\documentclass[amsmath,showpacs,aps,prl,reprint,longbibliography]{revtex4-1}
\usepackage{graphicx}
\usepackage{amsmath}
\usepackage{amssymb}
\usepackage{txfonts}
\usepackage[mathscr]{euscript}
\usepackage[pdftex,
	      pdfauthor={Pontus Laurell and Satoshi Okamoto},
	      pdftitle={Dynamical and thermal magnetic properties of the Kitaev spin liquid candidate $\alpha$-RuCl$_3$}]{hyperref}

\begin{document}
\title{Dynamical and thermal magnetic properties of the Kitaev spin liquid candidate \texorpdfstring{$\alpha$-RuCl$_3$}{α-RuCl3}}
\author{Pontus Laurell}
\email{laurellp@ornl.gov}
\affiliation{Center for Nanophase Materials Sciences, Oak Ridge National Laboratory, Oak Ridge, Tennessee 37831, USA}
\author{Satoshi Okamoto}
\affiliation{Materials Science and Technology Division, Oak Ridge National Laboratory, Oak Ridge, Tennessee 37831, USA}

\begin{abstract}
	What is the correct low-energy spin Hamiltonian description of $\alpha$-RuCl$_3$? The material is a promising Kitaev spin liquid candidate, but is also known to order magnetically, the description of which necessitates additional interaction terms. The nature of these interactions, their magnitudes and even signs, remain an open question. In this work we systematically investigate dynamical and thermodynamic magnetic properties of proposed effective Hamiltonians. We calculate zero-temperature inelastic neutron scattering (INS) intensities using exact diagonalization, and magnetic specific heat using a thermal pure quantum states method. We find that no single current model satisfactorily explains all observed phenomena of $\alpha$-RuCl$_3$. %
	In particular, we find that Hamiltonians derived from first principles can capture the experimentally observed high-temperature peak in the magnetic specific heat, while overestimating the magnon energy at the zone center. In contrast, other models reproduce important features of the INS data, but do not adequately describe the magnetic specific heat. To address this discrepancy we propose a modified ab initio model that is consistent with both magnetic specific heat and low-energy features of INS data.
\end{abstract}
\maketitle


\section*{Introduction}
Quantum spin liquids (QSL) are long-sought-after states of matter without magnetic order, but with nontrivial topological and potentially exotic properties \cite{Savary2017,RevModPhys.89.025003}. Much of the search has been focused on frustrated lattice systems \cite{Balents2010,RevModPhys.88.041002}, but in an important development in 2006 Kitaev \cite{Kitaev2006} introduced a novel exactly solvable paradigmatic QSL with bond-directional Ising terms on the bipartite honeycomb lattice. Importantly, this Kitaev model hosts anyonic excitations 
\cite{Hermanns2018}, which are of interest both for fundamental reasons and for their proposed application in topological quantum computing \cite{Kitaev2003,RevModPhys.80.1083}. 
It was realized that such interaction terms naturally appear {\cite{Khaliullin2005}} --- and can be large --- in Mott-insulating transition-metal systems with edge-sharing octahedra and strong spin-orbit coupling, such as in A$_2$IrO$_3$ (A=Na,Li) \cite{PhysRevLett.102.017205,PhysRevLett.105.027204}, $\alpha$-RuCl$_3$ \cite{PhysRevB.90.041112} and other materials \cite{Winter2017,Takagi2019}.
\begin{figure}[!thpb]
	\centering
	\includegraphics[width=\columnwidth]{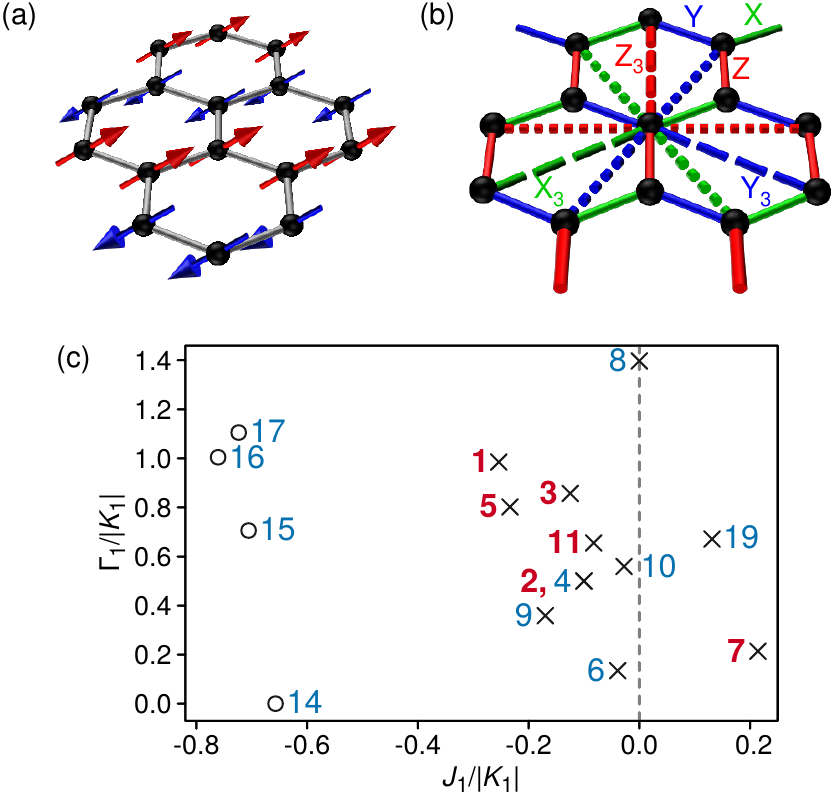}
	\caption{\label{fig:zigzagmodels}$\alpha$-RuCl$_3$. (a) The zigzag magnetic order. (b) The honeycomb lattice and its different bonds. Solid, dotted and dashed lines represent nearest, second-nearest and third-nearest neighbor bonds, respectively. (c) 
		The variability in two nearest neighbor (NN) parameters between various proposed spin Hamiltonians for $\alpha$-RuCl$_3$. The Hamiltonians marked by red, bold numbers (blue, roman) are discussed in the main text (Supplementa{ry} Information). Here $K_1$ and $J_1$ are the NN Kitaev and Heisenberg couplings, respectively, and $\Gamma_1$ is an NN symmetric off-diagonal interaction. Models with ferromagnetic (antiferromagnetic) $K_1$ are marked with crosses (open circles). Bond averaged values were used for anisotropic models.
	}
\end{figure}

However, the three mentioned materials are all found to order magnetically at low temperatures, and hence cannot be perfect realizations of the Kitaev model. Na$_2$IrO$_3$ \cite{PhysRevB.83.220403,PhysRevLett.108.127204,PhysRevB.85.180403} and $\alpha$-RuCl$_3$  \cite{PhysRevB.91.144420,PhysRevB.92.235119,PhysRevB.93.134423,Banerjee2016} both develop a zigzag order, as illustrated in Fig.~\ref{fig:zigzagmodels} (a), while Li$_2$IrO$_3$ displays an incommensurate spiral order \cite{PhysRevB.93.195158}. Despite the zigzag order, $\alpha$-RuCl$_3$ has emerged as a particularly promising Kitaev QSL candidate. Initial strong evidence came in the form of a strong and unusually stable scattering continuum at the zone center as observed in inelastic neutron {scattering} experiments (INS)  \cite{Banerjee2017,Banerjee2018,Do2017,PhysRevLett.118.107203}, which has been interpreted as evidence for the presence of fractional Majorana excitations. However, in an alternative picture it has been proposed that the continuum may consist of incoherent excitations due to {spontaneous} magnon decays \cite{10.1038/s41467-017-01177-0}. More recently, half-integer quantization of the thermal Hall conductivity was reported \cite{10.1038/s41586-018-0274-0}, also consistent with Majorana excitations. The quantization occurs in a narrow range of in-plane magnetic fields, where the magnetic order is melted, possibly uncovering an intermediate QSL state \cite{PhysRevLett.119.227208,PhysRevLett.119.037201,PhysRevB.95.180411,Banerjee2018,Balz2019}. Further evidence for Kitaev physics has been found in experiments reporting magnetic specific heat \cite{PhysRevB.96.041405,Do2017,PhysRevB.99.094415}, NMR \cite{PhysRevLett.119.037201,Jansa2018}, microwave absorption \cite{PhysRevB.98.184408}, Raman scattering \cite{PhysRevLett.114.147201,Du2018,Wang2018}, and THz spectroscopy results {\cite{PhysRevB.98.094425,Reschke2019,PhysRevLett.119.227201,PhysRevLett.119.227202}}.

Altogether, these experiments strongly suggest that $\alpha$-RuCl$_3$ can be described by a generalized Kitaev-Heisenberg Hamiltonian \cite{Winter2017,Takagi2019}, including off-diagonal and further-range interactions. {Theoretical} work leads to the same picture, whether the model is deduced from ab initio methods \cite{PhysRevB.100.075110,PhysRevB.90.041112,PhysRevB.91.241110,PhysRevB.93.155143,PhysRevB.93.214431,Yadav2016,PhysRevB.96.054410,PhysRevB.96.115103}, or from more phenomenological or ab initio inspired approaches \cite{10.1038/s41467-017-01177-0,PhysRevLett.120.077203,PhysRevB.98.060412,PhysRevB.97.134424}. Unfortunately, these different works, as well as experimental fits \cite{PhysRevLett.118.107203,PhysRevB.98.094425}, have led to a veritable zoo of proposed realistic spin Hamiltonian descriptions for $\alpha$-RuCl$_3$, and it's not currently clear which description is most accurate. Moreover, the proposed models disagree in terms of included spin-spin interaction terms, magnitudes of interaction parameters, and even signs. We illustrate this situation in Fig.~\ref{fig:zigzagmodels} (c) by a scatter plot of the values for just two relevant interaction terms, and in Table~\ref{table:models}.

\begin{figure*}
	\centering
	\includegraphics{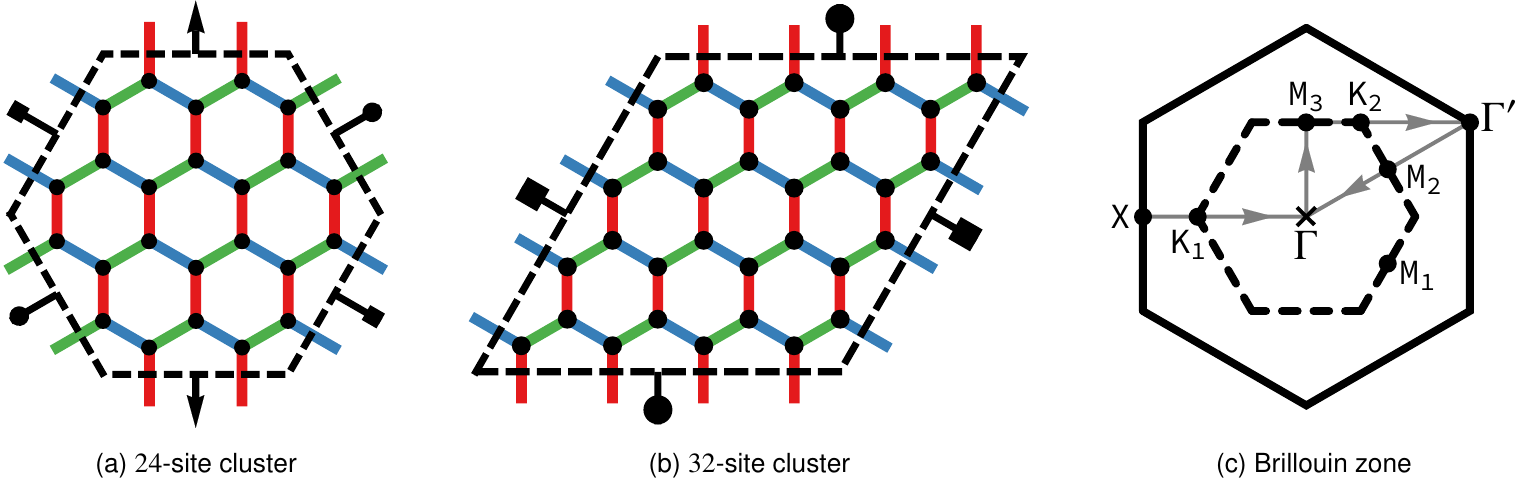}
	\caption{\label{fig:HoneycombClusters}Clusters. (a)-(b) Finite size clusters with periodic boundary conditions, and (c) the first and second Brillouin zones for the honeycomb lattice. Arrows indicate the high-symmetry path used for $I\left( \mathbf{q}, \omega\right)$ spectra.}
\end{figure*}

In this work, we adopt a systematic approach to address this uncertainty. We calculate static spin structure factors (SSF{s}) $S\left( \mathbf{q}\right)$ and INS intensities $I\left( \mathbf{q}, \omega\right)$ for all models listed in Table~\ref{table:models} using Lanczos exact diagonalization (ED) \cite{RevModPhys.66.763} on $24$-site clusters. {We also use ED to explore the evolution of the INS spectra away from the ferromagnetic Kitaev limit as new perturbations are introduced in a generalized Kitaev-Heisenberg model.} We {then} calculate the magnetic specific heat $C_\mathrm{mag}$ for the models using the thermal pure quantum state (TPQ) method \cite{PhysRevLett.108.240401} on the $24$- and $32$-site clusters shown in Fig.~\ref{fig:HoneycombClusters}. A few of the models considered here have previously been studied using similar methods in Refs.~\cite{10.1038/s41467-017-01177-0,PhysRevB.97.134424,Suzuki2018b}. For clarity we will restrict the discussion in the main text to six particularly relevant models. These models all have a ferromagnetic Kitaev coupling ($K_1<0$), which is expected in $\alpha$-RuCl$_3$ \cite{Winter2017,Takagi2019}. Results for the other models are included in the Supplementary Information.

Our key finding is that none of the studied models manages to fully capture the salient features of both the INS and magnetic specific heat data. The energy scales obtained in first principles approaches appear to be needed to reproduce a high-temperature peak in $C_\mathrm{mag}$, but the parameters proposed in the literature push the INS intensity at the $\Gamma$ point to higher energies. On the other hand, models obtained by fits to INS data run the risk of missing significant off-diagonal interactions, and fail to reproduce the experimentally observed temperature dependence of  $C_\mathrm{mag}$. {By modifying one of the ab initio models, we are able to find results consistent with both $C_\mathrm{mag}$ and low-energy features of the INS spectrum.} Our results thus provide important clues for an accurate and realistic description of of $\alpha$-RuCl$_3$.
\begin{table*}
	\caption{\label{table:models}\textbf{The spin Hamiltonians for $\alpha$-RuCl$_3$ considered in this work.} Dashes (--) indicate that the value is unavailable or negligible. The bolded models are considered in the main text, and results for the other models are given in the Supplementa{ry} Information. Asterisks in the `BA' column signify that the full Hamiltonian has different values for the X/Y bonds compared {with} the Z bonds, and that the parameter values given in the row have been bond averaged.}
	\footnotesize
	\begin{tabular}{lllllllllll}
		Reference           & Method   & $J_1$ & $K_1$ & $\Gamma_1$ & $\Gamma_1'$    & $J_2$                      & $K_2$   & $J_3$ & $K_3$ & BA       \\\hline
		\textbf{1}   \textbf{Winter et al. PRB} \cite{PhysRevB.93.214431}\footnote[1]{Using the proposed minimal model, which is bond averaged and neglects small $\Gamma_1'=-0.9$ meV. Values for the monoclinic ($C2/m$) crystal structure.}  & Ab initio (DFT + exact diag.)        & $-1.7$      & $-6.7$      & $+6.6$           & $-0.9$    & --  & --   & $+2.7$  & --    & $^\star$         \\
		\textbf{2}   \textbf{Winter et al. NC} \cite{10.1038/s41467-017-01177-0}   & Ab initio-inspired (INS fit)         & $-0.5$  & $-5.0$  & $+2.5$       & -- & -- & -- & $+0.5$  & --   &       \\
		\textbf{3}   \textbf{Wu et al.} \cite{PhysRevB.98.094425}           & THz spectroscopy fit                    & $-0.35$      & $-2.8$      & $+2.4$           & --  & --   & --    & $+0.34$      & -- &          \\
		4   Cookmeyer and Moore \cite{PhysRevB.98.060412} & Magnon thermal Hall (sign)                 & $-0.5$       & $-5.0$      & $+2.5$           & --        & --      & --     & $+0.1125$      & -- &          \\\hline
		\textbf{5}   \textbf{Kim and Kee} \cite{PhysRevB.93.155143}        & DFT + $t/U$ expansion  & $-1.53$      & $-6.55$      & $+5.25$            & $-0.95$     & --     & --     & -- & --   & $^\star$ \\
		6   Suzuki and Suga \cite{PhysRevB.97.134424,Suzuki2019}     & Magnetic specific heat               & $-1.53$      & $-24.4$      & $+5.25$           & $-0.95$  & --    & --    & --    & --  & $^\star$         \\\hline
		\textbf{7}   \textbf{Yadav et al.} \cite{Yadav2016}\footnote[2]{We use the sign convention in Refs.~\cite{PhysRevB.97.134424,PhysRevB.96.064430}.}       & Quantum chemistry (MRCI)  & $+1.2$      & $-5.6$      & $+1.2$  & $-0.7$   &   $+0.25$& --  & $+0.25$      & --     &    \\
		8   Ran et al. \cite{PhysRevLett.118.107203}         & Spin wave fit to INS gap   & --      & $-6.8$    & $+9.5$   & -- & -- & -- & -- & -- &      \\
		9   Hou et al. \cite{PhysRevB.96.054410}\footnote[3]{This work gives values for several values of $U$. Here we use the $U=3.5$eV parameters.}         & Constrained DFT +$U$   & $-1.87$ & $-10.7$      & $+3.8$  & -- & -- &  -- & $+1.27$ & $+0.63$     & $^\star$         \\
		10  Wang et al. \cite{PhysRevB.96.115103}\footnote[4]{Values for the C2 structure.}         & DFT + $t/U$ expansion  & $-0.3$      & $-10.9$      & $+6.1$      & -- & -- & -- & $+0.03$  & --      &          \\\hline
		\textbf{11}  \textbf{Eichstaedt et al.} \cite{PhysRevB.100.075110,Eichstaedt2019} {\footnote[5]{\label{fn:Eichstaedt}{These are the parameters from the preprint version in Ref.~\cite{Eichstaedt2019}. They were revised in the published version, Ref.~\cite{PhysRevB.100.075110}. In Supplementa{ry Note 4} we show that this slight modification does not affect our conclusions.}}}
		& Fully ab initio (DFT + cRPA + $t/U$) & $-1.4$       & $-14.3$      & $+9.8$           & $-2.23$  & --  & $-0.63$    & $+1.0$      & $+0.03$ & $^\star$ \\
		12  Eichstaedt et al. \cite{PhysRevB.100.075110,Eichstaedt2019}{\footnotemark[5]}  & Neglecting non-local Coulomb         & $-0.2$       & $-4.5$      & $+3.0$           &  $-0.73$ & -- & $-0.33$ & $+0.7$     & $+0.1$ &  $^\star$ \\
		13  Eichstaedt et al. \cite{PhysRevB.100.075110,Eichstaedt2019}{\footnotemark[5]}  & Neglecting non-local SOC             & $-1.3$      & $-13.3$      & $9.4$ & $-2.3$ & -- & $-0.67$ & $+1.0$      & $+0.1$& $^\star$\\\hline
		14	 Banerjee et al. \cite{Banerjee2016}  & Spin wave fit             & $-4.6$      & $+7.0$      & -- & -- & -- & -- & --      &--& \\
		15	 Kim et al. \cite{PhysRevB.91.241110,PhysRevB.96.064430}  & DFT + $t/U$ expansion             & $-12$      & $+17$      & $+12$ & -- & -- & -- & --      &--& \\
		16	 Kim and Kee\cite{PhysRevB.93.155143}\footnote[6]{Case 0, corresponding to P3 structure and weaker Hund's coupling than in Model 15.}  & DFT + $t/U$ expansion             & $-3.5$      & $+4.6$      & $+6.42$ & $-0.04$ & -- & -- & --      &--& \\
		17	 Winter et al. PRB \cite{PhysRevB.93.214431}\footnote[7]{Values for P3 structure.}  & Ab initio (DFT + exact diag.)             & $-5.5$      & $+7.6$      & $+8.4$ & $+0.2$ & -- & -- & $+2.3$      &--& \\\hline
		{18}	 {Ozel et al. PRB \cite{PhysRevB.100.085108}}  & {Spin wave fit / THz spectroscopy} & $-0.95$      & $+1.15$      & $+3.8$ & -- & -- & -- & --      &--& \\
		{19}	 {Ozel et al. PRB \cite{PhysRevB.100.085108}}  & {Spin wave fit / THz spectroscopy} & $+0.46$      & $-3.50$      & $+2.35$ & -- & -- & -- & --      &--& \\
	\end{tabular}
\end{table*}
%
%
%
\section*{Results}
Several of the Hamiltonians listed in Table~\ref{table:models} are special cases of a proposed minimal model \cite{PhysRevB.93.214431,Winter2017} for $\alpha$-RuCl$_3$,
\begin{align}
H_{J_1-K_1-\Gamma_1-J_3}	&=	\sum_{\langle i,j\rangle} \left[ J_1\mathbf{S}_i\cdot \mathbf{S}_j + K_1S_i^\gamma S_j^\gamma + \Gamma_1\left( S_i^\alpha S_j^\beta + S_i^\beta S_j^\alpha \right)\right] \nonumber\\
						&+ J_3\sum_{\langle\langle\langle i,j\rangle\rangle\rangle}\mathbf{S}_i\cdot\mathbf{S}_j,	\label{eq:HamiltonianMinimal}
\end{align}
where $\langle \dots\rangle$ and $\langle\langle\langle \dots\rangle\rangle\rangle$ denote nearest and third-nearest neighbors, respectively. $\gamma=${X,Y,Z} is the bond index shown in Fig.~\ref{fig:zigzagmodels} (b), and $\alpha,\beta$ are the two other bonds. The $\Gamma_1$ term is required to explain the moment direction, and $J_3>0$ helps stabilize the zigzag order. Ab initio and DFT studies also tend to report a sizable {symmetric} off-diagonal $\Gamma_1'$ interaction,
\begin{align}
	H_{\Gamma_1'}	&=	\Gamma_1' \sum_{\langle i,j\rangle} \sum_{\alpha\neq\gamma}\left[ S_i^\gamma S_j^\alpha + S_i^\alpha S_j^\gamma \right],	\label{eq:HamiltonianGammaPrime}
\end{align}
{which originates from trigonal distortion \cite{PhysRevLett.112.077204,Rau2014}. Since the crystal structure of $\alpha$-RuCl$_3$ features trigonal compression \cite{Winter2017,PhysRevB.92.235119,PhysRevB.93.134423,Do2017} perturbative calculations \cite{Rau2014,PhysRevB.93.214431} would predict that $\Gamma_1'<0$, which provides an additional mechanism to stabilize the zigzag order \cite{Rau2014,PhysRevB.99.064425}. In the absence of other crystal distortions, the most general nearest-neighbor Hamiltonian would thus be the $J_1-K_1-\Gamma_1-\Gamma_1'$  model. Combining these two proposed minimal models results in the $J_1-K_1-\Gamma_1-\Gamma_1'-J_3$ model.}

Further proposed extensions include second-nearest{-neighbor} Kitaev and Heisenberg terms, third-nearest{-neighbor} Kitaev terms, and additional symmetry-allowed anisotropies \cite{PhysRevB.93.214431,PhysRevB.100.075110}. In particular, $\alpha$-RuCl$_3$ does not have a perfect honeycomb lattice, which allows the parameters for the {Z} bond to deviate from those on the {X,Y} bonds. In Table~\ref{table:models} we have bond averaged such anisotropies for the sake of clarity, but we will use the full parameter sets in our calculations when appropriate.

\subsection{Spin structure factors and neutron scattering intensities}
\begin{figure*}[t]
	\centering
	\includegraphics{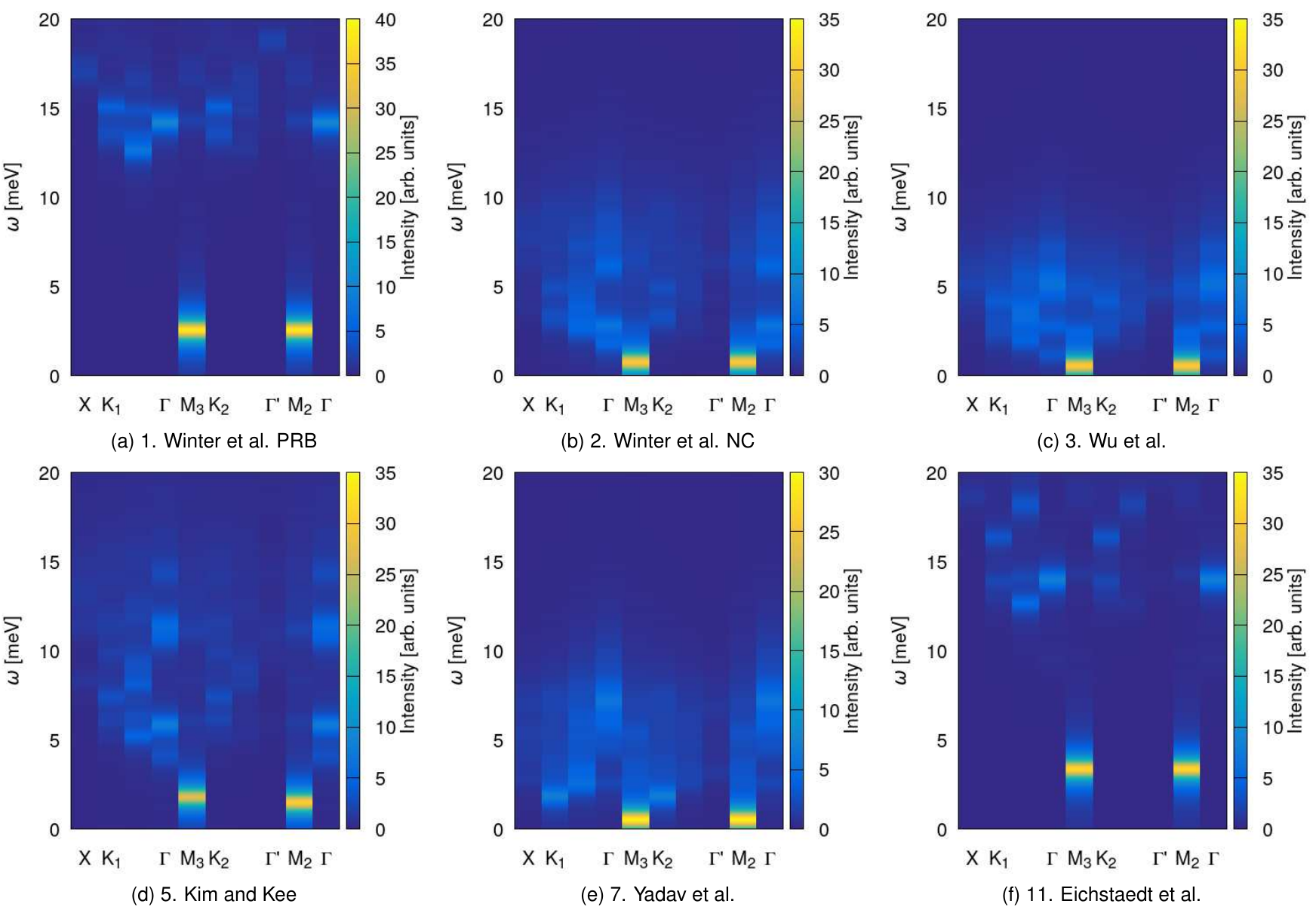}
	\caption{\label{fig:Iqomega}INS spectra. Inelastic neutron scattering intensities $I\left( \mathbf{q}, \omega\right)$ for the chosen models calculated at zero temperature using the $N=24$ site cluster. The model shown in (f) was bond averaged.}
\end{figure*}

Fig.~\ref{fig:Iqomega} shows predicted zero-temperature neutron scattering intensity spectra, $I\left( \mathbf{q}, \omega\right)$, for the six central models. All models feature sharp low-frequency peaks at the {M }points, indicating the zigzag order. The intensity at the {M} points is significantly higher than the intensity at the $\Gamma$ point, which is inconsistent with experimental observations \cite{Banerjee2018}. However, the {M} peaks could potentially be suppressed at finite temperatures \cite{PhysRevLett.120.077203}. The models in Fig.~\ref{fig:Iqomega} (b), (c), (d) and (e) all show clear signs of the scattering continuum at the $\Gamma$ point at frequencies comparable to the position of the {M} peak, whereas the two ab initio models in (a) and (f) display a sizable gap up to any noticeable scattering at the $\Gamma$ point.

\begin{figure*}[t]
	\centering
	\includegraphics[height=12.5cm]{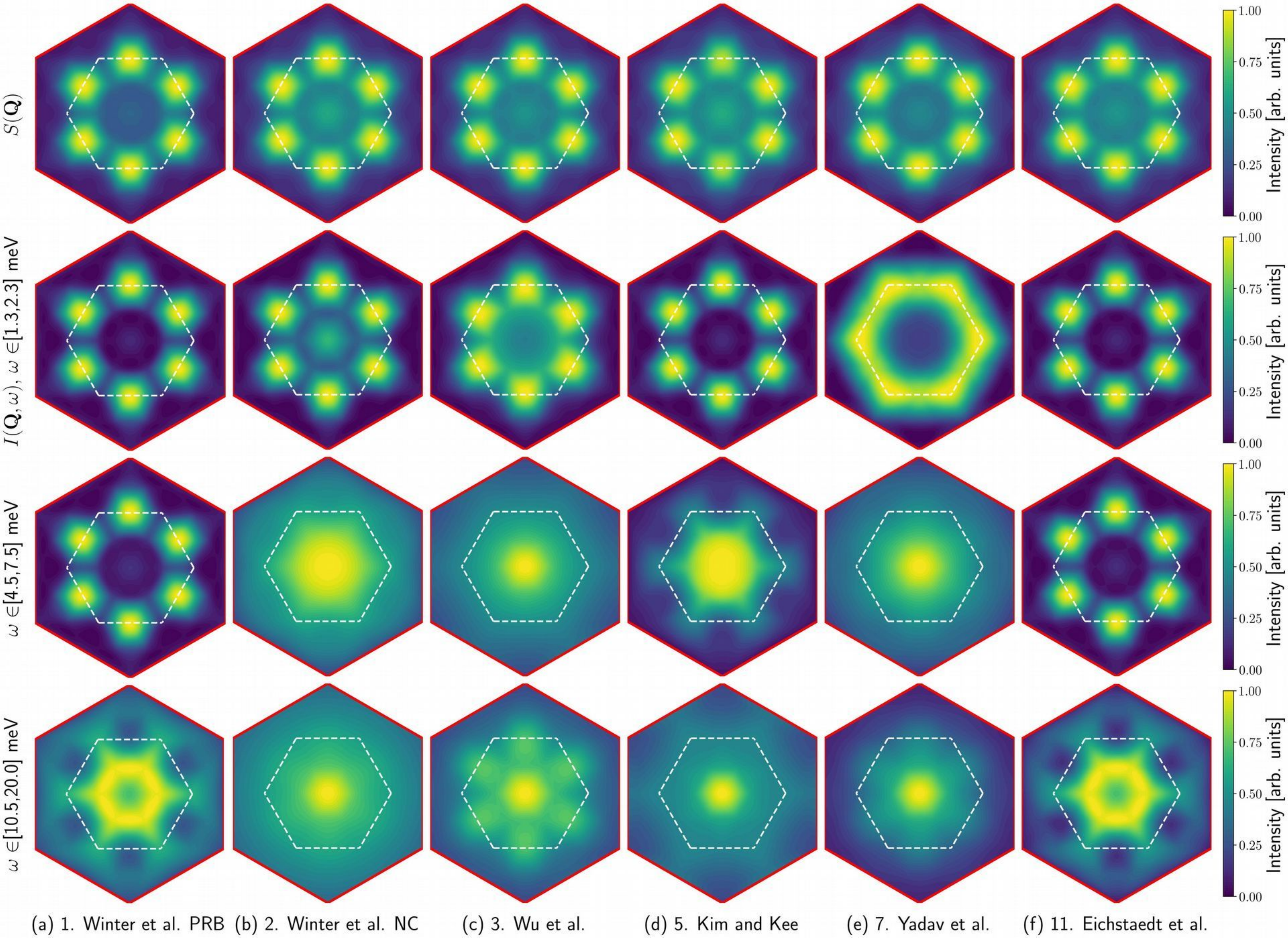}
	\caption{\label{fig:Sqomega}Additional inelastic neutron scattering intensity results {obtained on the $N=24$ site cluster}. The top row shows the static spin structure factors as a heat map over $k$-space. The dashed white hexagon marks the first Brillouin zone, and the outer red hexagon shows the second Brillouin zone. The three lower rows show the neutron scattering intensities $I\left( \mathbf{q}, \omega\right)$ integrated over representative energy windows i) $[1.3,2.3]$ meV, ii) $[4.5,7.5]$ meV, and iii) $[10.5,20.0]$ meV. Note that each heatmap is normalized separately, in order to showcase patterns in momentum space. Intensities in different heatmaps should not be compared.}
\end{figure*}

In the top row of Fig.~\ref{fig:Sqomega} we have plotted the static spin structure factors $S\left( \mathbf{q}\right)$. As shown, all six models are consistent with a zigzag ordering with some weight at the zone center. The model shown in Fig.~\ref{fig:Sqomega} (d) showcases how different interaction strengths for the {Z} bond results in weakly broken $C_3$ symmetry in the structure factors. The bottom three rows of Fig.~\ref{fig:Sqomega} shows integrated INS intensities for three different energy windows. Experimentally, a star-like pattern with strong weight at $\Gamma$ and arms extending to the {M} points was observed in the $\omega \in [4.5,7.5]$ meV energy window \cite{Banerjee2017}. We dub this pattern the {M} star. The only model displaying this pattern in the right energy window is the one due Yadav et al. \cite{Yadav2016}, in Fig.~\ref{fig:Sqomega} (e). In contrast, (b), (c) and (d) have star-like patterns where the arms extend towards the {K} points --- {K} star shapes. The two ab initio models in (a) and (f) do not capture the weight at $\Gamma$ at all, instead forming a flower-like shape that we would expect to see for lower frequencies, since the peak at $\Gamma$ is observed to be higher energy than the {M} point peak ($2.69\pm 0.11$ meV vs. $2.2\pm0.2$ meV from INS data \cite{Banerjee2018}). The high-energy window $\omega \in [10.5,20.0]$ meV is expected to be dominated by the continuum at $\Gamma$, which is consistent with (b), (c), (d), (e), but not the ``lotus root-like'' shapes in (a), and (f).

We summarize our computed neutron scattering intensity results in Table~\ref{table:results}, which also includes results for the models discussed in the Supplementa{ry} Information. The ab initio-inspired approach of Winter et al. \cite{10.1038/s41467-017-01177-0} was constructed to reproduce certain features in the INS spectrum, and thus does particularly well. It reproduces the $\Gamma$ point intensity profile well (see {Supplementary Figure 3}), and has an {M} star shape in the $[5.5,8.5]$ meV window, but not in $[4.5,7.5]$ meV.  It is thus natural to use this model as a starting point for INS data-compatible effective Hamiltonians, as done in the THz spectroscopy fit of Ref.~\cite{PhysRevB.98.094425}, and an analysis of the magnon thermal Hall conductivity in Ref.~\cite{PhysRevB.98.060412}. The latter work (for results see the Supplementa{ry Note 3}) proposes a particularly minor change --- only reducing the magnitude of $J_3$ from $0.5$ meV to $0.1125$ meV while keeping other parameters fixed --- which actually leads to an {M} star shape in the relevant window, but also significantly alters the intensity profile at the $\Gamma$ point. We mention this fact explicitly as an example of a more general observation: \emph{for these models even small changes to the parameters can result in significantly different spectra, while even significantly different models can produce very similar SSFs and the same magnetic order}. This difficulty calls for other methods to constrain the possible effective Hamiltonians, which is why we {will later study} the magnetic specific heat.

\begin{table*}[t]
	\caption{\label{table:results}\textbf{Summary of results, highlighting important features in the INS and magnetic specific heat predictions on the $24$-site cluster.} We focus on i) the positions $\omega_\Gamma$ and ${\omega_\mathrm{M}}$ of the initial spin wave peaks in the INS intensity at the $\Gamma$ and {M} points, respectively, ii) the shape of the neutron scattering intensity map (IM) in momentum space integrated over $[4.5,7.5]$ meV, and iii) the position of the high-temperature peak in the magnetic specific heat. {K (M)} star denotes a star-like shape pointing towards the {K (M)} points. {The bolded models are considered in the main text. Results for the other models are given in the Supplementary Information.}}
	\begin{ruledtabular}
	\begin{tabular}{lllll}
		Reference           & $\omega_\Gamma$ {[}meV{]} & ${\omega_\mathrm{M}}$ {[}meV{]} & IM shape      & $T_{{\mathrm{h}}}$ {[}K{]} \\\hline
		\textbf{1}   \textbf{Winter et al. PRB} \cite{PhysRevB.93.214431}  & $14.2$   & $2.55$ & Flower        & 54            \\
		\textbf{2}   \textbf{Winter et al. NC} \cite{10.1038/s41467-017-01177-0}   & $2.8$ & $0.75$    & {K} star      & 22            \\
		\textbf{3}   \textbf{Wu et al.} \cite{PhysRevB.98.094425}          & $1.18$  & $0.54$  & Dominated by $\Gamma$              & 17            \\
		4   Cookmeyer and Moore \cite{PhysRevB.98.060412} & $1.67$\footnote[1]{\label{fn:lowerpeak}There is also a clear, distinct peak at lower frequency, which would be hidden by the elastic scattering continuum.}  & $0.57$  & {M} star   &    24            \\
		\textbf{5}   \textbf{Kim and Kee} \cite{PhysRevB.93.155143}        & $4.12$     & $1.49$  & {K} star-like & 34            \\
		6   Suzuki and Suga \cite{PhysRevB.97.134424,Suzuki2019} & $2.05$ & $1.98$ & {K} star     & 94           \\
		\textbf{7}   \textbf{Yadav et al.} \cite{Yadav2016}       & $2.58$  & $0.58$   & {M} star      & 13            \\
		8   Ran et al.  \cite{PhysRevLett.118.107203} & $4.8$ & $1.37$  & {M} star      & 57            \\
		9   Hou et al. \cite{PhysRevB.96.054410} & $5.5$ & $1.87$  & {K} star      & 33            \\
		10  Wang et al. \cite{PhysRevB.96.115103} & $3.13$\footnotemark[1] & $1.03$  & {K} star      & 53            \\
		\textbf{11}  \textbf{Eichstaedt et al.} \cite{PhysRevB.100.075110}  & $11.2$  & $2.79$ & Flower / {M}$_3$ \footnote[2]{\label{fn:flower1}Flower shape when bond-averaged, otherwise dominated by {M}$_3$. Peak positions are given for the non-bond averaged case.}             & 66            \\
		12  Eichstaedt et al. \cite{PhysRevB.100.075110} & $5.1$  & $0.98$  & Lotus root    & 22            \\
		13  Eichstaedt et al. \cite{PhysRevB.100.075110} & $11.9$  & $2.43$  & Dominated by {M}$_3$  & 63            \\
		14  Banerjee et al. \cite{Banerjee2016}     &  $6.38$      & $1.21$ & Lotus root              & 21            \\
		15  Kim et al. \cite{PhysRevB.91.241110,PhysRevB.96.064430} & $6.52$  & $2.65$ & Flower              & 81            \\
		16  Kim and Kee \cite{PhysRevB.93.155143} & $4.39$    & $2.25$   & Dominated by $\Gamma$ and {K}              & 35            \\
		17  Winter et al. PRB \cite{PhysRevB.93.214431} & $7.35$ & $0.43$  & Ring     & 41       \\
		{18}	 {Ozel et al. PRB \cite{PhysRevB.100.085108}}  & $4.39$ & $0.31$ & Dominated by $\Gamma$ &16\\
		{19}	 {Ozel et al. PRB \cite{PhysRevB.100.085108}}  & $3.18 $& $3.92$& Dominated by $\Gamma$ &18
	\end{tabular}
	\end{ruledtabular}
\end{table*}

\subsection{Evolution of INS spectra}
Above we have provided results for spin Hamiltonians with multiple interaction parameters. To untangle the roles and effects of different interaction terms, we now focus on a minimal $J_1-K_1-\Gamma_1-\Gamma_1'-J_3$ model. We fix the energy scale $1=\sqrt{J_1^2+K_1^2+\Gamma_1^2+\left( \Gamma_1'\right)^2 + J_3^2}$, and use a hyperspherical parametrization where
\begin{align}
\Gamma_1 	&=	\cos \theta,\\
J_1			&=	\sin \theta \cos \phi,\\
K_1 		&=	\sin \theta \sin \phi \cos \chi, \\
\Gamma_1'	&= 	\sin \theta \sin \phi \sin \chi \cos \psi,\\
J_3 		&= 	\sin \theta \sin \phi \sin \chi \sin \psi
\end{align}
When $\chi=0$ this reduces to the notation used in Ref.~\cite{PhysRevLett.112.077204}. Following the typical hierarchy of interaction strengths in Table~\ref{table:models} ($|K_1|\ge |\Gamma_1|\ge |J_1| \sim |\Gamma_1'| \sim |J_3|$) we begin by assuming a dominant FM Kitaev interaction, and introduce other terms one by one. A representative selection of the resulting spectra is shown in Fig.~\ref{fig:IqomegaEvolution}. Additional parameter values are studied in Supplementa{ry Note 5}.

As Fig.~\ref{fig:IqomegaEvolution} (a) shows, the FM Kitaev limit has a flat spectrum with intensity peaked at the $\Gamma$ point, consistent with the exact theoretical result \cite{PhysRevLett.112.207203}. The spectral evolution away from this point can be qualitatively understood using previously obtained phase diagrams \cite{PhysRevLett.112.077204,Rau2014}. For $\Gamma_1/K_1=-1/2$ shown in Fig.~\ref{fig:IqomegaEvolution} (b), a sharp low-energy peak develops at {the center point between $\Gamma$ and K$_1$, ``K$_1/2$''}, 
signaling a tendency towards spiral order. The peak at the $\Gamma$ point remains strong, however, preventing a clear signal of the spiral phase in the static spin structure factor. We also note that the resolution of the spiral phase ordering vector is limited by the finite cluster size, and that some zigzag correlations remain at the {M} points. As $\Gamma_1$ is increased to $\Gamma_1/K_1=-1.0$ in Fig.~\ref{fig:IqomegaEvolution} (c) the intensities at {K}$_1/2$ and the {M} points become comparable in strength. As can be seen by comparison with the Kitaev limit, the presence of $\Gamma_1>0$ in (b) and (c) tends to produce a stronger excitation continuum, that stretches to higher frequencies.

We next introduce the nearest neighbor Heisenberg exchange $J_1$. Figs.~\ref{fig:IqomegaEvolution} (d) and (e) show the effect of adding antiferromagnetic and ferromagnetic $J_1$, respectively ($J_1/K_1=\mp 0.1$).  In this case $J_1>0$ produces a stronger peak at {K}$_1/2$ and weaker intensity at the $\Gamma$ point, signaling a stabilized spiral order phase. This is consistent with the classical phase diagram of Ref.~\cite{PhysRevLett.112.077204} and cluster mean field theory results of Ref.~\cite{PhysRevB.99.064425}. In contrast, $J_1<0$ sees the peaks at the $\Gamma$ point move down in frequency, a strengthening of the {M} peaks and significant reduction in intensity at {K}$_1/2$. These observations are consistent with moving into the regime of ferromagnetic ordering. We note that a strong continuum remains at the zone center. In the case of $\Gamma_1/K_1=0$ we would have had a weaker continuum, and a stronger low-energy peak at the zone center. Finally we also introduce $\Gamma_1'<0$ and $J_3>0$ in Figs.~\ref{fig:IqomegaEvolution} (f) and (g). These interactions both stabilize the zigzag order, while pushing the $\Gamma$ point peaks to higher frequency, and generally weakening the continuum nature of the excitation spectra. This suggests that the unrealistically large gaps at the $\Gamma$ points in Figs.~\ref{fig:Iqomega} (a) and (f) may be due to an overestimation of the $\Gamma_1'$ and $J_3$ parameters.
\begin{figure*}[t]
	\centering
	\includegraphics[height=10cm]{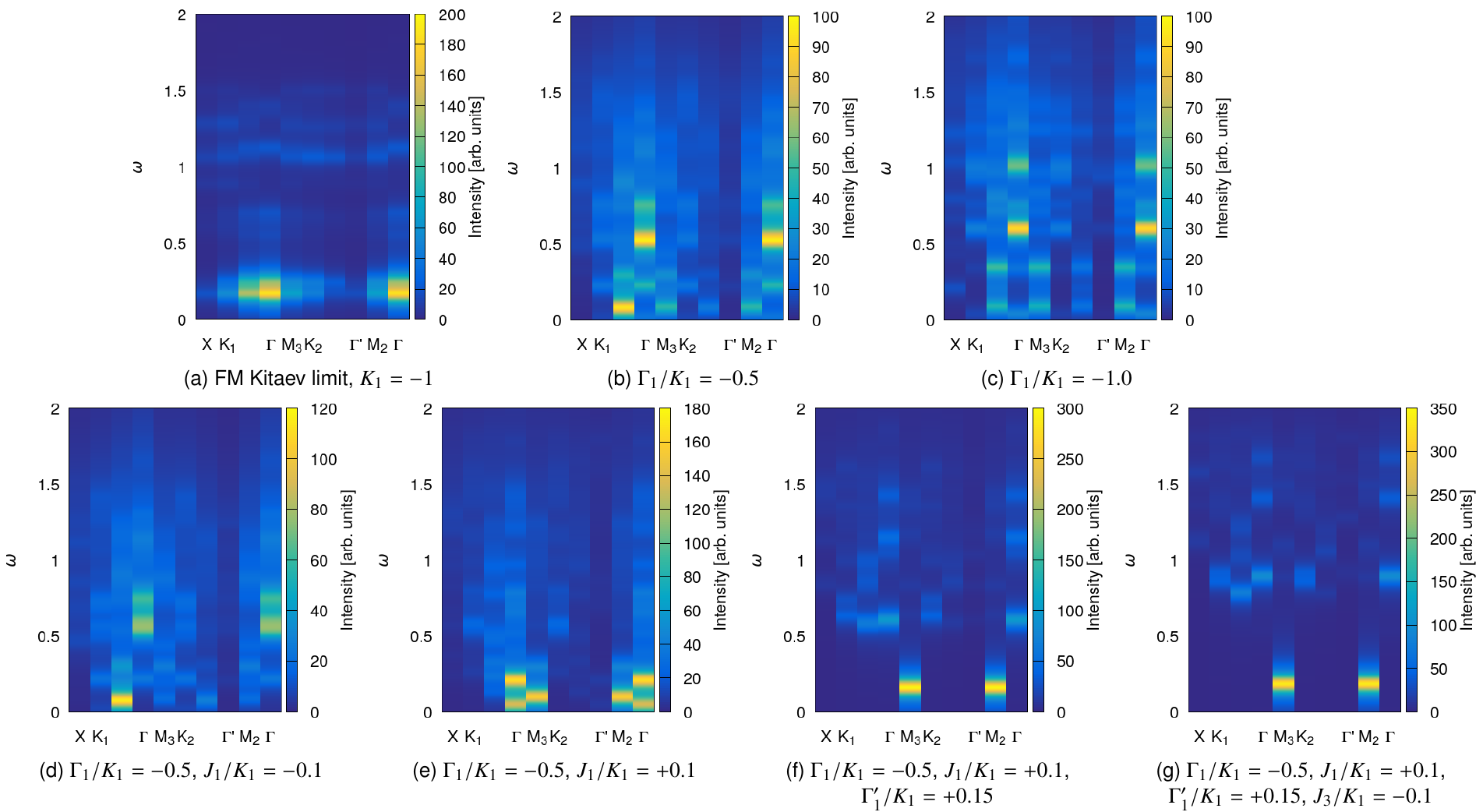}
	\caption{\label{fig:IqomegaEvolution}{
			Evolution of the INS intensity $I\left( \mathbf{q}, \omega\right)$ away from the ferromagnetic Kitaev limit. The limit is shown in (a). In (b), (c) a $\Gamma_1>0$ term is added. The cases of AFM $J_1$ and FM $J_1$ are considered in (d) and (e), respectively. Finally $\Gamma_1'<0$ and $J_3>0$ are introduced in (f) and (g). These results were obtained using the $24$-site cluster.}
	}
\end{figure*}

\subsection{Magnetic specific heat}
\begin{figure}
	\centering
	\includegraphics[width=\columnwidth]{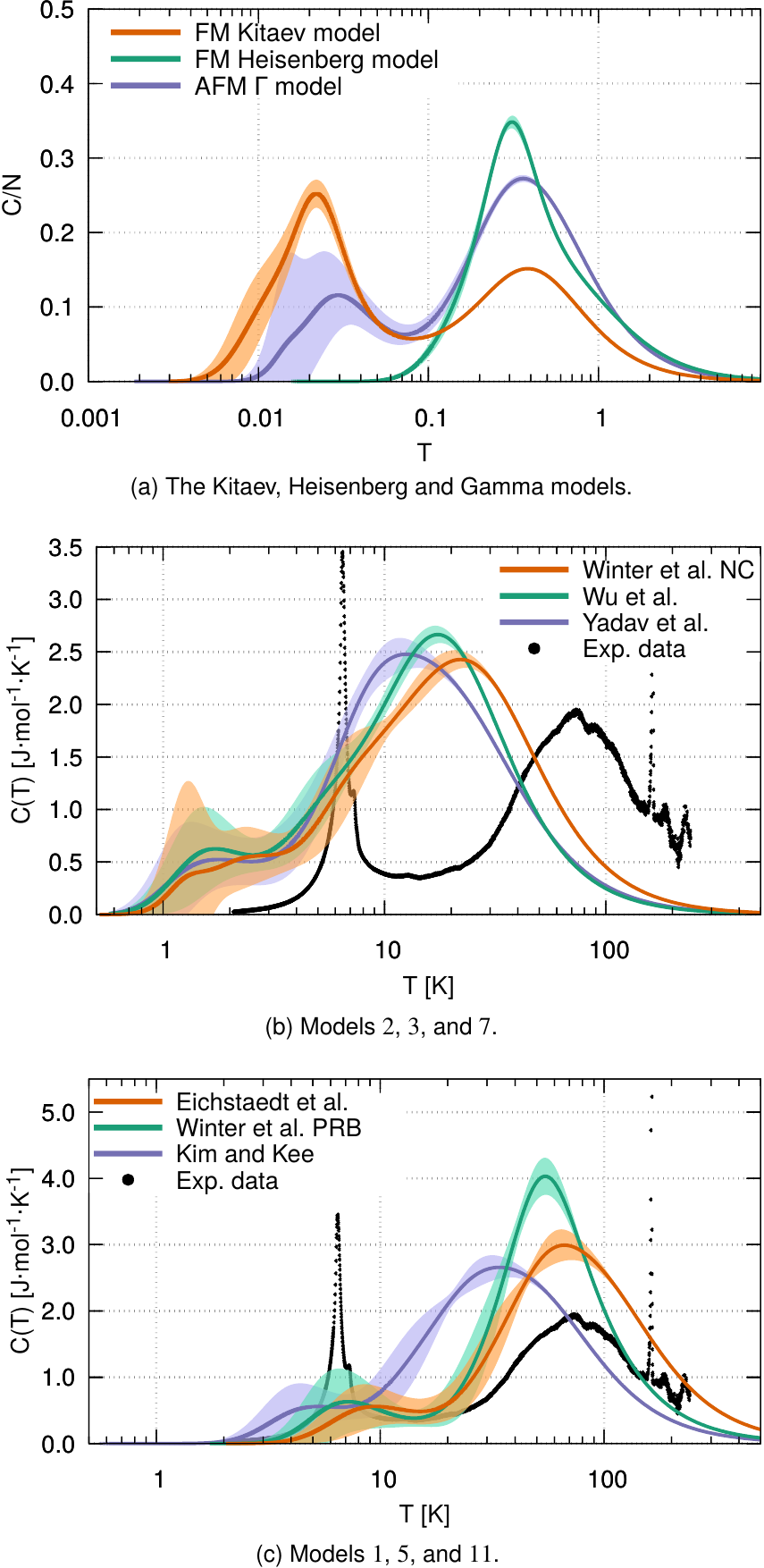}
	\caption{\label{fig:specHeat24}{Magnetic specific heat calculated using the TPQ method and the $N=24$ site cluster.} The solid lines show the average value over $15$ initial vectors {($100$ vectors were used for the $\Gamma$ model)}, and the shaded areas show the standard deviation. {(a) shows $C_\mathrm{mag}$ for the ferromagnetic Kitaev ($K_1=-1$), ferromagnetic Heisenberg ($J_1=-1$), and ``antiferromagnetic'' Gamma ($\Gamma_1=+1$) models. (b) and (c) shows $C_\mathrm{mag}$ for the six chosen models for $\alpha$-RuCl$_3$.} For comparison, the experimentally determined excess heat capacity from Ref.~\cite{PhysRevB.99.094415} is plotted using black dots. The peak in the experimental data near $~6.5$ K signals the magnetic ordering, and the strong peak at $170$ K is a structural transition, unrelated to the magnetic specific heat. Finally, the peak near $70$ K may correspond to itinerant Majorana quasiparticles \cite{PhysRevLett.113.197205,PhysRevB.92.115122,PhysRevB.99.094415}. The peak position is inconsistent with the models plotted in {(b)}, but consistent with the ab initio models plotted in {(c)}, with higher interaction strengths.
	}
\end{figure}

As shown in {Fig.~\ref{fig:specHeat24} (a)}, the magnetic specific heat of the pure Kitaev model features two characteristic, well-separated peaks at $T_{{\mathrm{l}}}$ and $T_{{\mathrm{h}}}$, where the low-$T$ one is due to thermal fluctuations of localized Majorana fermions, and the high-$T$ peak is related to itinerant Majoranas \cite{PhysRevLett.113.197205,PhysRevB.92.115122}. 
This two-peak structure appears to be stable to small perturbations away from the Kitaev point  \cite{PhysRevB.93.174425,Hermanns2018,Catuneanu2018}. Note that the presence of two peaks is not itself a unique signature of Kitaev physics \cite{PhysRevB.68.014424,PhysRevB.93.174425} and occurs also for e.g. the $\Gamma$ model \cite{Catuneanu2018,PhysRevB.98.045121}, see {Fig.~\ref{fig:specHeat24} (a)}.

A similar two-peak structure has been found in $\alpha$-RuCl$_3$ experiments, using both RhCl$_3$ \cite{PhysRevB.99.094415} and ScCl$_3$ \cite{Do2017,PhysRevB.95.241112,PhysRevB.91.094422} as nonmagnetic analogue compounds. In clean samples, a sharp low-$T$ peak representing the magnetic ordering occurs at $T_{{\mathrm{l}}} \approx 6.5$ K \cite{PhysRevLett.119.037201,PhysRevB.99.094415}, and then a broader peak occurs at a higher temperature $T_{{\mathrm{h}}}$, followed by a (non-magnetic) structural transition of $\alpha$-RuCl$_3$ near $165-170$ K \cite{Do2017,PhysRevB.99.094415}. So far, there is no clear consensus for the precise value of $T_{{\mathrm{h}}}$ (Widmann et al. report $T_{{\mathrm{h}}}\approx 70$ K \cite{PhysRevB.99.094415}, Do et al. find $T_{{\mathrm{h}}} \simeq 100$ K \cite{Do2017}, while Hirobe et al. \cite{PhysRevB.95.241112} and Kubota et al. \cite{PhysRevB.91.094422} find a broad maximum around $80-100$ K), but it appears to be an order of magnitude larger than $T_{{\mathrm{l}}}$. Whether or not the $T_{{\mathrm{h}}}$ peak can be attributed to fractionalized excitations due to a proximate Kitaev QSL, the feature appears to be real and ought to be captured by a realistic spin Hamiltonian. 
Two additional comments are in order. First, accurately determining the magnetic specific heat at higher temperatures is challenging, and sensitive to details of the analysis. This may partly explain the range of $T_{{\mathrm{h}}}$ values mentioned above. Second, an optical spectroscopy study \cite{PhysRevB.94.195156} extracted a crossover temperature for magnetic correlations, $T^\star\sim 35$ K, which, in an analysis relying on the pure Kitaev model, was equated to $T_{{\mathrm{h}}}$. In the following we will rely mainly on data from Widmann et al. \cite{PhysRevB.99.094415}, but there is clearly some uncertainty to the value of $T_{{\mathrm{h}}}$.

In Fig{s}.~\ref{fig:specHeat24} {(b)--(c)} we plot the magnetic specific heat, $C_\mathrm{mag}(T)$, for the six considered models on $24$-site clusters, along with the excess heat capacity determined in Ref.~\cite{PhysRevB.99.094415}. We note that the finite-size clusters are far from the thermodynamic limit, so we cannot expect to numerically observe a sharp magnetic transition, but the location of the peaks can provide useful information. We see that the models plotted in {(b)} are clearly inconsistent with the experimental data. However, the two ab initio models in {(c)} (which did not capture the INS data) actually have peak positions that are consistent with the data. In fact, the model fully determined from first-principles in Eichstaedt et al. \cite{PhysRevB.100.075110} has a peak at $T_{{\mathrm{h}}}\approx 66$ K, while the experimental data is centered around $~70$ K, with a peak at $68$ K.

\begin{figure}
	\centering
	\includegraphics[width=\columnwidth]{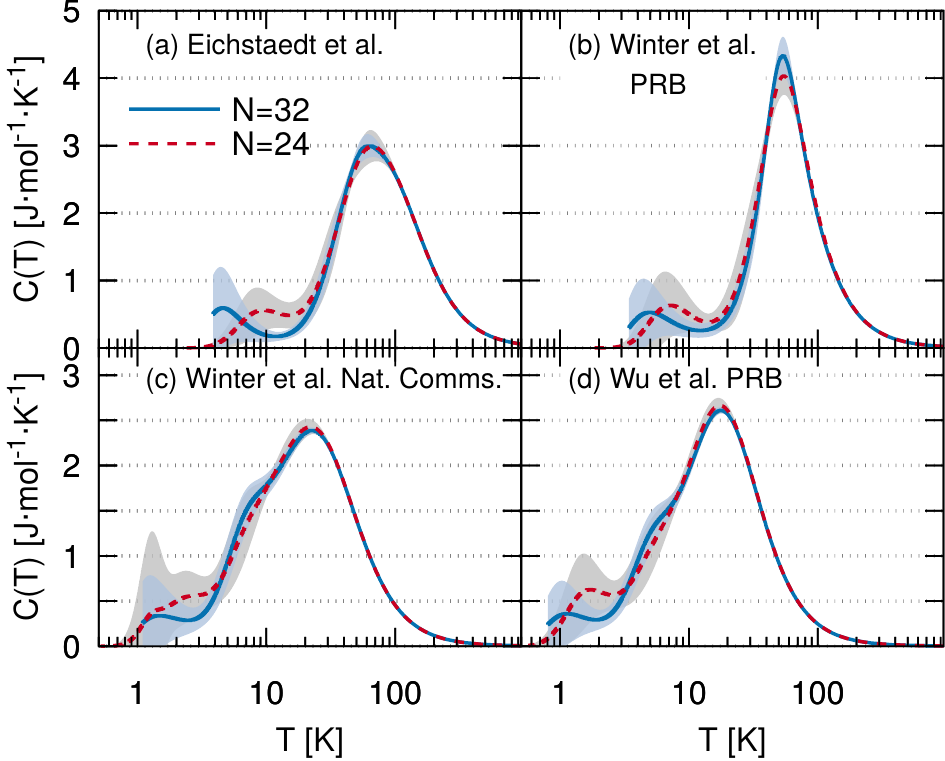}
	\caption{\label{fig:specHeat:FiniteSize}Finite size scaling of specific heat for four select Hamiltonians (models $1$, $2$, $3$, and $11$). The data is averaged over $15$ initial TPQ vectors, and the shaded regions show the standard deviations. We generically observe a two-peak structure, the high-$T$ peak of which appears to be well converged. The lower temperature peak corresponds to magnetic ordering, and may be quite sensitive to finite size effects, or to the difference in symmetry between the 24- and 32-site clusters.}
\end{figure}

In Fig.~\ref{fig:specHeat:FiniteSize} we provide $32$-site cluster TPQ results for a subset of the models, and show the finite size scaling tendencies. The two cluster sizes have different symmetry properties, which could explain part of the differences. Unfortunately, going to even larger cluster sizes (for better scaling or to preserve symmetries) using the TPQ method becomes computationally prohibitive. We find that the position of the high-temperature peak changes only marginally (see Supplementa{ry Note 2} for details), while the low-temperature behavior is much less well-converged.  We thus conclude that the two ab initio models describe the magnetic specific heat better than the other models.

\subsection{Modified ab initio model}
Having established that the two ab initio models are consistent with the experimental specific heat, we now ask whether they can be modified to better describe the INS data. As discussed in the section on evolution of INS spectra, the large gaps at the zone centers in Figs.~\ref{fig:Iqomega} (a) and (f) may be due to overestimated $\Gamma_1'$ or $J_3$ values. Since $\Gamma_1'$ is sensitive to the degree of trigonal distortion, it can be expected to vary between crystal samples. It is thus likely the parameter with the highest degree of uncertainty. For these reasons, we consider the effect of reducing $|\Gamma_1'|$ in the model of Ref.~\cite{PhysRevB.100.075110}, while leaving other parameters ($J_3$ included) unchanged. We use bond-averaged interaction parameters.

\begin{figure}
	\includegraphics[width=\columnwidth]{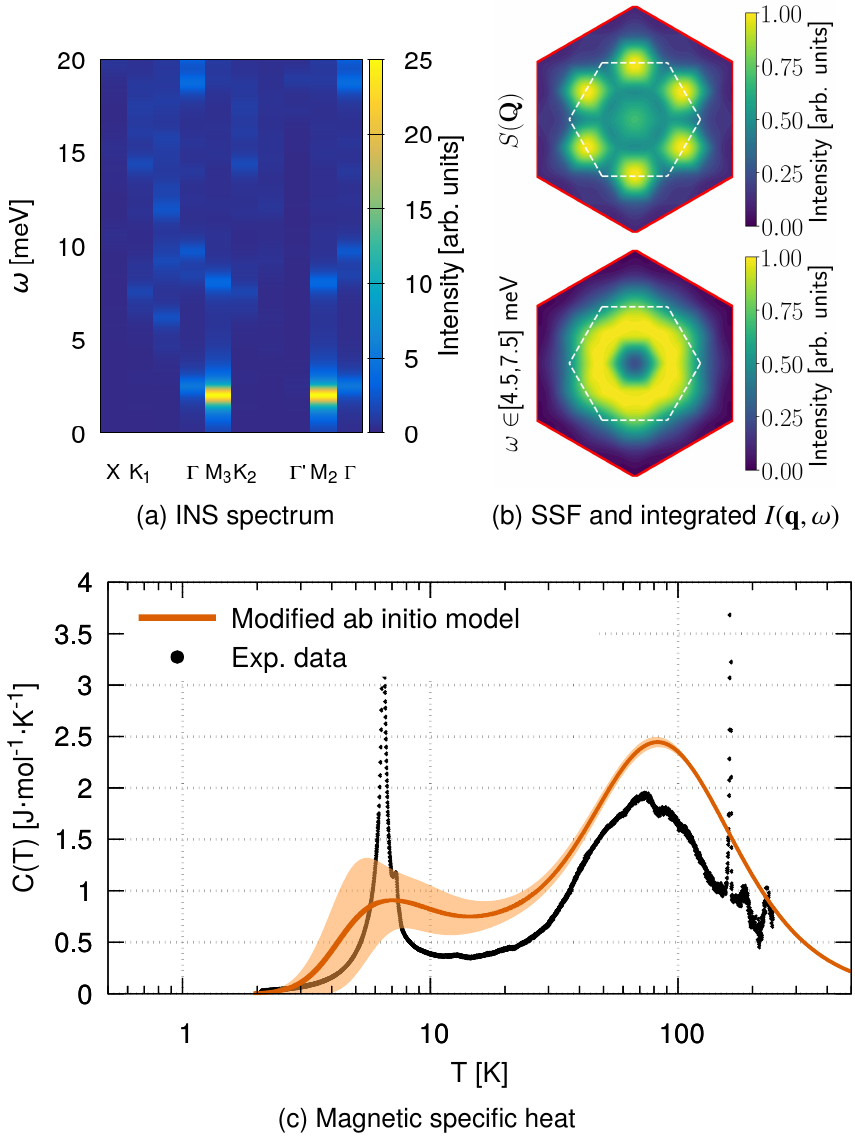}
	\caption{\label{fig:modifiedmodel}The modified ab initio model. Shown in (a) is the $I(\mathbf{q},\omega)$ spectrum for the modified ab initio model, calculated using ED. We find low-energy peaks at the {M} and $\Gamma$ points consistent with peaks in experimental INS data. The SSF is shown in the top panel of (b), clearly consistent with a zigzag order. The bottom panel of (b) shows $I(\mathbf{q},\omega)$ integrated over $[4.5,7.5]$ meV, with less intensity at the zone center than is experimentally observed. (c) shows the magnetic specific heat calculated using TPQ and $15$ random initial vectors compared {with} the experimental data from Ref.~\cite{PhysRevB.99.094415}. {The shaded region shows the standard deviation.} The calculations were all done using the $24$-site cluster.}
\end{figure}

We again find that the spin wave gap at the $\Gamma$ point depends strongly on the value of $\Gamma_1'$, and that the low-energy features of the INS spectrum can be well explained when $|\Gamma_1'|$ is significantly reduced but still finite. Specifically, we take $\Gamma_1' \rightarrow 0.05\Gamma_1'$. (The full parameter set is given in {Supplementary Table 2}.) The spectrum for this case is shown in Fig.~\ref{fig:modifiedmodel} (a). In this case, we find $\omega_\Gamma = 2.5$ meV, ${{\omega_\mathrm{M}}}= 2$ meV, close to the values obtained in Ref.~\cite{Banerjee2018}: $\omega_\Gamma = 2.69\pm 0.11$ meV, ${\omega_\mathrm{M}}= 2.2\pm 0.2$ meV. We note that our value for $\omega_\Gamma$ is consistent with the the low-energy magnon energy of $2.5$ meV observed using THz spectroscopy \cite{PhysRevB.98.094425,Reschke2019}. As we noted earlier, the relative intensity of the $\Gamma$ point peak may be enhanced at finite temperature \cite{PhysRevLett.120.077203}.

Fig.~\ref{fig:modifiedmodel} (b) shows the SSF, which remains consistent with zigzag order, and $I(\mathbf{q},\omega)$ integrated over the $[4.5,7.5]$ meV range. From this integrated intensity and Fig.~\ref{fig:modifiedmodel} (a) we see that there is a lack of intensity at the zone center within the chosen energy range, unlike the notable star shape in Ref.~\cite{Banerjee2017}. The source of this discrepancy is not clear, and calls for further study and parameter refinement. Finally, in (c) we show the magnetic specific heat calculated for this modified model. We do find $T_{{\mathrm{h}}}\approx 83$ K, which is higher than the $\approx 70$ K reported in Ref.~\cite{PhysRevB.99.094415}, yet consistent with the broader range of values proposed ($70-100$ K). 


\section*{Discussion}
We have found that there is a considerable qualitative difference between proposed spin Hamiltonians that describe the INS data well, and realistic models derived using ab initio methods, which are consistent with the reported magnetic specific heat observations. This difference is accompanied by a significant discrepancy in overall energy scales. The specific heat measurements probe the energy density of states, and should represent a good guide to the energy scale, provided the phonon background is handled adequately. In light of our results we thus expect the Kitaev and off-diagonal couplings strengths to be larger, and that $\alpha$-RuCl$_3$ may be closer to the QSL regime than previously believed. (We note that recent anisotropic susceptibility \cite{PhysRevB.98.100403} and THz spectroscopy experiments \cite{Reschke2019} are also consistent with higher Kitaev strengths than in Model 2.) In contrast, the calculated dynamical spin structure factors and INS intensities are much more sensitive to the relative strengths of different interaction terms. They are particularly useful probes for models with fewer degrees of freedom{, such as the $J_1-K_1-\Gamma_1-\Gamma_1'-J_3$ model we study}. 
At the same time, static properties such as magnetic order or SSFs, are clearly insufficient to fully constrain the $\alpha$-RuCl$_3$ spin Hamiltonians. In this respect, properties in the presence of magnetic fields, such as phase transitions and the magnon thermal Hall effect \cite{PhysRevB.98.060412}, present a particularly promising direction for both theory and experiment.

By using one of the ab initio models (Eichstaedt et al. \cite{PhysRevB.100.075110}) as a starting point and reducing the magnitude of $\Gamma_1'$, we are able to identify a set of parameters (full values are given in {Supplementary Table 2}) that partly resolve the discrepancy between the two classes of models mentioned above. We find low-energy peaks in the INS spectrum and a high-temperature peak in the magnetic specific heat that are consistent with experiment. However, this should not yet be considered a fully accurate model, as there is an unexplained lack of intensity at the zone center at intermediate frequencies. Instead, we consider it a new starting point.

With these results in mind, we now return to the question we posed in the abstract, about the nature of the correct spin Hamiltonian for $\alpha$-RuCl$_3$. From a variety of ab initio and DFT calculations, we expect a minimal model to include ferromagnetic nearest-neighbor Kitaev and Heisenberg couplings, $\Gamma_1>0$, a $\Gamma_1'<0$ term, and a small $J_3>0$. {Our results for the modified ab initio model further suggest that $\Gamma_1'$ should be small, but finite.} Since {both $\Gamma_1'$ and} $J_3$ act to stabilize the zigzag order, {small values are} consistent with the fact that a relatively weak {in-plane} magnetic field can take $\alpha$-RuCl$_3$ out of the ordered phase. Alternatively, $\alpha$-RuCl$_3$ might be close to a quantum critical point \cite{PhysRevB.96.041405,PhysRevLett.120.077203,Balz2019}, which would be a very exciting scenario. Anisotropic susceptibility measurements \cite{PhysRevB.98.100403} point towards significant off-diagonal $\Gamma_1$ and $\Gamma_1'$ terms, which may also help stabilize the purported spin liquid phase at finite magnetic fields \cite{Catuneanu2018,Gordon2019,PhysRevB.99.224409}. At this point it is not clear whether anisotropies between bonds or the interlayer coupling play a qualitative role, but they are also expected in a full model.

We hope that our results can help guide further theory development and interpretation of experimental results going forward{, both for $\alpha$-RuCl$_3$ and other Kitaev spin liquid candidate materials}. Since we found the INS predictions to be particularly sensitive to small parameter changes, it would be very useful to consider additional modeling techniques and additional observables. For example, machine learning methods may be a promising way to efficiently handle the high-dimensional parameter space. In addition, further experiments in applied magnetic fields can help constrain the Hamiltonian by suppressing fluctuations.


\section*{Methods}
We use the $\mathcal{H}\Phi$ \cite{Kawamura2017} library for numerical calculations on finite-size systems. We employ a $24$-site cluster with $C_3$ symmetry, and a rhombic $32$-site cluster, see Fig.~\ref{fig:HoneycombClusters}. {The momenta compatible with the finite-size clusters are shown in {Supplementary Note 1}.} Finite-temperature specific heat is computed using the microcanonical thermal pure quantum state (TPQ) method \cite{PhysRevLett.108.240401,PhysRevB.93.174425,Sugiura2017}, and averaged over $\geq{}15$ random initial vectors. The key idea behind the TPQ method is that a quantum system at thermal equilibrium can be reliably described by a single, iteratively constructed state. Utilizing this fact allows for a significant reduction in computational cost compared {with} finite-temperature exact diagonalization methods.

Zero temperature properties are calculated using the Lanczos exact diagonalization method, and the continued fraction expansion (CFE) \cite{RevModPhys.66.763} is used to compute the dynamical quantities. A total of $500$ Lanczos steps are used to calculate the CFE. We take $1$ meV as a representative value for the experimental energy resolution at full-width/half-maximum \cite{Banerjee2017,Banerjee2018}, and emulate it in the exact diagonalization calculations by using a Lorentzian broadening of $0.5$ meV. {For the INS spectral evolution calculations we used a Lorentzian broadening of $0.05$ in the fixed energy scale.} The neutron scattering intensity $I\left( \mathbf{q}, \omega\right)$ is defined \cite{10.1038/s41467-017-01177-0,PhysRevLett.120.077203}
\begin{align}
I\left( \mathbf{q}, \omega\right)	&\propto f^2(q) \sum_{\mu,\nu}  \left[ \delta_{\mu\nu}-\frac{q_\mu q_\nu}{q^2}\right] S^{\mu\nu} \left( \mathbf{q},\omega\right),
\end{align}
where 
$f(q)$ is the magnetic form factor, $q^a$ is the projection of the momentum vector onto the spin components in the local cubic coordinate system also used for the spin Hamiltonian, and $S^{\mu\nu} \left( \mathbf{q}, \omega\right)$ is the dynamical spin structure factor at momentum $\mathbf{q}$ and frequency $\omega$,
\begin{align}
	S^{\mu\nu}\left( \mathbf{q},\omega\right)	&= \int \sum_{i,j}\left\langle S^\mu_i \left(t\right) S^\nu_j \left( 0\right) \right\rangle {\mathrm{e}}^{-i\mathbf{q}\cdot \left(\mathbf{r}_i - \mathbf{r}_j \right)} {\mathrm{e}}^{-i\omega t} \,\mathrm{d}t 	\label{eq:Smunu}
\end{align}
Note that the off-diagonal elements of $S^{\mu\nu} \left( \mathbf{q}, \omega\right)$ contribute significantly for most models studied here, due to the presence of $\Gamma_1$ and $\Gamma_1'$ interactions. The static spin structure factor,  $S\left(\mathbf{q}\right)=\int S\left(\mathbf{q}, \omega\right)\, \mathrm{d}\omega$, is evaluated separately. {We note that neutron scattering experiments probe the magnetization $\mathbf{M}$, while the Hamiltonians in Eqs.~\eqref{eq:HamiltonianMinimal} and \eqref{eq:HamiltonianGammaPrime} are expressed in terms of the pseudospin $\mathbf{S}$. Hence the form of Eq.~\eqref{eq:Smunu} amounts to an assumption that $\mathbf{M}$ and $\mathbf{S}$ are approximately parallel. However, trigonal distortion can induce an angle between the two vectors, and a resulting $g$-factor anisotropy \cite{PhysRevB.94.064435}. While there have been conflicting reports about the degree of anisotropy \cite{PhysRevB.91.180401,PhysRevB.91.094422,Yadav2016}, a more recent X-ray absorption spectroscopy study \cite{PhysRevB.96.161107} found $\alpha$-RuCl$_3$ to have only weak trigonal distortion and a nearly isotropic $g$-factor. In light of this result and the relatively weak $\Gamma_1'$ interactions in Table~\ref{table:models} we assume that $\mathbf{M}$ and $\mathbf{S}$ are indeed approximately parallel in $\alpha$-RuCl$_3$.
}

$f(q)$ is assumed to be isotropic, which is justified for small scattering wave numbers \cite{Banerjee2017}. The magnetic form factor for Ru$^{3+}$ was calculated using DFT in the Supplementa{ry} Material of Ref.~\cite{Do2017}. By fitting their data to a Gaussian we have obtained the analytical approximation
\begin{align}
f(q) &= \exp\left(-\frac{q^2}{\left(2\pi*0.25\right)^2}\right).
\end{align}
We integrate over the momentum direction perpendicular to the honeycomb plane, following the experiment \cite{Banerjee2017}. Since the ED calculation is necessarily two-dimensional we assume that $S^{\mu\nu} \left( \mathbf{q}, \omega\right)$ is constant along the perpendicular direction during the integration step. We expect this to be a reasonable approximation due to the strong two-dimensionality of $\alpha$-RuCl$_3$ \cite{PhysRevB.98.205110}, and the relatively small interlayer coupling \cite{Balz2019}.

\section*{Data availability statement}
The data that support the findings of this study are available from the corresponding author upon reasonable request.


\section*{Acknowledgments} 
We thank C. Balz, A. Banerjee, T. Berlijn, S. E. Nagler, A. M. Samarakoon, and D. A. Tennant for helpful discussions, and A. Loidl for providing the magnetic specific data. We thank Y. Yamaji both for useful discussions and assistance with $\mathcal{H}\Phi$. We are also grateful to an anonymous referee for suggesting a systematic approach to study the spectral evolution away from the Kitaev limit, which lead us to the improved interaction parameters proposed in this work.

The research by P.L. and S.O. was supported by the Scientific Discovery through Advanced Computing (SciDAC) program funded by the US Department of Energy, Office of Science, Advanced Scientific Computing Research and Basic Energy Sciences, Division of Materials Sciences and Engineering. This research used resources of the Oak Ridge Leadership Computing Facility, which is a DOE Office of Science User Facility supported under Contract DE-AC05-00OR22725, and of the Compute and Data Environment for Science (CADES) at the Oak Ridge National Laboratory, which is managed by UT-Battelle and supported by the Office of Science of the U.S. Department of Energy under Contract No. DE-AC05-00OR22725. {An award of computer time was provided by the INCITE program. A portion of the work was conducted at the Center for Nanophase Materials Sciences, which is a DOE Office of Science User Facility.}


\section*{Competing interests} The authors declare that there are no competing interests.

\section*{Author contributions}
P.L. performed the ED and TPQ calculations. S.O. supervised the study. Both authors contributed to the writing of the manuscript.



\end{document}


\title{Supplementary Information for\texorpdfstring{\\}{ }Dynamical and thermal magnetic properties of the Kitaev spin liquid candidate \texorpdfstring{$\alpha$-RuCl$_3$}{α-RuCl3}}
	\author{Pontus Laurell}
	\affiliation{Center for Nanophase Materials Sciences, Oak Ridge National Laboratory, Oak Ridge, Tennessee 37831, USA}
	\author{Satoshi Okamoto}
	\affiliation{Materials Science and Technology Division, Oak Ridge National Laboratory, Oak Ridge, Tennessee 37831, USA}
	\date{\today}
	\maketitle
	\thispagestyle{plain}
	\pagestyle{plain}
	
	\renewcommand{\figurename}{Supplementary Figure}
	\renewcommand{\tablename}{Supplementary Table}
	
	In this Supplement, we present results for the Hamiltonians not covered extensively in the main text, as well as more details for the models covered in the main text. In particular, we include comparisons of the predicted inelastic neutron scattering (INS) profiles at the $\Gamma$ and M points {with} experimental data from Ref.~\cite{Banerjee2018} for all models, as well as additional results for the magnetic specific heat and structure factors. {All results reported in the Supplement, except where otherwise explicitly noted, were obtained using the $N=24$ cluster.}
	
\section{{Supplementary Note 1:} Finite-size cluster momenta}
	
	The finite-size clusters are compatible with the momenta shown in Supplementary Fig.~\ref{fig:ClusterMomenta}. To plot the static spin structure factors (SSFs), and energy slices of the INS intensity, $I\left( \mathbf{q}, \omega\right)$, over reciprocal space, we use cubic interpolation over a hexagonal grid centered in the small hexagons.
	\begin{figure}[tbp]
		\centering
		\subfloat[$24$-site cluster momenta]{
			\includegraphics[width=.48\columnwidth]{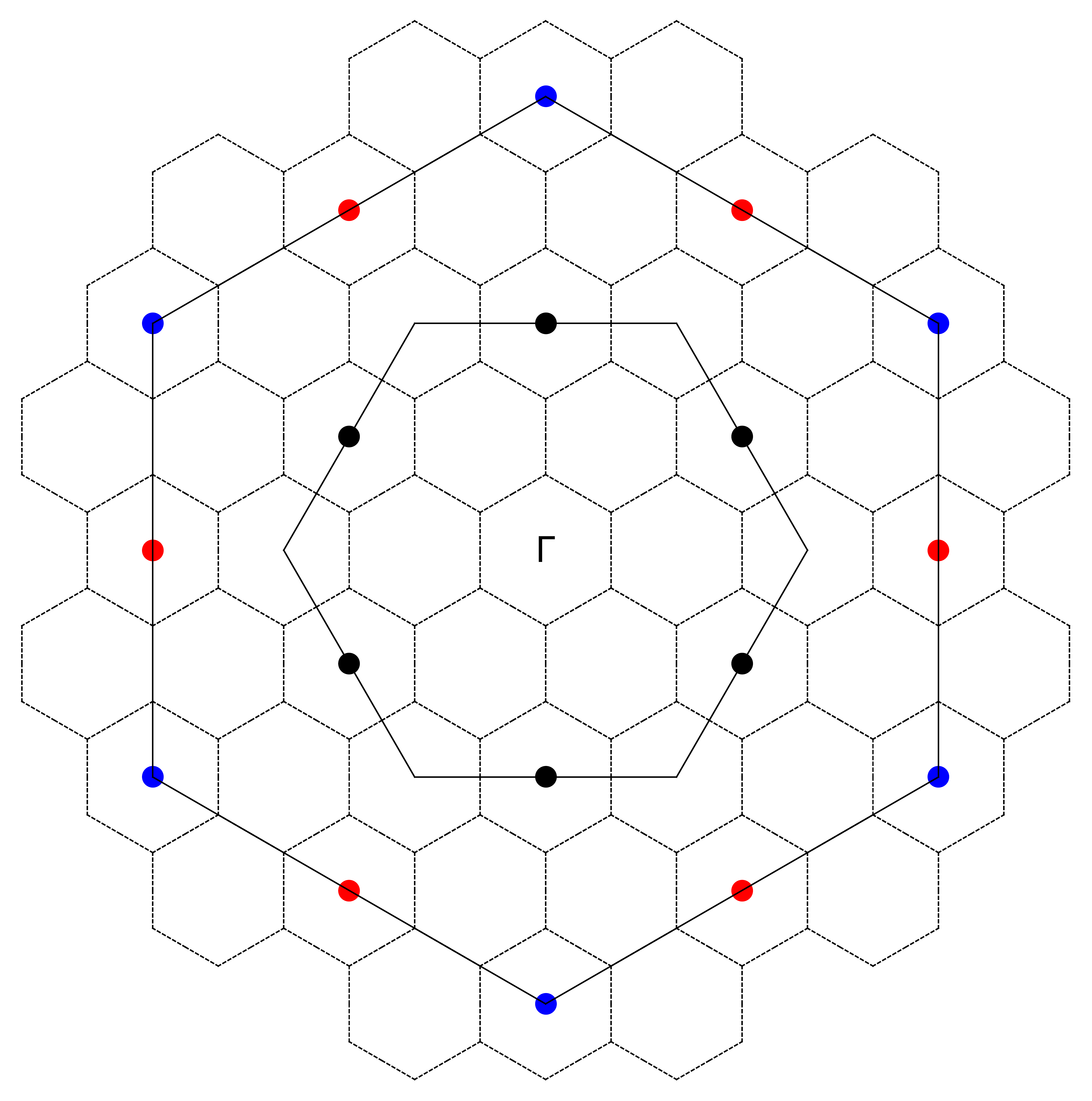}}
		\subfloat[$32$-site cluster momenta]{	\includegraphics[width=.48\columnwidth]{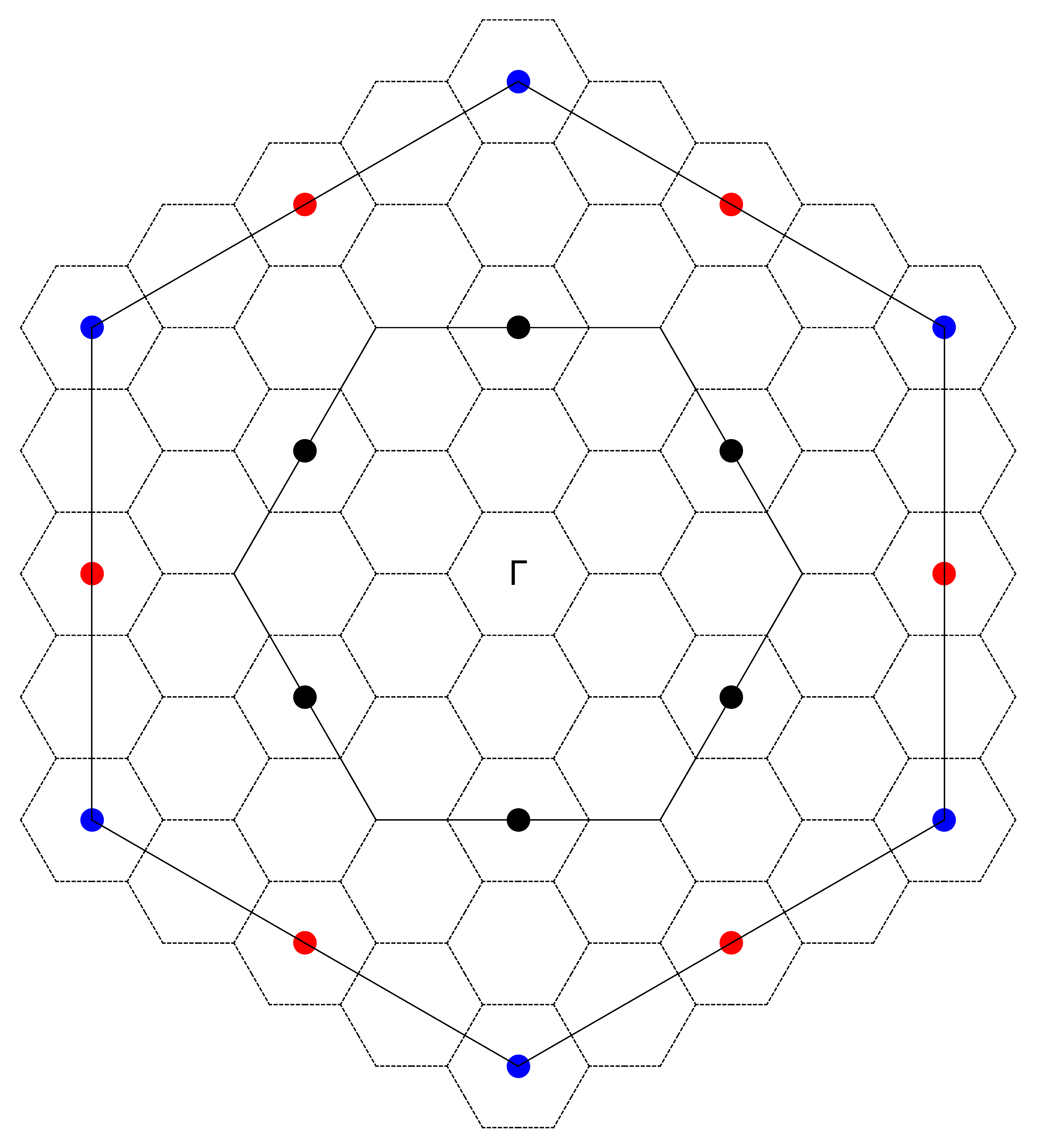}}
		\caption{\label{fig:ClusterMomenta}Cluster momenta. Allowed momenta for the a) $24$-site, and b) $32$-site clusters {are represented by the center points of the small dashed hexagons}. The inner (outer) large solid hexagons represent the first (second) Brillouin zones. Black, red, and blue disks represent M, X and $\Gamma$' points, respectively.}
	\end{figure}
	
\section{Supplementary Note 2: Additional magnetic specific heat results}
	
\begin{figure*}[tbp]
		\centering
		\subfloat[Models 2, 3, and 4 from Refs.~\cite{10.1038/s41467-017-01177-0,PhysRevB.98.060412,PhysRevB.98.094425}.]{
			\includegraphics[width=\columnwidth]{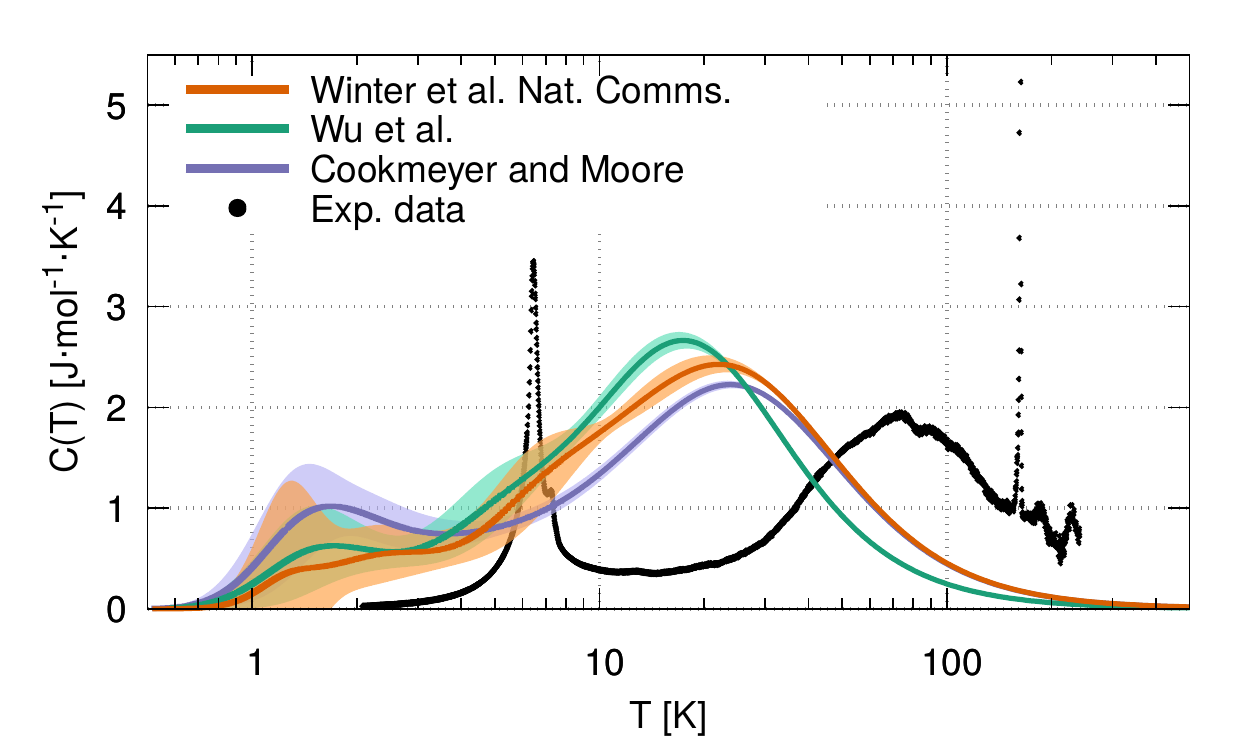}
		}%
		\subfloat[Models 6, 8, 9, and 10 from Refs.~\cite{PhysRevB.97.134424,PhysRevB.96.115103,PhysRevB.96.054410,PhysRevLett.118.107203}.]{	\includegraphics[width=\columnwidth]{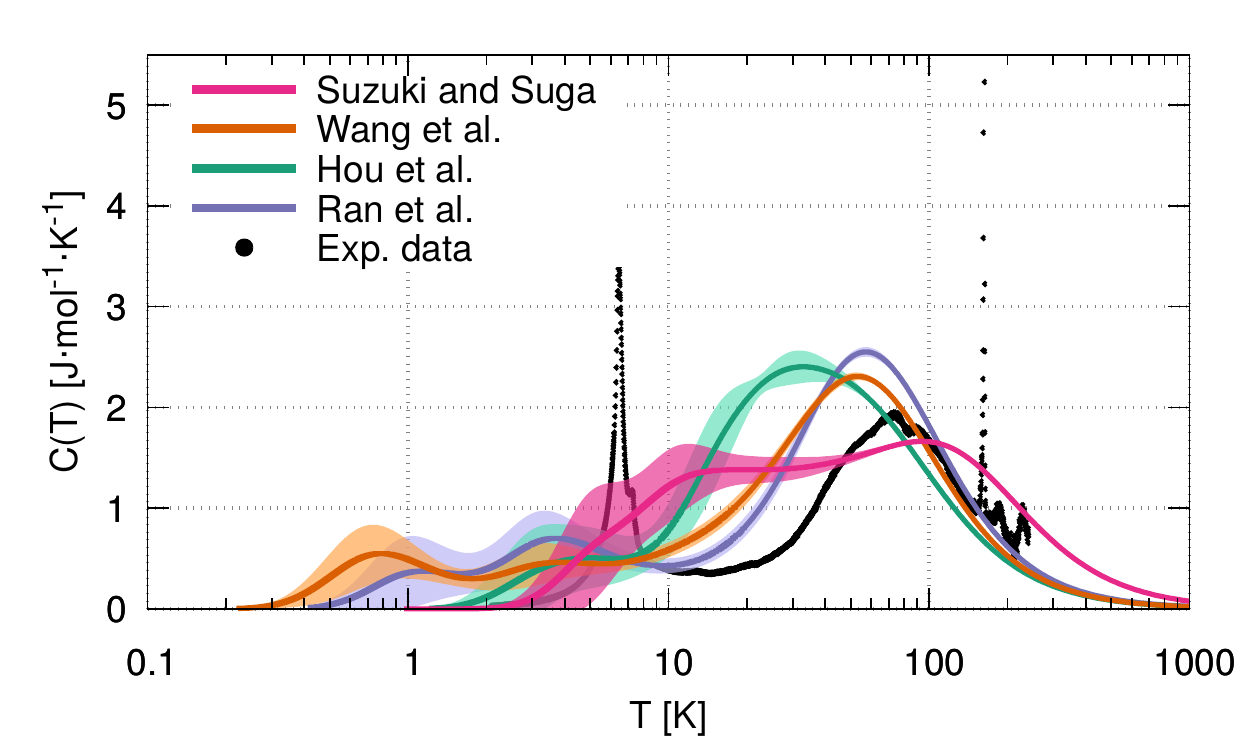}
		}\\\vspace{-0.2375cm}
		\subfloat[Models 11, 12, and 13 from Ref.~\cite{Eichstaedt2019}.]{
			\includegraphics[width=\columnwidth]{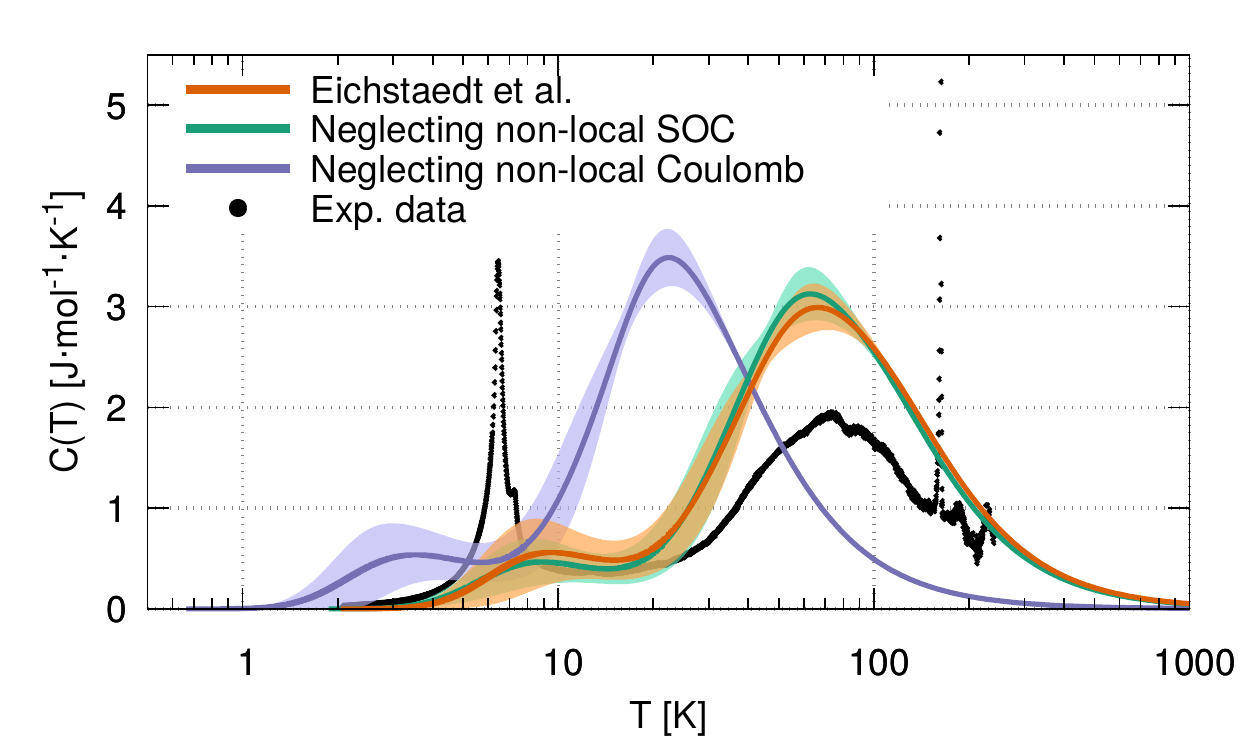}
		}%
		\subfloat[Models 14, 15, 16, and 17 from Refs.~\cite{PhysRevB.93.214431,PhysRevB.93.155143,PhysRevB.91.241110,Banerjee2016}.]{
			\includegraphics[width=\columnwidth]{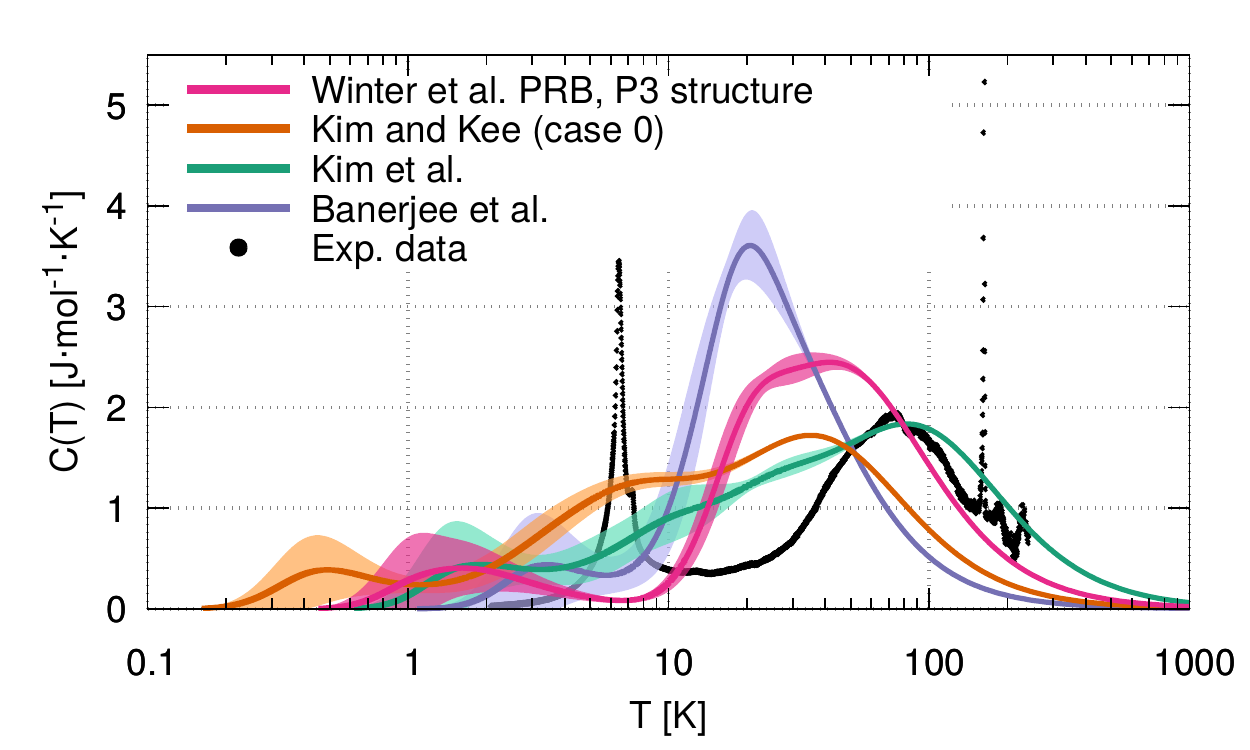}
		}\\\vspace{-0.2375cm}
		\subfloat[{Models 18 and 19 from Ref.~\cite{PhysRevB.100.085108}.}]{
			\includegraphics[width=\columnwidth]{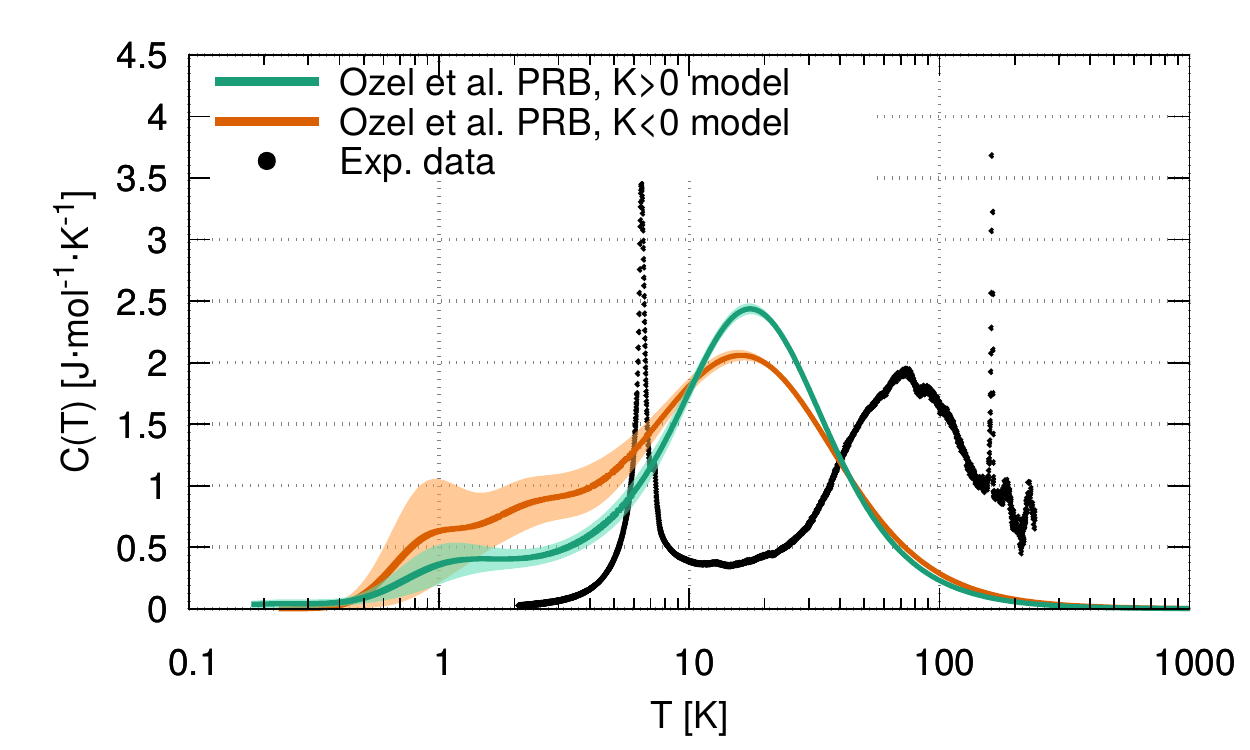}
		}
		\caption{\label{fig:supp:specHeat24}Magnetic specific heat calculated using the TPQ method for various proposed $\alpha$-RuCl$_3$ Hamiltonians, compared with experimentally determined excess heat capacity from Ref.~\cite{PhysRevB.99.094415}. The model numbers match those in Table I of the main text. The solid lines show the calculated average value over $15$ initial vectors, and the shaded areas show the standard deviation.
		}
	\end{figure*}

	Supplementary Fig.~\ref{fig:supp:specHeat24} shows the magnetic specific heat for the models not considered in the main text. Two of the panels also include comparisons to a few models that were discussed in the main text. This is the case for (a), where the Winter et al. Nat. Comms. model \cite{10.1038/s41467-017-01177-0} is shown along with two related models, due to Cookmeyer and Moore \cite{PhysRevB.98.060412}, and Wu et al. \cite{PhysRevB.98.094425}, respectively. In (c) the different cases considered in Eichstaedt et al. \cite{Eichstaedt2019} are shown. Both these panels clearly indicate that, for related models, the high-temperature peak $T_\mathrm{h}$ and the $C(T)$ profile vary mostly smoothly with the overall energy scale of the model. In (a) we also note that the Cookmeyer and Moore parameters produce a more prominent low-temperature peak than the Winter et al. Nat. Comms. model. We interpret this as an effect of the weaker third-nearest Heisenberg exchange, $J_3$, which results in a less stable zigzag ordering and puts the system closer to the Kitaev limit, for which a two-peak structure is expected.
	
	In Supplementary Fig.~\ref{fig:supp:specHeat24} (b), the models of Ran et al.~\cite{PhysRevLett.118.107203} and Wang et al. \cite{PhysRevB.96.115103} both have high-temperature peaks near $\approx 55$ K, which is relatively close to the experimental value ($\approx 70$ K) obtained in Ref.~\cite{PhysRevB.99.094415}. Meanwhile, the Suzuki and Suga model  \cite{PhysRevB.97.134424,Suzuki2019} has its peak at $\approx 94$ K, consistent with the peak position of $\simeq 100$ K obtained in Ref.~\cite{Do2017}. In (d) results for four Hamiltonians with $K_1>0$ are shown. We note that the Kim et al. model \cite{PhysRevB.91.241110,PhysRevB.96.064430} has its peak at a reasonable position ($T_\mathrm{h}=81$ K), but a rather flat temperature dependence. Supplementary Fig.~\ref{fig:supp:specHeat24} (e) shows the magnetic specific heat for the two spin wave fits to THz spectroscopy data obtained by Ozel et al.~\cite{PhysRevB.100.085108}. The fit with $K_1>0$ has $T_\mathrm{h} \approx 16$ K, and the fit with $K_1<0$ has $T_\mathrm{h}=17.5$ K. Both models are inconsistent with the magnetic specific heat data, but we note that these $T_\mathrm{h}$ values are very consistent with the similarly obtained model of Wu et al.~\cite{PhysRevB.98.094425}, which has $T_\mathrm{h}\approx 17$ K.
		
	In Fig. 7 of the main text we contrasted the $24$- and $32$-site cluster results for four models, arguing that the high-$T$ peak appears to be fairly stable. To quantify this, Supplementary Table~\ref{supp:table:SpecHeatThTmeps} lists the numerical positions for the high-$T$ peaks. Overall, the relative difference $\Delta T_\mathrm{h}$ between the two clusters considered is on the order of a few percent, despite the cluster changing both size and shape. This relatively small difference suggests that the TPQ method is a useful numerical tool, particularly at high temperatures, despite being limited to small clusters. This is in line with previous results for nearest-neighbor honeycomb Kitaev-Heisenberg \cite{PhysRevB.93.174425} and kagome Heisenberg \cite{PhysRevLett.111.010401} models.
	
	\begin{table}[t]
		\caption{\label{supp:table:SpecHeatThTmeps}Cluster dependence on the location of the high-temperature peak in the magnetic specific heat calculated using the TPQ method. Both cluster size and symmetry properties may influence the results. The larger difference for model 11 may be due to the fact that the full ab initio Hamiltonian, without $C_3$ symmetry was used, including bond anisotropies and second nearest neighbor interactions. The relative difference is defined $\Delta T_\mathrm{h}=(T_\mathrm{h}^{N=32}-T_\mathrm{h}^{N=24})/T_\mathrm{h}^{N=24}$.}
		\begin{ruledtabular}
			\begin{tabular}{llrrr}
				& Reference         & $T_\mathrm{h}^{N=24}$ {[}K{]} & $T_\mathrm{h}^{N=32}$ {[}K{]}  & $\Delta T_\mathrm{h}$ {[}\%{]} \\\hline
				1  & Winter et al. PRB & $54.5$	& $53.6$	& $-1.7$ \\
				2  & Winter et al. NC  & $22.1$	& $22.6$	& $+2.4$ \\
				3  & Wu et al.         & $17.4$ & $17.7$	& $+1.9$ \\
				11 & Eichstaedt et al. & $66.4$ & $62.8$ 	& $-5.4$
			\end{tabular}
		\end{ruledtabular}
	\end{table}
	
\section{Supplementary Note 3: Additional INS results}

	The $T=0$ INS intensity profiles at the $\Gamma$ and M$_1$ points for the six models considered in the main text are plotted in Supplementary Fig.~\ref{fig:supp:INSprofile1} as a function of energy. The ED calculation is compared with experimental data from Ref.~\cite{Banerjee2018}, in which the positions of the first spin-wave peaks were estimated to be $2.69\pm 0.11$meV at the $\Gamma$ point, and $2.2\pm 0.2$meV at the M point.
	
	The ab initio inspired Winter et al. Nat. Comms. parameters reproduce the intensity profile at the $\Gamma$ point particularly well. This is no surprise, given that the parameters were chosen to reproduce broad features of the INS spectrum, especially near the $\Gamma$ point. In contrast, from panels (a) and (f), we see that the two models that most accurately predicted the high-temperature peak in the magnetic specific heat have the intensity at the $\Gamma$ point shifted to much higher frequencies.
	\begin{figure*}[tbp]
		\centering
		\subfloat[1. Winter et al. PRB]{
			\includegraphics[width=0.66\columnwidth]{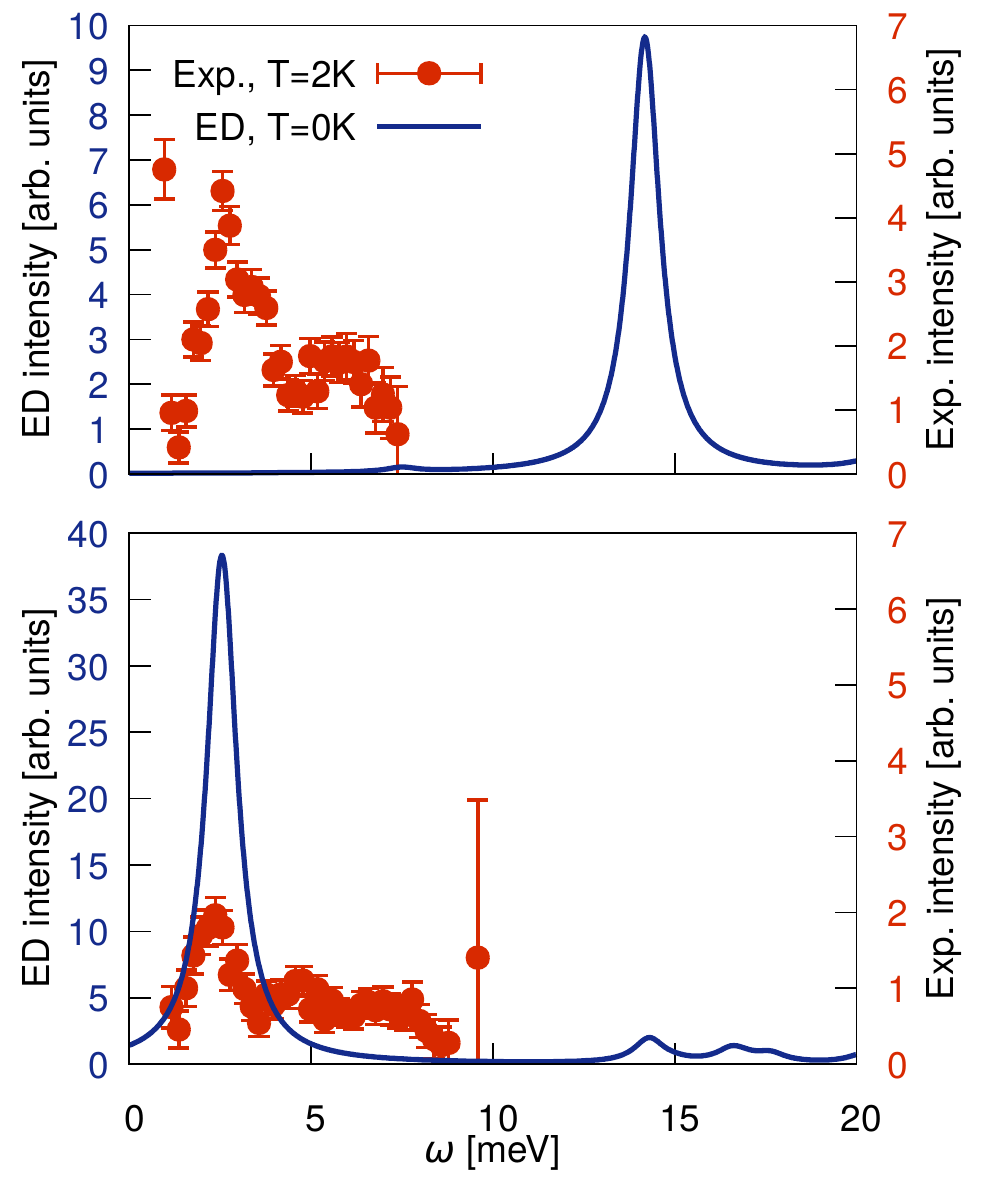}
		}\hspace{2cm}%
		\subfloat[2. Winter et al. Nat. Comms.]{
			\includegraphics[width=.66\columnwidth]{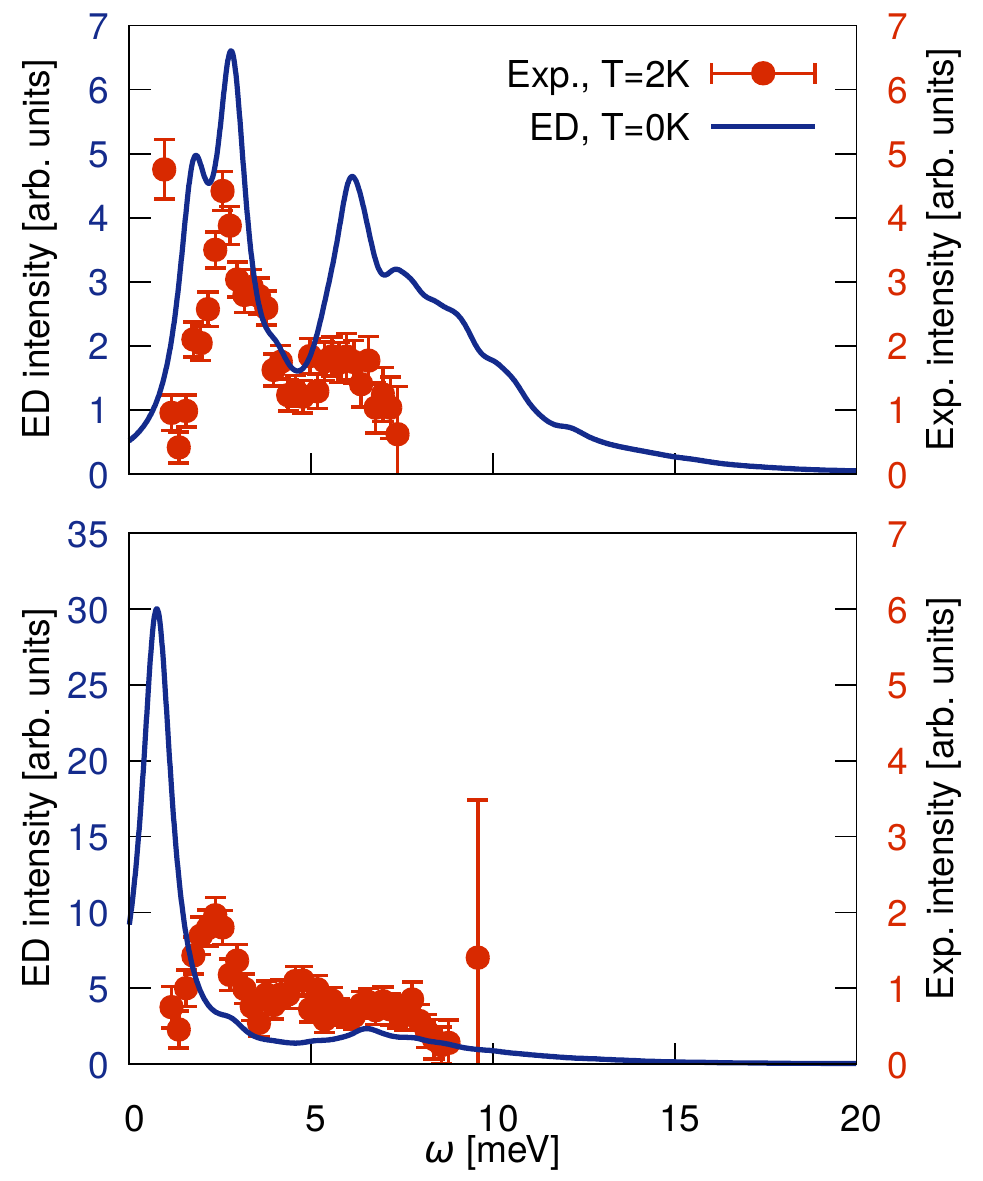}
		}\\\vspace{-0.305cm}
		\subfloat[3. Wu et al.]{				
			\includegraphics[width=.66\columnwidth]{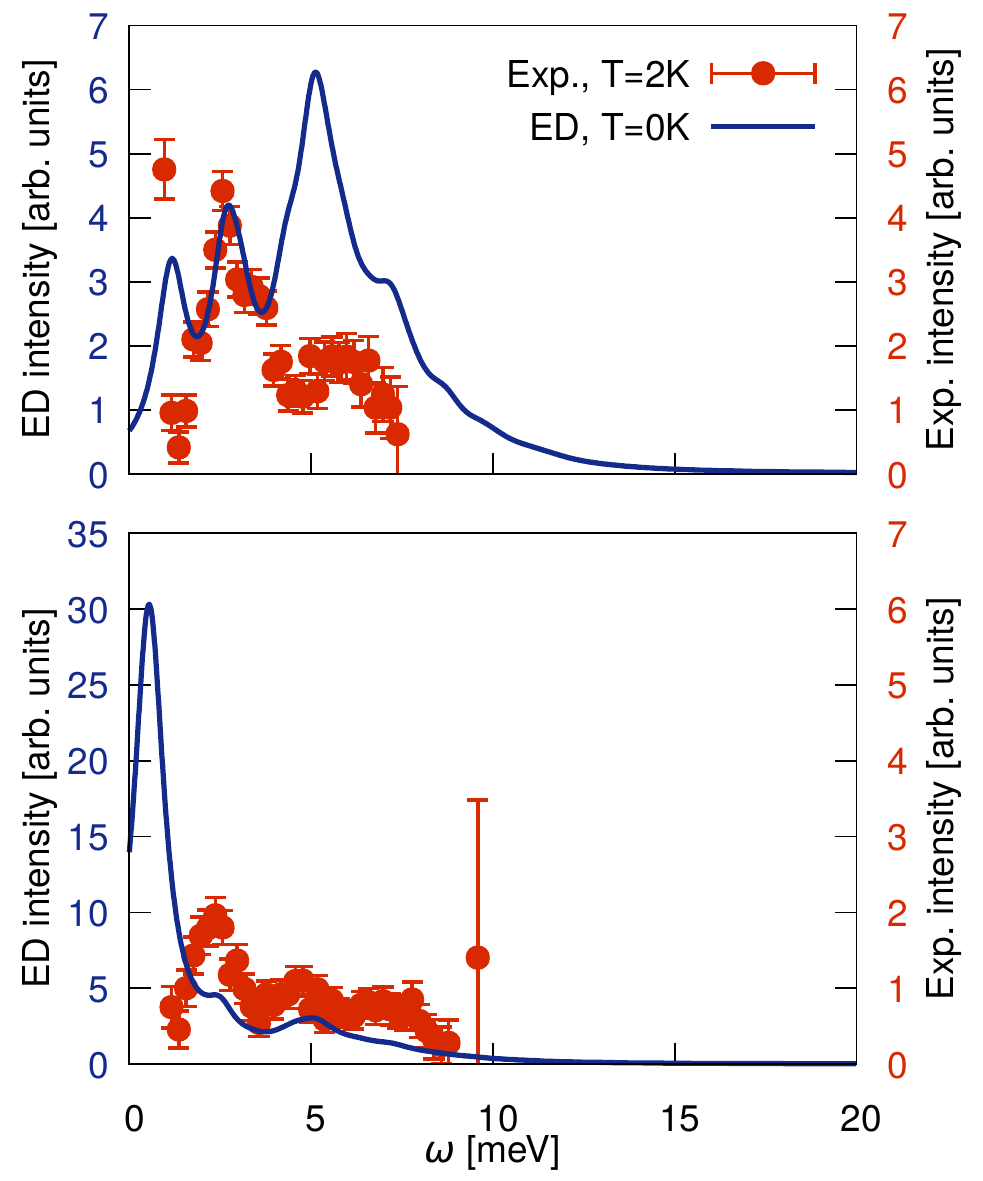}
		}\hspace{2cm}%
		\subfloat[5. Kim and Kee]{
			\includegraphics[width=0.66\columnwidth]{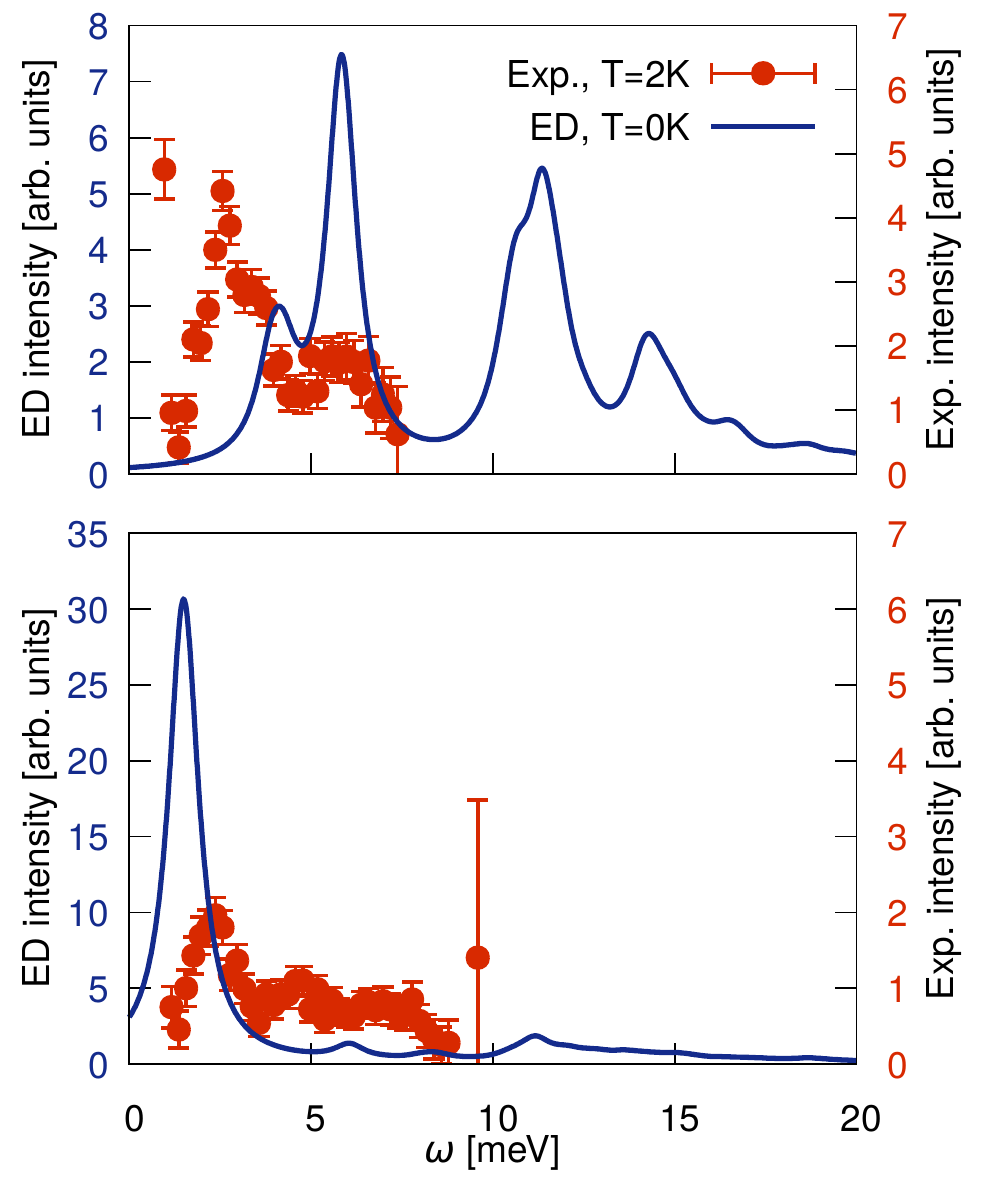}
		}\\\vspace{-0.305cm}
		\subfloat[7. Yadav et al.]{				
			\includegraphics[width=.66\columnwidth]{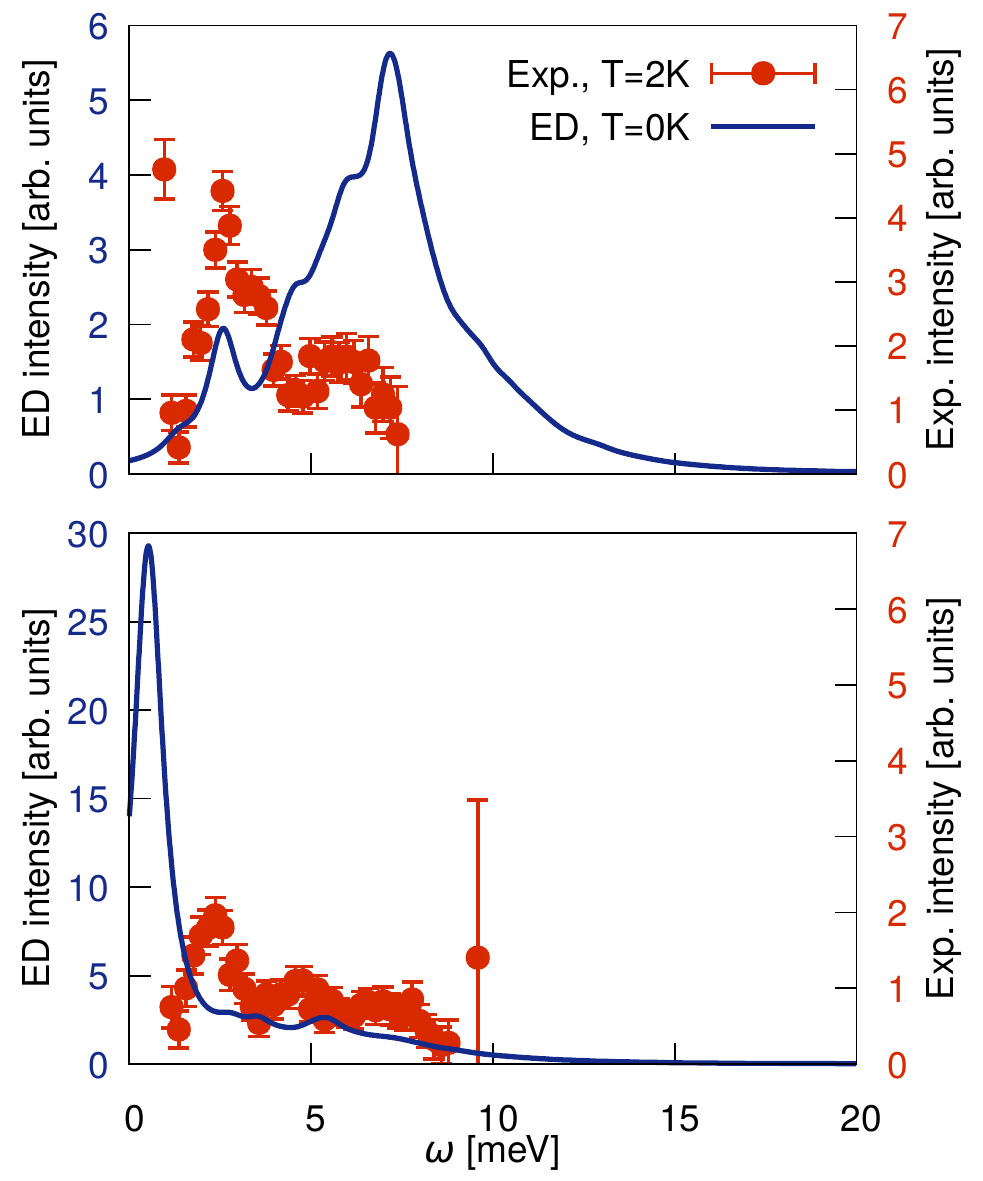}
		}\hspace{2cm}%
		\subfloat[11. Eichstaedt et al.]{
			\includegraphics[width=.66\columnwidth]{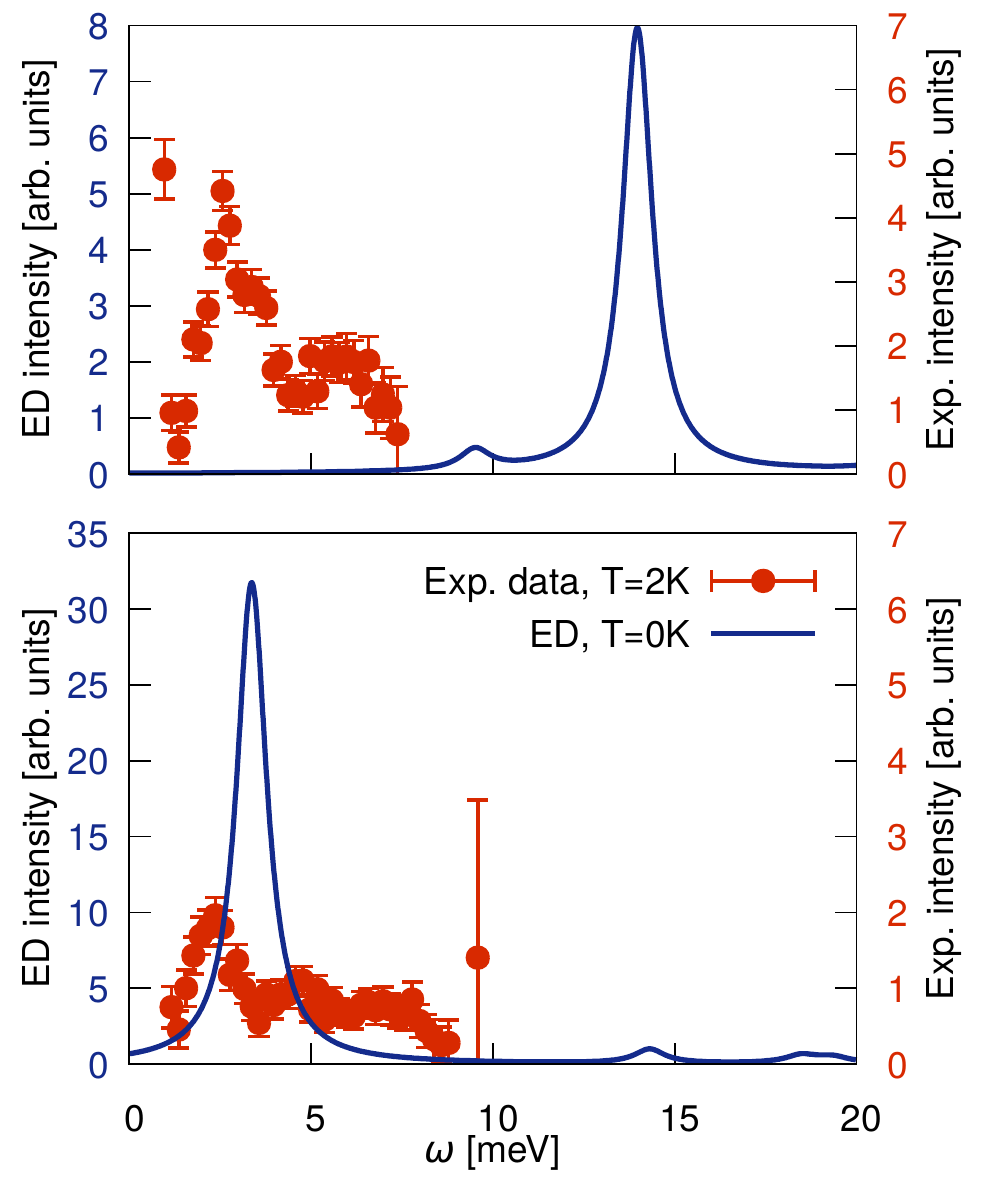}
		}
	\caption{\label{fig:supp:INSprofile1}$I\left( \mathbf{q}, \omega\right)$ at the $\Gamma$ (top panel) and M$_1$ points (bottom panel) for the six models studied in the main text. Experimental data from Ref.~\cite{Banerjee2018}, with error bars representing one standard deviation assuming Poisson counting statistics.}
	\end{figure*}
	
	We next study the fully ab initio-determined parameters from Eichstaedt et al. \cite{Eichstaedt2019} more closely. These parameters were derived using density functional theory, constrained RPA, and perturbation theory ($t/U$ expansion), and the calculations included nonlocal Coulomb interaction and spin-orbit coupling terms. This fact makes it interesting to consider approximate versions of the full Hamiltonian, to better understand the role of different effects. The $I\left( \mathbf{q}, \omega\right)$ spectra are shown in Supplementary Fig.~\ref{fig:supp:IqomegaEichtaedt}, while the static spin structure factor $S\left( \mathbf{q}\right)$ and energy-integrated slices of $I\left( \mathbf{q}, \omega\right)$ are shown in Supplementary Fig.~\ref{fig:supp:SqomegaEichtaedt}. The full Hamiltonian is bond-anisotropic, which results in inequivalent M peaks. Bond averaging makes the model $C_3$ symmetric, and the M points equivalent. As can be seen in Table I of the main text, the non-local SOC only has a small effect on the spin interaction parameters, which leads to minor changes to $I\left( \mathbf{q}, \omega\right)$. Finally, the INS intensity profiles for the four cases are plotted in Supplementary Fig.~\ref{fig:supp:INSprofileEichstaedt}.
	\begin{figure*}
		\centering
		\subfloat[Full ab initio model]{
			\includegraphics[height=4.3cm]{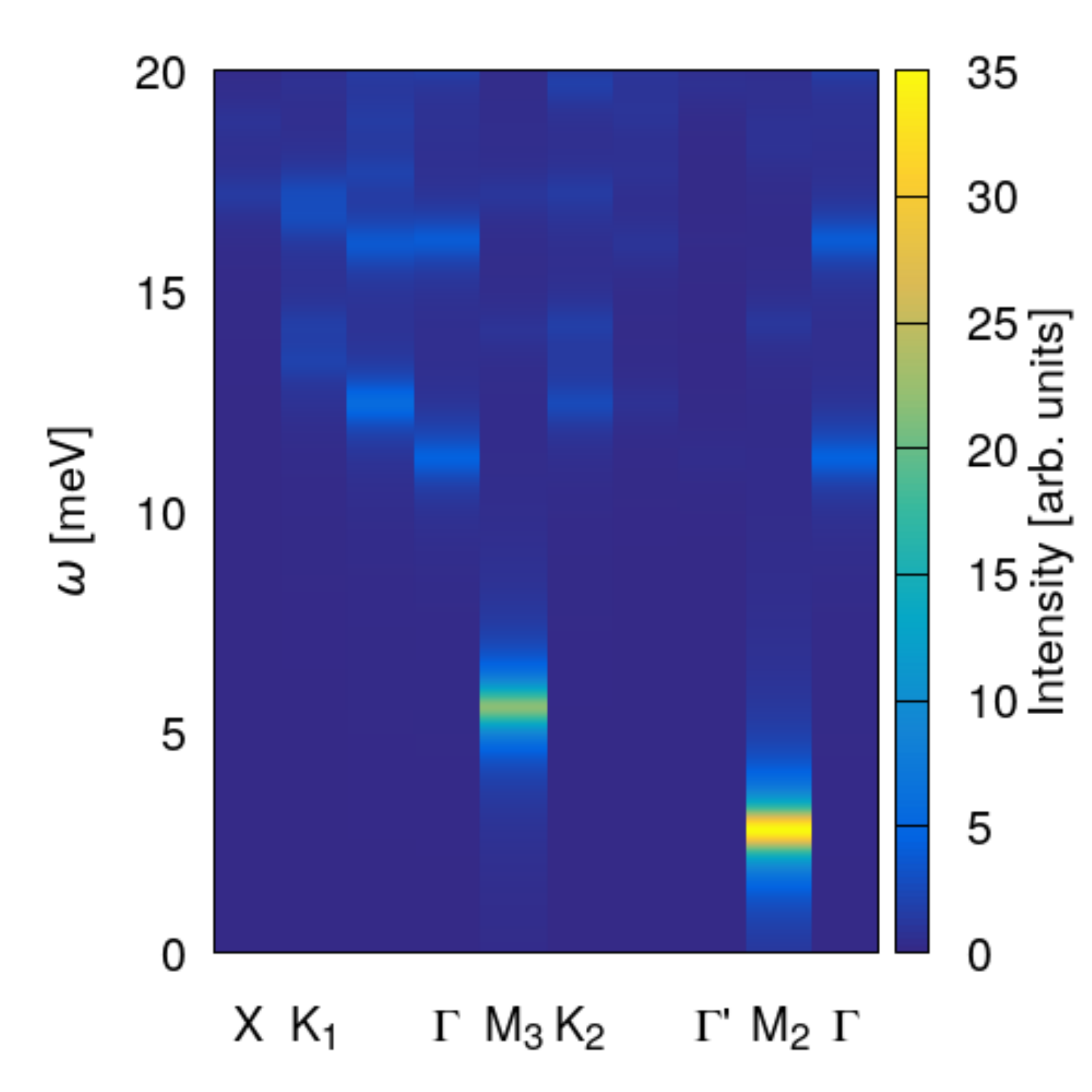}
		}%
		\subfloat[Bond-averaged full model]{
			\includegraphics[height=4.3cm]{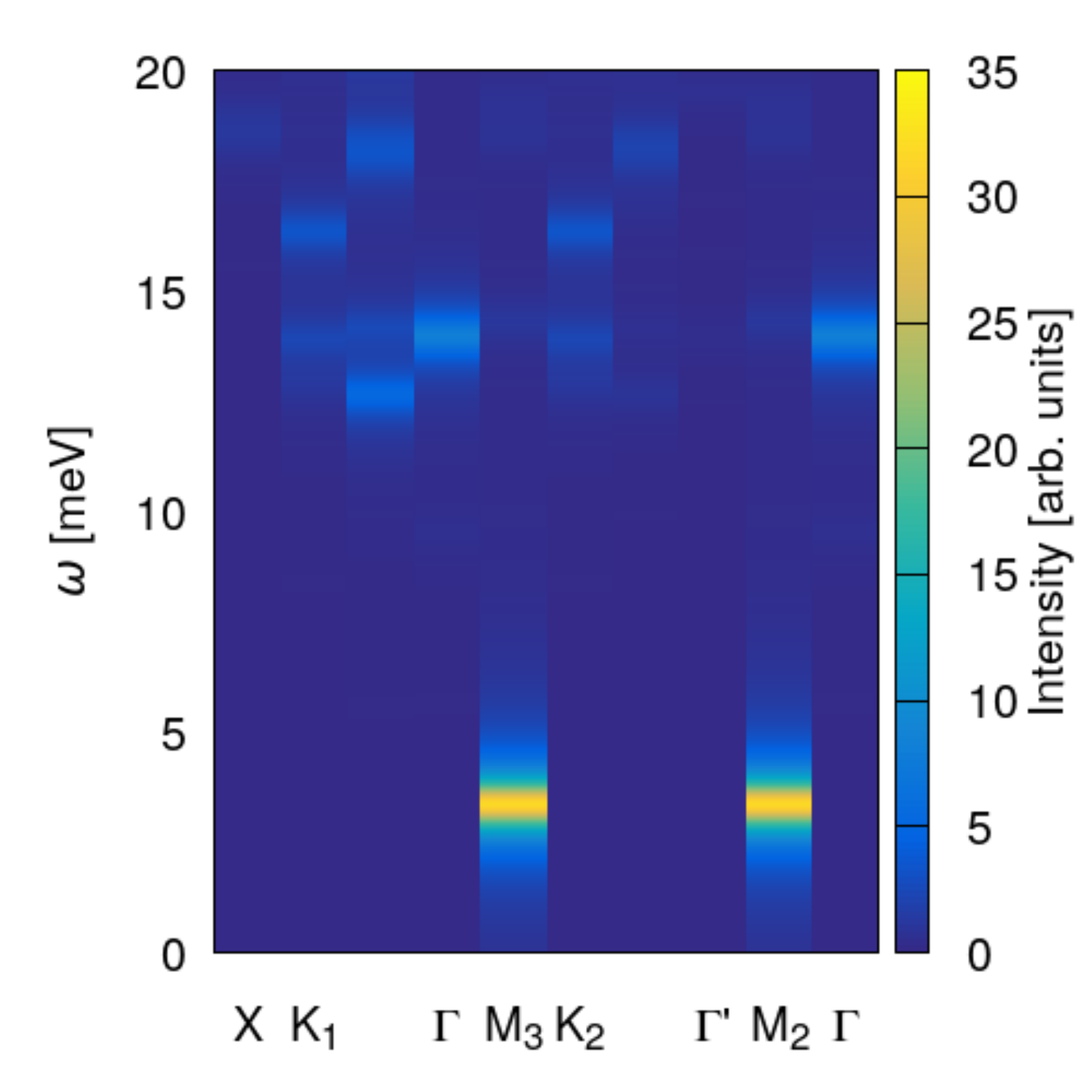}
		}%
		\subfloat[Neglecting nonlocal Coulomb]{
			\includegraphics[height=4.3cm]{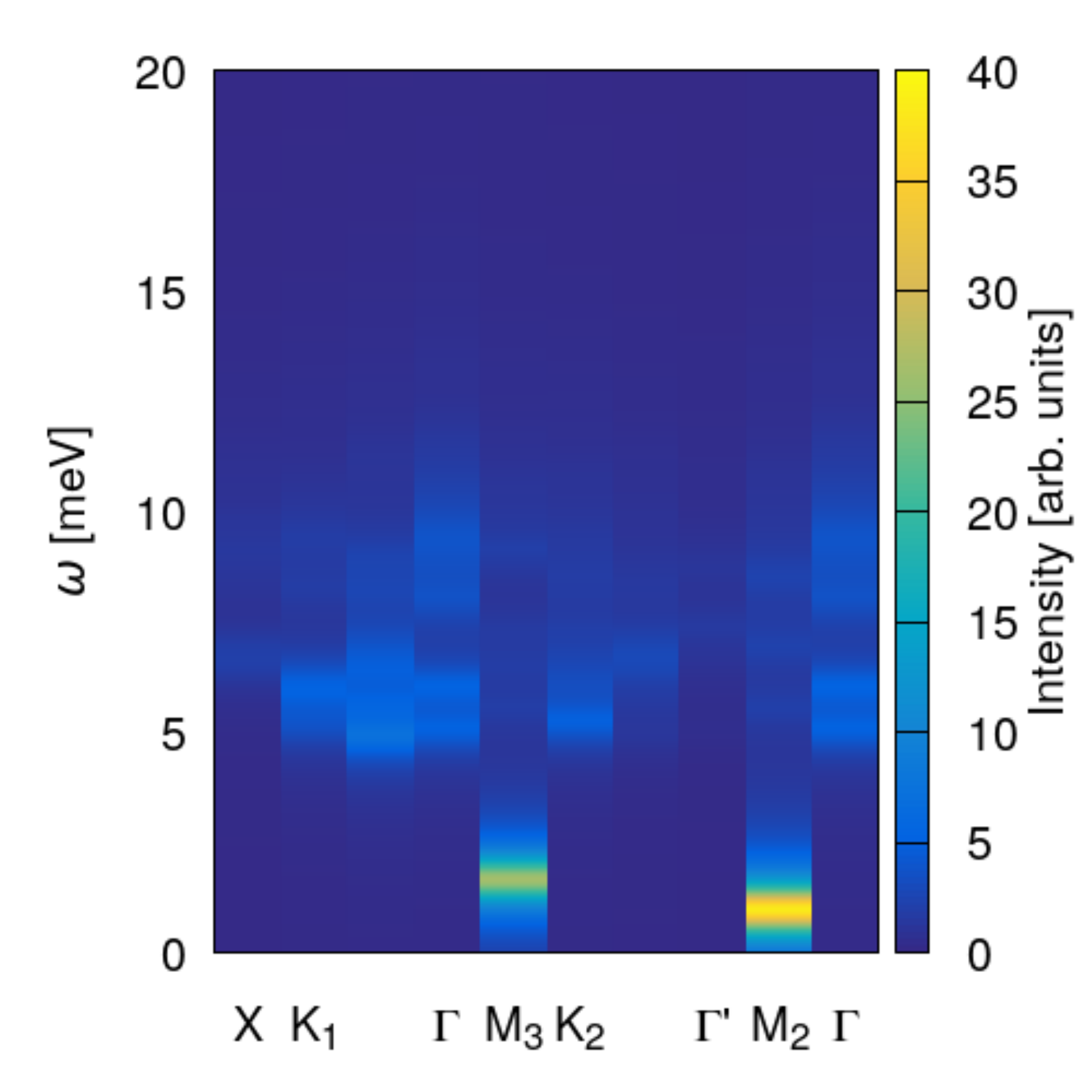}
		}
		\subfloat[Neglecting nonlocal SOC]{
			\includegraphics[height=4.3cm]{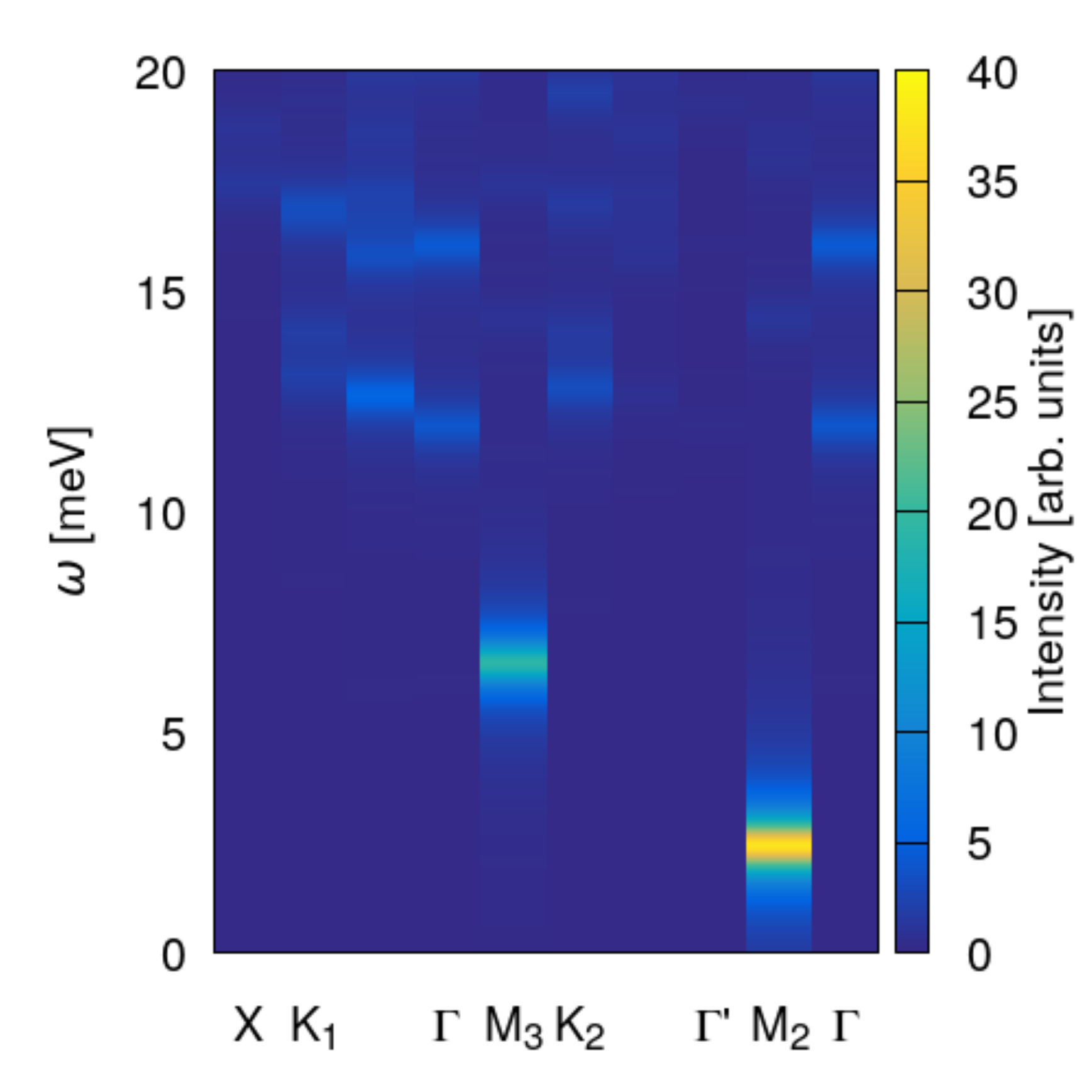}
		}
		\caption{\label{fig:supp:IqomegaEichtaedt}Inelastic neutron scattering intensities $I\left( \mathbf{q}, \omega\right)$ for ab initio parameters from Ref.~\cite{Eichstaedt2019}. (a) shows the spectrum for the full, bond-anisotropic model, while the results for the bond-averaged parameters are shown in (b). (c) and (d) correspond to cases $2$ and $3$ in Ref.~\cite{Eichstaedt2019}, and have not been bond averaged.}
	\end{figure*}
	\begin{figure*}[tbp]
		\centering
		\includegraphics[height=12.5cm]{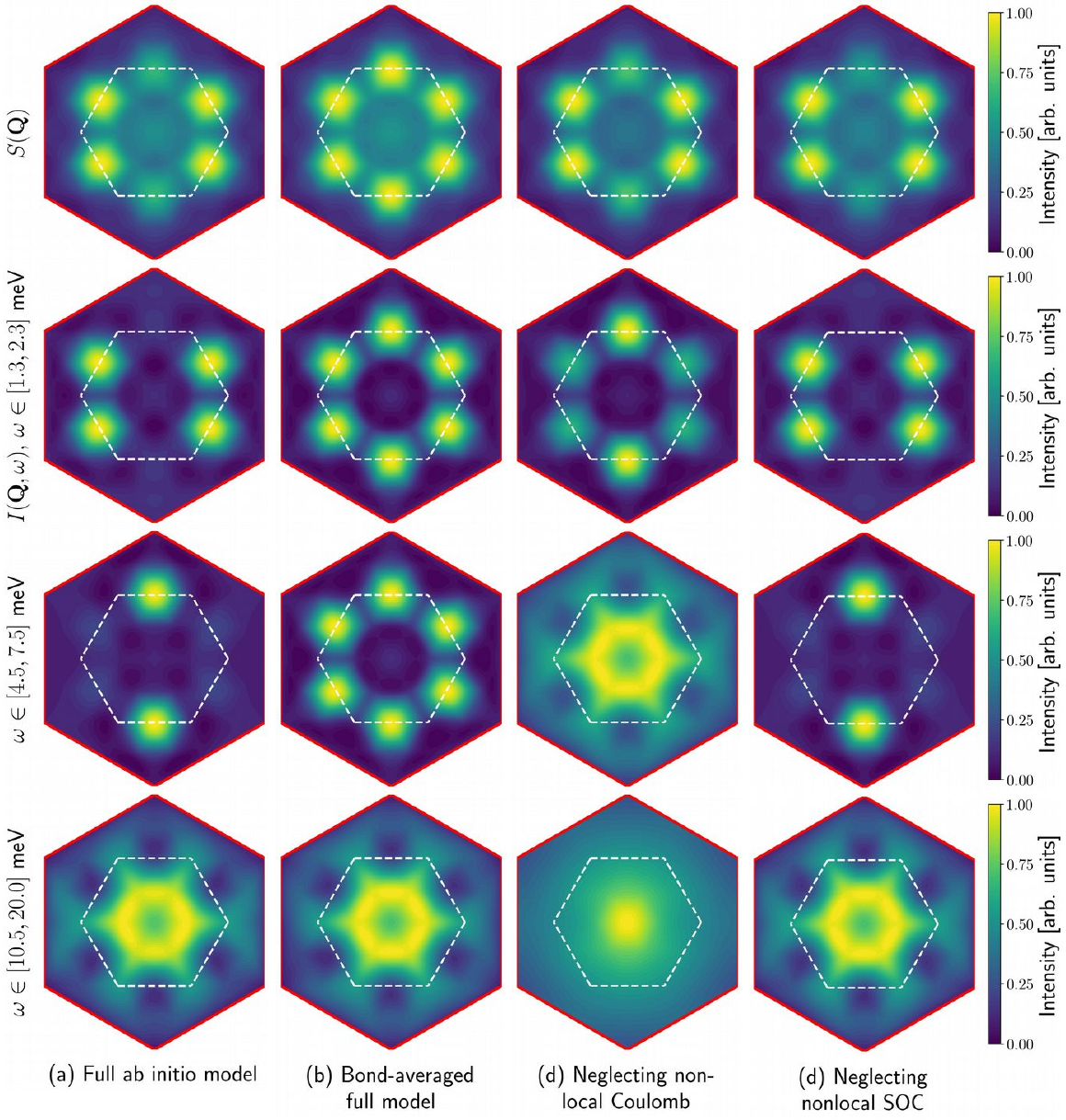}
		\caption{\label{fig:supp:SqomegaEichtaedt}Ab initio parameters from Ref.~\cite{Eichstaedt2019}. The top row shows the static spin structure factors, while the three lower rows show the neutron scattering intensities $I\left( \mathbf{q}, \omega\right)$ integrated over representative energy windows i) $[1.3,2.3]$ meV, ii) $[4.5,7.5]$ meV, and iii) $[10.5,20.0]$ meV. Note that each heatmap is normalized separately, in order to showcase patterns in momentum space. Intensities in different heatmaps should not be compared.}
	\end{figure*}
	\begin{figure*}[tbp]
		\centering
		\subfloat[Full ab initio model]{
			\includegraphics[width=.66\columnwidth]{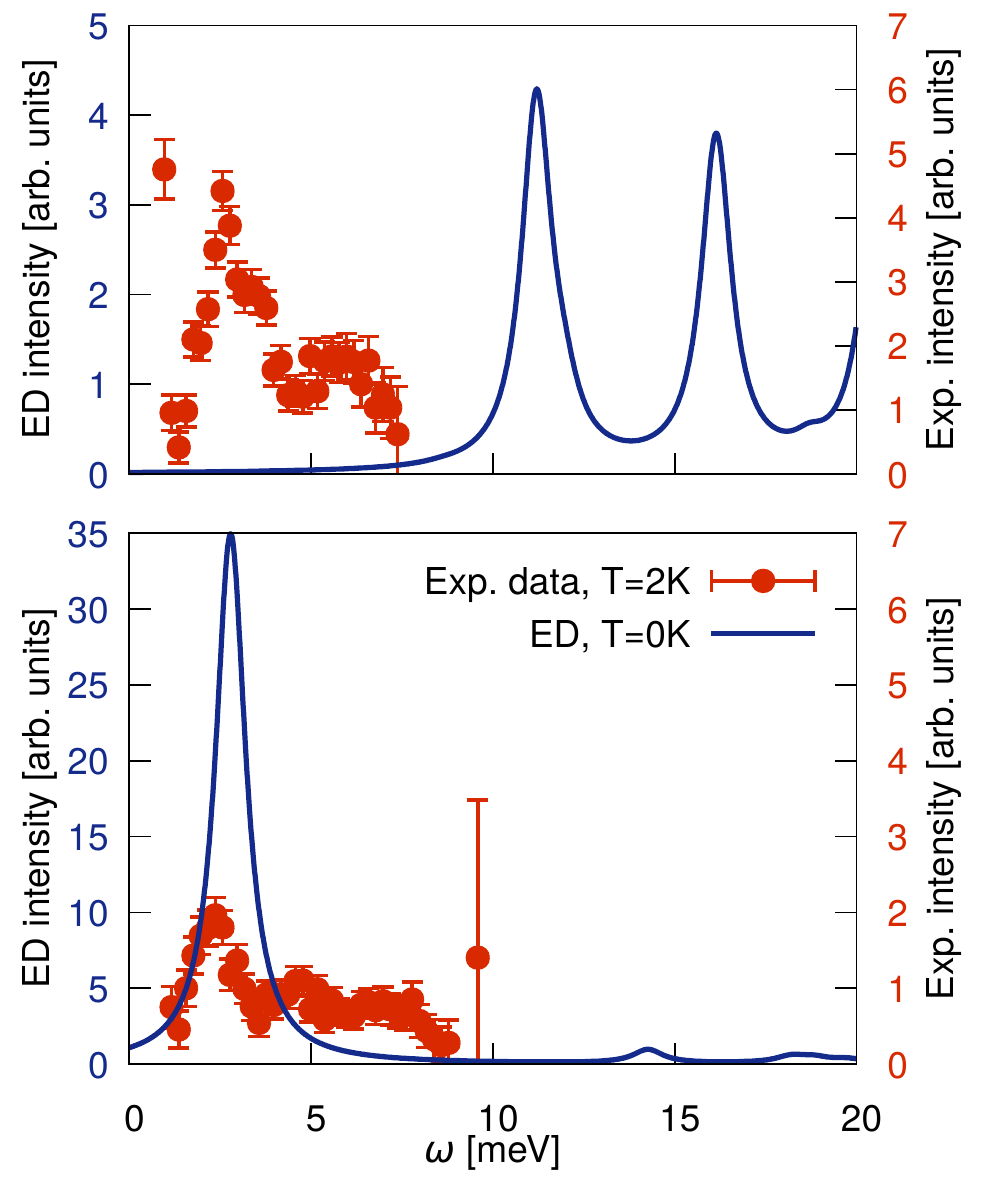}
		}\hspace{2cm}%
		\subfloat[Bond-averaged full model]{				
			\includegraphics[width=.66\columnwidth]{SuppFigures/Supp_IntensityProfile_Eichstaedt_Full_BA.pdf}
		}\\\vspace{-0.305cm}
		\subfloat[Neglecting nonlocal Coulomb]{				
			\includegraphics[width=.66\columnwidth]{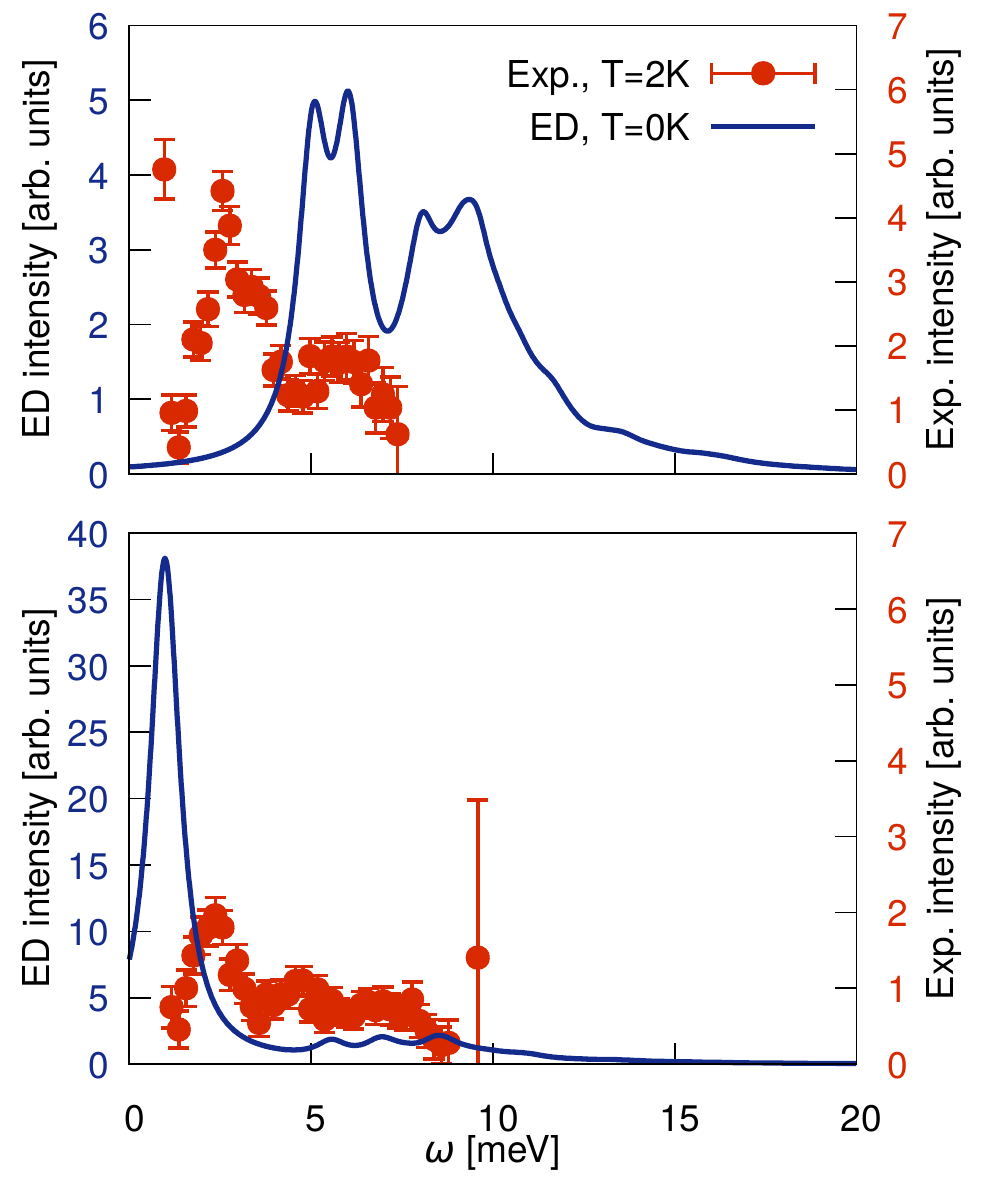}
		}\hspace{2cm}%
		\subfloat[Neglecting nonlocal SOC]{
			\includegraphics[width=.66\columnwidth]{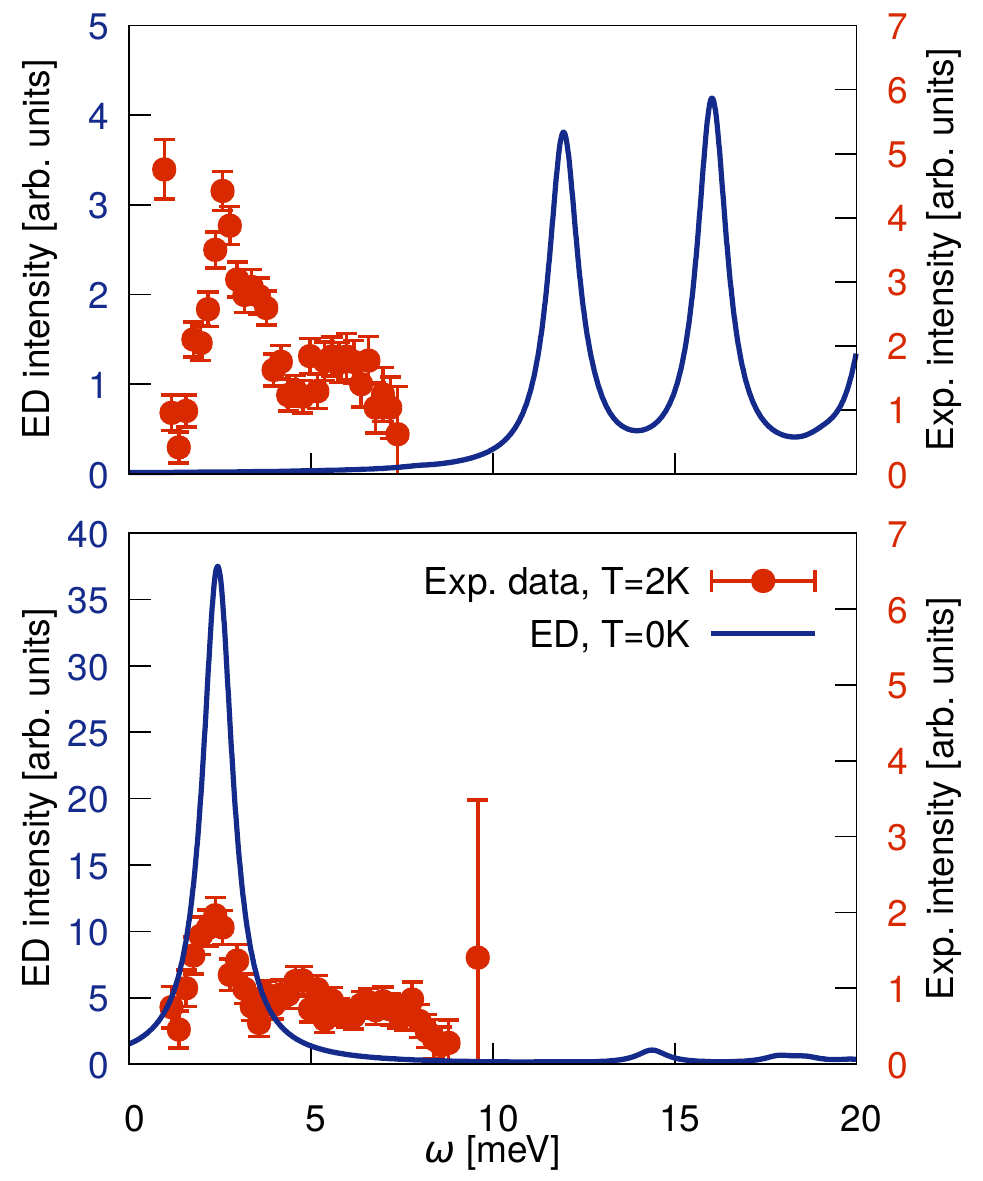}
		}
		\caption{\label{fig:supp:INSprofileEichstaedt}Intensity profiles for the ab initio parameters from Ref.~\cite{Eichstaedt2019} at the $\Gamma$ (top panel) and M$_1$ points (bottom panel). The experimental data is from Ref.~\cite{Banerjee2018}, with error bars representing one standard deviation assuming Poisson counting statistics.
		}
	\end{figure*}
	
	We turn next to the nine of the remaining Hamiltonians. Their respective $I\left( \mathbf{q}, \omega\right)$ spectra are shown in Supplementary Fig.~\ref{fig:supp:Iqomega1}. Five (four) of these models have ferromagnetic (antiferromagnetic) nearest-neighbor Kitaev interactions. Further results for models with $K_1<0$ are shown in Supplementary Figs.~\ref{fig:supp:Sqomega2} and \ref{fig:supp:INSprofile2}. See Supplementary Figs.~\ref{fig:supp:SqomegaPositiveK} and \ref{fig:supp:INSprofilePositiveK} for the models with $K_1>0$. 
	
	Among the first five models, we note that the Cookmeyer and Moore parameters \cite{PhysRevB.98.060412} produce the expected SSF and intensity map in the $[4.5,7.5]$ meV window. In this sense it improves on the Winter et al. Nat. Comms. parameters \cite{10.1038/s41467-017-01177-0}. However, the latter model yields intensity profiles at the $\Gamma$ and M points that better fit the experiment. The SSF for Suzuki and Suga's model \cite{PhysRevB.97.134424,Suzuki2019} is dominated by the $\Gamma$ point peak, due to a pronounced scattering continuum extending to large energies. However, note that Supplementary Fig.~\ref{fig:supp:Iqomega1} (b) shows that the M point peak in $I\left( \mathbf{q}, \omega\right)$ occurs at lower energy than the $\Gamma$ point peak, and is stronger in intensity, which is consistent with the zigzag order.
	
	The Kitaev-$\Gamma$ model proposed by Ran et al. \cite{PhysRevLett.118.107203}, model 8, displays strong low-energy peaks at both the M points, and the center between $\Gamma$ and K points. This spectral structure is similar to that of the pure ``antiferromagnetic'' $\Gamma$ model. Model 9 from Hou et al. \cite{PhysRevB.96.054410} produces a $\Gamma$ point peak at too high $\omega$, similar to the ab initio parameters for models 1 and 11. Model 10 from Wang et al. \cite{PhysRevB.96.115103} is an interesting case. While the strongest $I\left( \mathbf{q}, \omega\right)$ peak occurs at the M point, it also has a low-lying peak at the $\Gamma$ point. Such a peak could potentially be hidden by the elastic continuum in an INS experiment, but we note that THz spectroscopy experiments down to $0.3-0.4$meV find no signs of such a peak \cite{PhysRevLett.119.227201,PhysRevB.98.094425}. The SSF for this model has higher intensity at the $\Gamma$ point than the M points. However, we note that the trace over diagonal components of the SSF, $S^{\mu\mu}\left( \mathbf{q}\right)$, has the opposite pattern, while the off-diagonal components are weaker. This suggests that the pattern in momentum space for this model may be quite sensitive to the radial dependence of the magnetic form factor.
	
	We next turn to the four models with antiferromagnetic NN Kitaev coupling. Models 14 and 16 \cite{Banerjee2017,PhysRevB.93.214431} both have $I\left( \mathbf{q}, \omega\right)$ dominated by low-energy M point peaks. The $\Gamma$ point peaks are too high in energy, and we do not observe extended scattering minima. Models 15 and 16 \cite{PhysRevB.91.241110,PhysRevB.93.155143} are found to have strong weight at the K points at low frequencies, as well as in the SSF. This shows that they order in the $120^\circ$ configuration, which is consistent with the phase diagrams obtained in Ref.~\cite{PhysRevLett.112.077204}.
	\begin{figure*}
		\centering
		\subfloat[4. Cookmeyer and Moore]{
			\includegraphics[height=5cm]{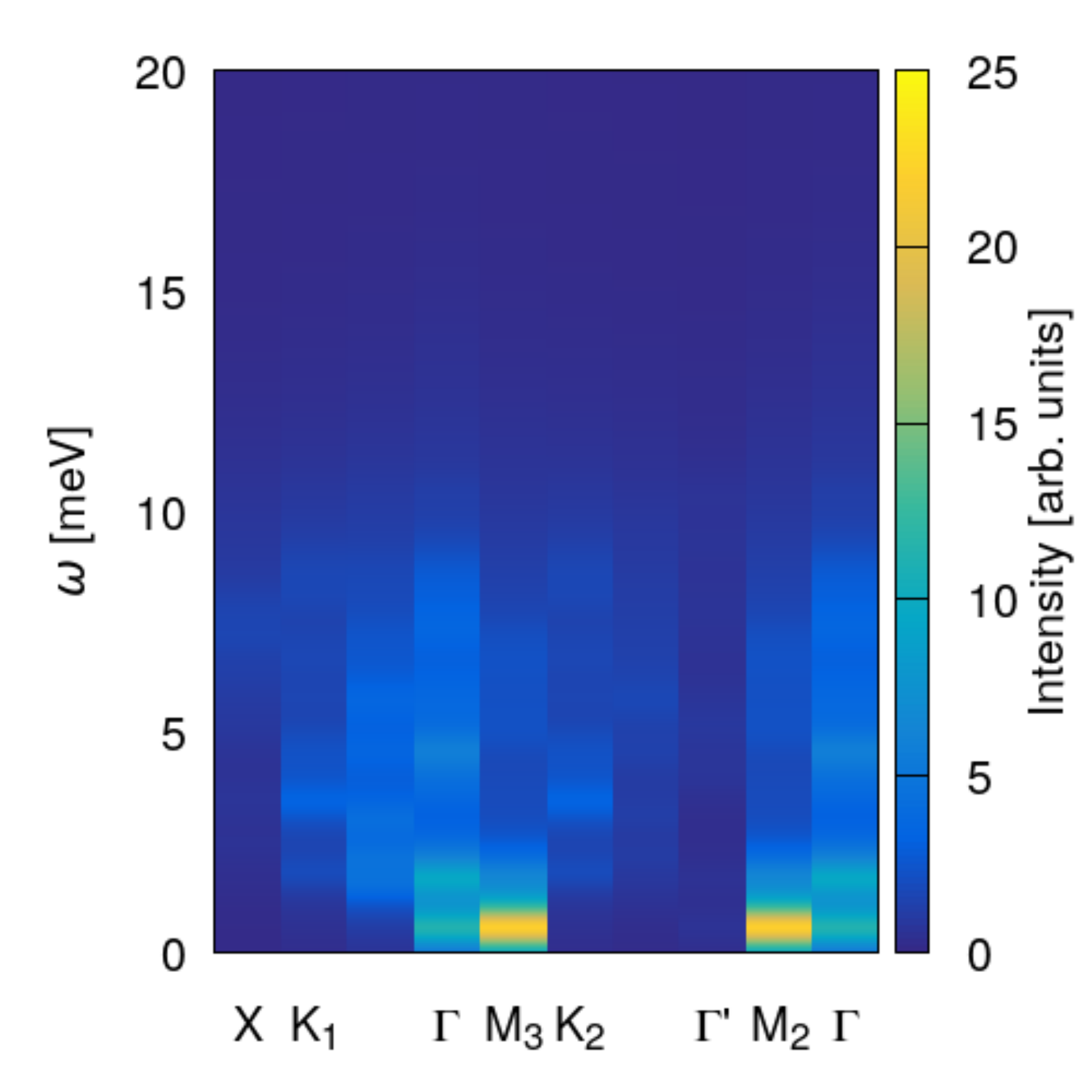}
		}%
		\subfloat[6. Suzuki and Suga]{
			\includegraphics[height=5cm]{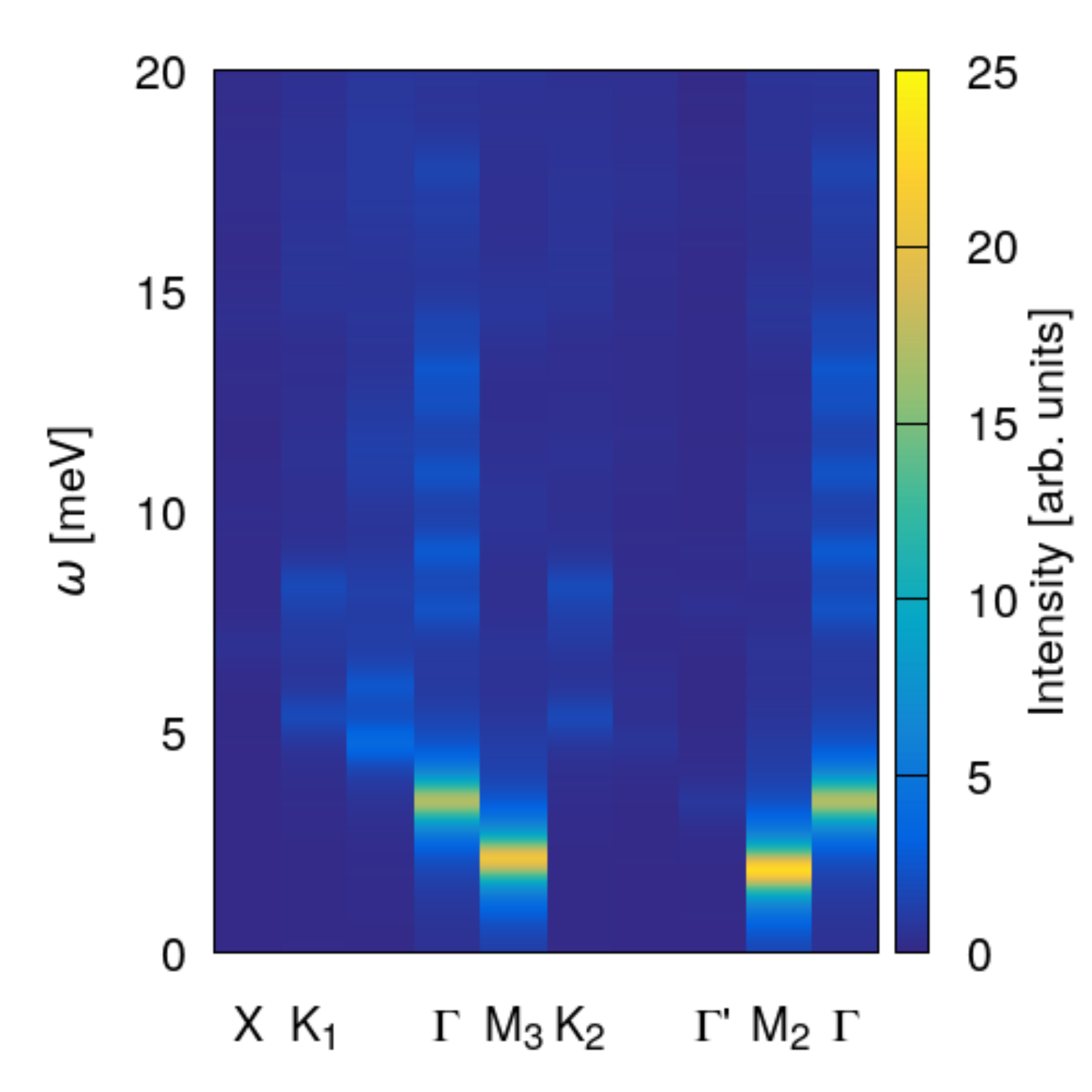}
		}%
		\subfloat[8. Ran et al.]{
			\includegraphics[height=5cm]{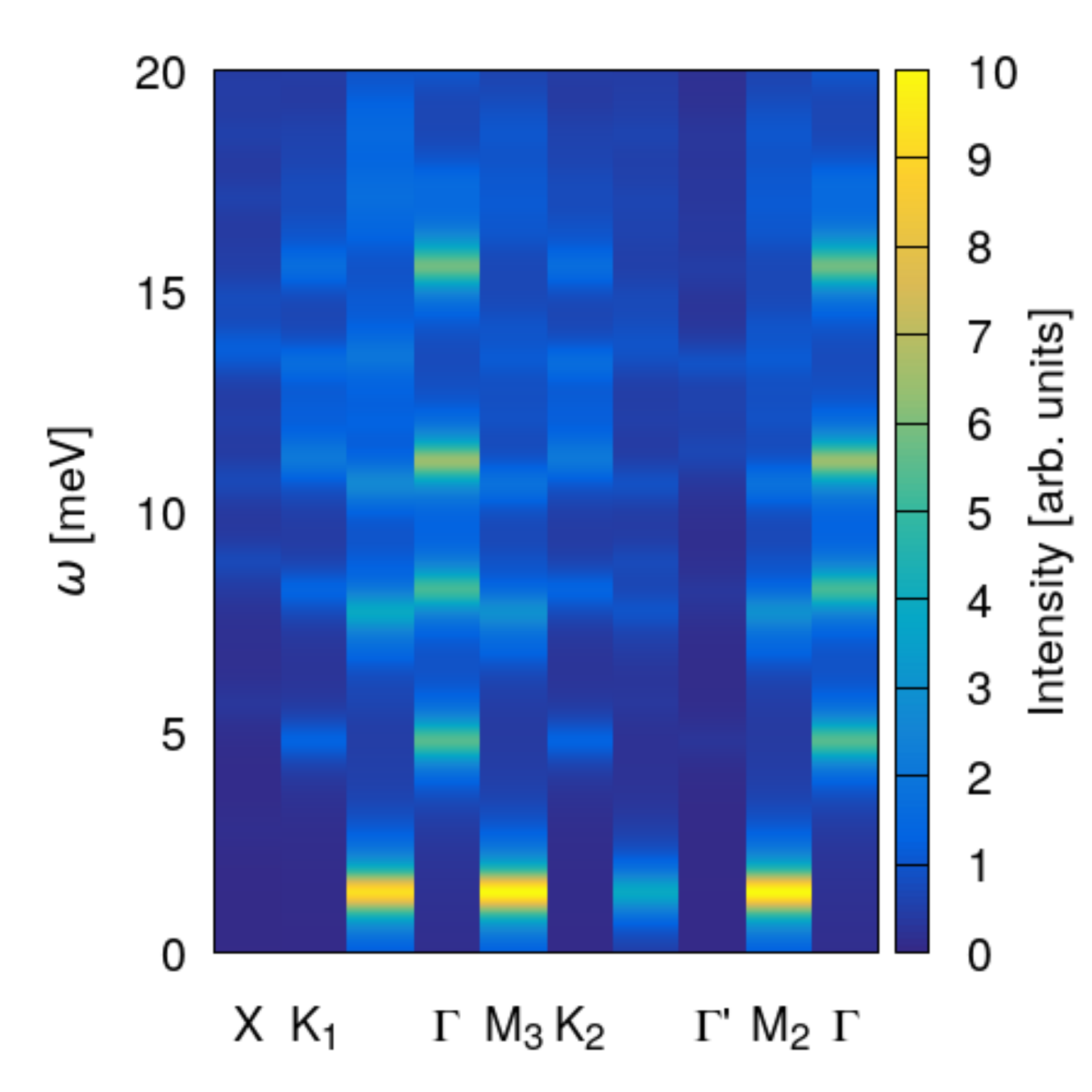}
		}\\
		\subfloat[9. Hou et al.]{
			\includegraphics[height=5cm]{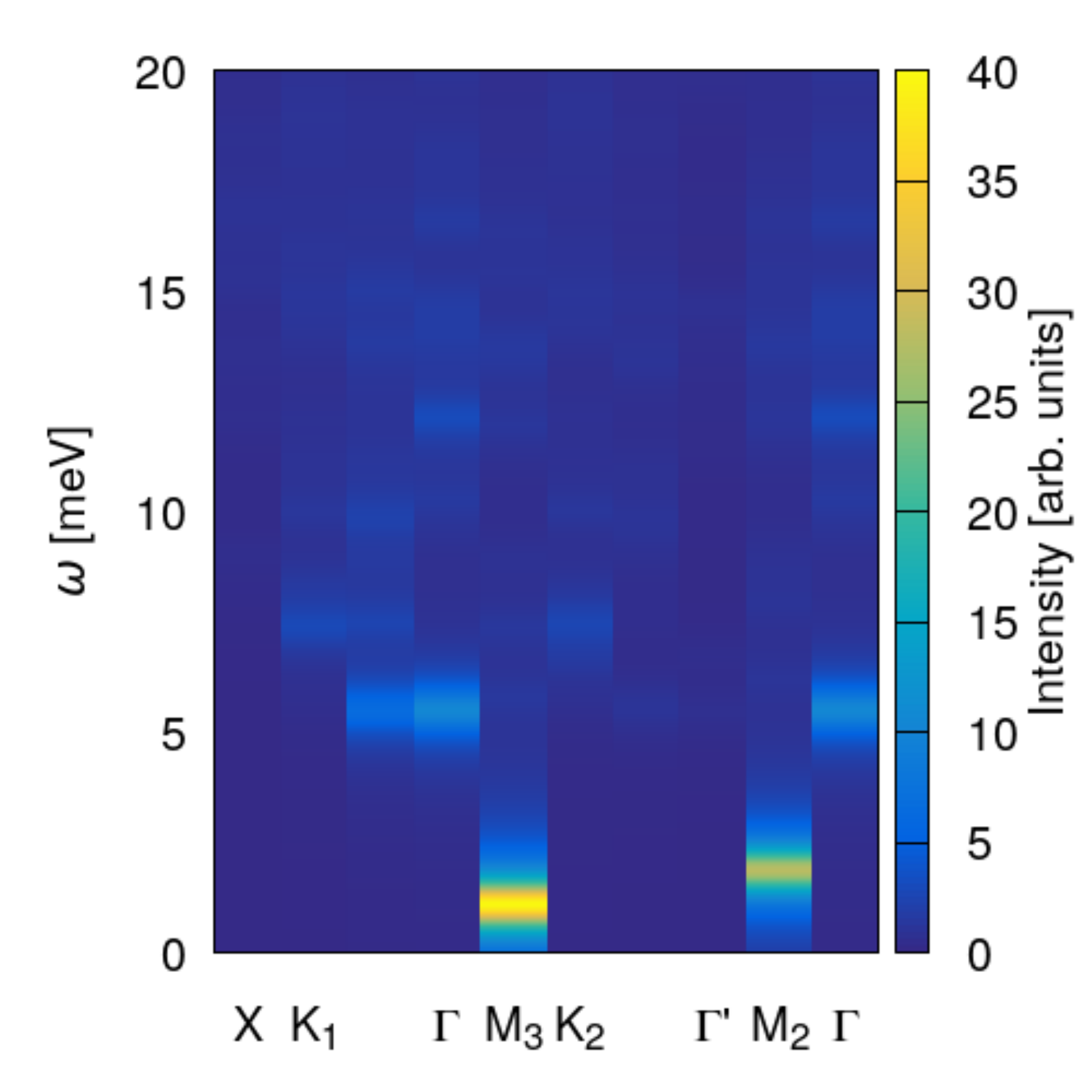}
		}%
		\subfloat[10. Wang et al.]{
			\includegraphics[height=5cm]{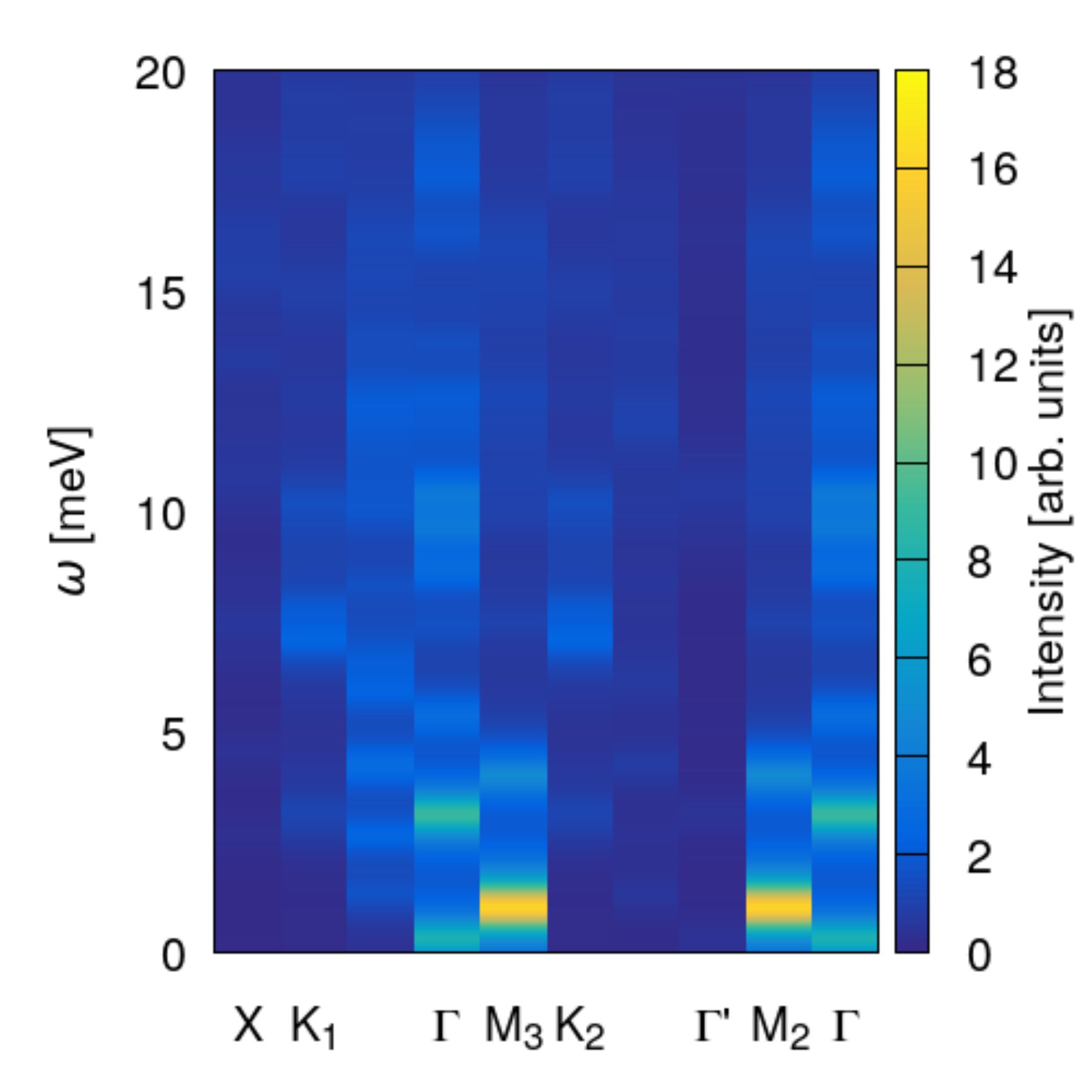}
		}
		\subfloat[14. Banerjee et al.]{
			\includegraphics[height=5cm]{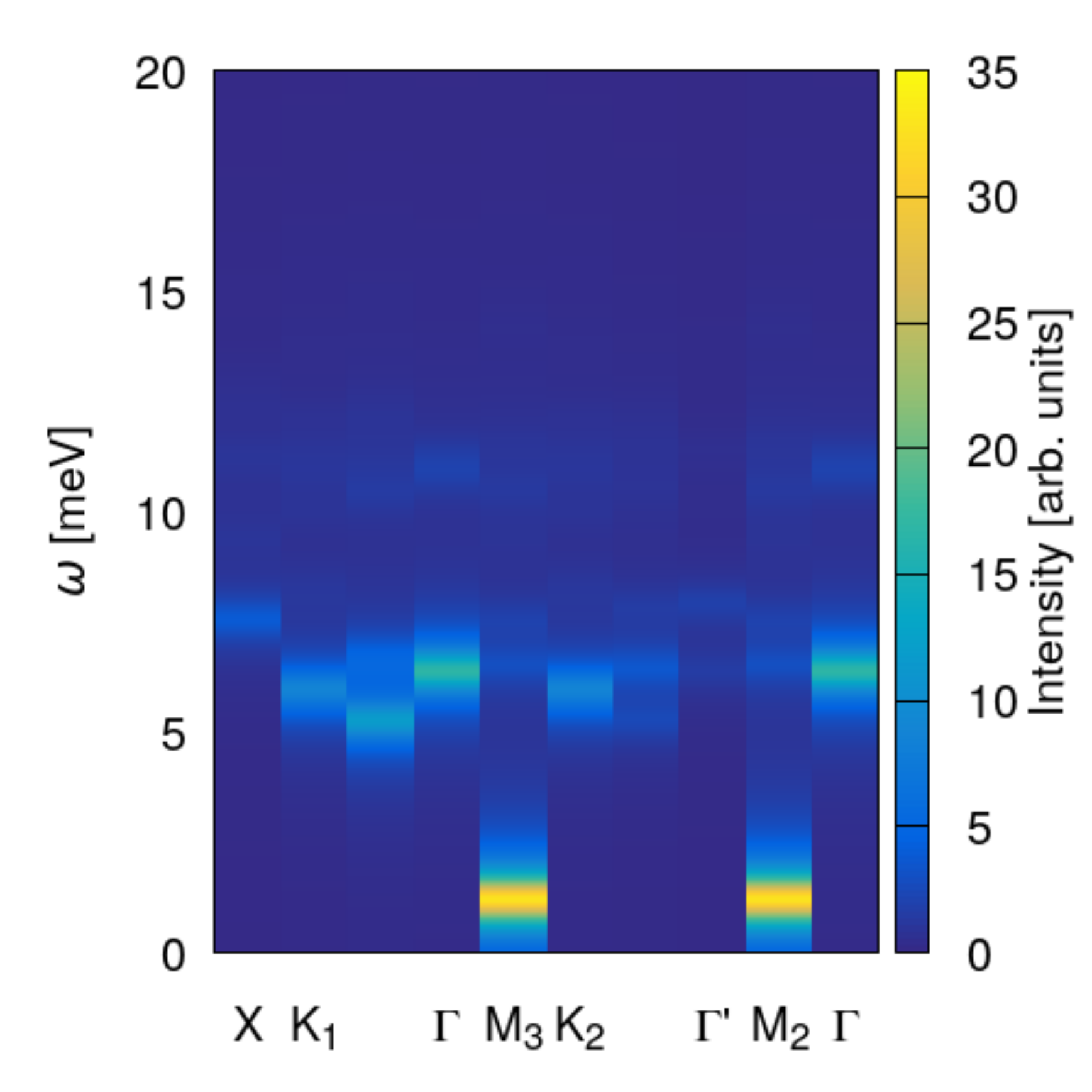}
		}\\
		\subfloat[15. Kim et al.]{
			\includegraphics[height=5cm]{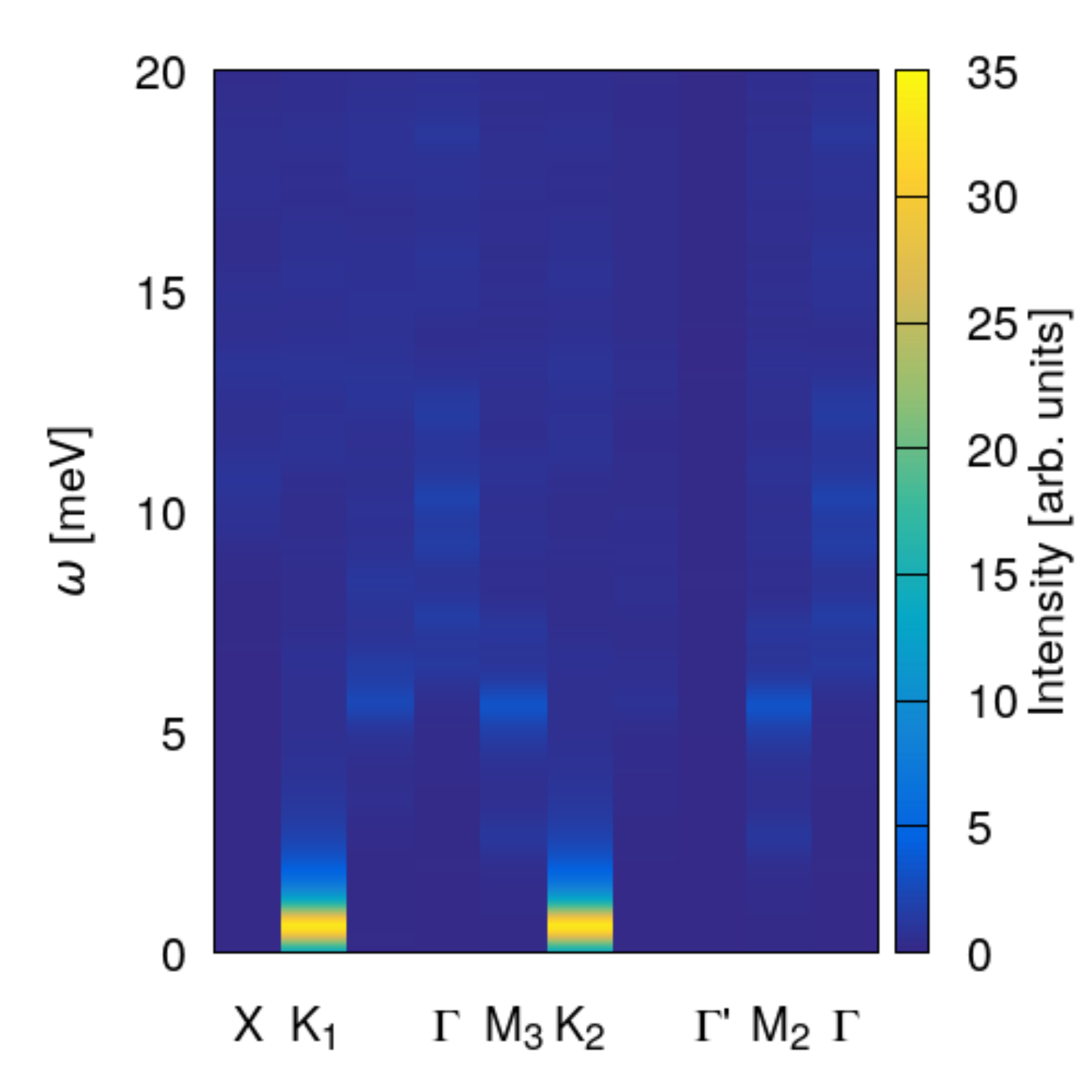}
		}%
		\subfloat[16. Kim and Kee (Case 0)]{
			\includegraphics[height=5cm]{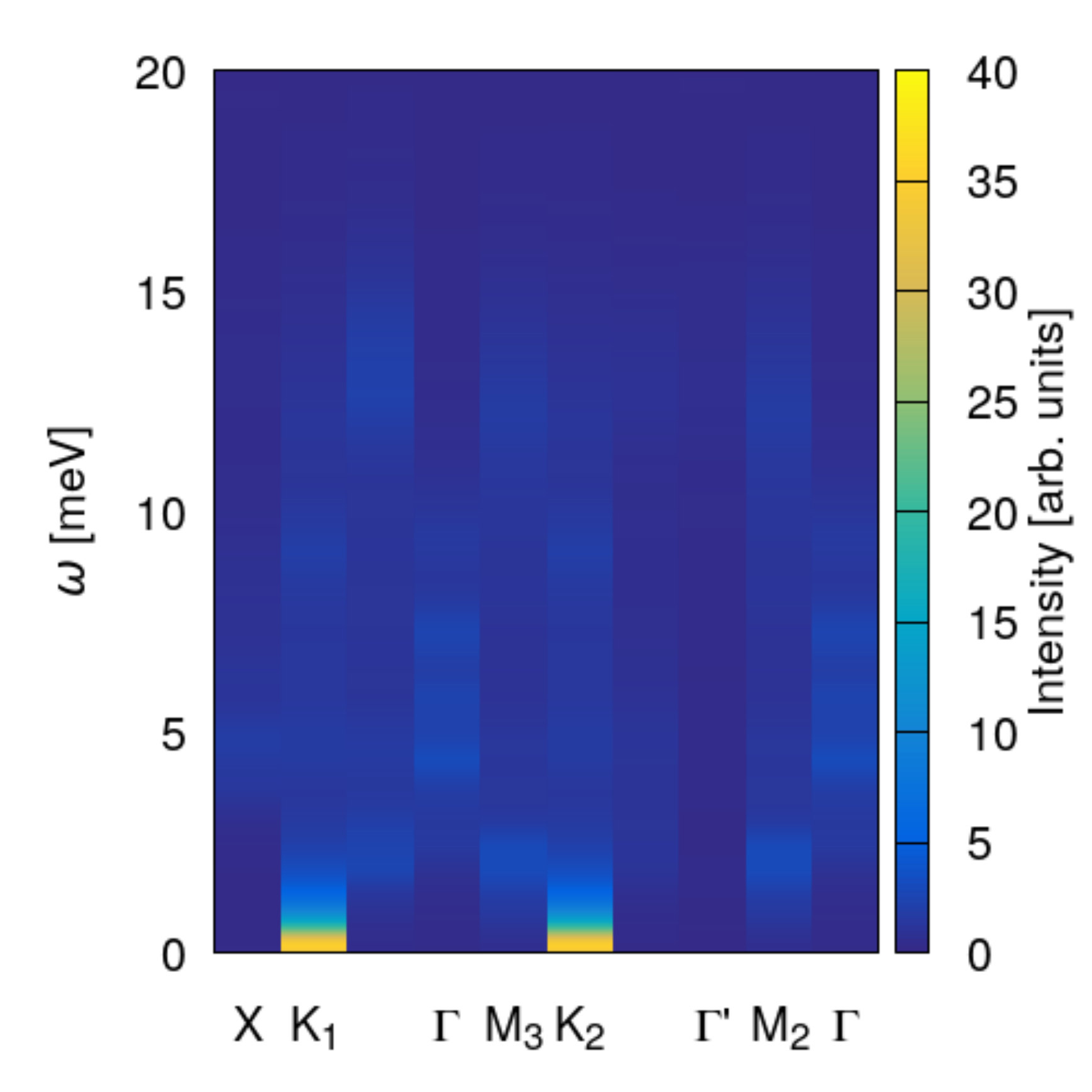}
		}
		\subfloat[17. Winter et al. PRB (P3)]{
			\includegraphics[height=5cm]{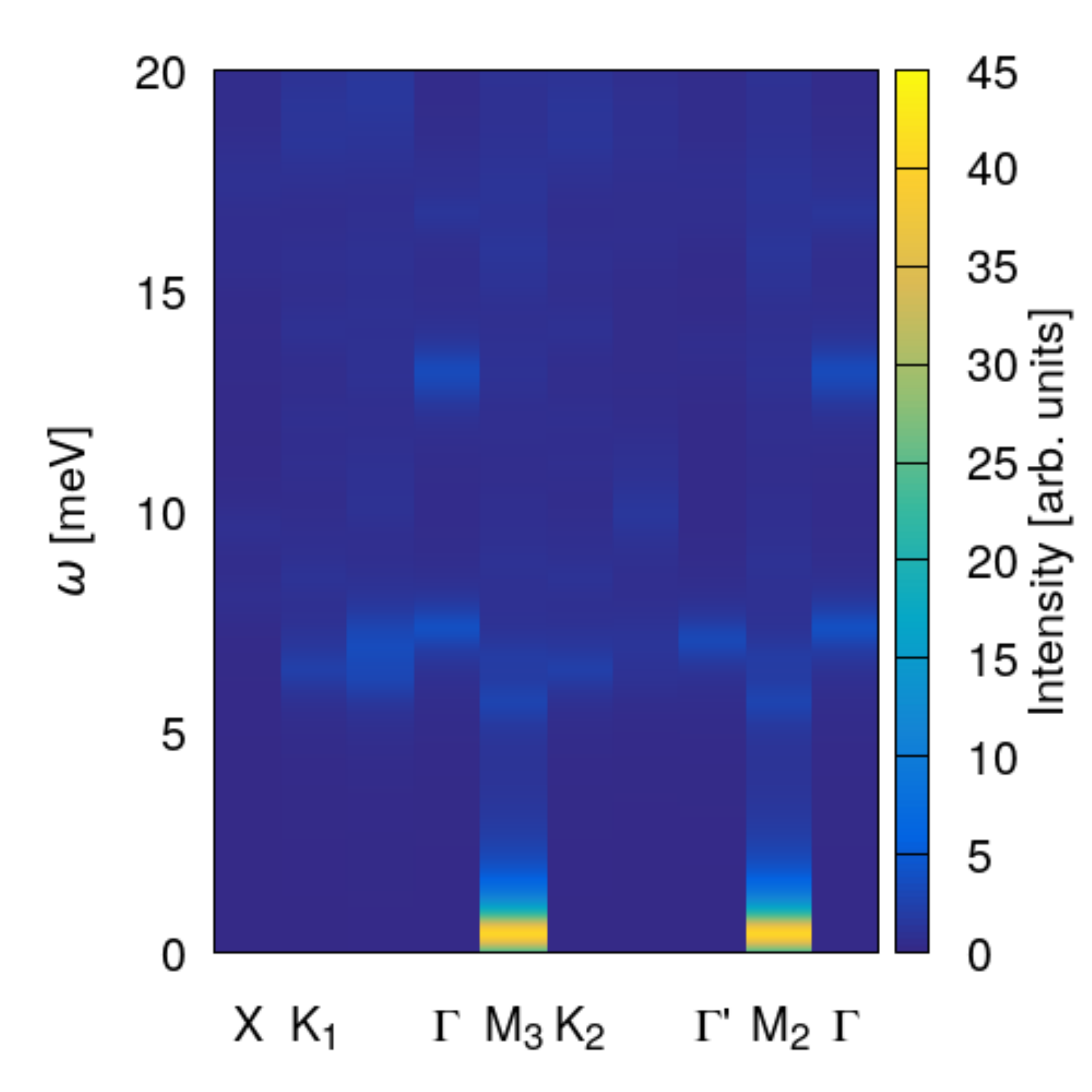}
		}
		\caption{\label{fig:supp:Iqomega1}Inelastic neutron scattering intensities $I\left( \mathbf{q}, \omega\right)$ calculated at zero temperature using $N=24$ sites for nine of the models not discussed in the main text.}
	\end{figure*}
	%
	\begin{figure*}[t]
		\centering
		\includegraphics[height=12.5cm]{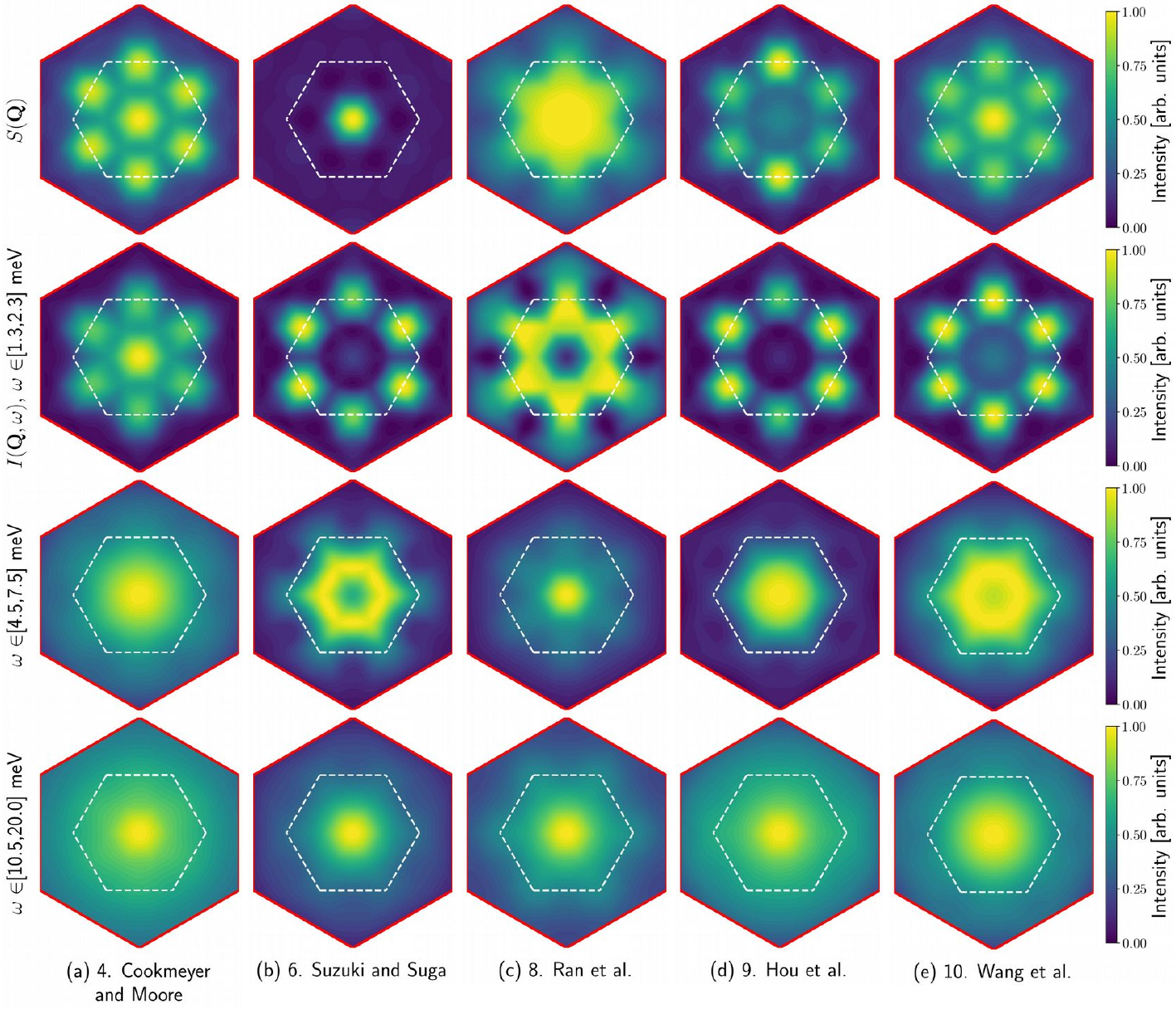}
		\caption{\label{fig:supp:Sqomega2}Static spin structure factors, $S\left( \mathbf{q}\right)$, and energy-integrated neutron scattering intensities, $I\left( \mathbf{q}, \omega\right)$, for five of the models not considered in the main text}
	\end{figure*}
	%
	\begin{figure*}[tbp]
		\centering
		\subfloat[4. Cookmeyer and Moore]{
			\includegraphics[width=.66\columnwidth]{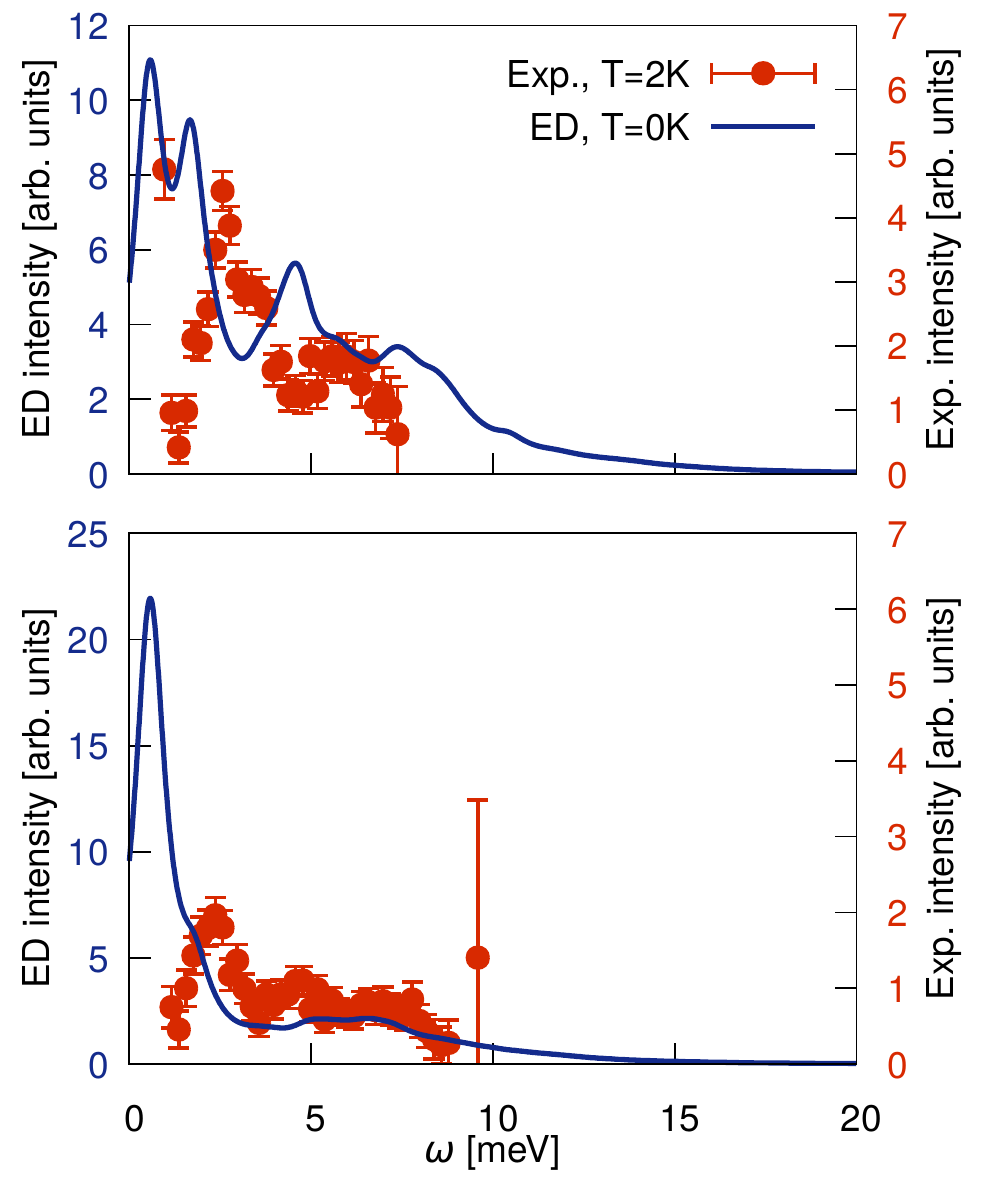}
		}\hspace{2cm}%
		\subfloat[6. Suzuki and Suga]{				
			\includegraphics[width=.66\columnwidth]{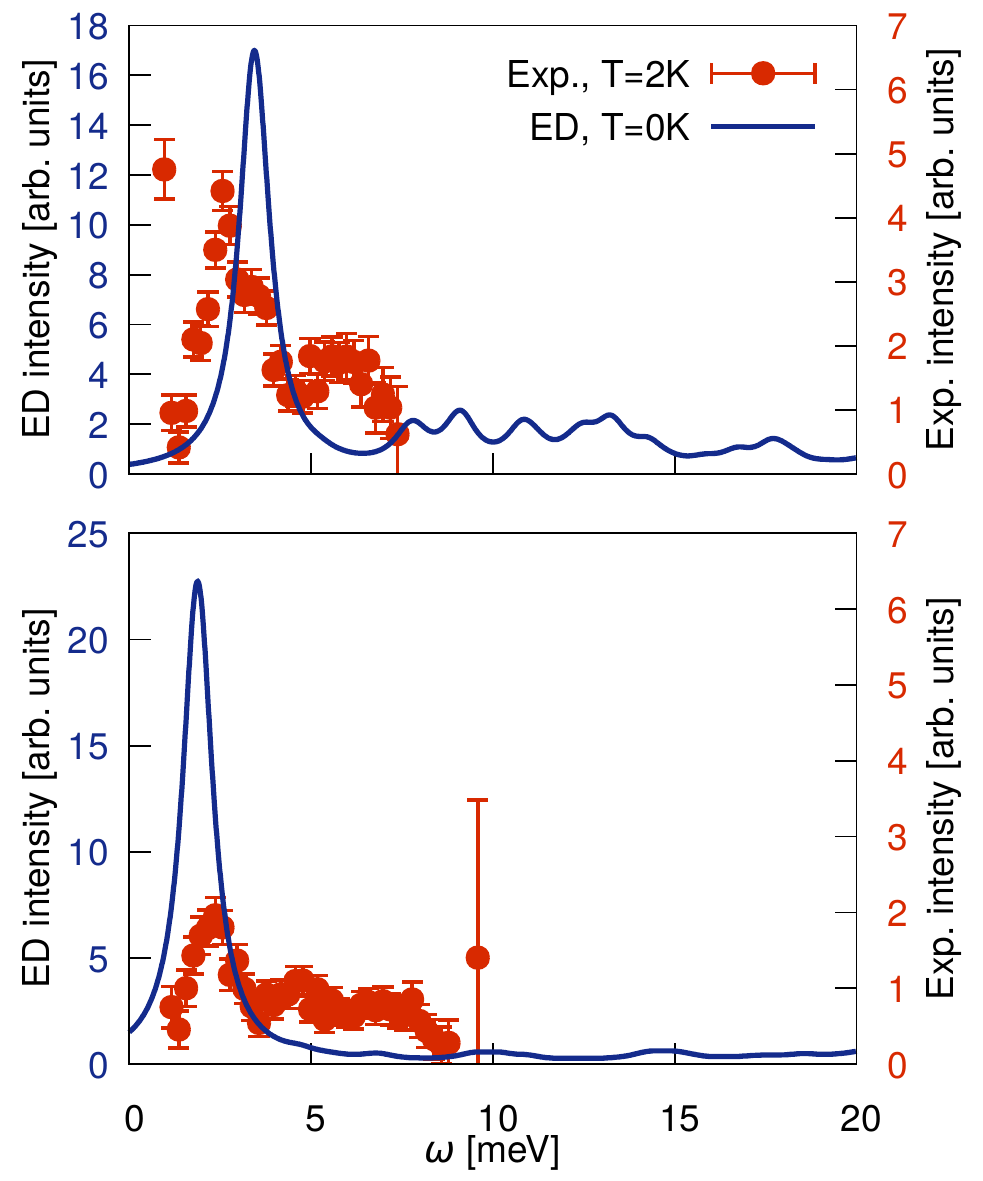}
		}\\\vspace{-0.305cm}
		\subfloat[8. Ran et al.]{				
			\includegraphics[width=.66\columnwidth]{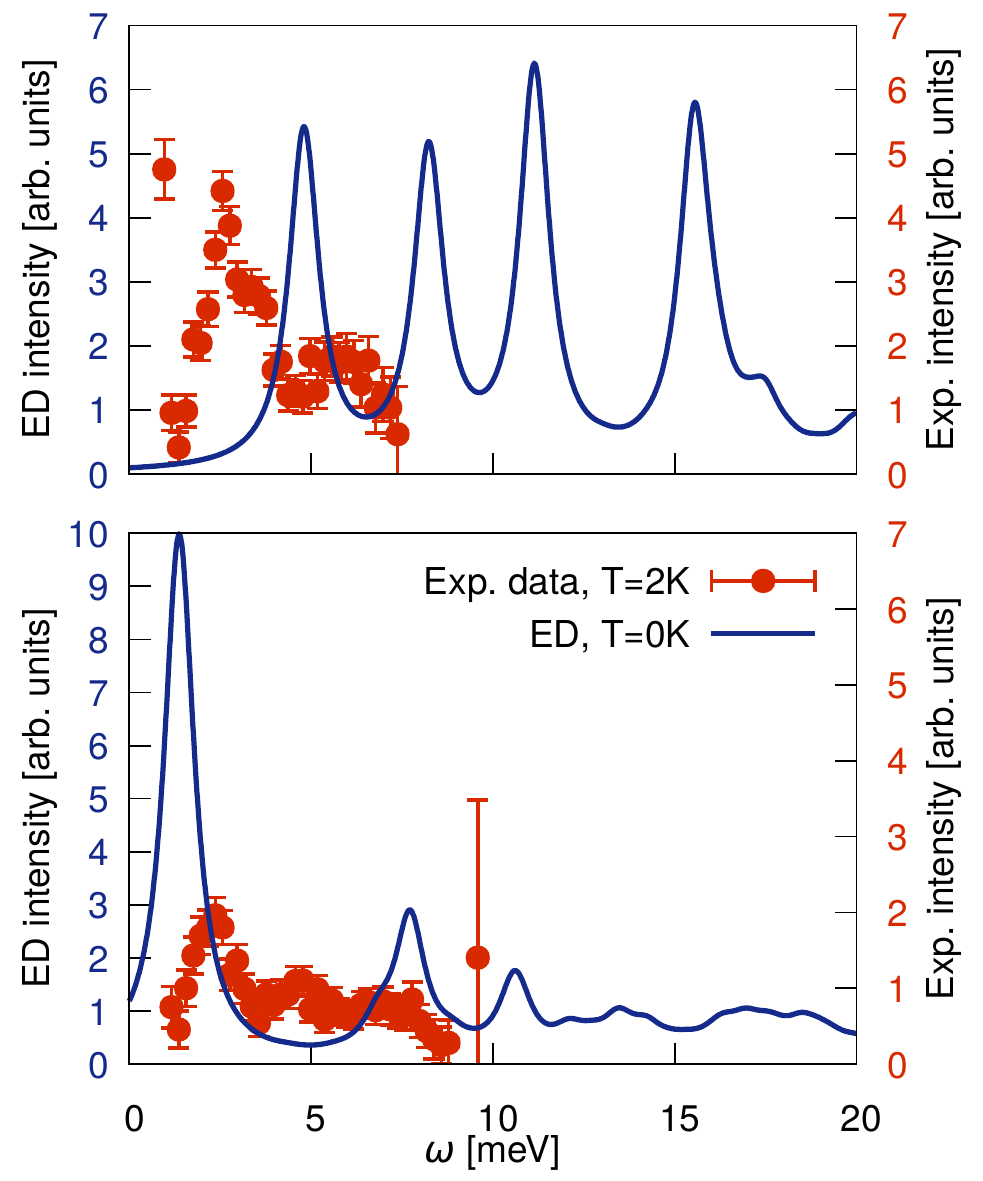}
		}\hspace{2cm}%
		\subfloat[9. Hou et al.]{
			\includegraphics[width=.66\columnwidth]{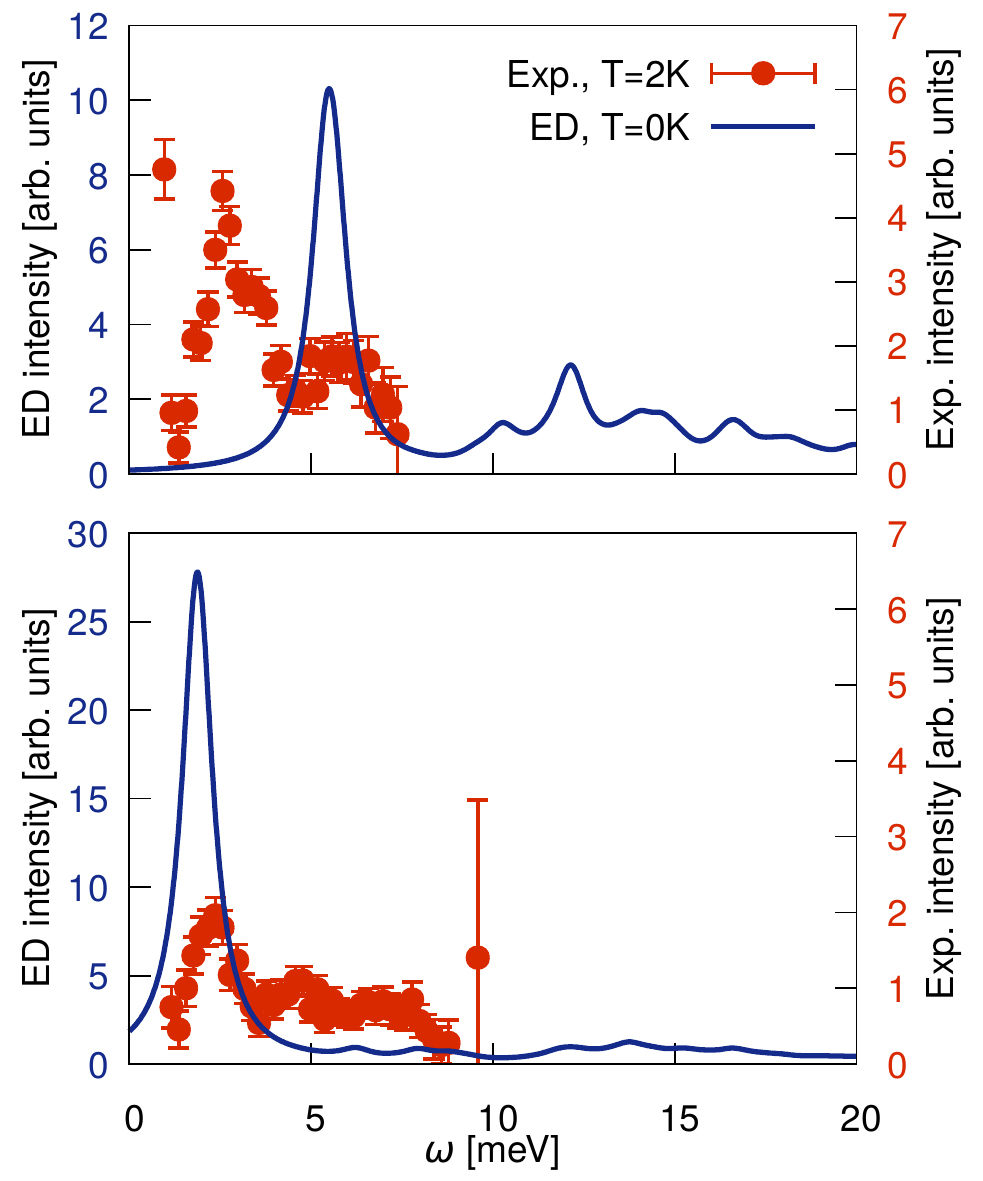}
		}\\\vspace{-0.305cm}
		\subfloat[10. Wang et al.]{
			\includegraphics[width=0.66\columnwidth]{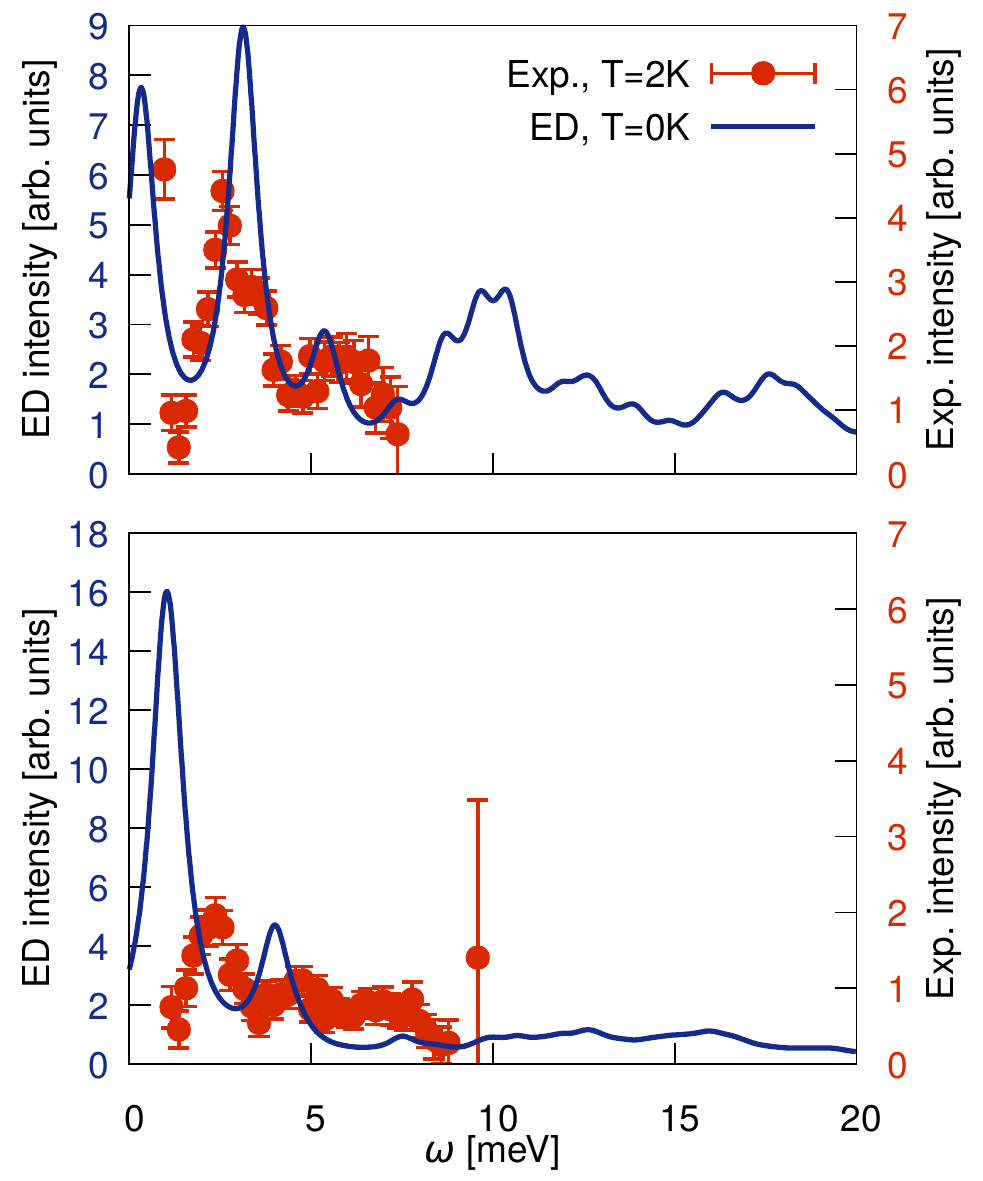}
		}
		\caption{\label{fig:supp:INSprofile2}$I\left( \mathbf{q}, \omega\right)$ at the $\Gamma$ (top panel) and M$_1$ points (bottom panel) for five of the models not considered in the main text. Experimental data from Ref.~\cite{Banerjee2018}, with error bars representing one standard deviation assuming Poisson counting statistics.
		}
	\end{figure*}
	%
	\begin{figure*}[t]
		\centering
		\includegraphics[height=12.5cm]{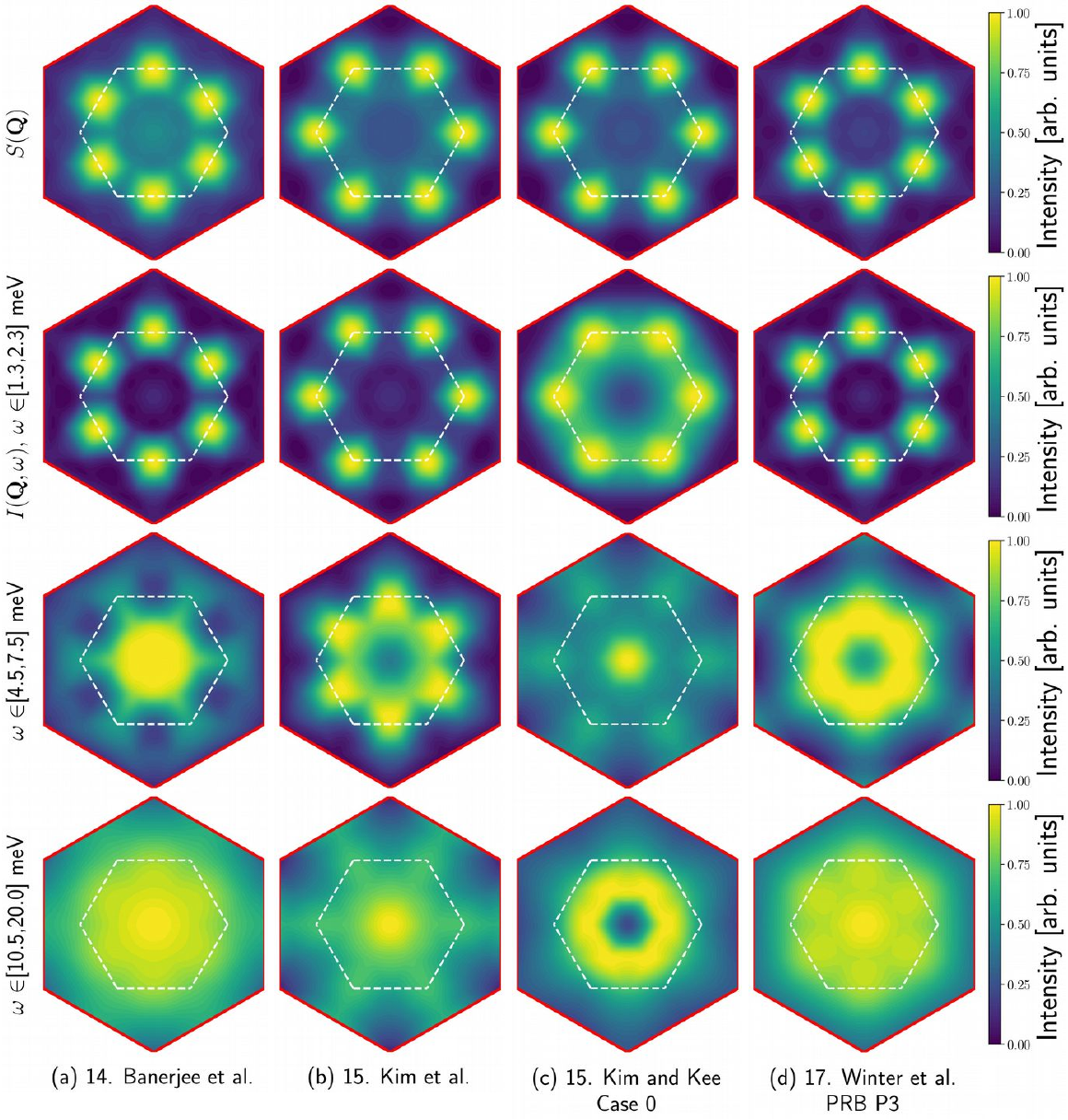}
		\caption{\label{fig:supp:SqomegaPositiveK}Static spin structure factors, $S\left( \mathbf{q}\right)$, and energy-integrated neutron scattering intensities, $I\left( \mathbf{q}, \omega\right)$, for the four models with antiferromagnetic Kitaev coupling.}
	\end{figure*}
	%
	\begin{figure*}[tbp]
		\centering
		\subfloat[14. Banerjee et al.]{
			\includegraphics[width=.70\columnwidth]{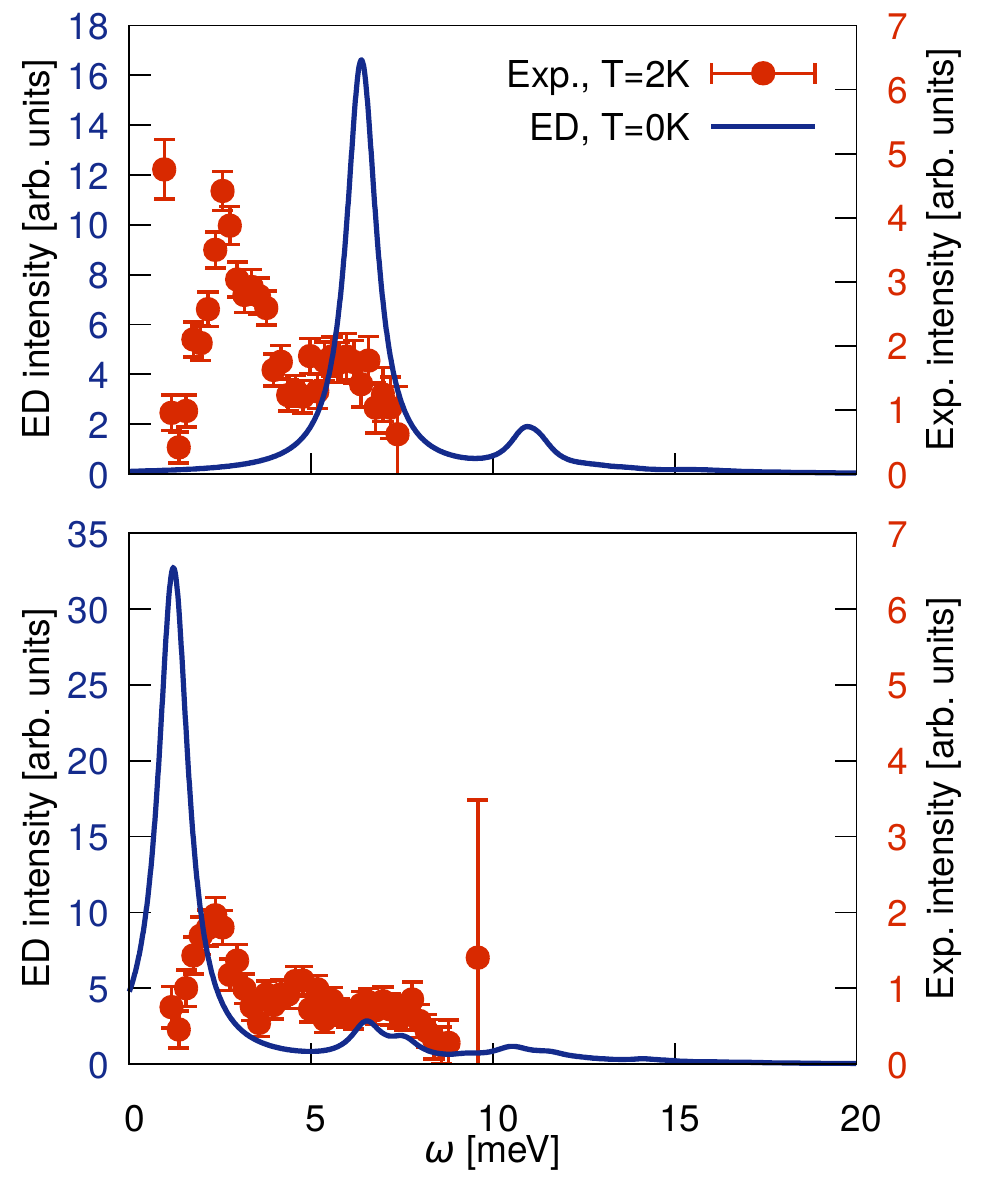}
		}\hspace{2cm}%
		\subfloat[15. Winter et al. P3 structure]{				
			\includegraphics[width=.7\columnwidth]{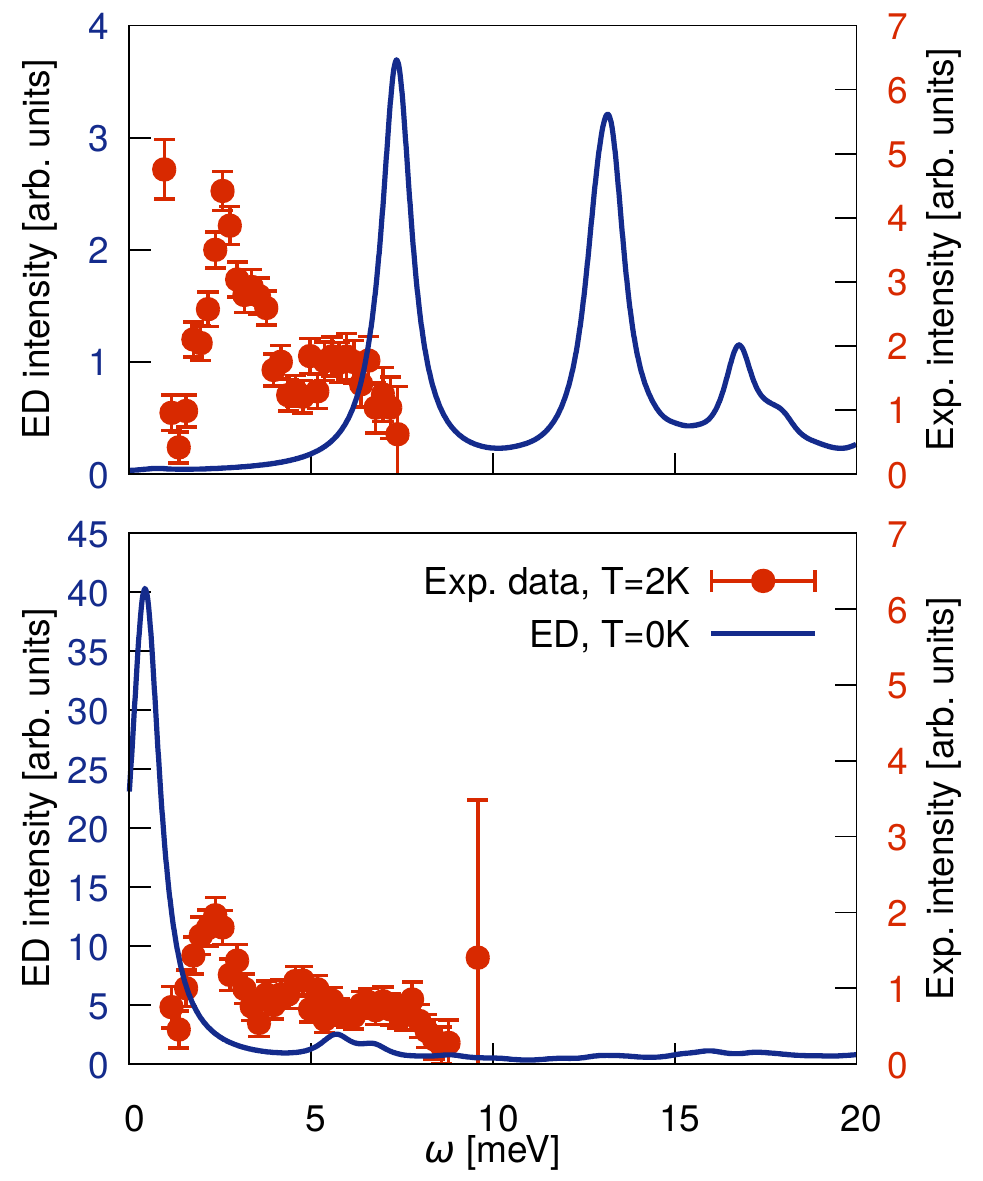}
		}\\
		\subfloat[16. Kim et al.]{				
			\includegraphics[width=.7\columnwidth]{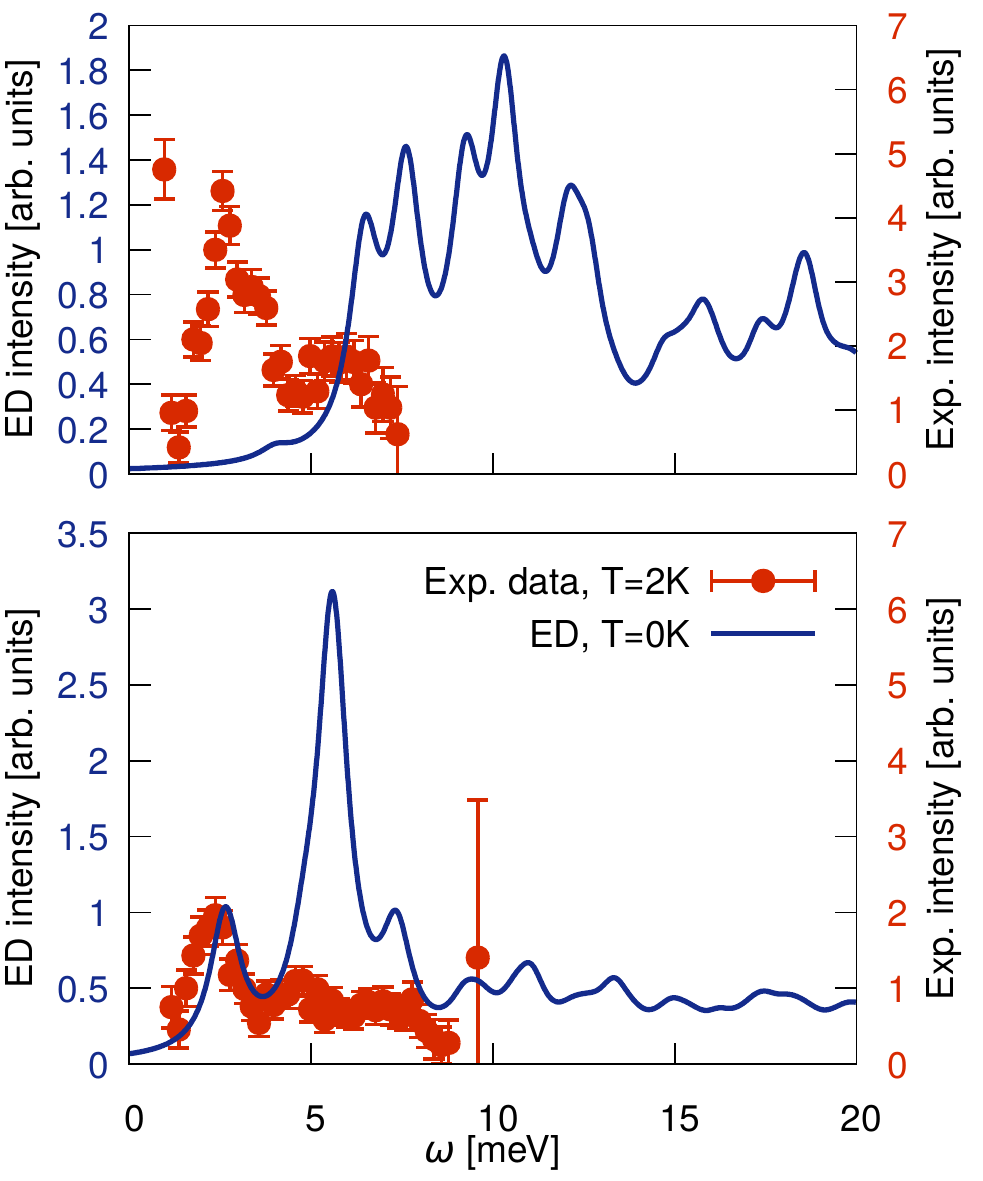}
		}\hspace{2cm}%
		\subfloat[17. Kim and Kee Case 0]{
			\includegraphics[width=.7\columnwidth]{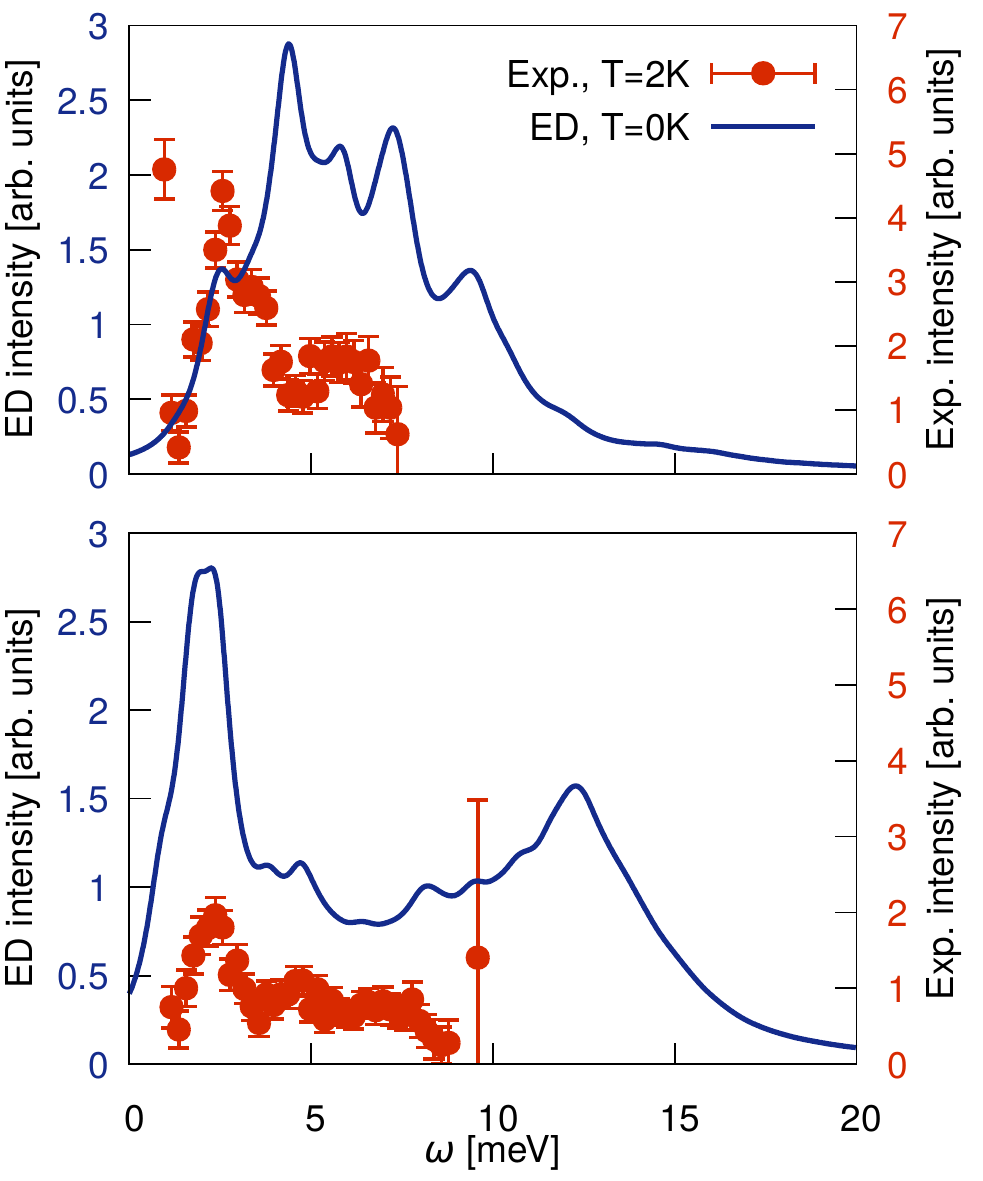}
		}
		\caption{\label{fig:supp:INSprofilePositiveK}$I\left( \mathbf{q}, \omega\right)$ at the $\Gamma$ (top panel) and M$_1$ points (bottom panel) for the four models with antiferromagnetic Kitaev coupling. Experimental data from Ref.~\cite{Banerjee2018}, with error bars representing one standard deviation assuming Poisson counting statistics.
		}
	\end{figure*}
	
	Finally we turn to two models obtained by Ozel et al. \cite{PhysRevB.100.085108} by linear spin wave fits to THz spectroscopy data. Their spectra and intensity profiles are shown in Supplementary Fig.~\ref{fig:supp:INSOzel}, and the static spin structure factors and energy slices of their spectra are shown in Supplementary Fig.~\ref{fig:supp:SqomegaOzel}. Since THz spectroscopy is a probe of the physics at the $\Gamma$ point, we may expect good agreement there. We do indeed find peaks that reasonably resemble the shape of the experimental INS data. The frequency shifts in the peak position may be explained by quantum renormalization. The M point intensities are, however, not captured. Despite prominent low-energy peaks at the M and K$_1/2$ points, respectively, the static spin structure factors both suggest ferromagnetic phases.
	%
	\begin{figure*}
		\centering
		\subfloat[18. Ozel et al. PRB, $K_1>0$ model]{
			\raisebox{1.25cm}{\includegraphics[height=5cm]{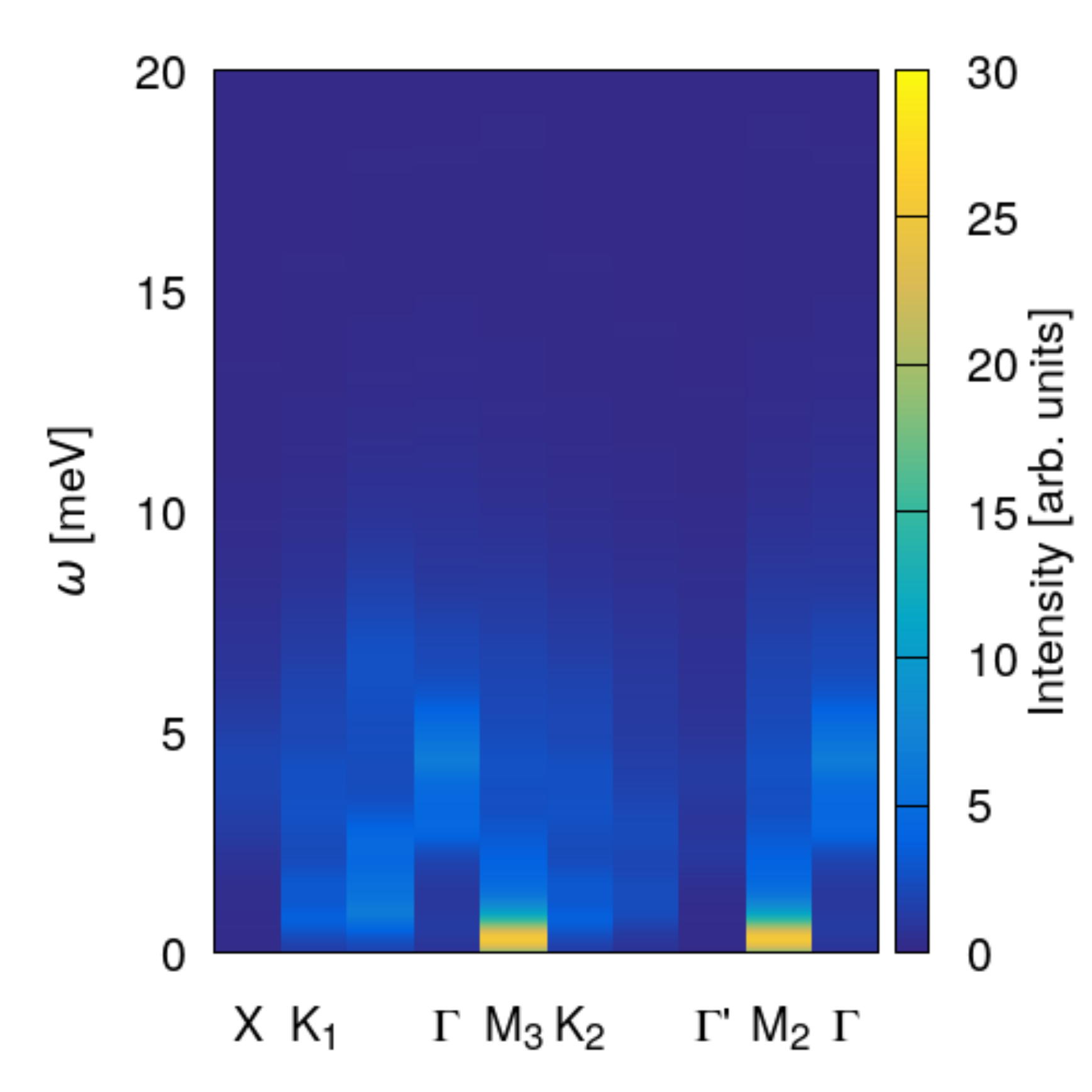}}\hspace{1cm}
			\includegraphics[width=.7\columnwidth]{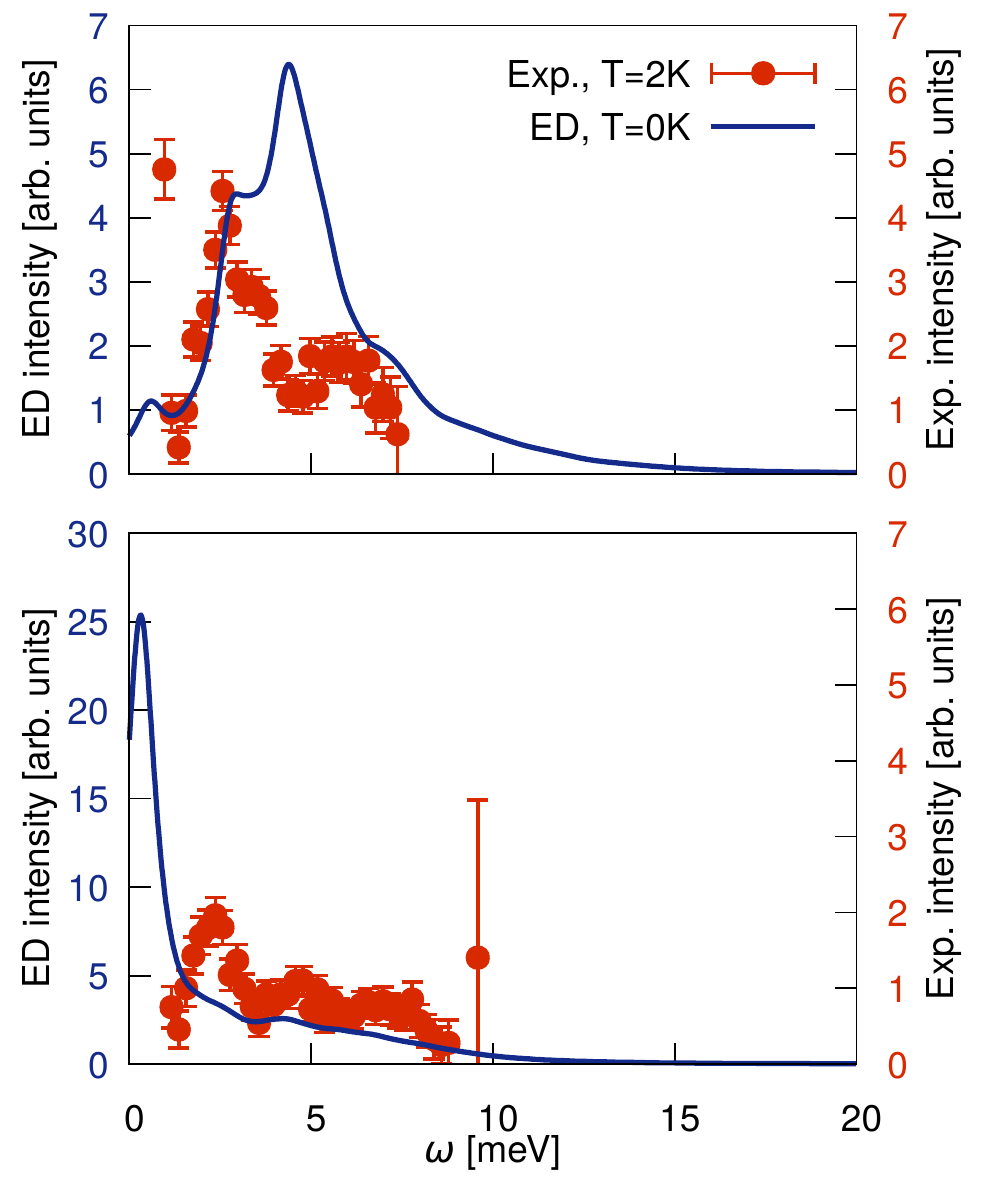}
		}\\
		\subfloat[19. Ozel et al. PRB, $K_1<0$ model]{
			\raisebox{1.25cm}{\includegraphics[height=5cm]{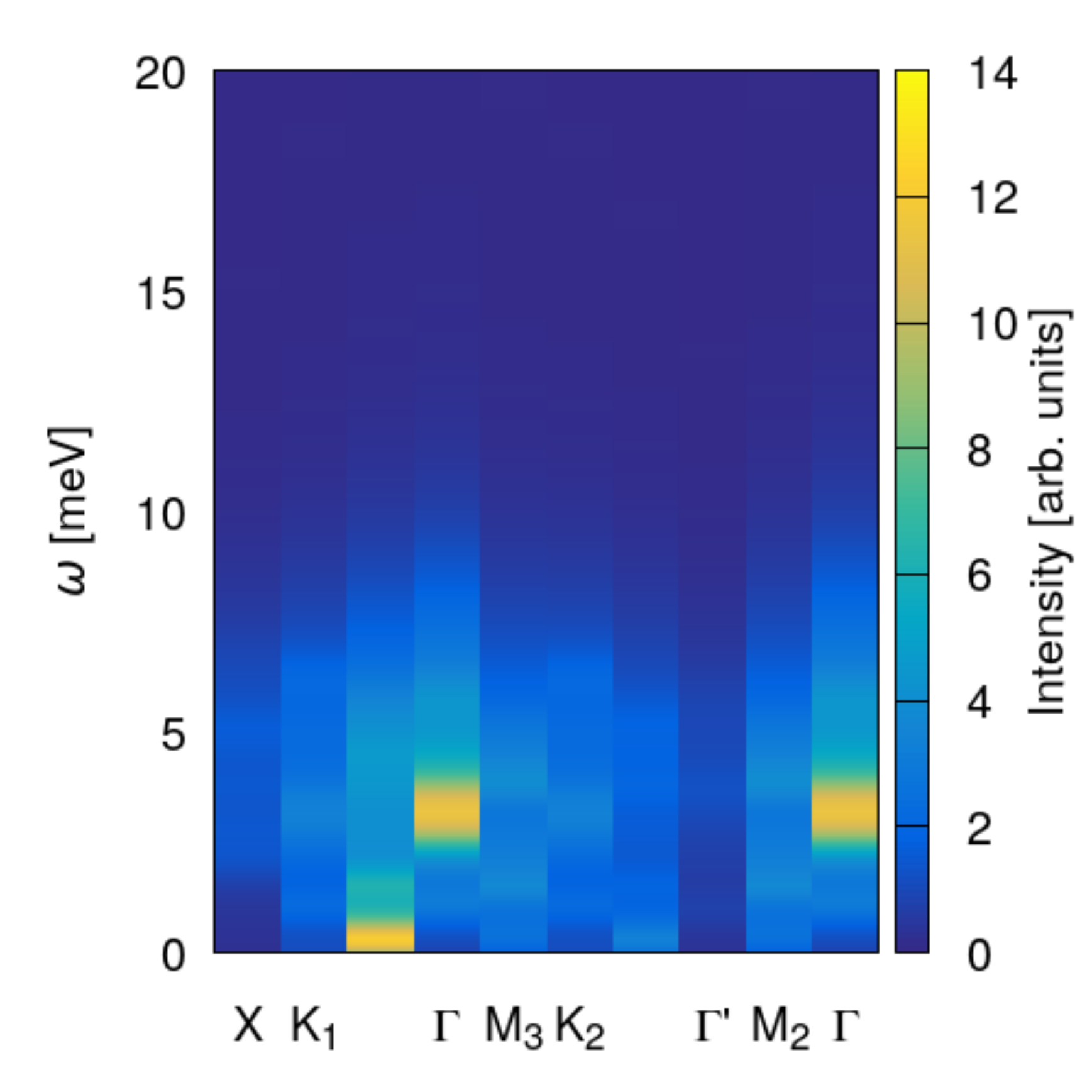}}\hspace{1cm}
			\includegraphics[width=.7\columnwidth]{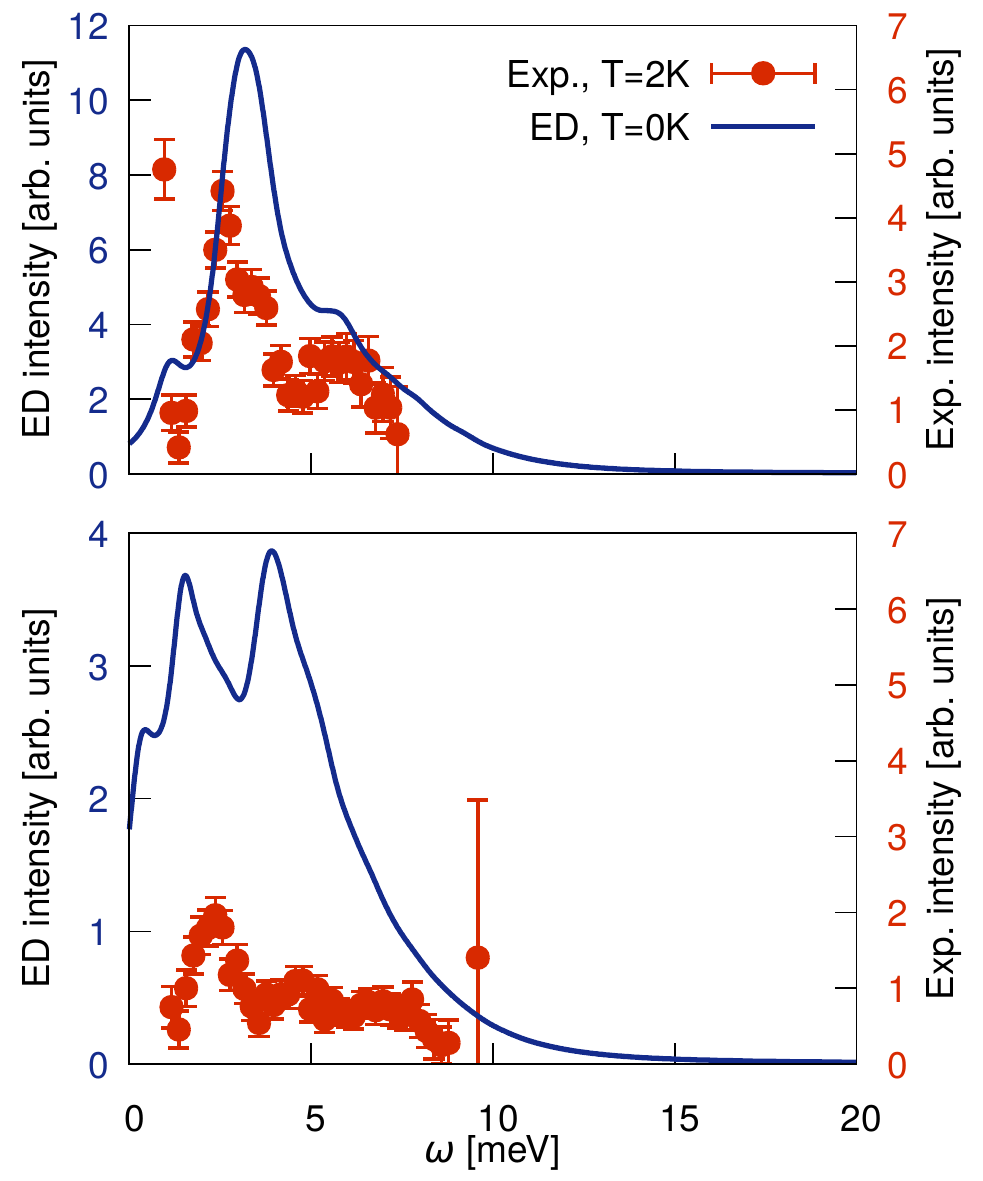}
		}    
		\caption{\label{fig:supp:INSOzel}{Inelastic neutron scattering intensities $I\left( \mathbf{q}, \omega\right)$ calculated at zero temperature using $N=24$ sites for the two models of Ozel et al.} The experimental data is from Ref.~\cite{Banerjee2018}, with error bars representing one standard deviation assuming Poisson counting statistics.}
	\end{figure*}
	\begin{figure}
		\centering
		\subfloat{
			\includegraphics[height=12cm]{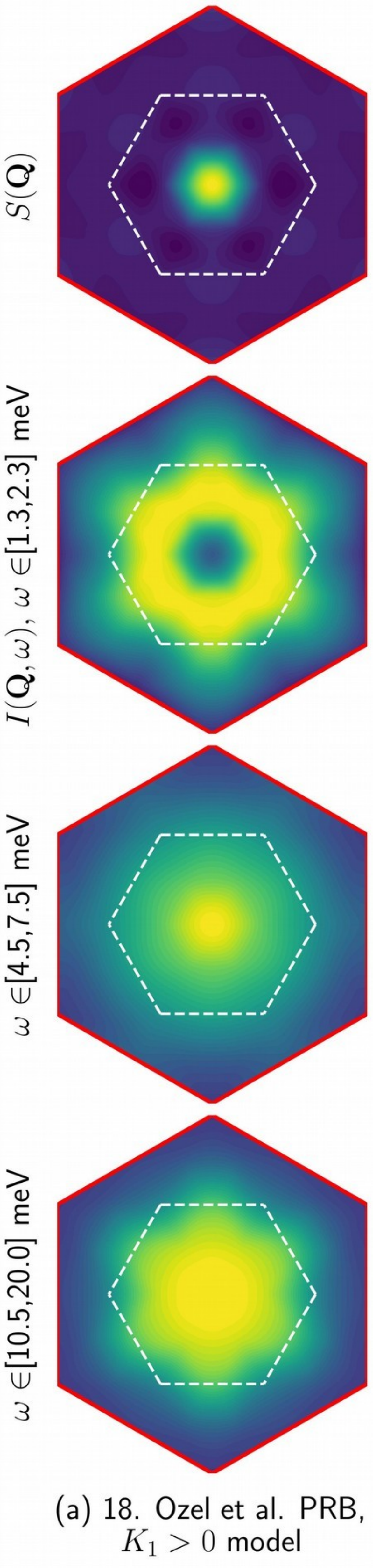}
		}%
		\subfloat{
			\includegraphics[height=12cm]{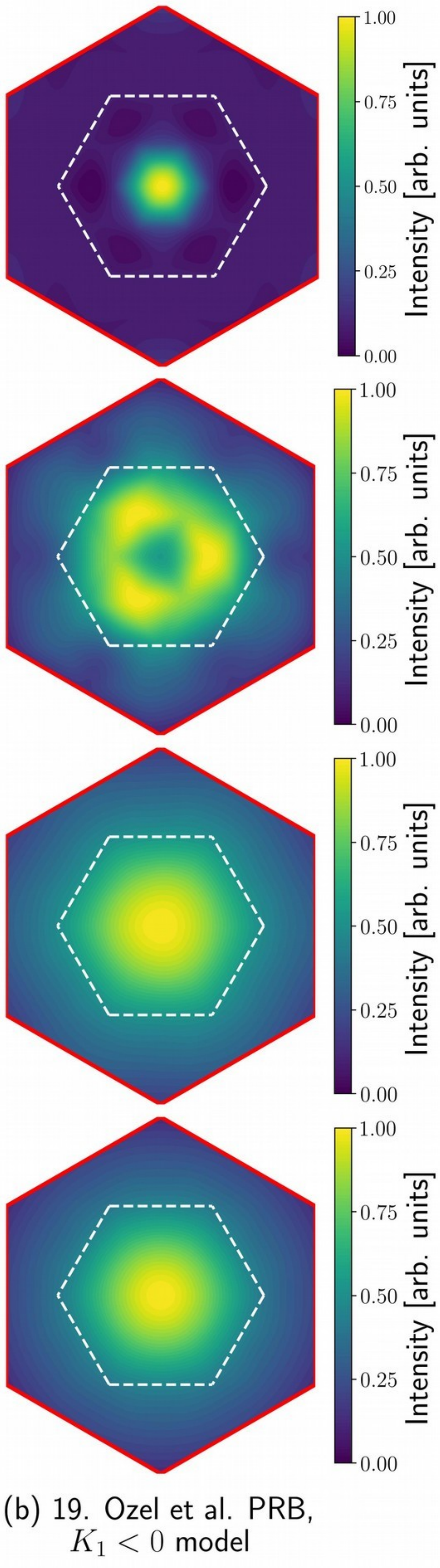}
		}
		\caption{\label{fig:supp:SqomegaOzel}{Static spin structure factors, $S\left( \mathbf{q}\right)$, and energy-integrated neutron scattering intensities, $I\left( \mathbf{q}, \omega\right)$, for the two models due to Ozel et al.}}
	\end{figure}
\section{Supplementary Note 4: Comparing the updated and original Eichstaedt et al. parameters}

	As mentioned in Table~I of the main text, the Eichstaedt et al. interaction parameters were revised in the published version \cite{PhysRevB.100.075110}. Above and in the main text (except for the case of the modified ab initio model) we have used the preliminary values from the preprint version \cite{Eichstaedt2019}. As Supplementary Table~\ref{table:Eichstaedt} shows these are relatively small changes. In this section we compare results for the old and updated parameters, showing that the results do not crucially depend on this modification. The magnetic specific heat curves are shown in Supplementary Fig.~\ref{fig:supp:Crev_Eichstaedt}, showing minimal differences, which is reflected in the positions of the high-temperature peak. The revised (original) full model has $T_\mathrm{h}\approx 68.6$ K ($T_\mathrm{h}\approx 66.4$ K). Similarly the revised (original) parameters for the cases neglecting non-local SOC and Coulomb interactions have $T_\mathrm{h}\approx 63.1$ ($T_\mathrm{h}\approx 62.9$) and $T_\mathrm{h}\approx 22.9$ K ($T_\mathrm{h}\approx 22.4$ K), respectively.
	
	We also find visually near-identical INS spectra. We plot the intensity profiles at the $\Gamma$ and M$_1$ points in Supplementary Fig.~\ref{fig:supp:INSprofile_Echstaedt_Revised}, from which it is clear that the strong low-energy features in the spectrum are nearly unchanged. We thus conclude that there is no qualitative difference between the revised and original interaction parameters.
	\begin{table*}[bt]
		\caption{\label{table:Eichstaedt}{The original Eichstaedt et al. parameters from Ref.~\cite{Eichstaedt2019} considered in the main text (in bold), compared with the revised, published parameters from Ref.~\cite{PhysRevB.100.075110}. The differences are only minor. The last row shows the parameters for the modified ab initio model, which is based on the revised, fully ab initio model. Dashes (--) indicate that the value is unavailable or negligible. Asterisks in the `BA' column signify that the full Hamiltonian has different values for the X/Y bonds compared to the Z bonds, and that the parameter values given in the row have been bond averaged.
		}}
		\begin{ruledtabular}
			\begin{tabular}{llllllllllll}
				& Reference           & Method   & $J_1$ & $K_1$ & $\Gamma_1$ & $\Gamma_1'$    & $J_2$                         & $K_2$   & $J_3$ & $K_3$ & BA       \\\hline                  
				\textbf{11} & \textbf{Eichstaedt et al.} \cite{Eichstaedt2019} 
				& Fully ab initio (DFT + cRPA + $t/U$) & $-1.4$       & $-14.3$      & $+9.8$           & $-2.23$  & --  & $-0.63$    & $+1.0$      & $+0.03$ & $^\star$ \\
				& Eichstaedt et al. \cite{PhysRevB.100.075110} 
				& Fully ab initio (DFT + cRPA + $t/U$) & $-1.3$       & $-15.1$      & $+10.1$           & $-2.35$  & --  & $-0.68$    & $+0.9$      & $+0.13$ & $^\star$ \\\hline
				\textbf{12} & \textbf{Eichstaedt et al.} \cite{Eichstaedt2019}  & Neglecting non-local Coulomb         & $-0.2$       & $-4.5$      & $+3.0$           &  $-0.73$ & -- & $-0.33$ & $+0.7$     & $+0.1$ &  $^\star$ \\
				& Eichstaedt et al. \cite{PhysRevB.100.075110}  & Neglecting non-local Coulomb         & $-0.2$       & $-4.8$      & $+3.1$           &  $-0.75$ & -- & $-0.38$ & $+0.7$     & $+0.1$ &  $^\star$ \\\hline
				\textbf{13} & \textbf{Eichstaedt et al.} \cite{Eichstaedt2019}  & Neglecting non-local SOC             & $-1.3$      & $-13.3$      & $9.4$ & $-2.3$ & -- & $-0.67$ & $+1.0$      & $+0.1$& $^\star$\\
				& Eichstaedt et al. \cite{PhysRevB.100.075110}  & Neglecting non-local SOC             & $-1.3$      & $-13.3$      & $9.4$ & $-2.3$ & -- & $-0.70$ & $+1.0$      & $+0.1$& $^\star$\\\hline
				& Modified ab initio model
				&Reducing $\Gamma'_1$ (this work) & $-1.3$       & $-15.1$      & $+10.1$           & $-0.1175$  & --  & $-0.68$    & $+0.9$      & $+0.13$ &  \\\hline
			\end{tabular}
		\end{ruledtabular}
	\end{table*}
	%
	\begin{figure}
		\centering
		\includegraphics[width=\columnwidth]{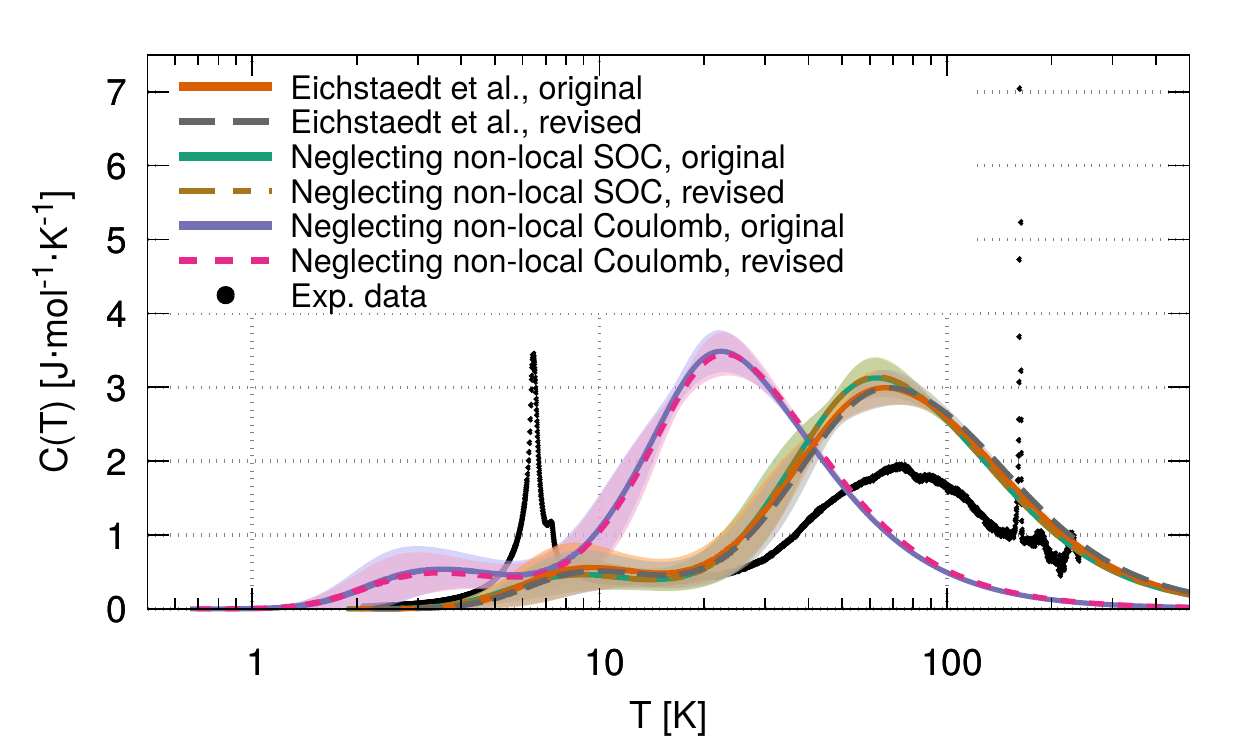}
		\caption{\label{fig:supp:Crev_Eichstaedt}Magnetic specific heat calculated using the TPQ method for original \cite{Eichstaedt2019} and revised \cite{PhysRevB.100.075110} models 11, 12, and 13. The solid lines show the calculated average value over $15$ initial vectors, and the shaded areas show the standard deviation.}
	\end{figure}
	\begin{figure}
		\centering
		\subfloat[Full ab initio model]{
			\includegraphics[width=.625\columnwidth]{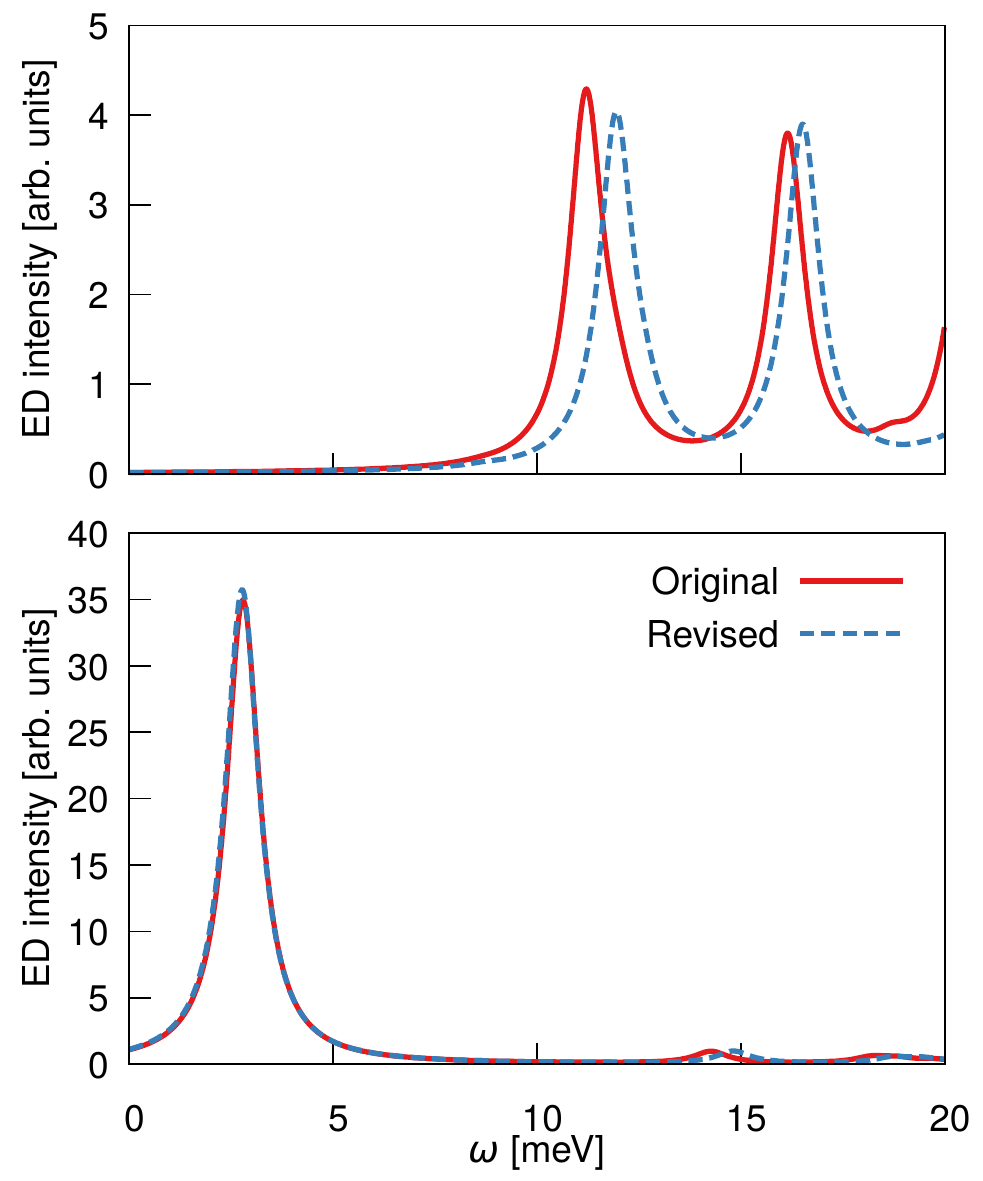}
		}\\
		\subfloat[Neglecting nonlocal Coulomb]{				
			\includegraphics[width=.625\columnwidth]{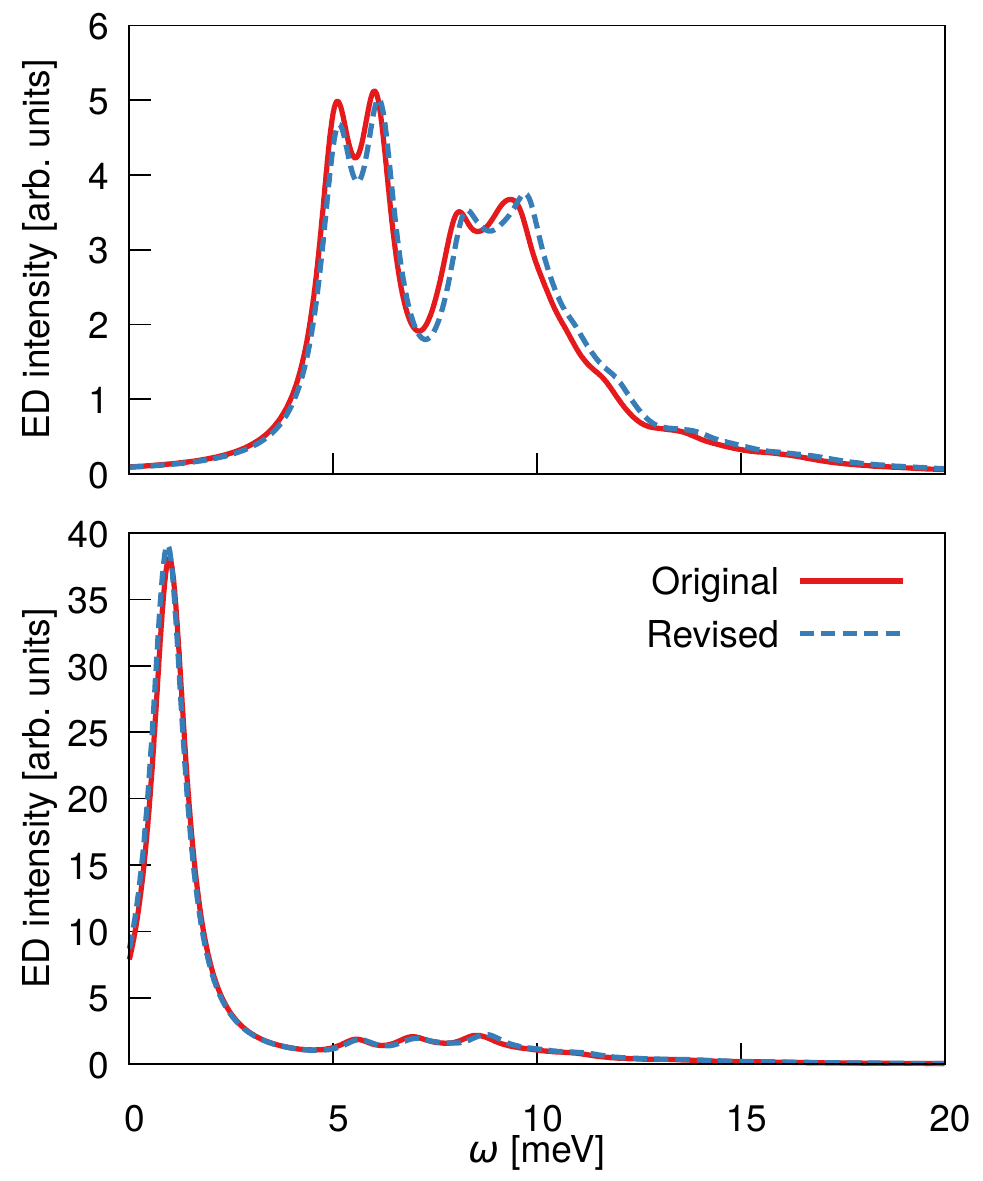}
		}\\
		\subfloat[Neglecting nonlocal SOC]{				
			\includegraphics[width=.625\columnwidth]{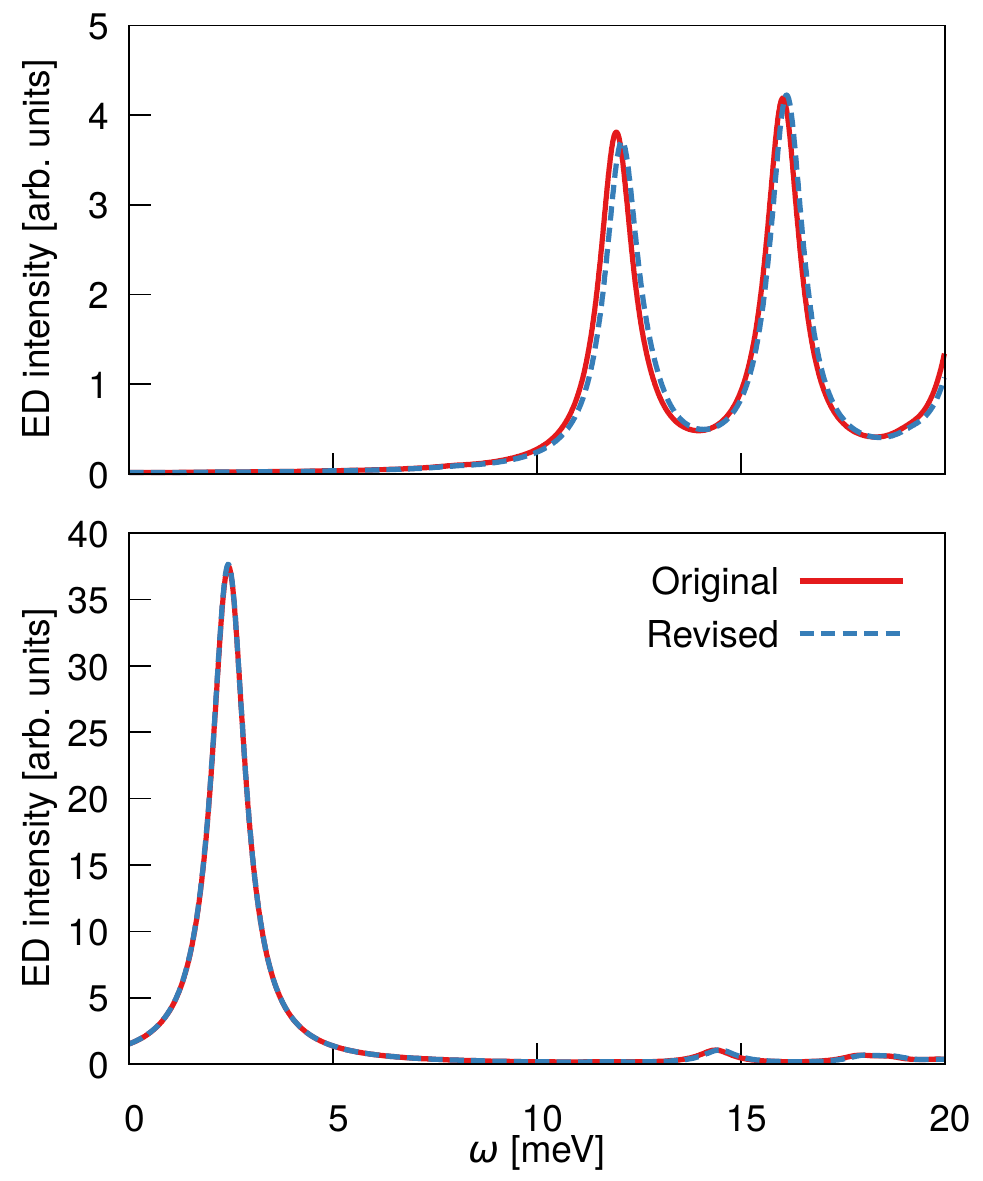}
		}
		\caption{\label{fig:supp:INSprofile_Echstaedt_Revised}{$I\left( \mathbf{q}, \omega\right)$ at the $\Gamma$ (top panel) and M$_1$ points (bottom panel) for the original and revised Eichstaedt et al. parameters.
			}
		}
	\end{figure}%
\section{Supplementary Note 5: Evolution of INS spectra away from the Kitaev limit, and modified ab initio model}
\begin{figure*}
		\begin{tabular}{llllll} 
			& $\quad\quad\Gamma_1/K_1=0.00$                                                                                  & $\quad\quad\Gamma_1/K_1=-0.25$                                                                                 & $\quad\quad\Gamma_1/K_1=-0.50$                                                                                 & $\quad\quad\Gamma_1/K_1=-0.75$                                                                                 & $\quad\quad\Gamma_1/K_1=-1.00$                                                                                 \\
			\raisebox{1.5cm}{$J_1=\Gamma_1'=J_3=0$}                                                           & \includegraphics[height=3cm]{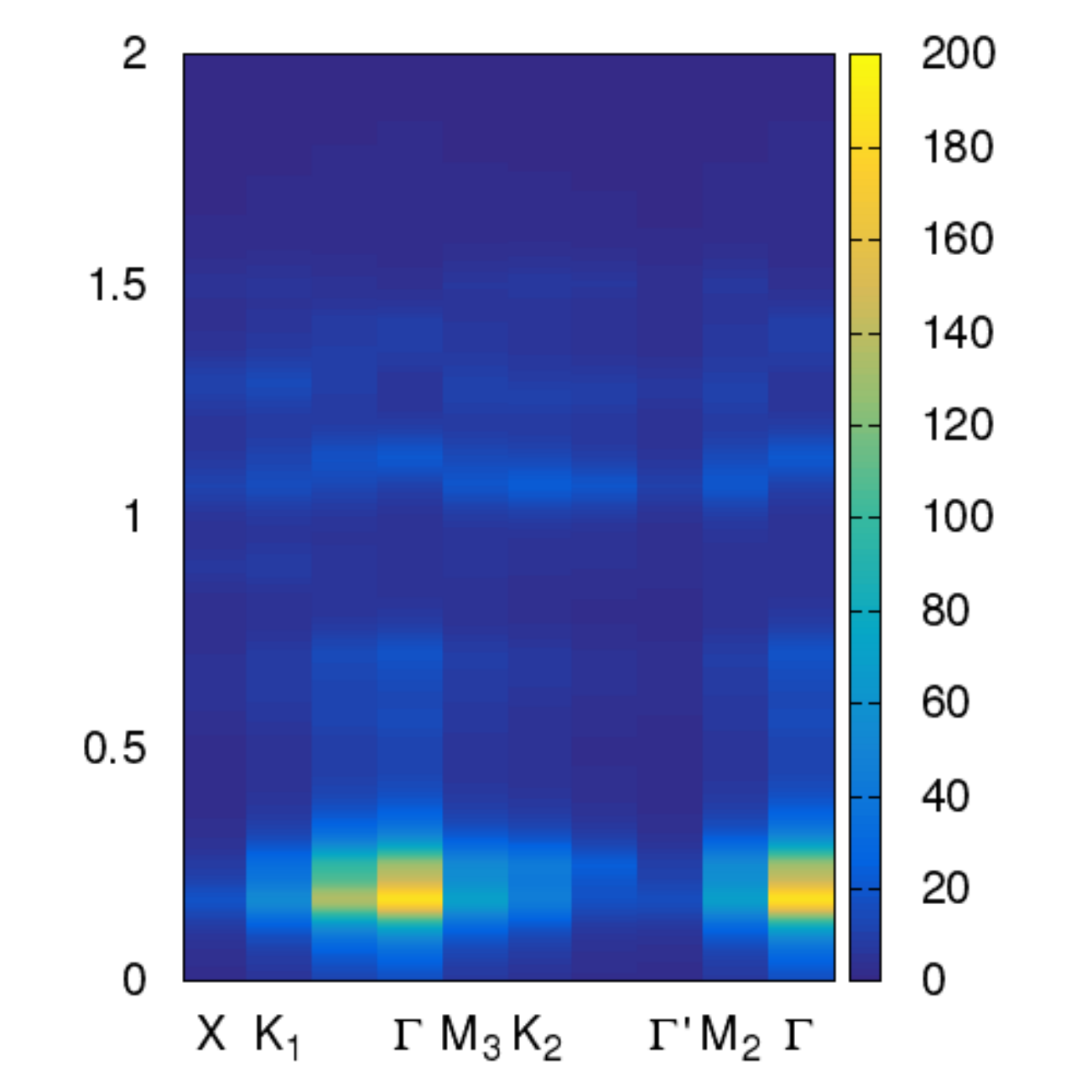}                                & \includegraphics[height=3cm]{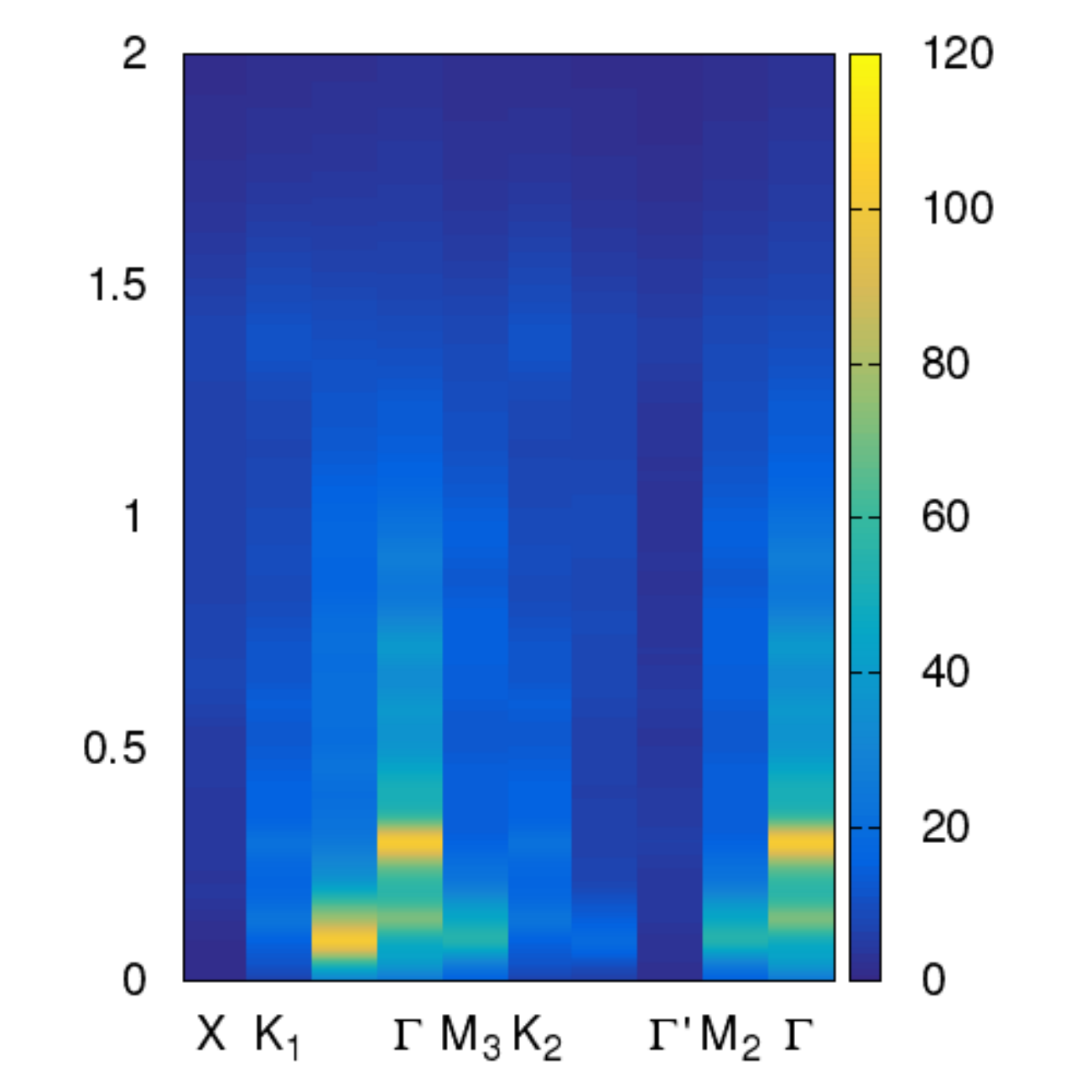}                               & \includegraphics[height=3cm]{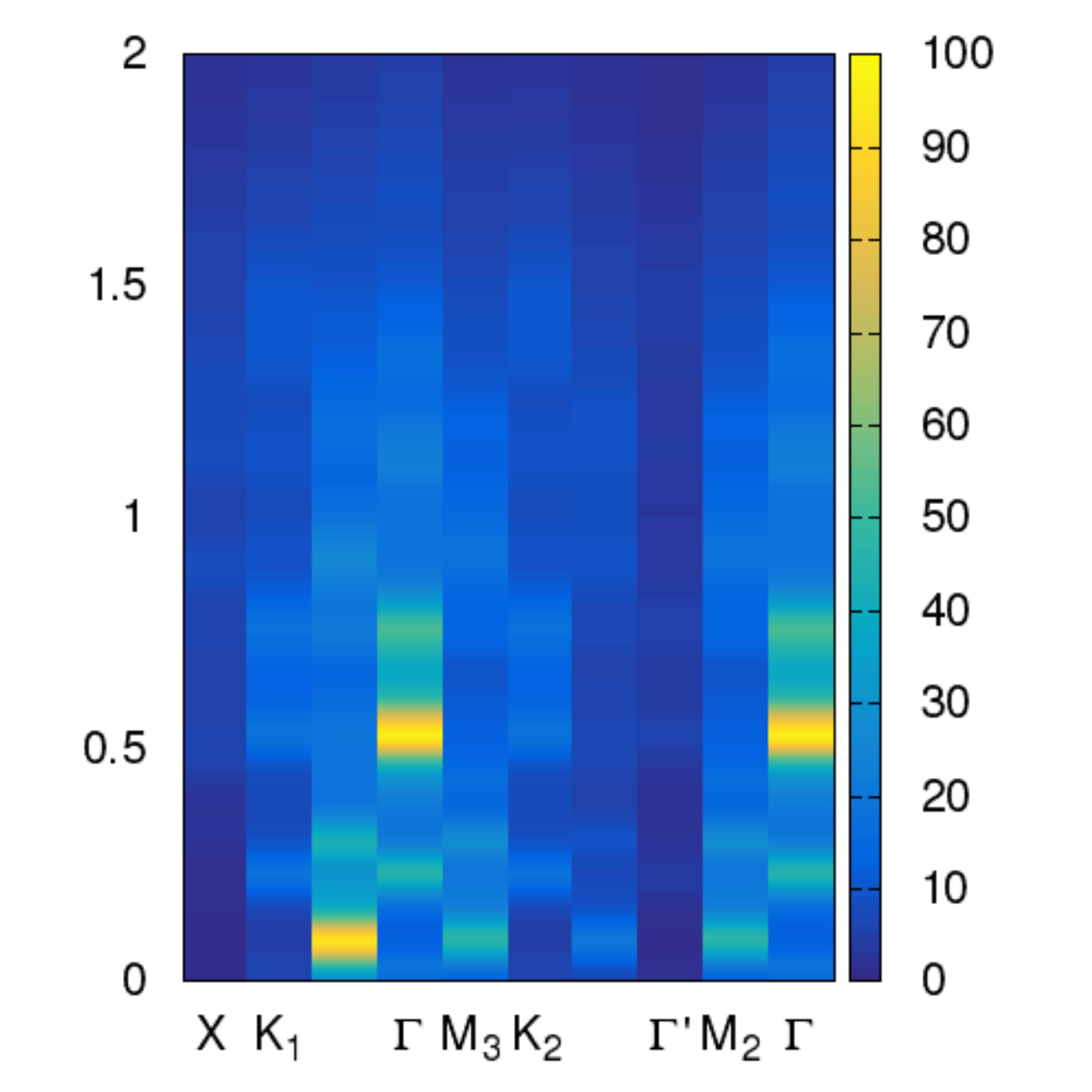}                               & \includegraphics[height=3cm]{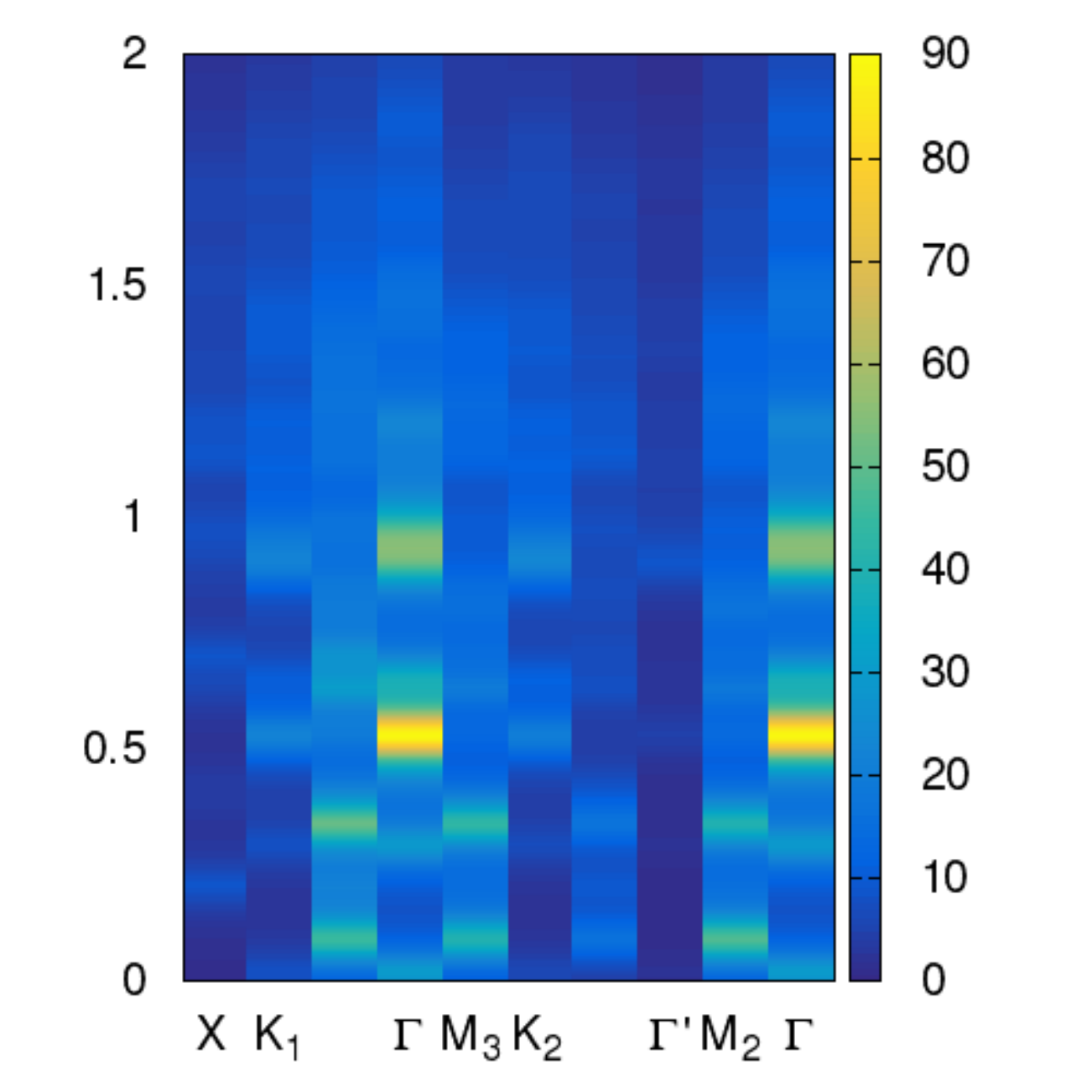}                               & \includegraphics[height=3cm]{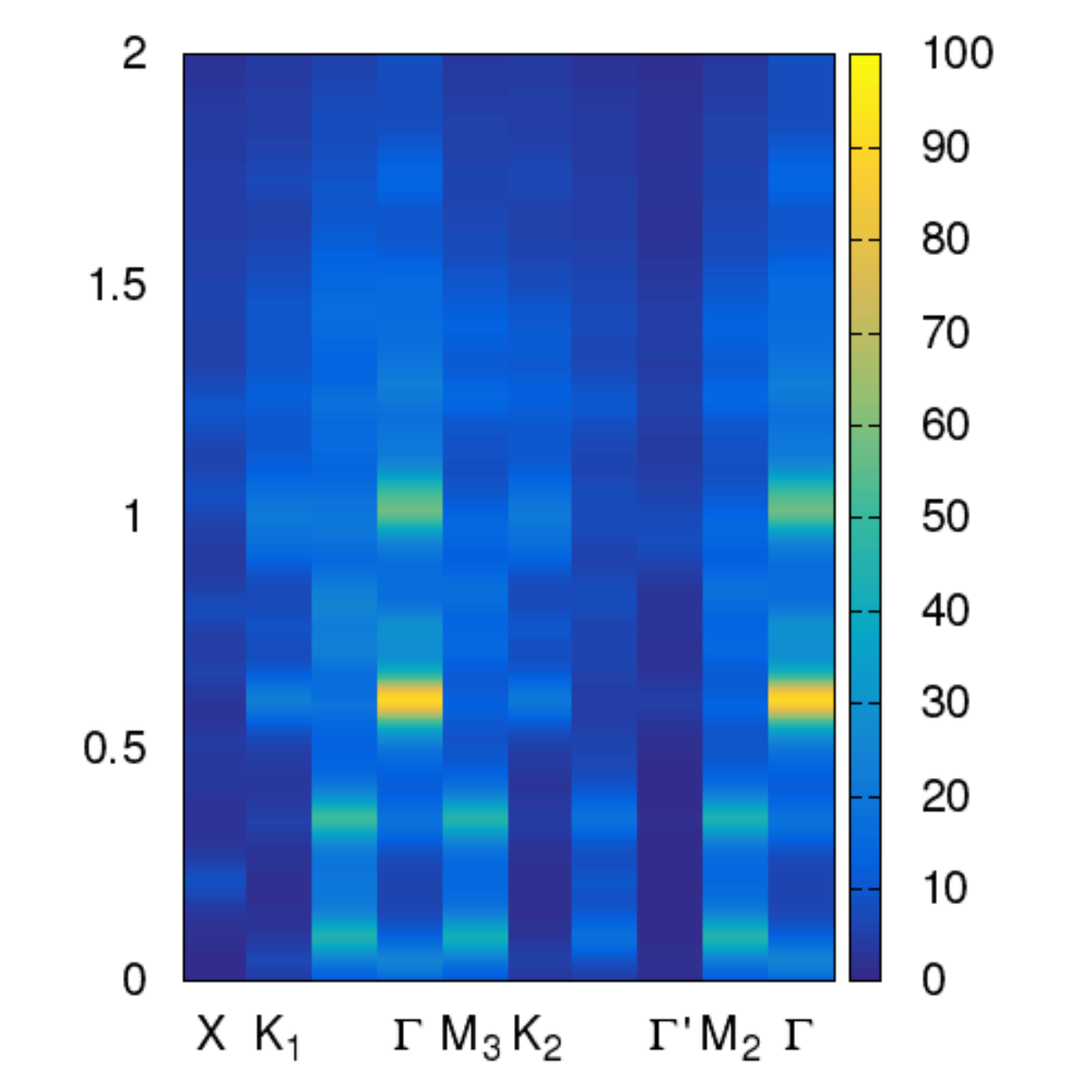}                               \\
			\raisebox{1.5cm}{\begin{tabular}{l}$J_1/K_1=-0.10$,\\$\Gamma_1'=J_3=0$\end{tabular}}              & \includegraphics[height=3cm]{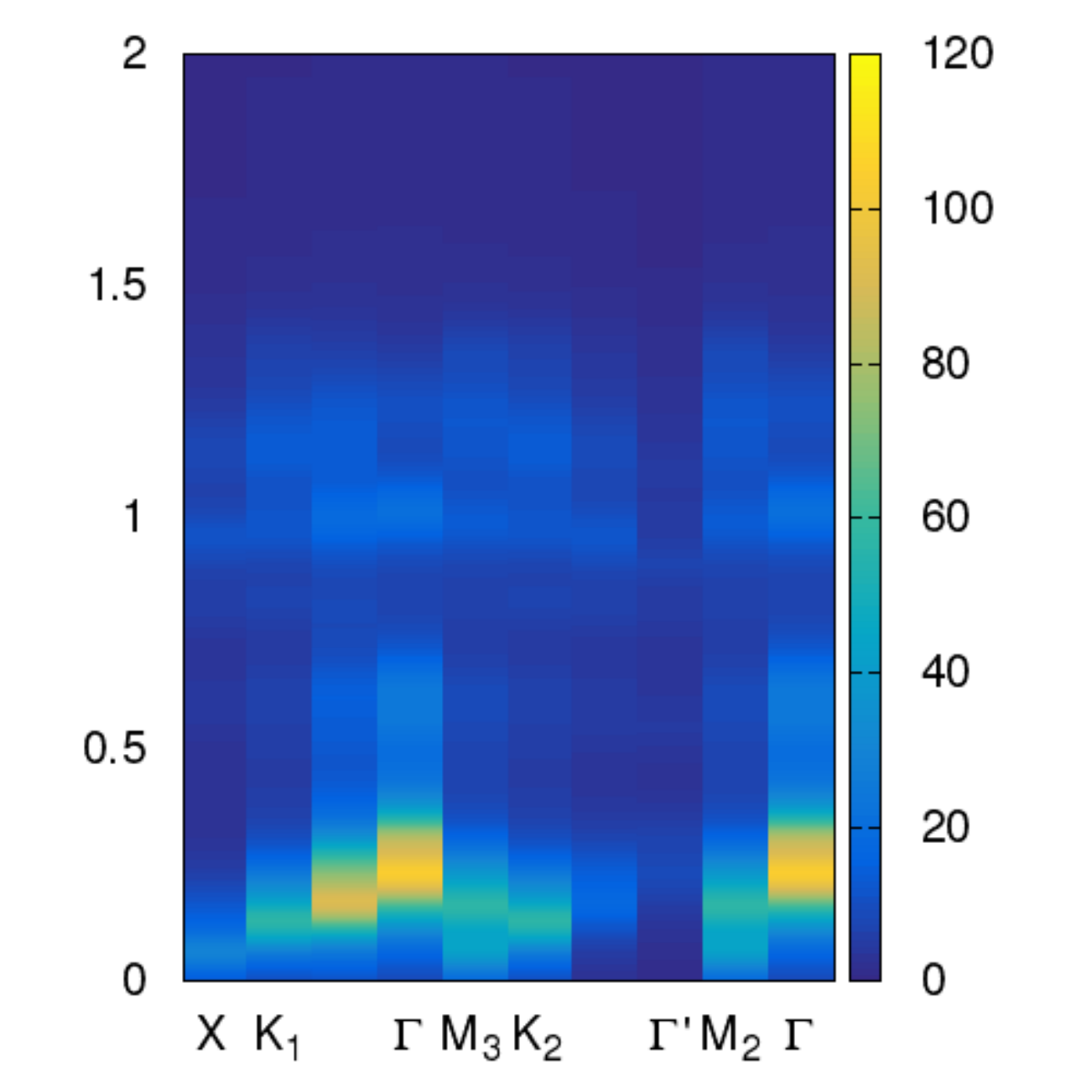}                     & \includegraphics[height=3cm]{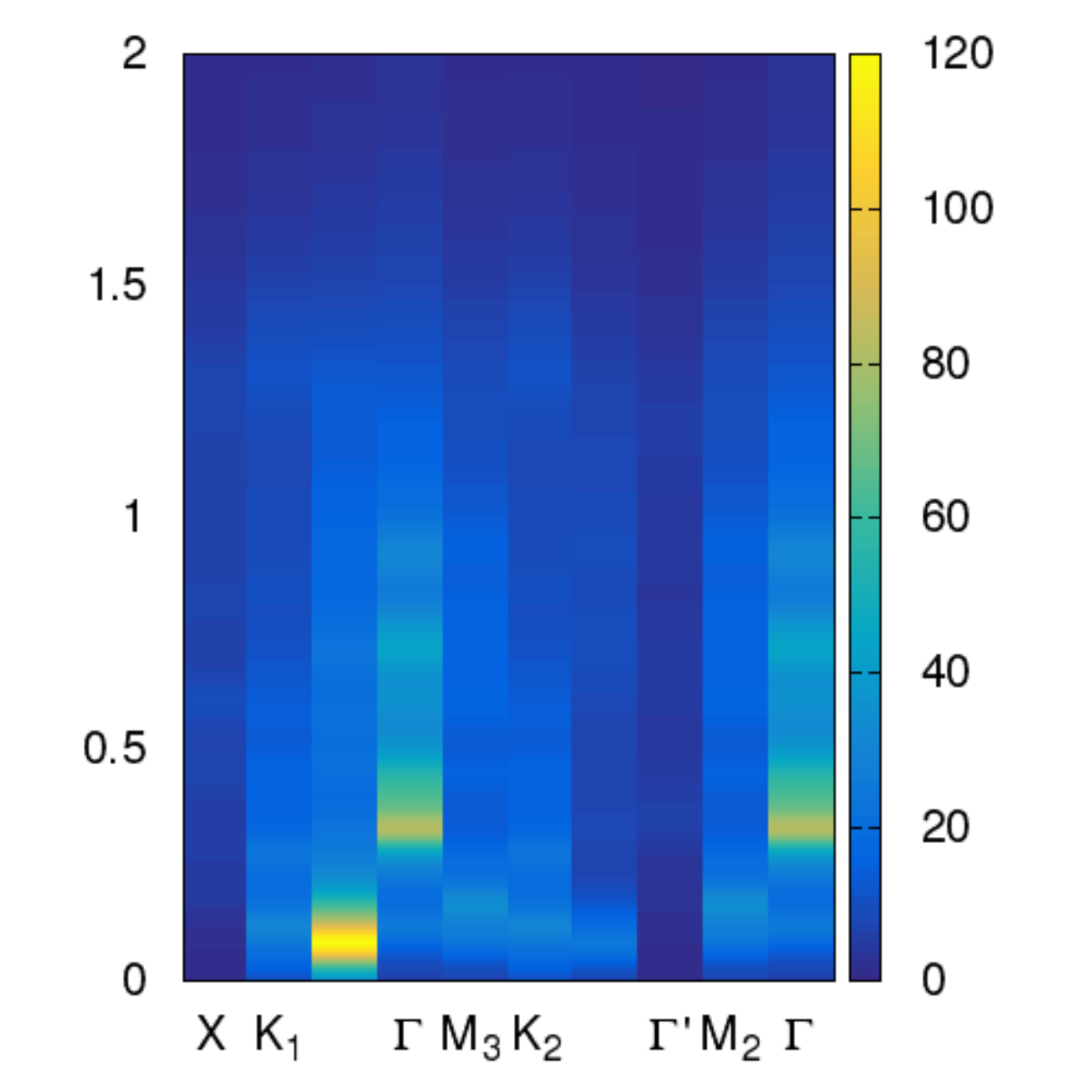}                     & \includegraphics[height=3cm]{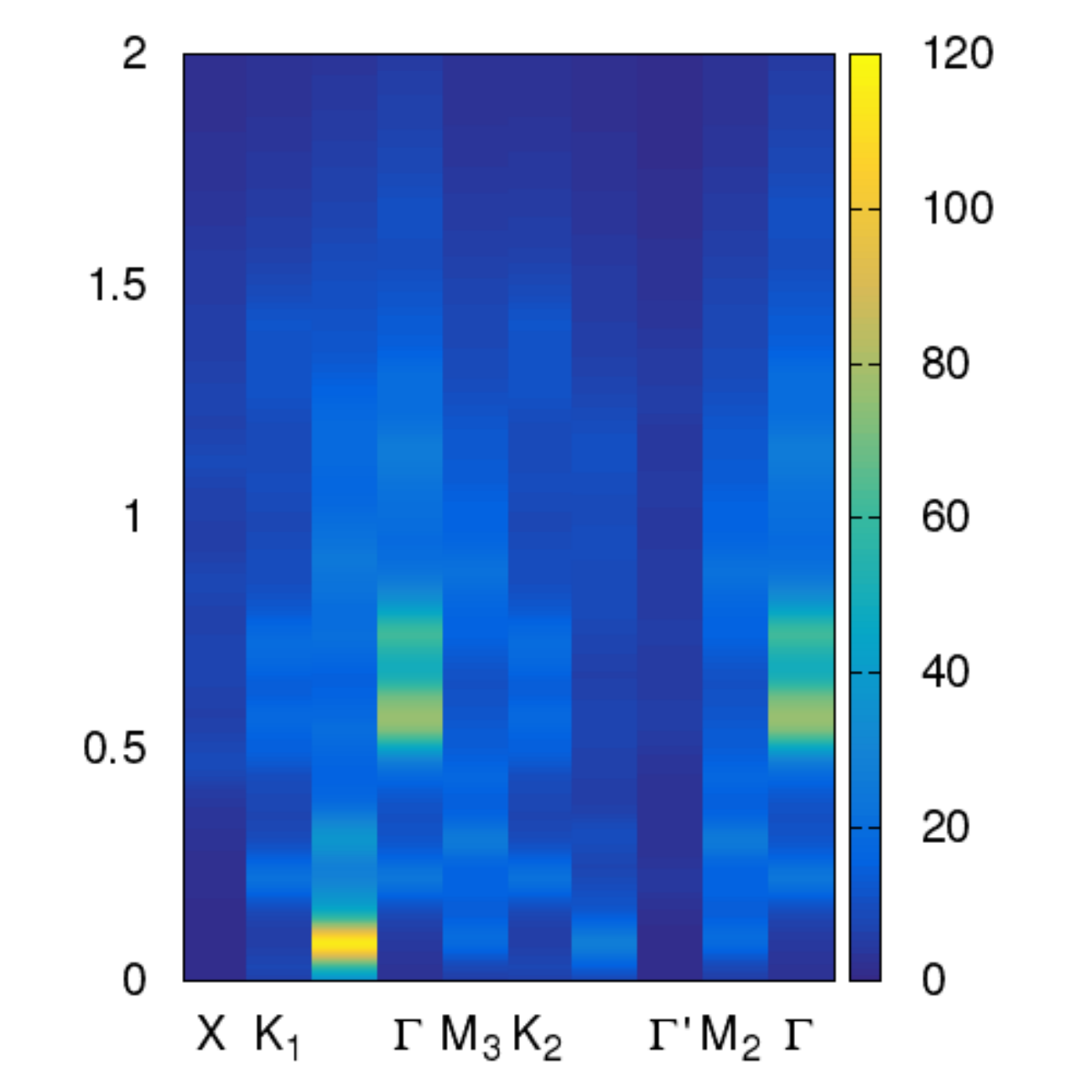}                     & \includegraphics[height=3cm]{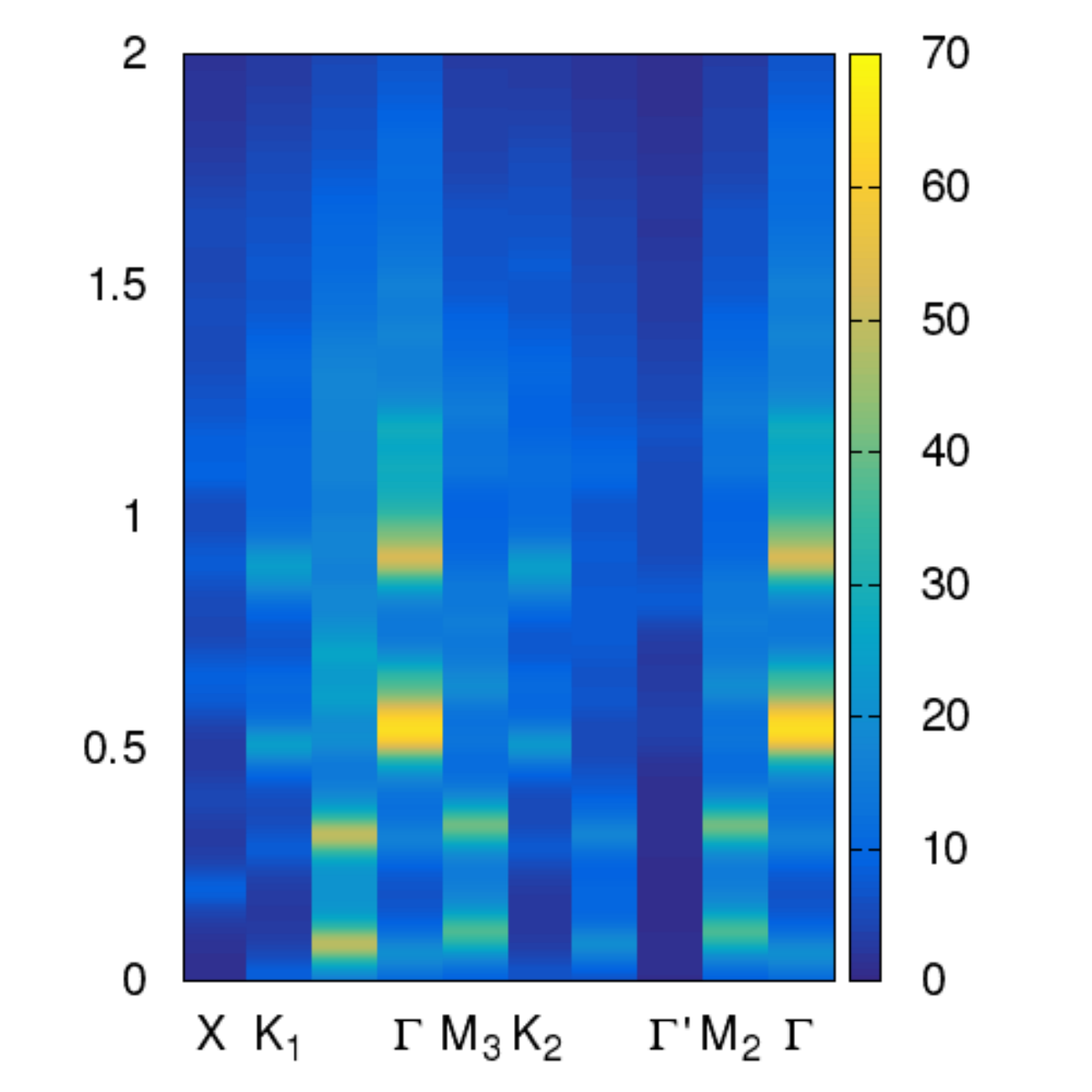}                     & \includegraphics[height=3cm]{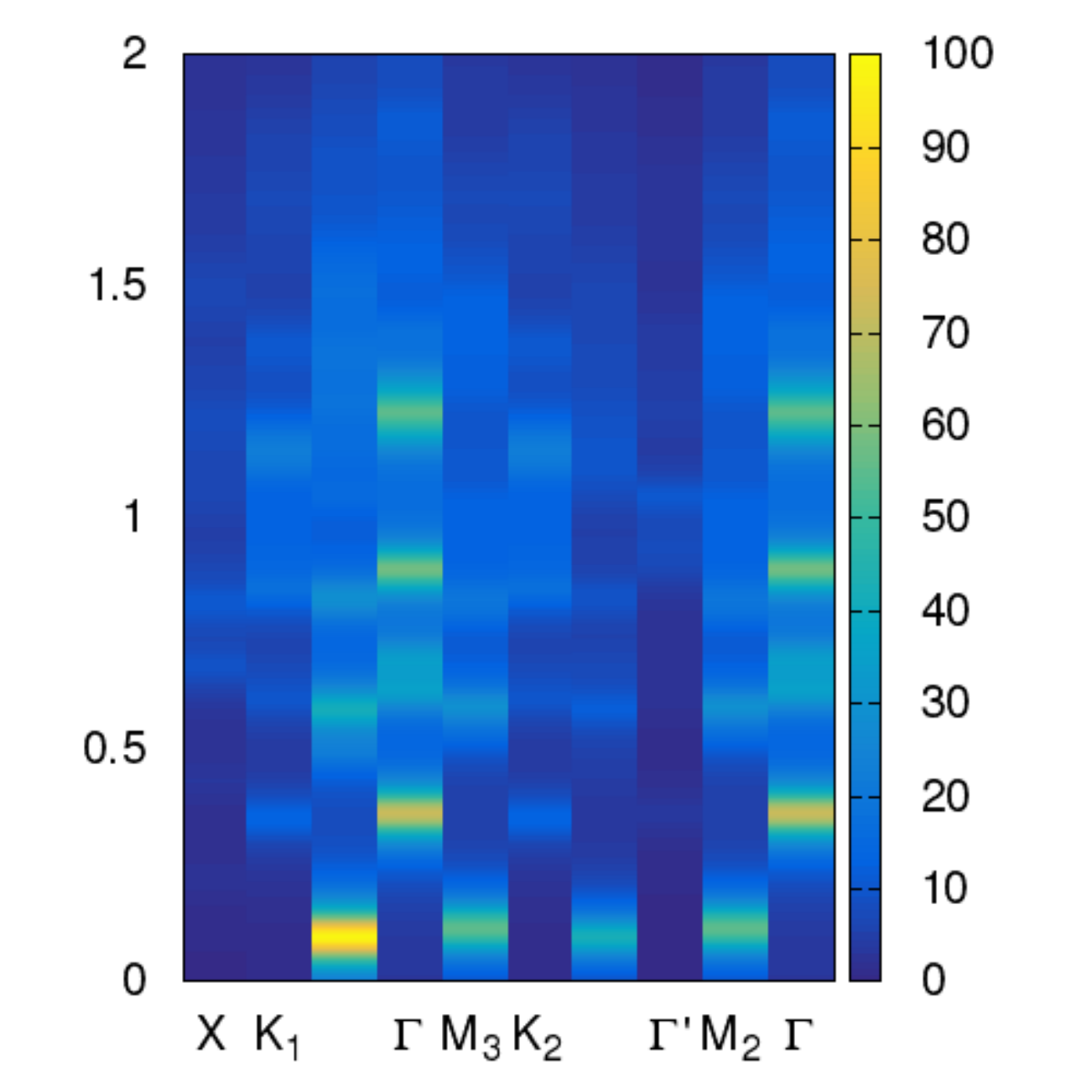}                     \\
			\raisebox{1.5cm}{\begin{tabular}{l}$J_1/K_1=+0.10$,\\$\Gamma_1'=J_3=0$\end{tabular}}              & \includegraphics[height=3cm]{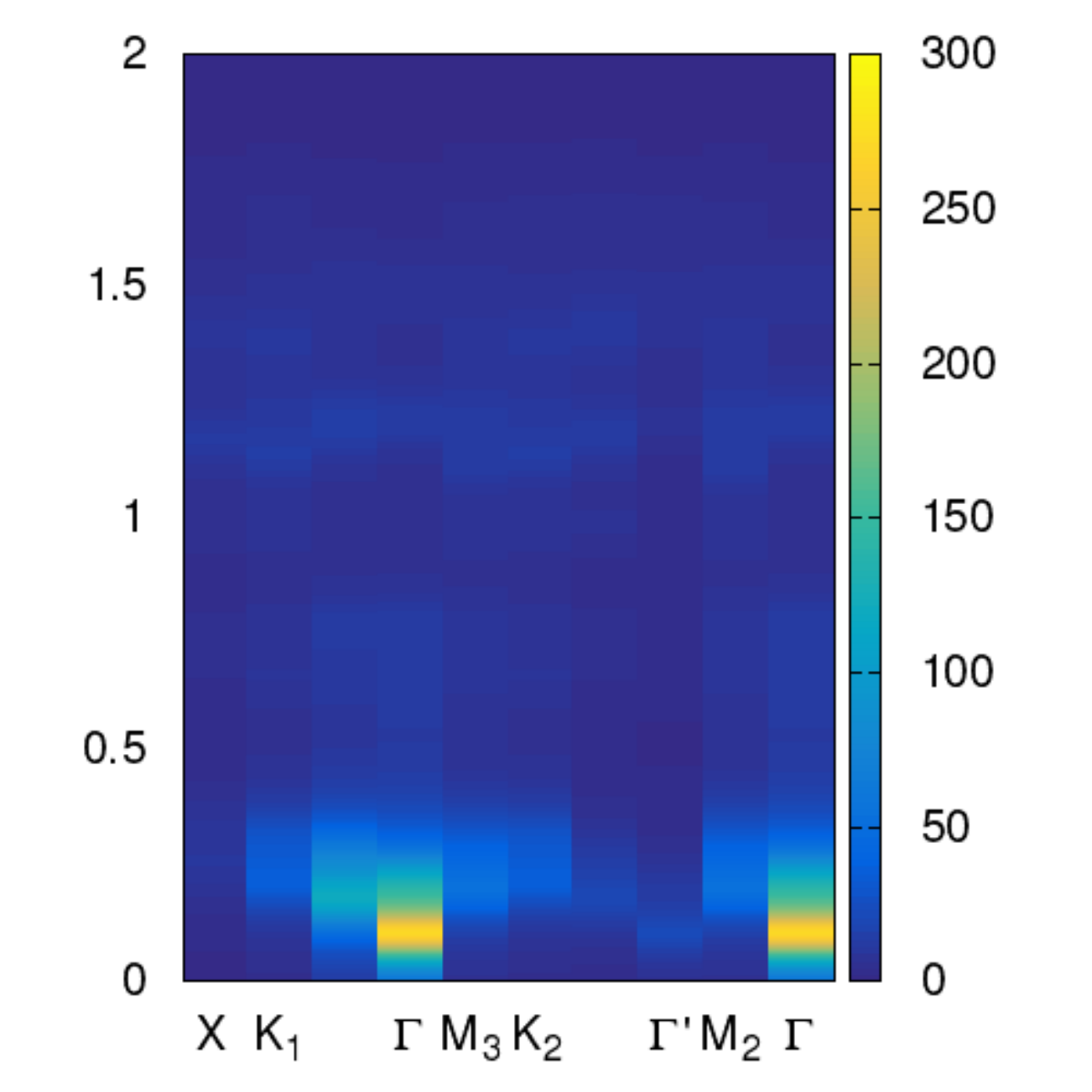}                      & \includegraphics[height=3cm]{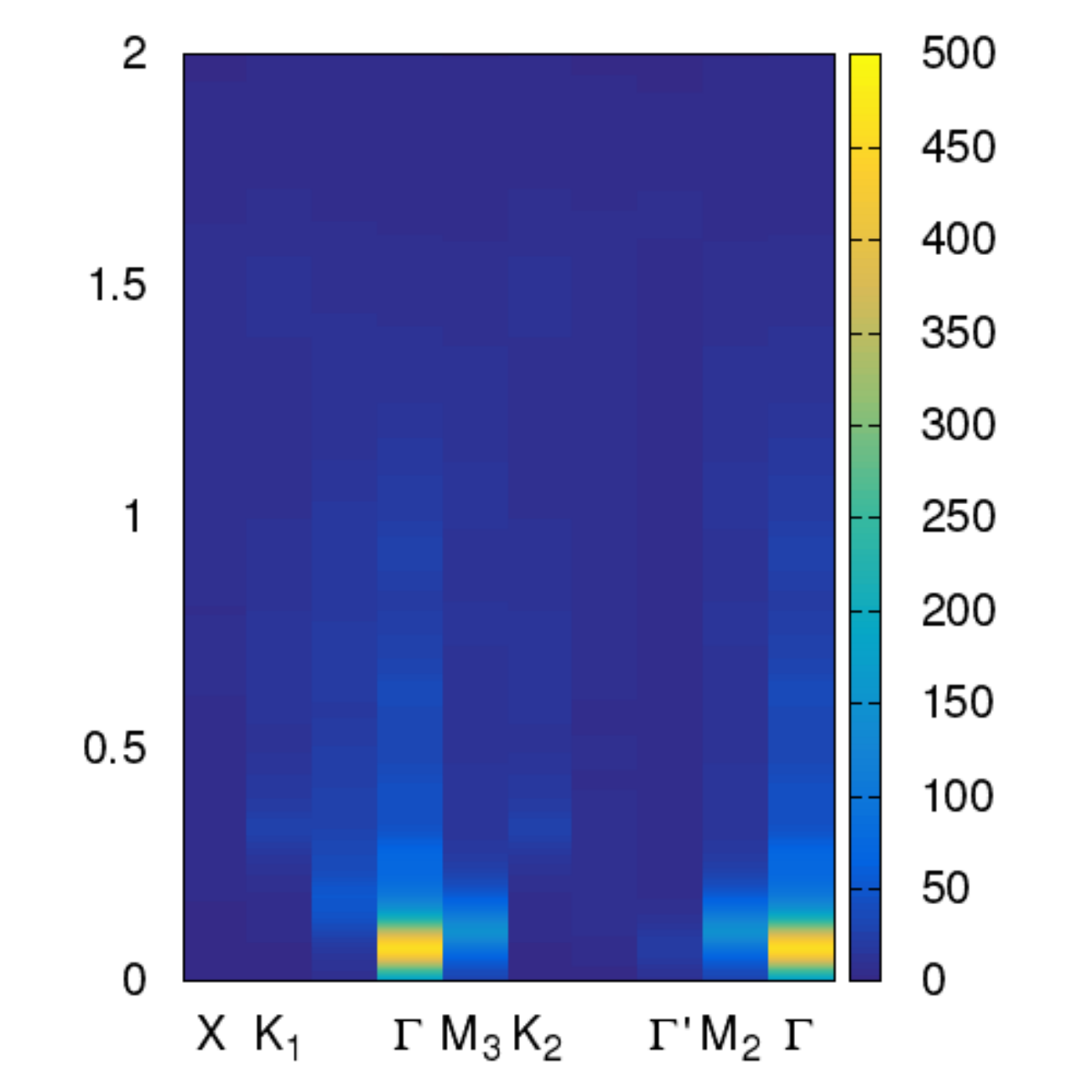}                      & \includegraphics[height=3cm]{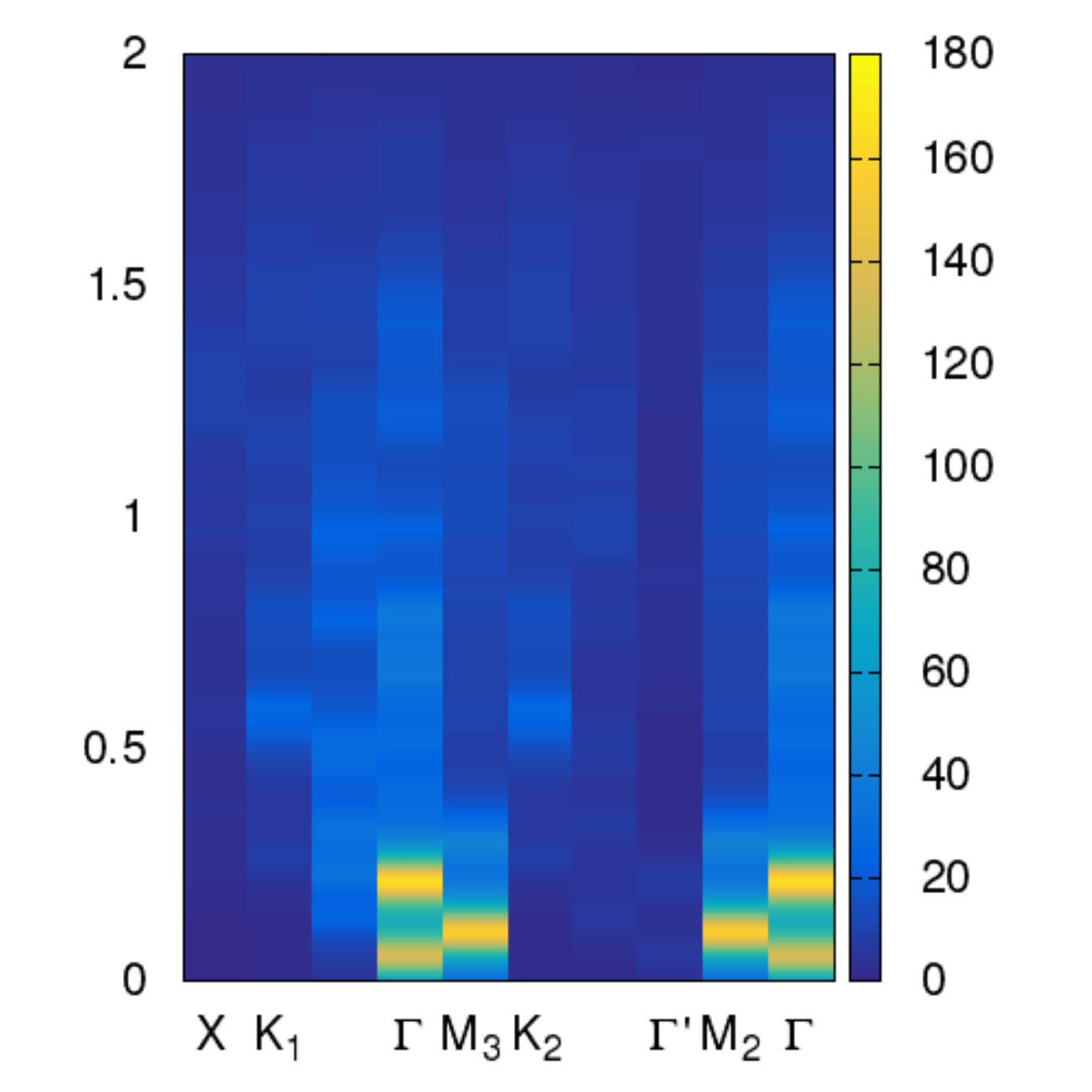}                      & \includegraphics[height=3cm]{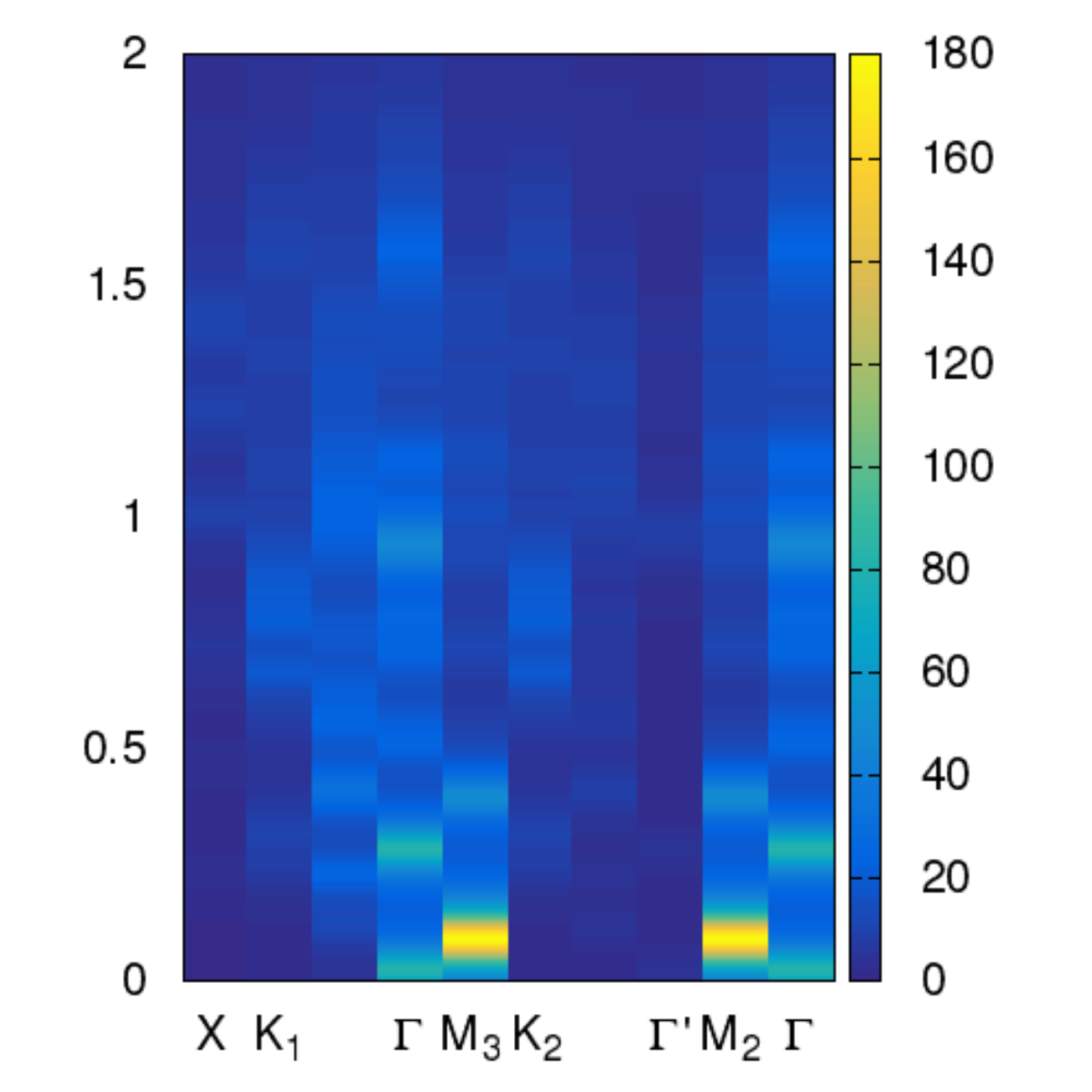}                      & \includegraphics[height=3cm]{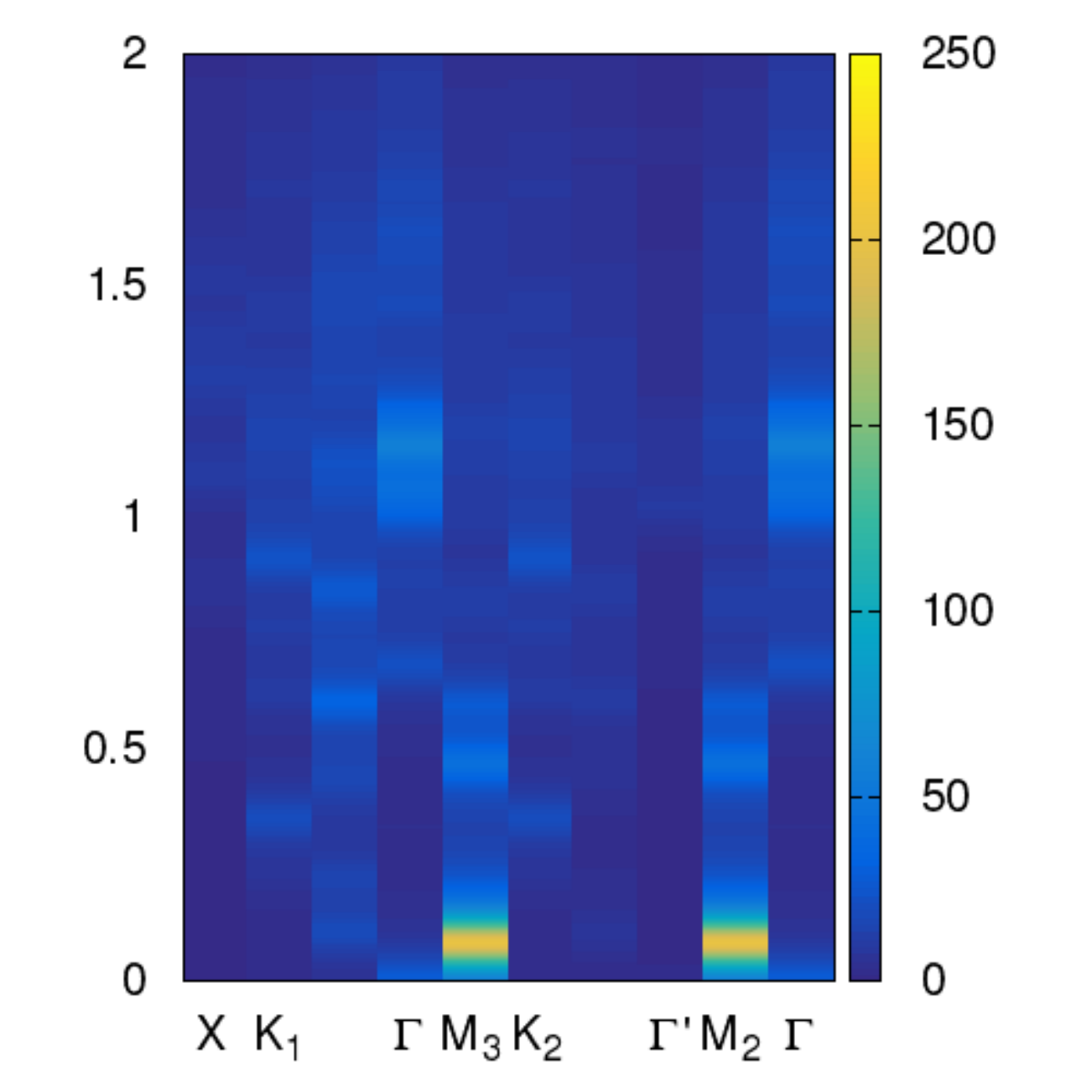}                      \\
			\raisebox{1.5cm}{\begin{tabular}{l}$J_1/K_1=+0.25$,\\$\Gamma_1'=J_3=0$\end{tabular}}              & \includegraphics[height=3cm]{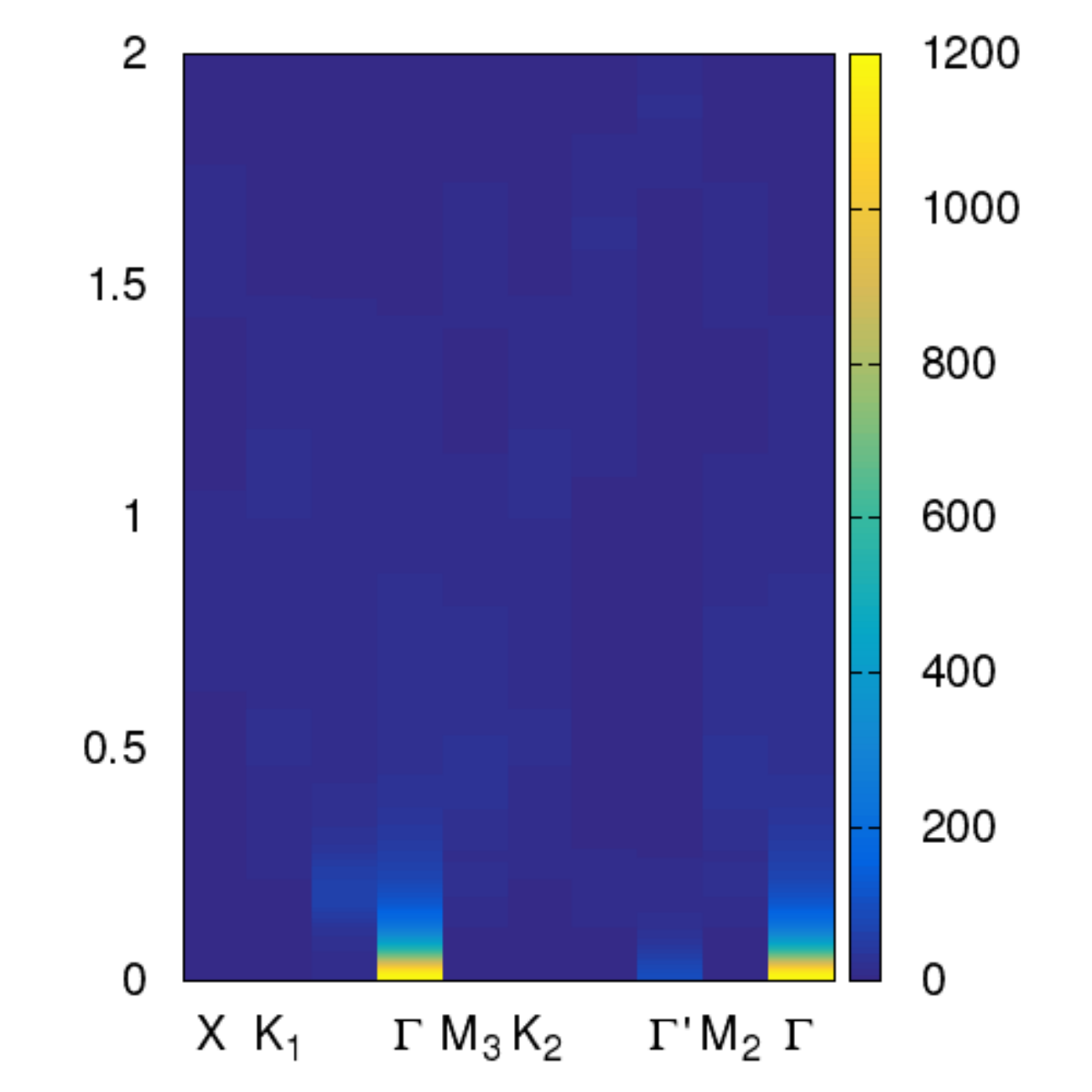}                      & \includegraphics[height=3cm]{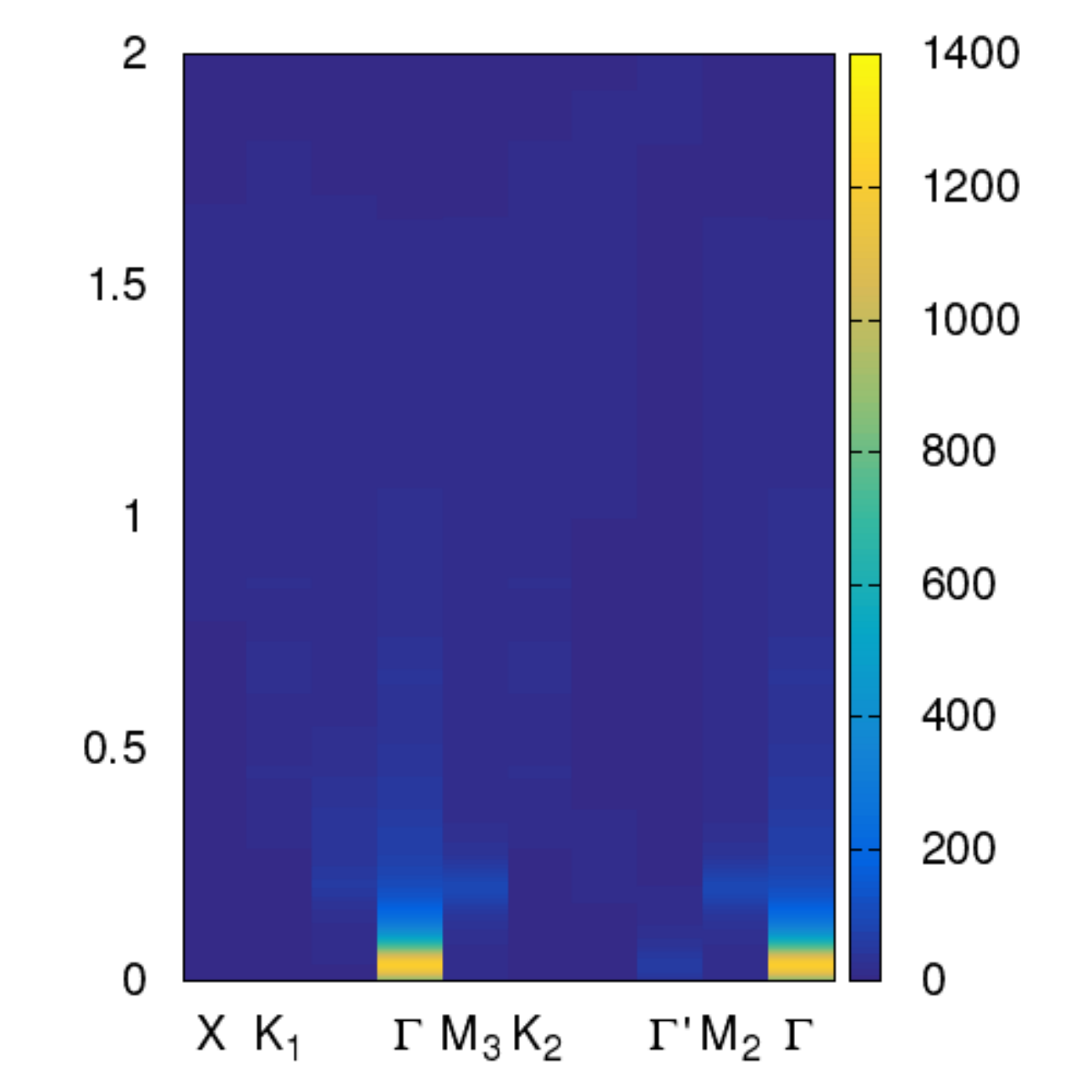}                      & \includegraphics[height=3cm]{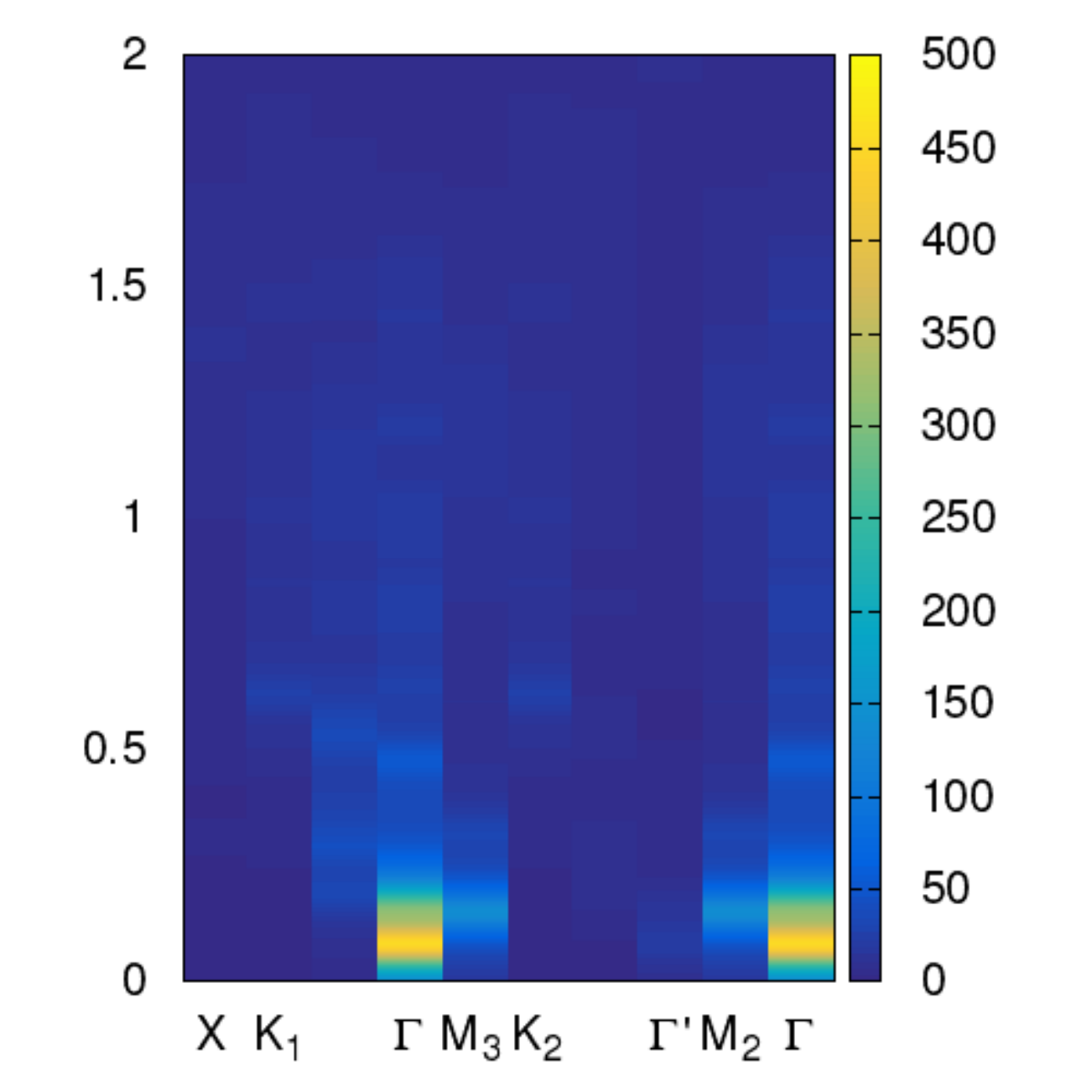}                      & \includegraphics[height=3cm]{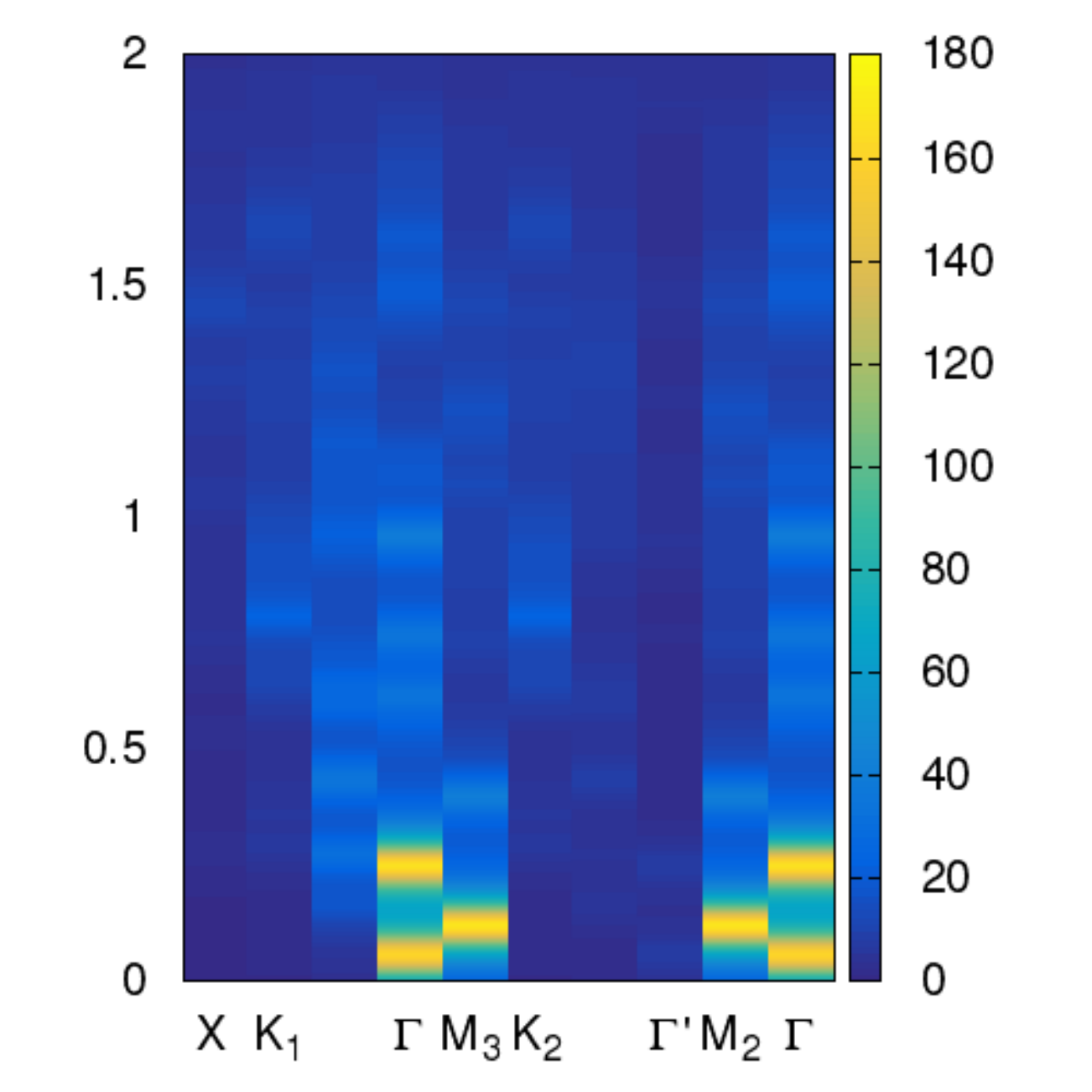}                      & \includegraphics[height=3cm]{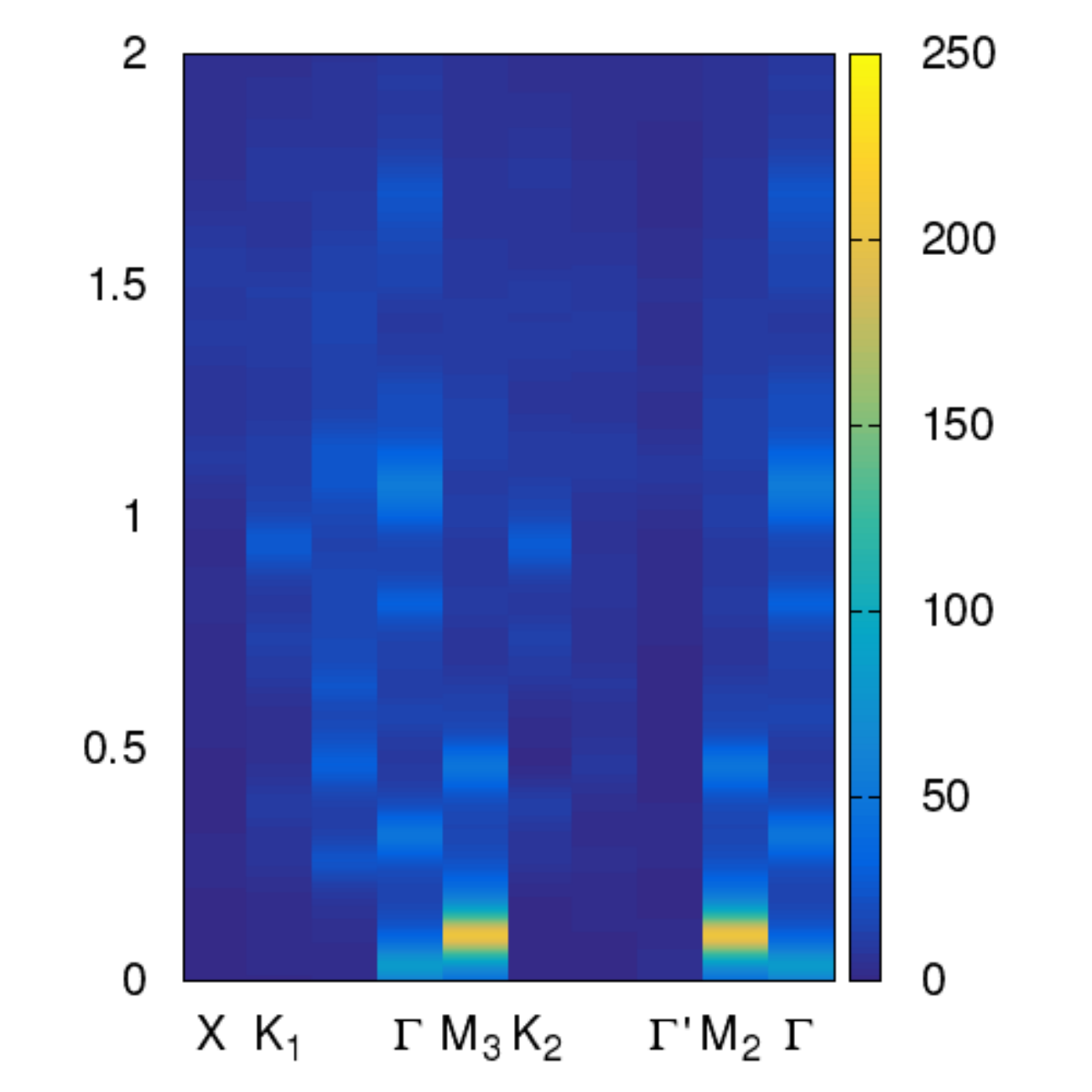}                      \\
			\raisebox{1.5cm}{\begin{tabular}{l}$J_1/K_1=+0.10$,\\$\Gamma_1'=0.15,$\\$J_3=0$\end{tabular}}     & \includegraphics[height=3cm]{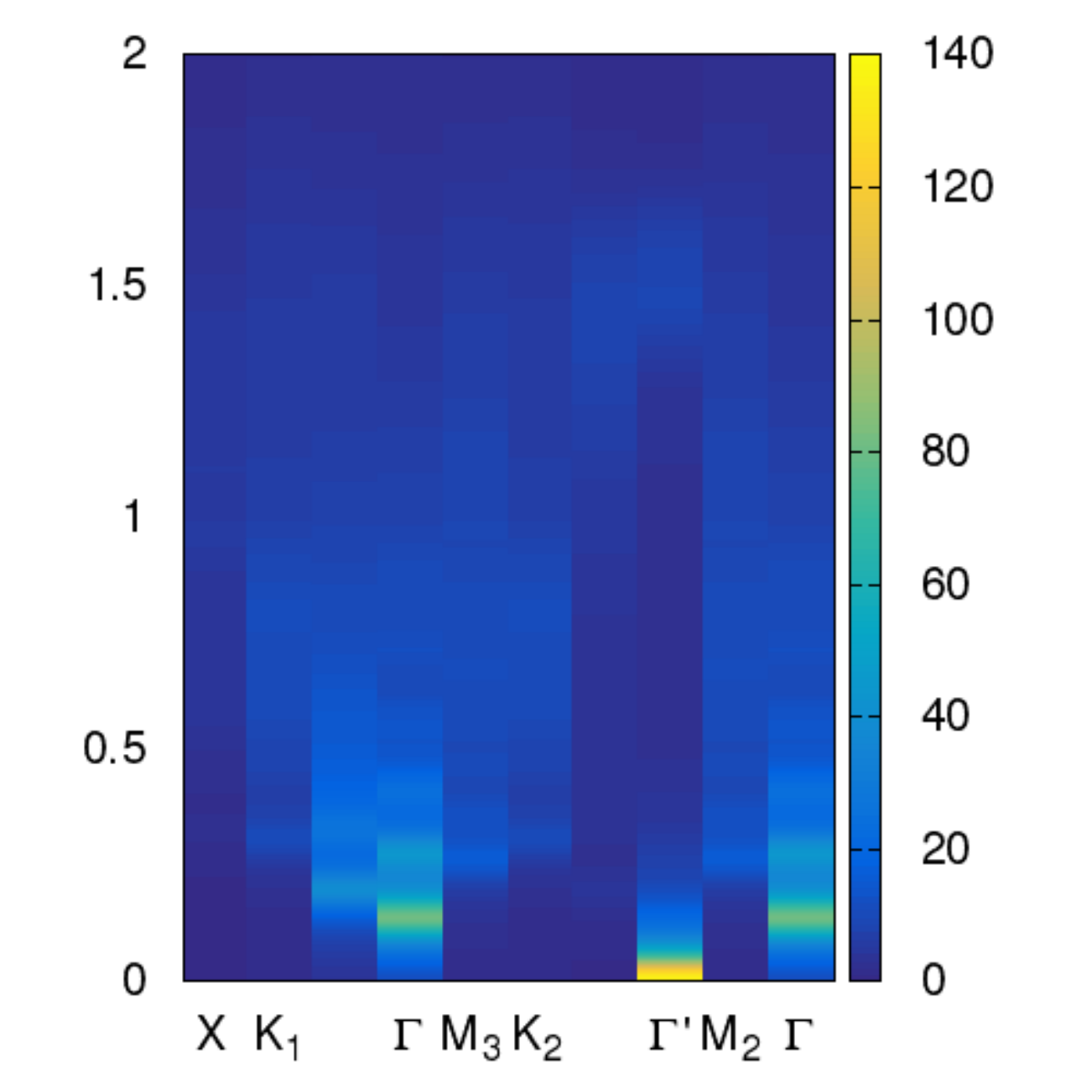}            & \includegraphics[height=3cm]{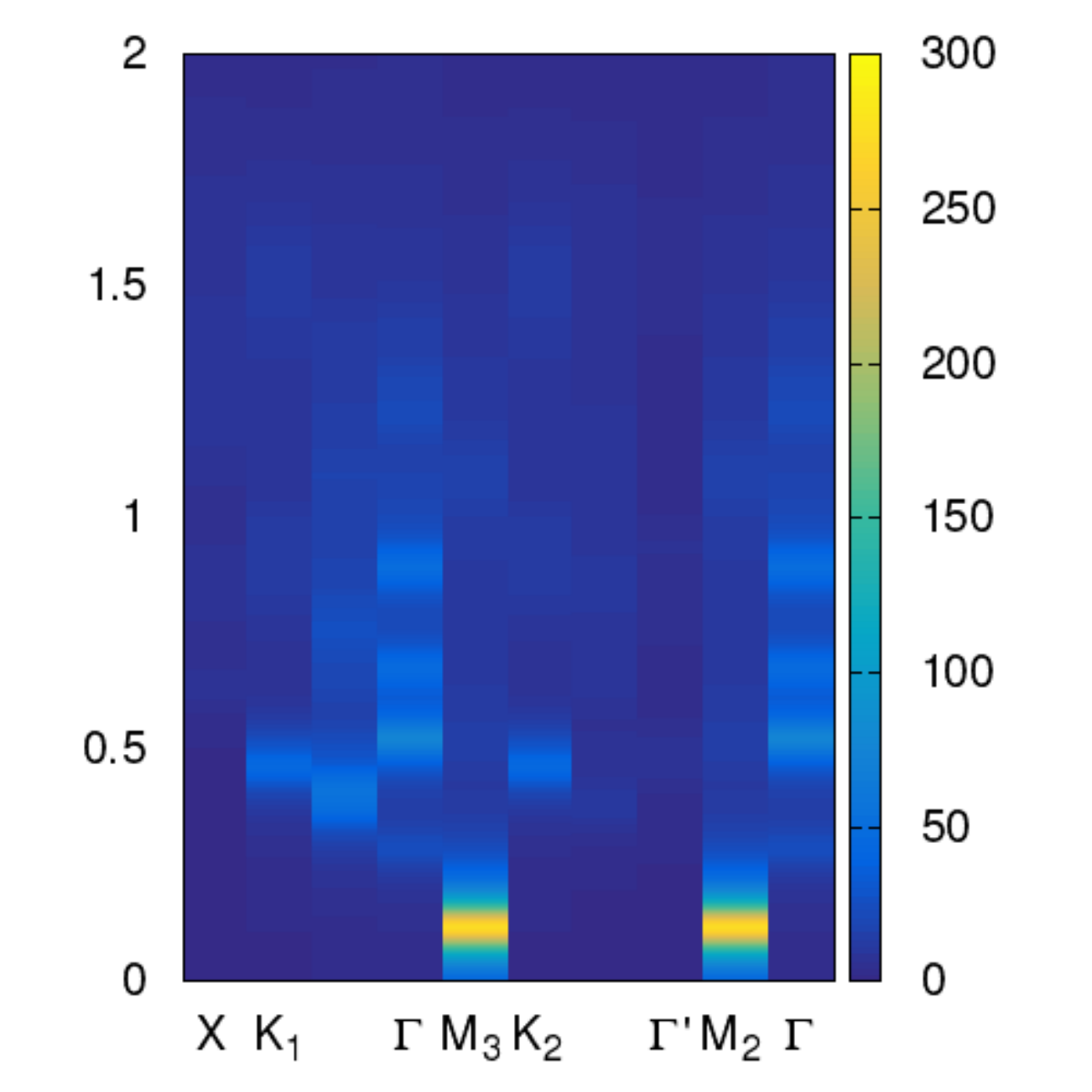}            & \includegraphics[height=3cm]{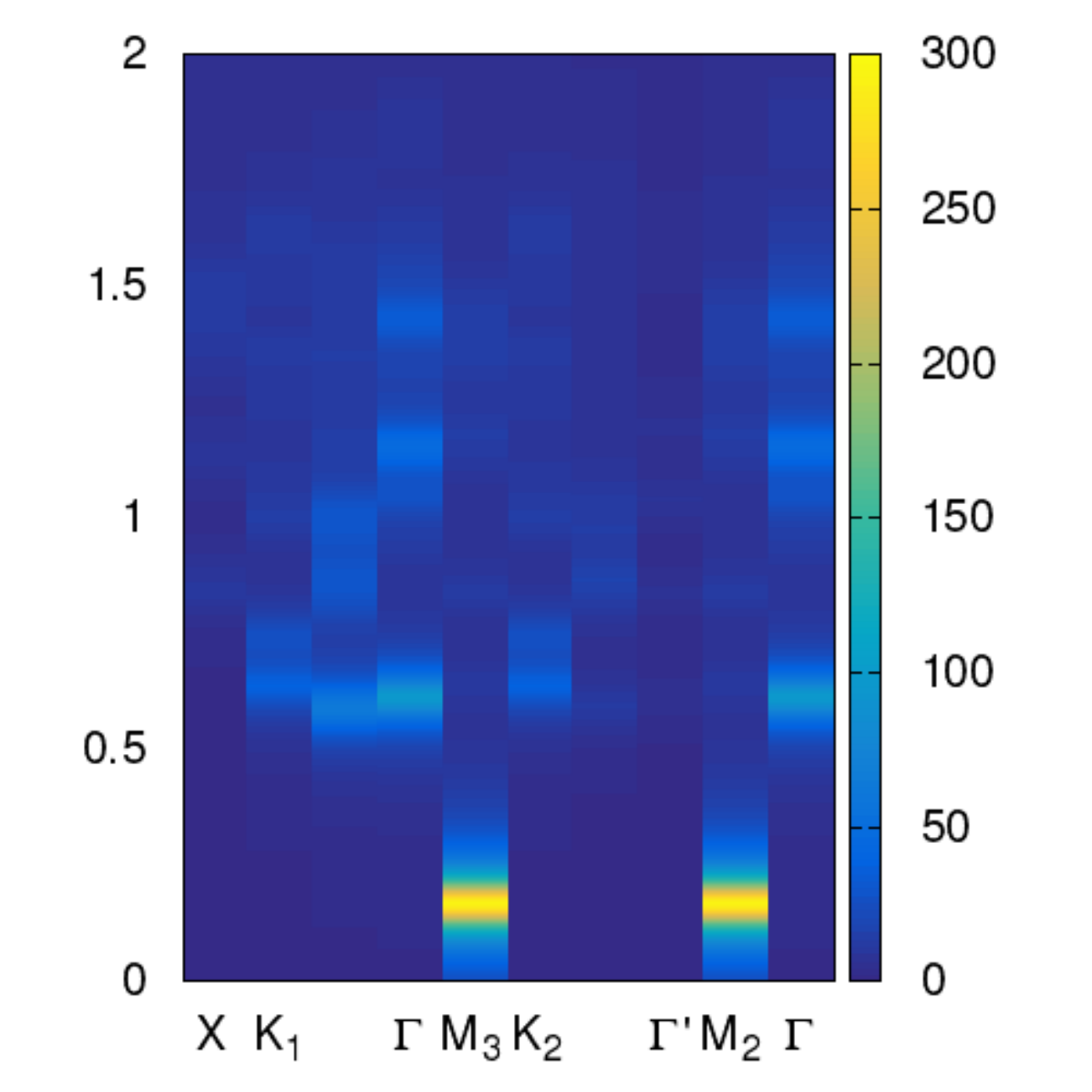}            & \includegraphics[height=3cm]{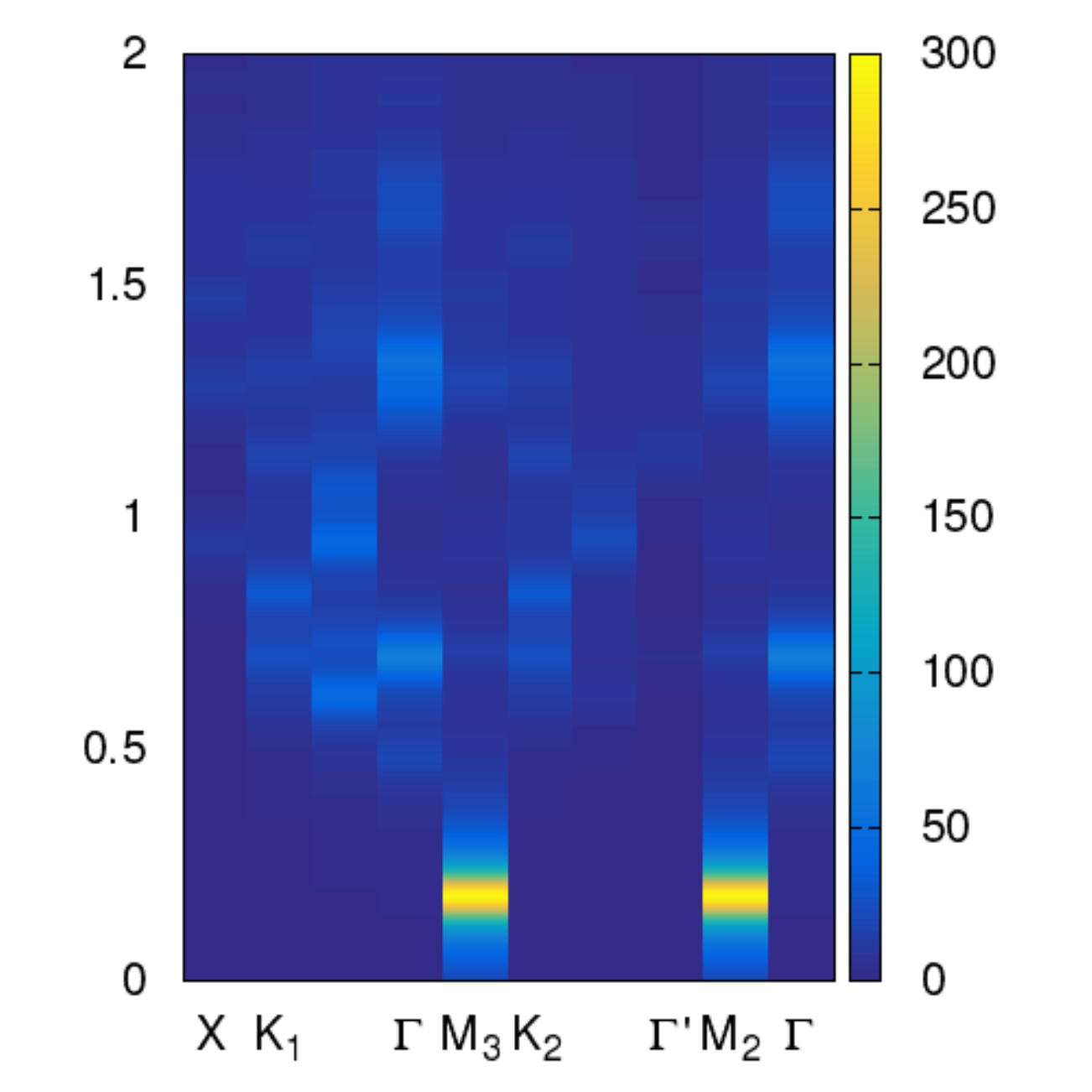}            & \includegraphics[height=3cm]{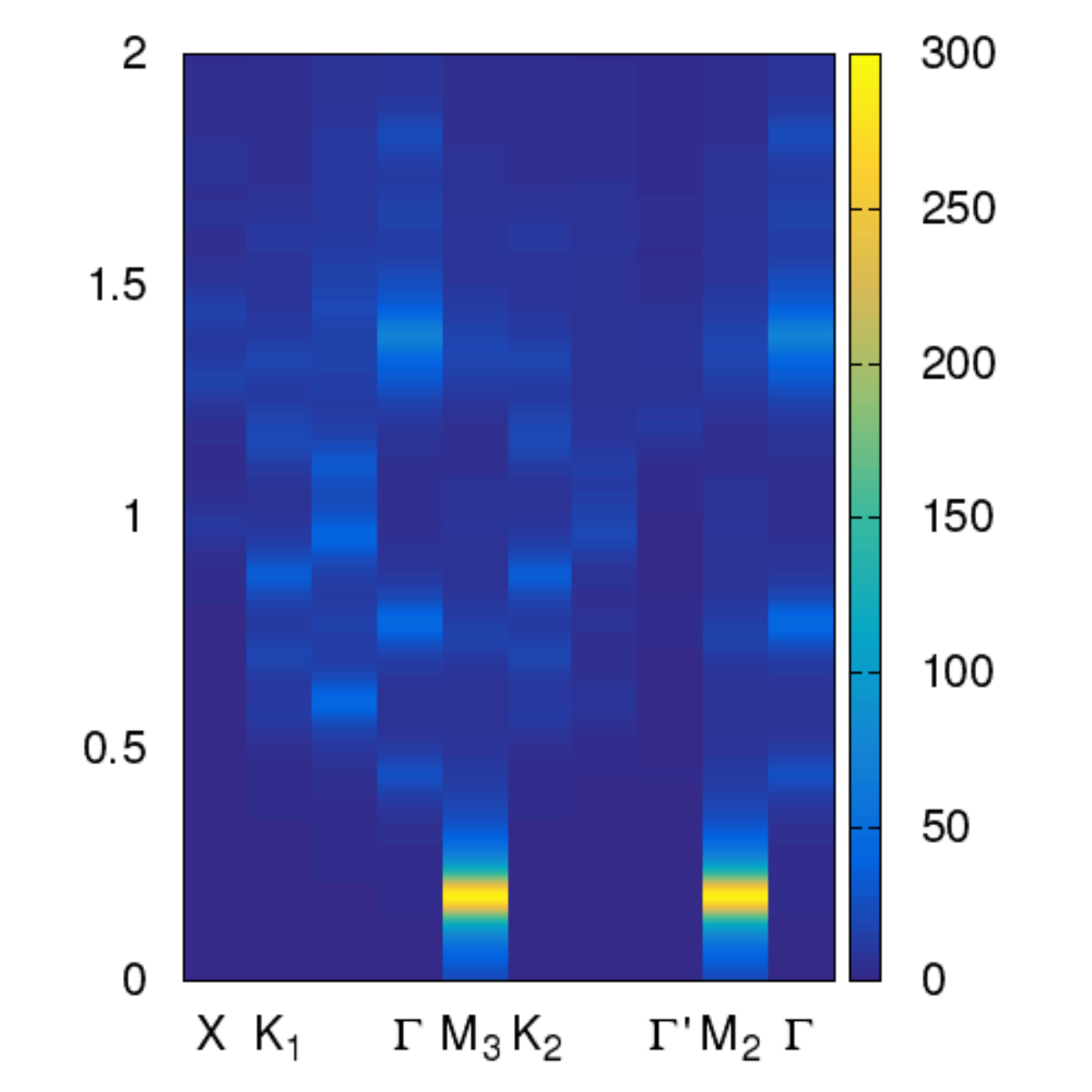}            \\
			\raisebox{1.5cm}{\begin{tabular}{l}$J_1/K_1=+0.10$,\\$\Gamma_1'=0.15,$\\$J_3=-0.10$\end{tabular}} & \includegraphics[height=3cm]{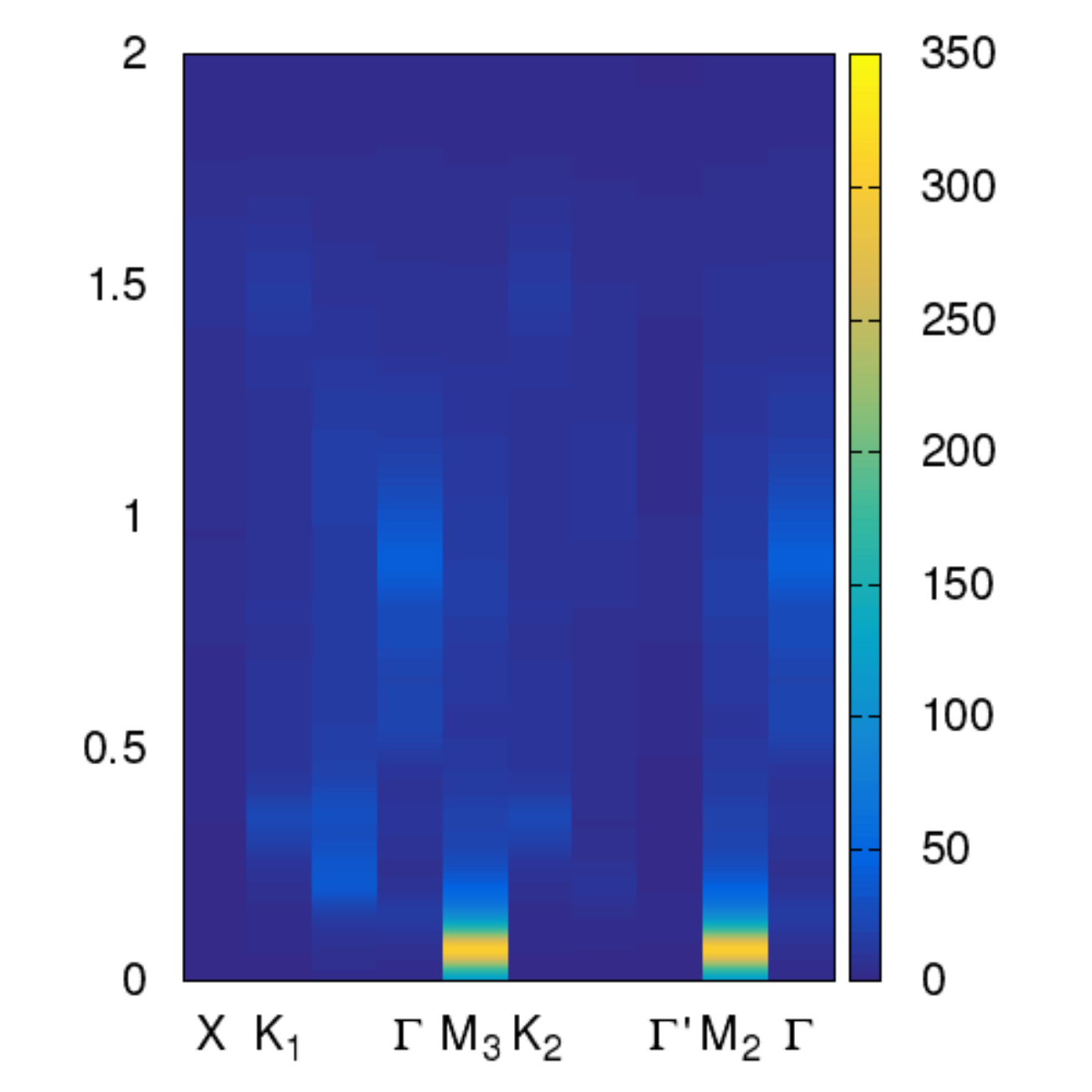} & \includegraphics[height=3cm]{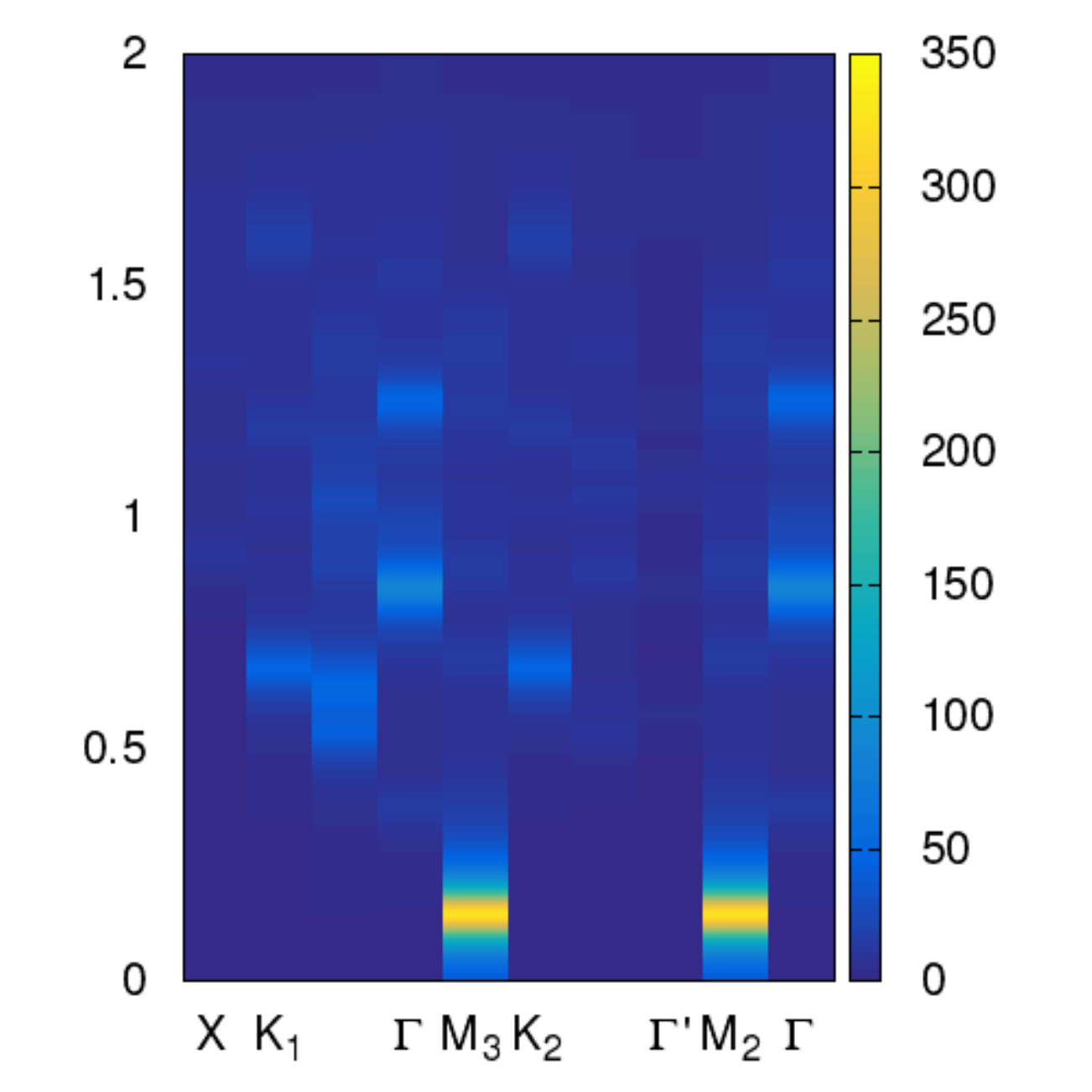} & \includegraphics[height=3cm]{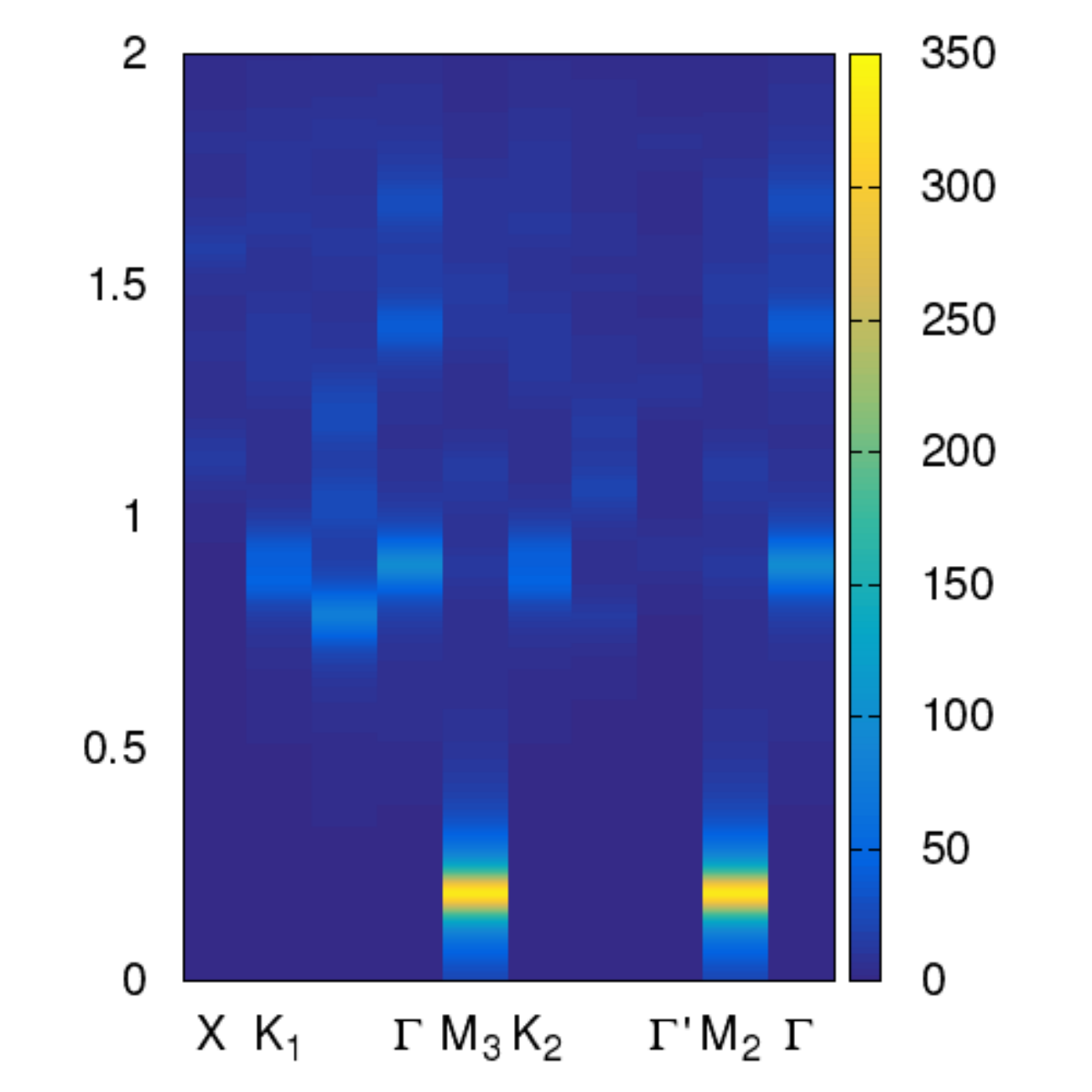} & \includegraphics[height=3cm]{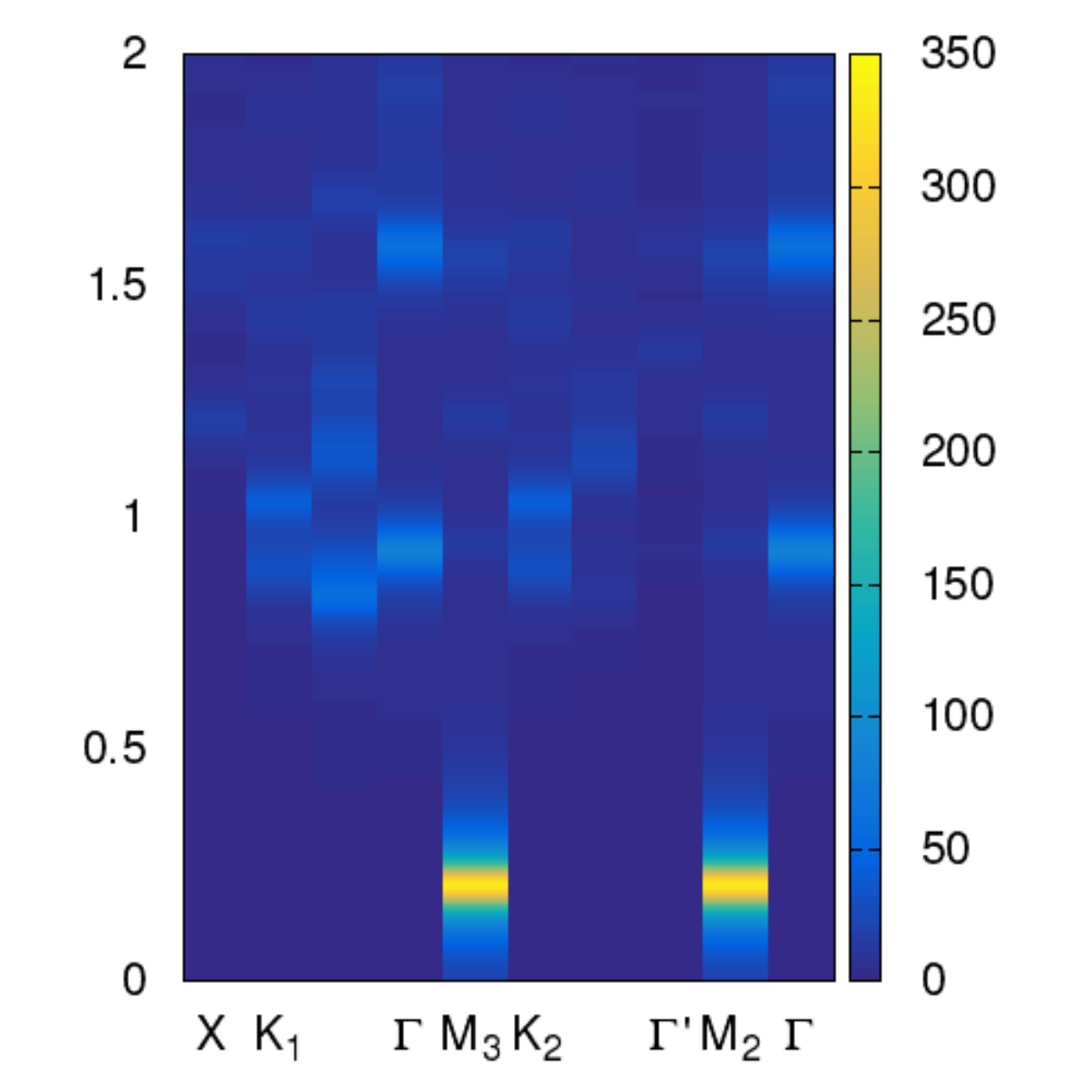} & \includegraphics[height=3cm]{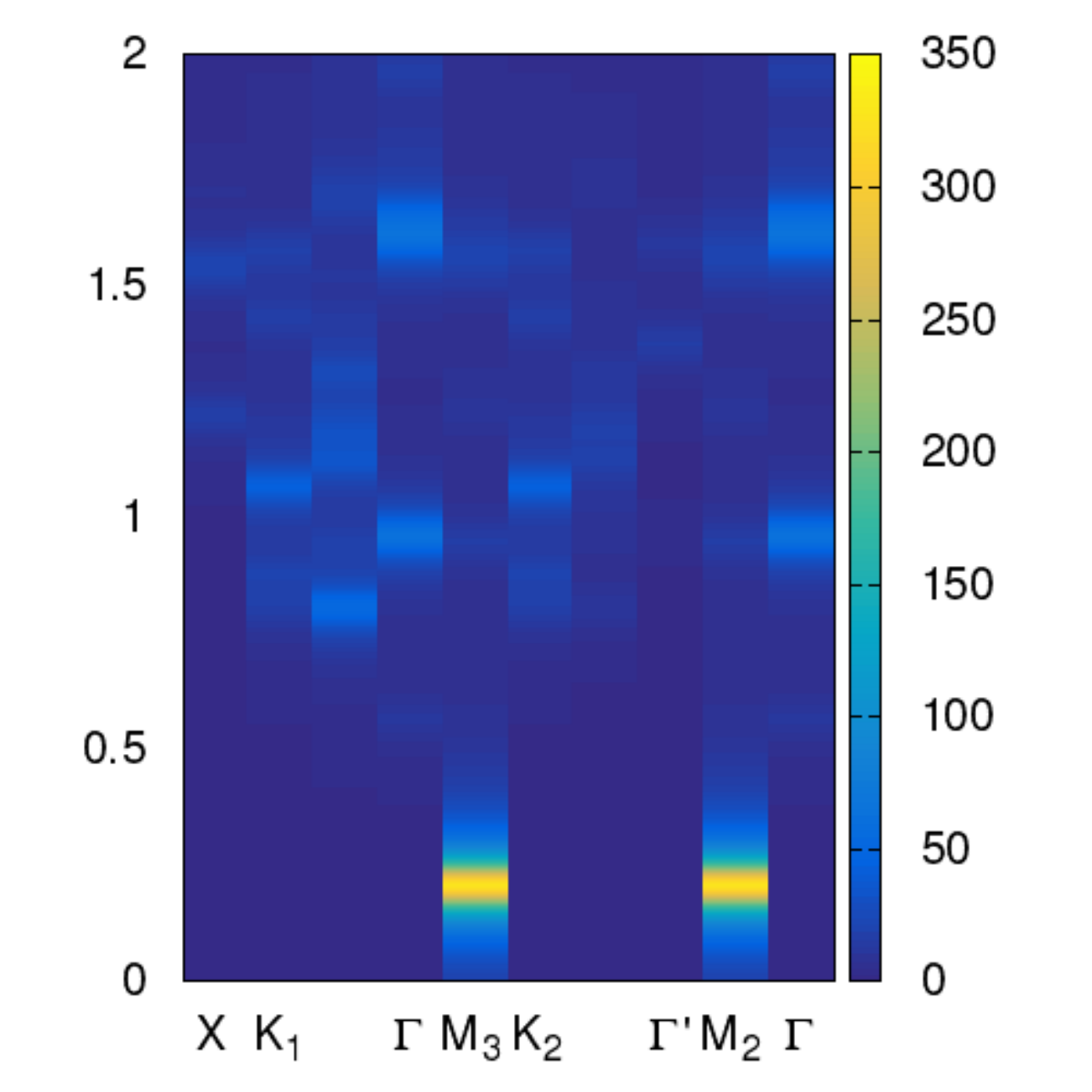} \\
			\raisebox{1.5cm}{\begin{tabular}{l}$J_1/K_1=+0.10$,\\$\Gamma_1'=0.15,$\\$J_3=-0.40$\end{tabular}} & \includegraphics[height=3cm]{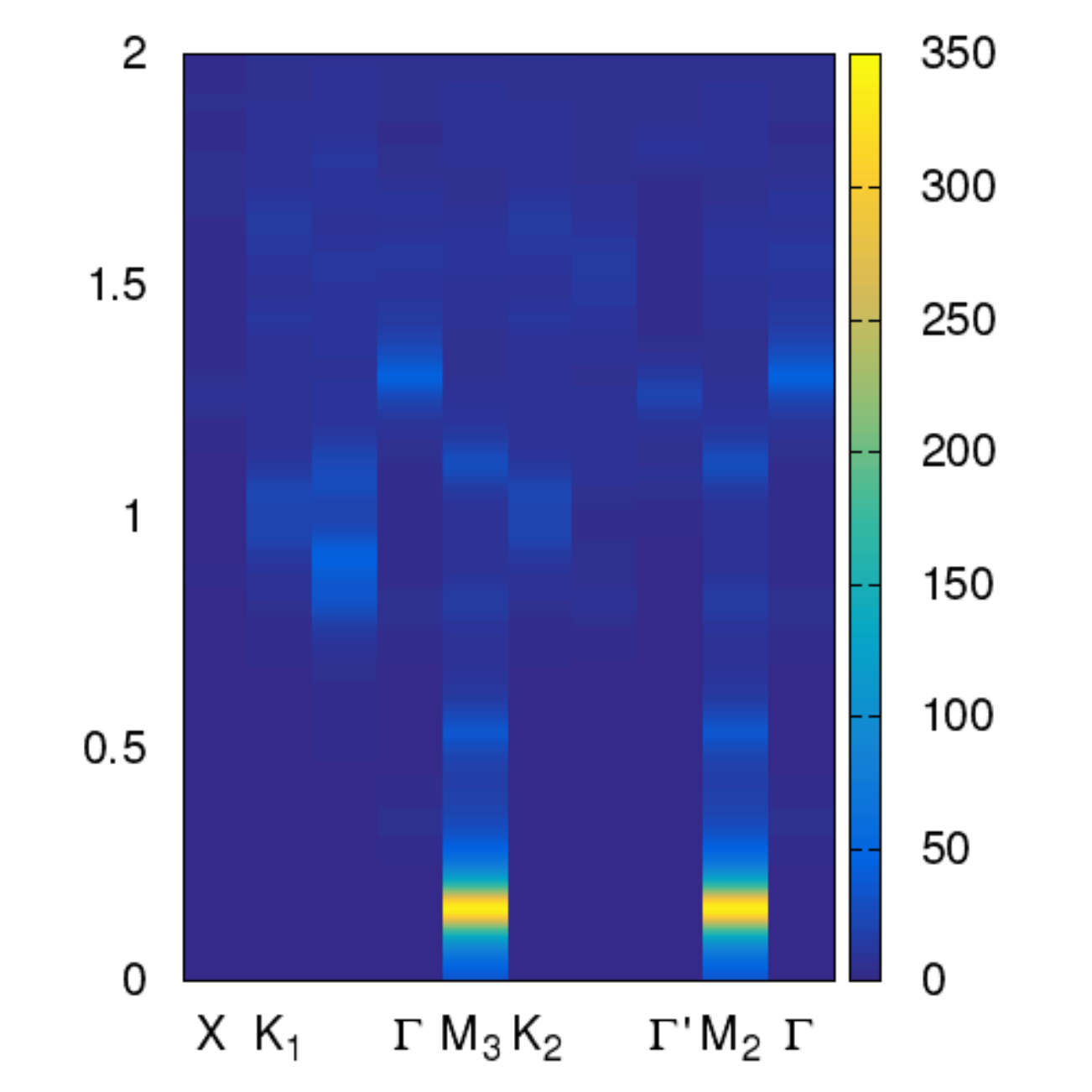} & \includegraphics[height=3cm]{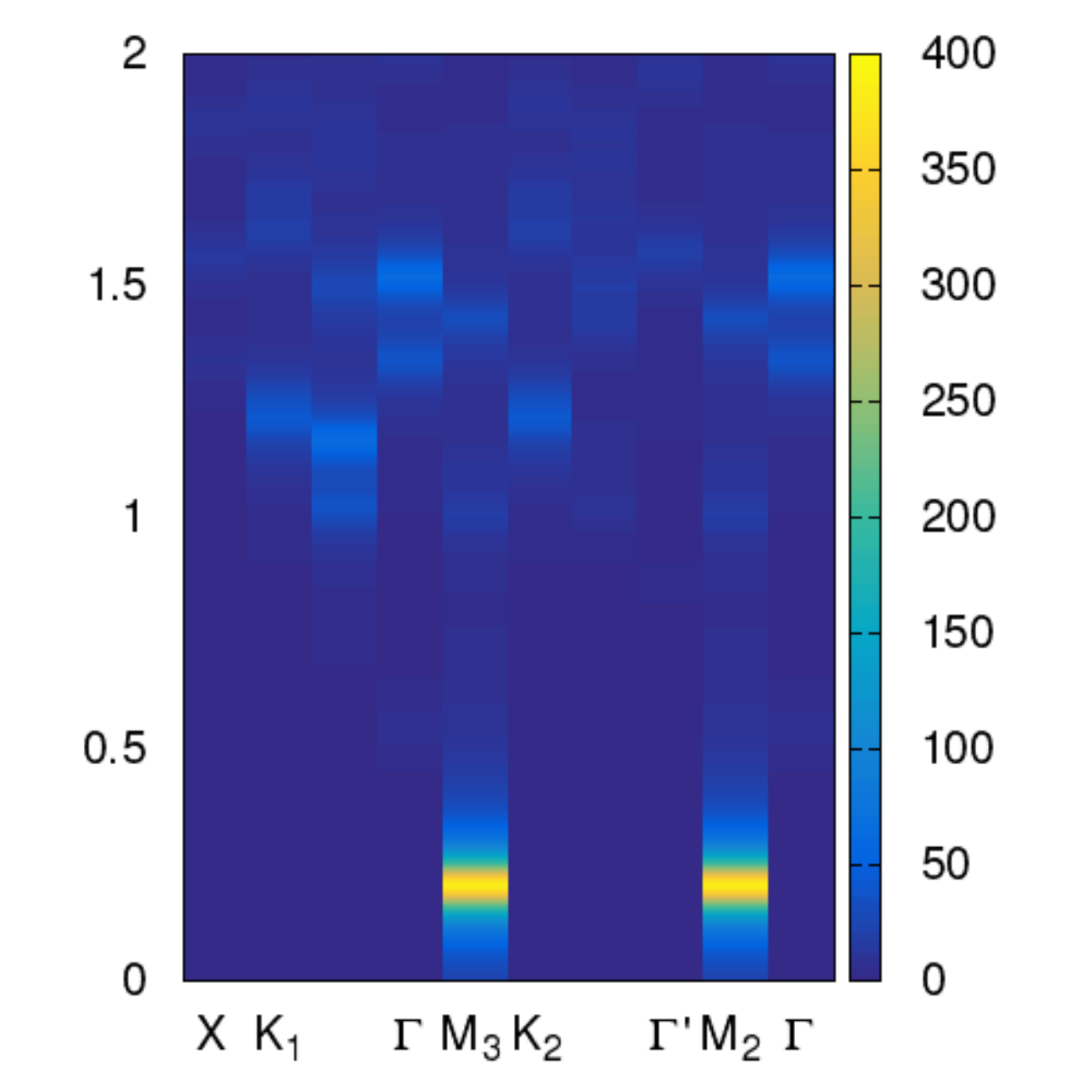} & \includegraphics[height=3cm]{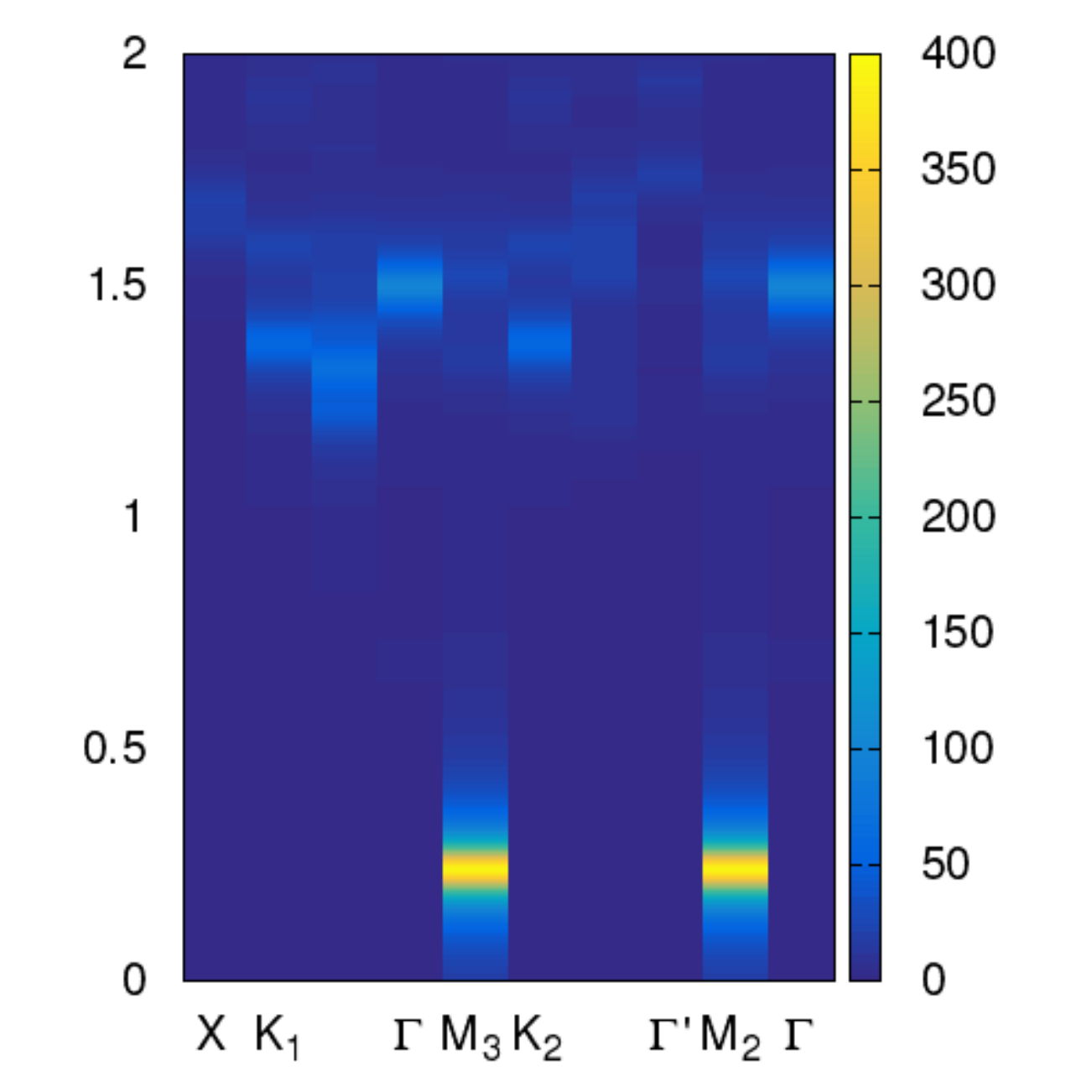} & \includegraphics[height=3cm]{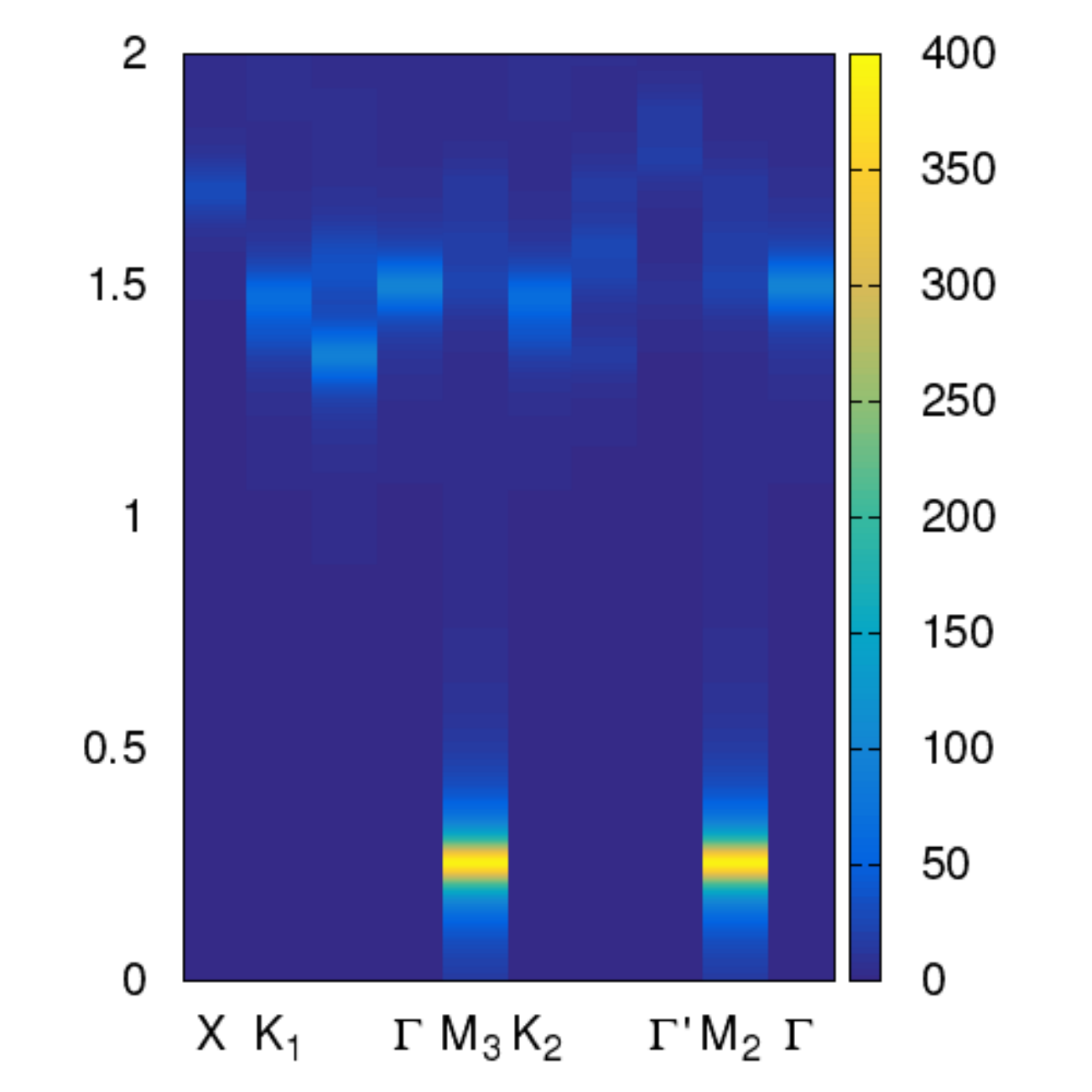} & \includegraphics[height=3cm]{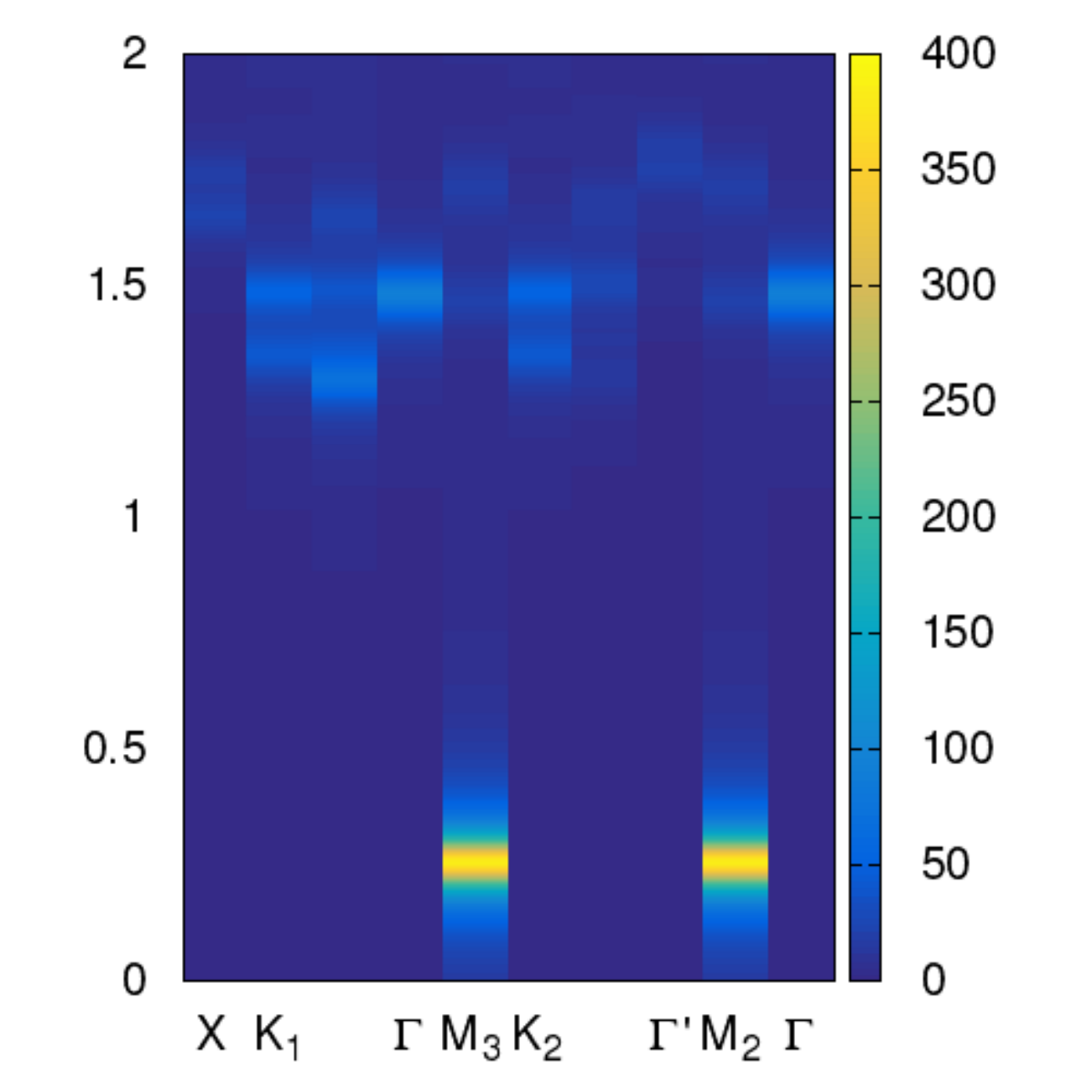}
		\end{tabular}
		\caption{\label{fig:supp:IqomegaEvolution}
				Evolution of the INS intensity spectrum away from the ferromagnetic Kitaev limit ($K_1=-1$). All color scales show intensity in arbitrary units.
		}
	\end{figure*}

\begin{figure*}[tbp]
		\centering
		\subfloat[$J_1=\Gamma_1'=J_3=0$]{
			\includegraphics[width=.66\columnwidth]{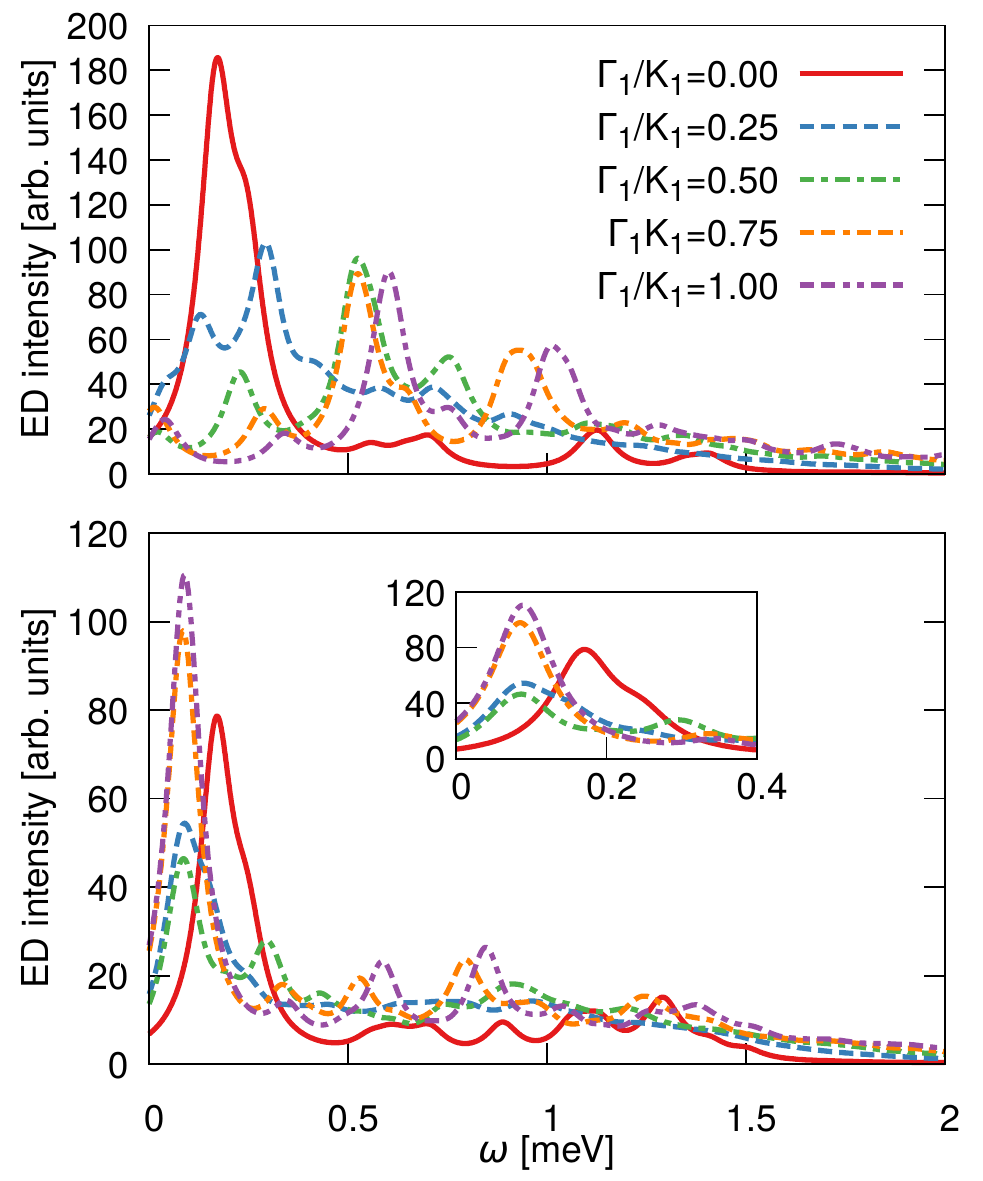}
		}\\\vspace{-0.305cm}
		\subfloat[$\Gamma_1/K_1=-0.5$, $\Gamma_1'=J_3=0$]{				
			\includegraphics[width=.66\columnwidth]{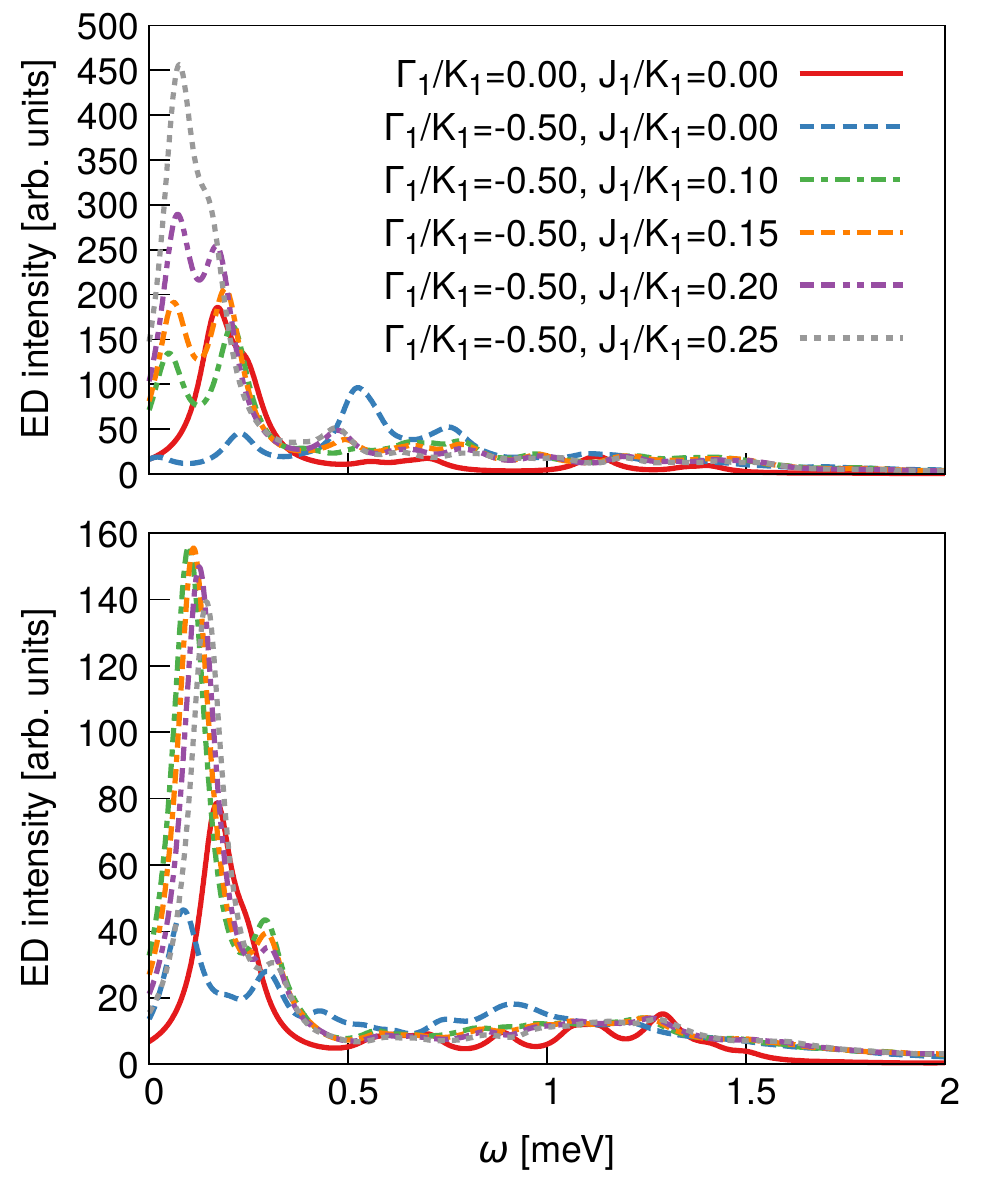}
		}\hspace{2cm}%
		\subfloat[$\Gamma_1/K_1=-1.0$, $\Gamma_1'=J_3=0$]{				
			\includegraphics[width=.66\columnwidth]{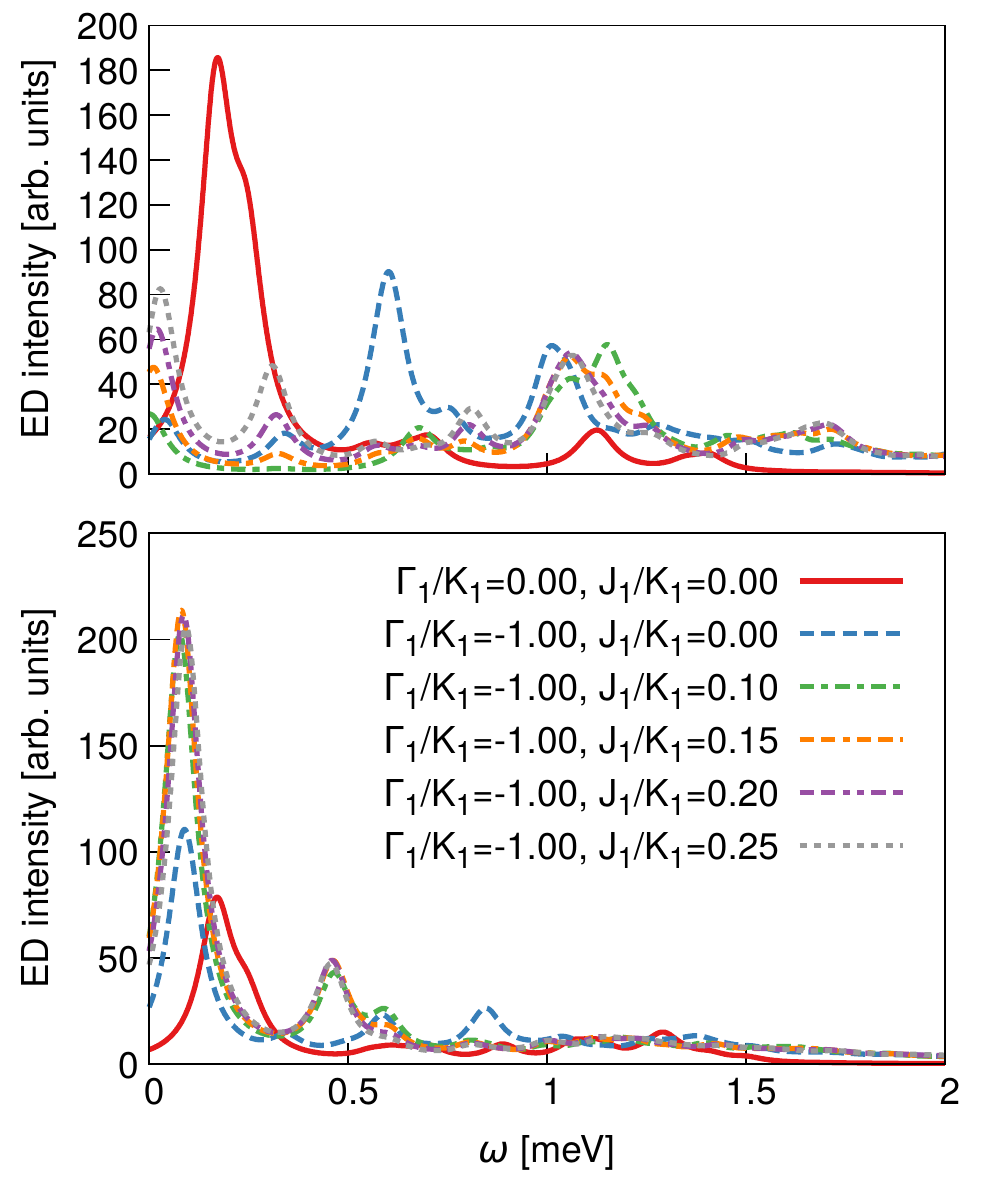}
		}\\\vspace{-0.305cm}
		\subfloat[$\Gamma_1/K_1=-0.5$, $J_1=0.10$, $\Gamma_1'/K_1=0.15$]{
			\includegraphics[width=.66\columnwidth]{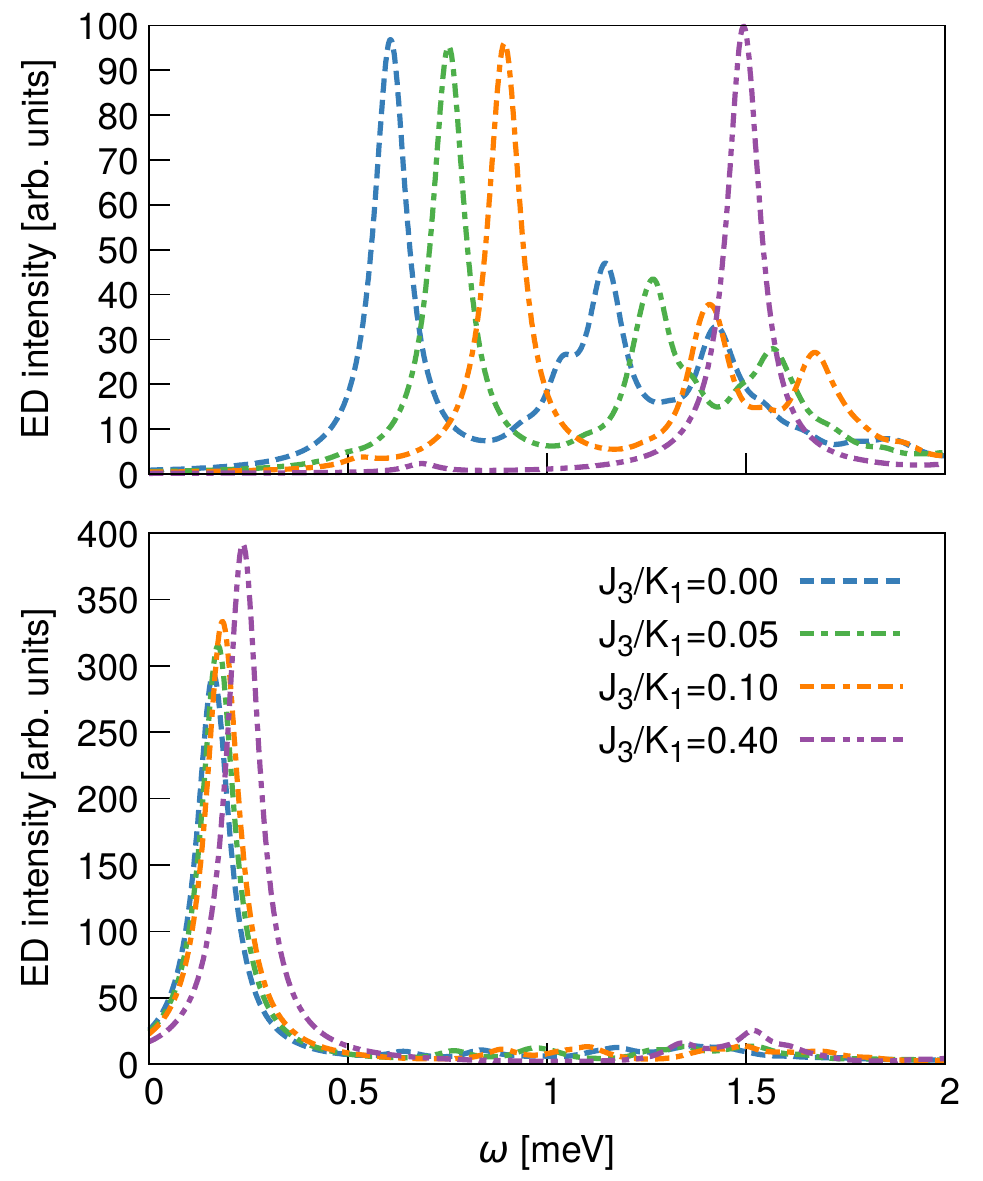}
		}\hspace{2cm}%
		\subfloat[$\Gamma_1/K_1=-1.0$, $J_10.10$, $\Gamma_1'/K_1=0.15$]{
			\includegraphics[width=0.66\columnwidth]{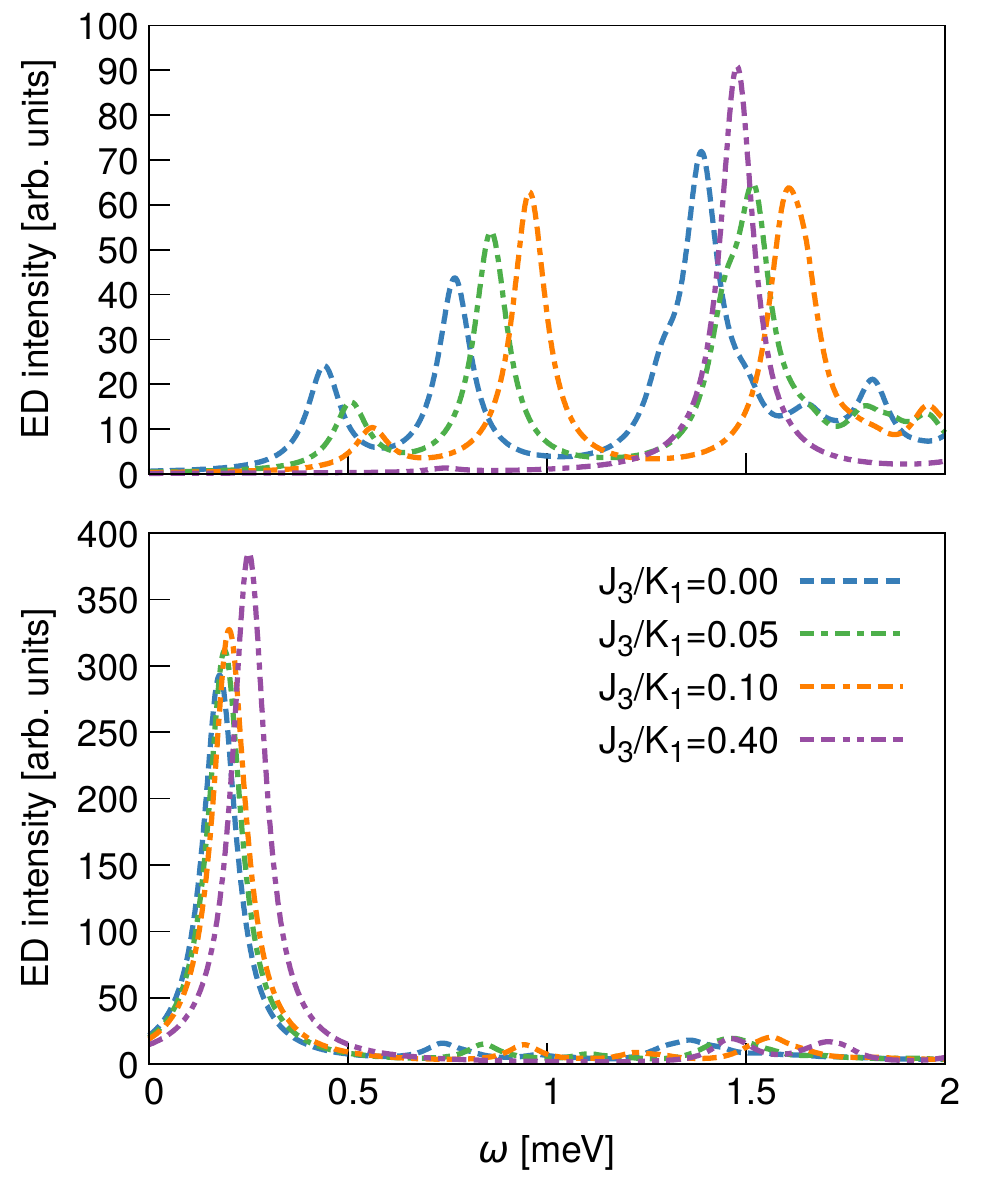}
		}
		\caption{\label{fig:supp:INSprofileEvolution}{$I\left( \mathbf{q}, \omega\right)$ at the $\Gamma$ (top panel) and M$_1$ points (bottom panel) for parameters away from the ferromagnetic Kitaev limit ($K_1=-1$).}
		}
	\end{figure*}

	Supplementary Fig.~\ref{fig:supp:IqomegaEvolution} shows INS spectra of the energy-normalized $J_1-K_1-\Gamma_1-\Gamma_1'-J_3$ model for additional parameter values. Supplementary Fig.~\ref{fig:supp:INSprofileEvolution} shows the evolution of the $\Gamma$ and M$_1$ point intensities as the $\Gamma_1$, $J_1$ and $J_3$ values are varied.
	
	The modified ab initio model proposed in the main text uses the revised Eichstaedt et al. parameters \cite{PhysRevB.100.075110} as its starting point. The full parameters of the modified model are given in the last row of Supplementary Table~\ref{table:Eichstaedt}. The values are the same as for the model in the second row of the table, but with $\Gamma_1'=-0.1175$ meV. The intensity profile, SSF, and integrated $I(\mathbf{q},\omega)$ results are shown in Supplementary Fig.~\ref{fig:supp:modifiedabinitio}.
	
	%
	\clearpage
	\begin{figure}
		\subfloat[Intensity profile]{
			\raisebox{1.75cm}{\includegraphics[width=.55\columnwidth]{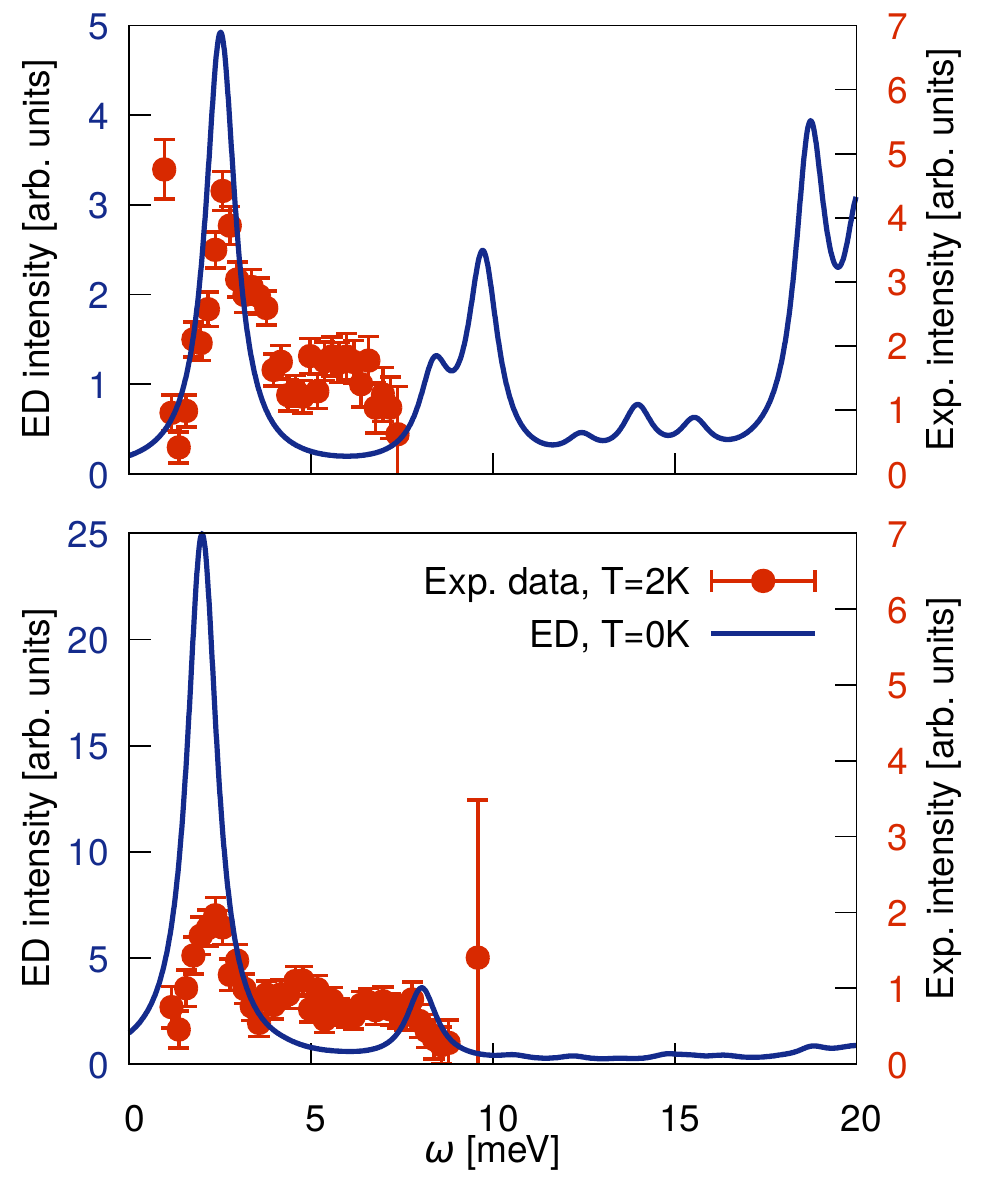}}
		}\hfill%
		\subfloat[SSF and integrated $I(\mathbf{q},\omega)$]{
			\includegraphics[height=10cm]{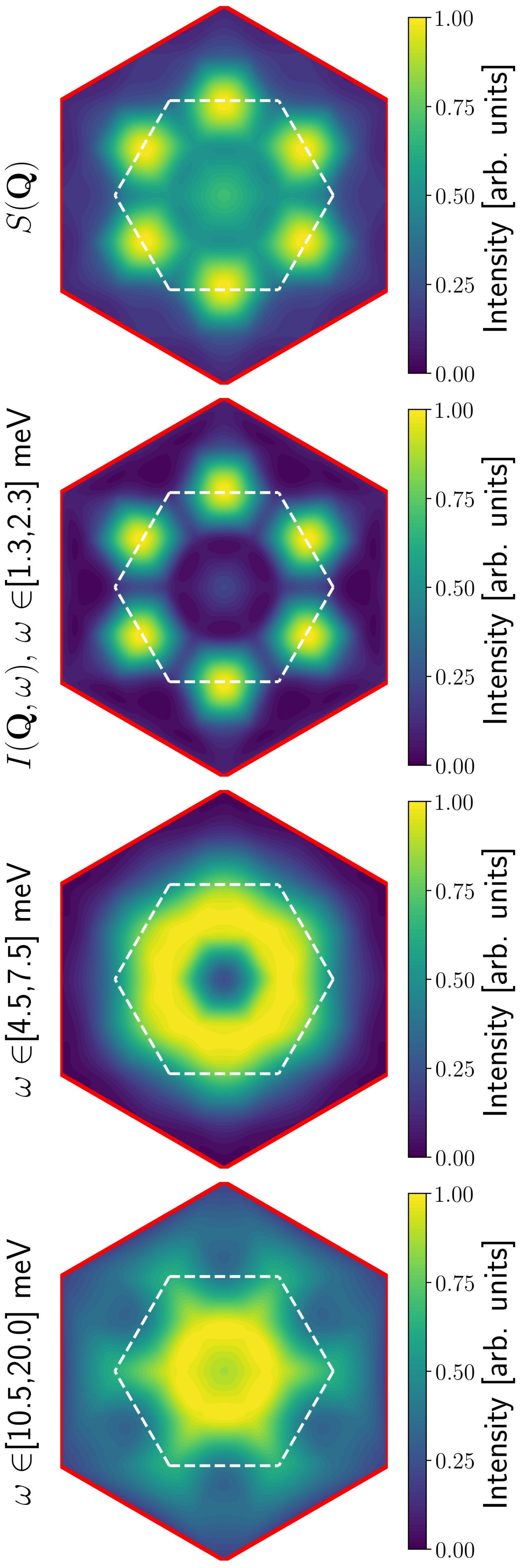}	
		}
		\caption{\label{fig:supp:modifiedabinitio}Modified ab initio model. In (a) the intensity profile of $I(\mathbf{q},\omega)$ is shown for the $\Gamma$ (top) and M$_1$ points (bottom). (b) SSF and additional integrated energy cuts of $I(\mathbf{q},\omega)$. The experimental data is from Ref.~\cite{Banerjee2018}, with error bars representing one standard deviation assuming Poisson counting statistics.}
	\end{figure}
		
\section{Supplementary References}
	\def\bibsection{}